\documentclass[preprint,3p,12pt]{./styles/elsarticle}
\pdfoutput=1
\usepackage[german,english]{babel}
\usepackage[latin1]{inputenc}
\usepackage[T1]{fontenc}
\usepackage{./styles/epsfig}
\usepackage[centertags]{amsmath}
\usepackage{amssymb,amscd}
\usepackage{ulem}
\usepackage{tabularx}
\usepackage{eurosym}
\usepackage{upgreek}
\usepackage{units}
\usepackage{paralist}
\usepackage{./styles/myxspace}
\usepackage{./styles/mysidecaption}
\sidecaptionraggedtrue
\def\zero{{\scriptscriptstyle 0}}

\providecommand{\coll}{Coll.\xspace}

\providecommand{\etal}{et al.\xspace}

\catcode`\@=11 
\@ifundefined{euro}%
{\providecommand{\Euro}{Euro\xspace}
 }%
{\providecommand{\Euro}{\euro\xspace}
 }
\catcode`\@=12 

\def\Z0{{Z^\zero}}
\def\eVdist{\kern-0.06667em}
\def\Ev{{\text{e}\eVdist\text{V\/}}}

\def\Mev{{\text{Me}\eVdist\text{V\/}}}
\def\Gev{{\text{Ge}\eVdist\text{V\/}}}
\def\Tev{{\text{Te}\eVdist\text{V\/}}}
\def\Pev{{\text{Pe}\eVdist\text{V\/}}}
\def\Eev{{\text{Ee}\eVdist\text{V\/}}}
\def\Zev{{\text{Ze}\eVdist\text{V\/}}}
\def\ev{{\,\text{e}\eVdist\text{V\/}}}
\def\kev{{\,\text{ke}\eVdist\text{V\/}}}
\def\mev{{\,\text{Me}\eVdist\text{V\/}}}
\def\gev{{\,\text{Ge}\eVdist\text{V\/}}}
\def\tev{{\,\text{Te}\eVdist\text{V\/}}}
\def\pev{{\,\text{Pe}\eVdist\text{V\/}}}
\def\eev{{\,\text{Ee}\eVdist\text{V\/}}}

\def\met{\,\text{m}}
\def\Met{\text{m}}
\def\cm{\,\text{cm}}
\def\Cm{\text{cm}}
\def\mm{\,\text{mm}}
\def\km{\,\text{km}}
\def\Km{\text{km}}

\def\nm{\,\text{nm}}

\def\Hz{\,\text{Hz}}
\def\kHz{\,\text{kHz}}

\def\scnd{\,\text{s}}
\def\Scnd{\text{s}}
\def\musec{\,\upmu\text{s}}
\def\ns{\,\text{ns}}

\def\gramm{\,\text{g}}

\def\Kelvin{\,\text{K}}

\def\erg{\,\text{erg}}

\def\sr{\,\text{sr}}
\def\flunit{\gev\cm^{-2}\scnd^{-1}\sr^{-1}}

\def\D{{\rm d}}

\def\IP{{\rm I$\kern-0.01667em$P}\xspace}

\def\Ptlj{{\not{\kern-0.55ex P}}_t\ell j}
\def\Ptmiss{{\not{\kern-0.55ex P}}_t}

\def\rnge{{\,\text{--}\,}}

\mathchardef\qsm=63
\mathchardef\pls=43
\mathchardef\mns=512
\mathchardef\plm=518
\mathchardef\eql=61
\mathchardef\smallleft=300
\mathchardef\smallright=301
\mathchardef\perslsh=47
\mathchardef\les=316
\mathchardef\gre=318
\mathchardef\leq=532
\mathchardef\grq=533
\chardef\amp=38
\chardef\usc=95
\chardef\til=126
\chardef\Lslash=138
\def\intl{\int\limits}

\def\sqr#1#2#3{{\vcenter{\hrule height.#3ex\hbox{\vrule width.#2ex height#1ex
    \kern#1ex\vrule width.#3ex}\hrule height.#2ex}}}

\def\angleto{\vrule width.035em height2.1ex depth-.56ex\unskip\kern-.6ex\to}
\def\perchc#1{{\raise.4ex\hbox{$\mkern4mu#1{\it\perslsh}_
             {\mkern-5mu\scriptscriptstyle{{\rm o}\!{\rm o}}}^
             {\mkern-12.8mu\scriptscriptstyle{\rm o}}$}}}

\def\widebar#1{\mkern1.5mu\overline{\mkern-1.5mu#1\mkern-1.mu}\mkern1.mu}
\catcode`\@=11 
\def\parenbar{\mathpalette\p@renb@r}
\def\p@renb@r#1#2{\vbox{%
  \ifx#1\scriptscriptstyle \dimen@.7em\dimen@ii.2em\else
  \ifx#1\scriptstyle \dimen@.8em\dimen@ii.25em\else
  \dimen@1em\dimen@ii.4em\fi\fi \offinterlineskip
  \ialign{\hfill##\hfill\cr
    \vbox{\hrule width\dimen@ii}\cr
    \noalign{\vskip-.3ex}%
    \hbox to\dimen@{$\mathchar300\hfil\mathchar301$}\cr
    \noalign{\vskip-.3ex}%
    $#1#2$\cr}}}
\catcode`\@=12 
\def\nuan{\parenbar{\nu}}

\def\dbar{\widebar{d}}
\def\ubar{\widebar{u}}

\def\bbar{\widebar{b}}

\def\nubar{\widebar{\nu}}

\newbox\struttbox
\setbox\struttbox=\hbox{\vrule height1.65ex depth.485ex width0pt}
\def\strutt{\relax\ifmmode\copy\struttbox\else\unhcopy\struttbox\fi}
\def\stru#1#2{\relax\ifmmode\hbox{\vrule height#1 depth#2 width0pt}
\else\vrule height#1 depth#2 width0pt\fi}
\def\uline#1{$\underline{\hbox{#1\strutt}}$}
\def\ronum#1{\uppercase\expandafter{\romannumeral#1}}
\def\ronuml#1{\expandafter{\romannumeral#1}}

\def\cbk{\kern-0.5em}
\newcommand{\pcite}[1]{{\protect\cite{#1}}}

\newcommand{\linebox}[2][3.ex]{\uline{\hbox to #2{\stru{#1}{0.pt}\hfil}}}

\newcounter{seqnum}
\DeclareMathAlphabet{\mathbf}{OT1}{cmr}{bx}{n}

\DeclareMathAlphabet{\mathbfs}{OT1}{lcmss}{bx}{sl}

\newcommand{\PreserveBackslash}[1]{\let\temp=\\#1\let\\=\temp}
\newlength\listtextwidth

\catcode`\@=11 
\newlength{\@tabfninsert}
\newlength{\@tabfnwidth}
\newcommand{\tabfootnote}[2]{%
  \setlength{\@tabfninsert}{0.8em}
  \setlength{\@tabfnwidth}{\textwidth}
  \addtolength{\@tabfnwidth}{-\@tabfninsert}
  \addtolength{\@tabfnwidth}{-0.4em}
  \noindent\makebox[\@tabfninsert][r]{\footnotesize$^{#1}$\hfil}\hfill%
  \parbox[t]{\@tabfnwidth}{\footnotesize #2\hfill}}
\catcode`\@=12 
\newcommand{\boldarrayrulewidth}{1pt}

\catcode`\@=11 
\let\tab@penalty\relax
\newcount\tab@state
\def\bcline#1{%
  \noalign{\kern-.5\arrayrulewidth\tab@penalty}%
  \omit%
  \global\tab@state\@ne%
  \ranges\bcline@i{#1}%
  \cr%
  \noalign{\kern-.5\arrayrulewidth\tab@penalty}%
}
\def\bcline@i#1#2{%
  \ifnum#1<\tab@state\relax%
    \tab@@cr%
    \noalign{\kern-\arrayrulewidth\tab@penalty}%
    \omit%
    \global\tab@state\@ne%
  \fi%
  \@whilenum\tab@state<#1\do{%
    \hfil\tab@@tab@omit%
    \global\advance\tab@state\@ne%
  }%
  \ifnum\tab@state>\@ne%
    \kern-\arrayrulewidth%
  \fi%
  \@whilenum\tab@state<#2\do{%
    \tab@@span@omit%
    \global\advance\tab@state\@ne%
  }%
  \leaders\hrule\@height\boldarrayrulewidth\hfill%
}
\def\ranges#1#2{%
  \gdef\ranges@temp{#1}%
  \begingroup%
  \ranges@i#2 \q@delim%
}
\def\ranges@i{%
  \@ifnextchar\q@delim\ranges@done{\afterassignment\ranges@ii\count@}%
}
\def\ranges@ii{%
  \@ifnextchar-\ranges@iii{\ranges@do\count@\count@\ranges@v}%
}
\def\ranges@iii-{\afterassignment\ranges@iv\@tempcnta}
\def\ranges@iv{\ranges@do\count@\@tempcnta\ranges@v}
\def\ranges@v{%
  \@ifnextchar,%
    \ranges@vi%
    {%
      \@ifnextchar\q@delim%
        \ranges@done%
        {\tab@err@range\ranges@vi,}%
    }%
}
\def\ranges@vi,{\afterassignment\ranges@ii\count@}
\def\ranges@do#1#2{%
  \ifnum#1>#2\else%
    \expandafter\endgroup%
    \expandafter\ranges@temp%
    \expandafter{%
    \the\expandafter#1%
    \expandafter}%
    \expandafter{%
    \the#2%
    }%
    \begingroup%
  \fi%
}
\def\ranges@done\q@delim{\endgroup}
\def\ifinrange#1#2{%
  \@tempswafalse%
  \count@#1%
  \ranges\ifinrange@i{#2}%
  \if@tempswa%
    \expandafter\@firstoftwo%
  \else%
    \expandafter\@secondoftwo%
  \fi%
}
\def\ifinrange@i#1#2{%
  \ifnum\count@<#1 \else\ifnum\count@>#2 \else\@tempswatrue\fi\fi%
}
\def\tab@@cr{\cr}
\def\tab@@tab@omit{&\omit}
\def\tab@@span@omit{\span\omit}
\def\tab@checkrule#1{%
  \count@#1\relax%
  \expandafter\ifinrange%
  \expandafter\count@%
  \expandafter{\tab@xcols}%
    {\tab@checkrule@i}%
    {}%
}
\def\bhline{\noalign{\ifnum0=`}\fi\hrule \@height  
\boldarrayrulewidth \futurelet \@tempa\@xhline}
\def\@xhline{\ifx\@tempa\hline\vskip \doublerulesep\fi
      \ifnum0=`{\fi}}
\catcode`\@=12 

\catcode`\@=11 
\newcounter{pict@width}
\newcounter{pict@height}
\newlength{\pict@scale}
\newcommand{\psfigadd}[4]{%
\setcounter{pict@width}{1*\ratio{#2+\pict@scale/2}{\pict@scale}}
\setcounter{pict@height}{1*\ratio{#3+\pict@scale/2}{\pict@scale}}
\setlength{\unitlength}{\pict@scale}
\hbox to #2{\hspace{-\fill}\begin{picture}(\thepict@width,\thepict@height)
\put(0,0){\psfig{figure=#1,width=#2,height=#3,clip=}}
\SetScale{0.283466457}
\SetWidth{1.763889}
{#4}
\end{picture}}
}
\newcounter{pict@widthfst}
\newcounter{pict@widthscd}
\newcounter{pict@widthtot}
\newcommand{\psfigaddtwo}[7]{%
\setcounter{pict@widthfst}{1*\ratio{#2+\pict@scale/2}{\pict@scale}}
\setcounter{pict@widthscd}{1*\ratio{#2+#4+\pict@scale/2}{\pict@scale}}
\setcounter{pict@widthtot}{1*\ratio{#2+#4+#6+\pict@scale/2}{\pict@scale}}
\setcounter{pict@height}{1*\ratio{#3+\pict@scale/2}{\pict@scale}}
\setlength{\unitlength}{\pict@scale}
\hbox{\hspace{-\fill}\begin{picture}(\thepict@widthtot,\thepict@height)
\put(0,0){\psfig{figure=#1,width=#2,height=#3,clip=}}
\put(\thepict@widthscd,0){\psfig{figure=#5,width=#6,height=#3,clip=}}
\SetScale{0.283466457}
\SetWidth{1.763889}
{#7}
\end{picture}}
}
\newcommand{\psfigror}[4]{%
\setcounter{pict@width}{1*\ratio{#2+\pict@scale/2}{\pict@scale}}
\setcounter{pict@height}{1*\ratio{#3+\pict@scale/2}{\pict@scale}}
\setlength{\unitlength}{\pict@scale}
\hbox{\begin{picture}(\thepict@width,\thepict@height)
\put(0,\thepict@height){\psfig{figure=#1,width=#3,height=#2,clip=,angle=270}}
\SetScale{0.283466457}
\SetWidth{1.763889}
{#4}
\end{picture}}
}
\newcommand{\psfigrol}[4]{%
\setcounter{pict@width}{1*\ratio{#2+\pict@scale/2}{\pict@scale}}
\setcounter{pict@height}{1*\ratio{#3+\pict@scale/2}{\pict@scale}}
\setlength{\unitlength}{\pict@scale}
\hbox{\begin{picture}(\thepict@width,\thepict@height)
\put(0,0){\psfig{figure=#1,width=#3,height=#2,clip=,angle=90}}
\SetScale{0.283466457}
\SetWidth{1.763889}
{#4}
\end{picture}}
}
\catcode`\@=12 

\biboptions{square,comma,sort&compress}
\journal{Progress in Particle and Nuclear Physics}
\begin{document}
\begin{frontmatter}
\title{High-Energy Neutrino Astrophysics:\\ 
       Status and Perspectives}

\author[labeluk]{U.F.~Katz\fnref{fnuk}}
\author[labelcs]{Ch.~Spiering\corref{fncs}}
\address[labeluk]{ECAP, University of Erlangen, Erwin-Rommel-Str.~1, 91058 Erlangen, Germany}
\address[labelcs]{DESY, Platanenallee 6, 15738 Zeuthen, Germany}
\cortext[fncs]{Email: csspier@ifh.de}
\fntext[fnuk]{Email: katz@physik.uni-erlangen.de}

\begin{abstract}
Neutrinos are unique cosmic messengers. Present attempts are directed to extend
the window of cosmic neutrino observation from low energies (Sun, supernovae) to
much higher energies. The aim is to study the most violent processes in the
Universe which accelerate charged particles to highest energies, far beyond the
reach of laboratory experiments on Earth. These processes must be accompanied by
the emission of neutrinos. Neutrinos are electrically neutral and interact only
weakly with ordinary matter; they thus propagate through the Universe without
absorption or deflection, pointing back to their origin. Their feeble
interaction, however, makes them extremely difficult to detect. The years
2008-2010 have witnessed remarkable steps in developing high energy neutrino
telescopes. In 2010, the cubic-kilometre neutrino telescope IceCube at the
South Pole has been completed. In the Mediterranean Sea the first-generation
neutrino telescope ANTARES takes data since 2008, and efforts are directed
towards KM3NeT, a telescope on the scale of several cubic kilometres. The next
years will be key years for opening the neutrino window to the high energy
Universe. With an instrumented volume of a cubic kilometre, IceCube is entering
a region with realistic discovery potential. Discoveries or non-discoveries of
IceCube will have a strong impact on the future of the field and possibly mark a
``moment of truth''. In this review, we discuss the scientific case for neutrino
telescopes, describe the detection principle and its implementation in first-
and second-generation installations and finally collect the existing physics
results and the expectations for future detectors. We conclude with an outlook
to alternative detection methods, in particular for neutrinos of extremely high
energies.
\end{abstract}

\end{frontmatter}
\clearpage
\pagenumbering{roman}
\setcounter{page}{2}
\tableofcontents
\clearpage
\pagenumbering{arabic}
\section{Introduction}
\label{sec-int}

High-energy neutrinos, with energies much larger than $100\mev$, must be emitted
as a by-product of collisions of charged cosmic rays with matter; in fact, only
neutrinos provide incontrovertible evidence for hadronic acceleration. Since
they can escape much denser celestial environments than light, they can be
tracers of processes which stay hidden to traditional astronomy. At the same
time, however, their extremely low reaction probability makes their detection
extraordinarily difficult.

First ideas to search for cosmic neutrinos other than those from the Sun date
back to the late fifties. In 1960, K.\,Greisen proposed a 3000\,ton underground
Cherenkov detector to record neutrinos emitted by the Crab nebula
\cite{Greisen-1960}. He was flanked by F.\,Reines \cite{Reines-1960}, who
realised, however, that ``the cosmic neutrino flux cannot be usefully
predicted'' -- i.e.\ that a mass of 3000\,tons may be far too small. In the same
year, M.\,Markov made his ground-breaking proposal ``to install detectors deep
in a lake or in the sea to determine the direction of charged particles with the
help of Cherenkov radiation'' \cite{Markov-1960}. In the decades since then it
was realised that high-energy neutrino astronomy requires detectors of a cubic
kilometre or larger that can indeed only be implemented in open media. Actually,
the first project of that size, IceCube at the South Pole, has just been completed. 
Two others, KM3NeT in the Mediterranean Sea and GVD in Lake Baikal, are in their
preparatory phases. No doubt, we are entering an exciting era of opportunity.

\begin{figure}[ht]
\sidecaption
\epsfig{file=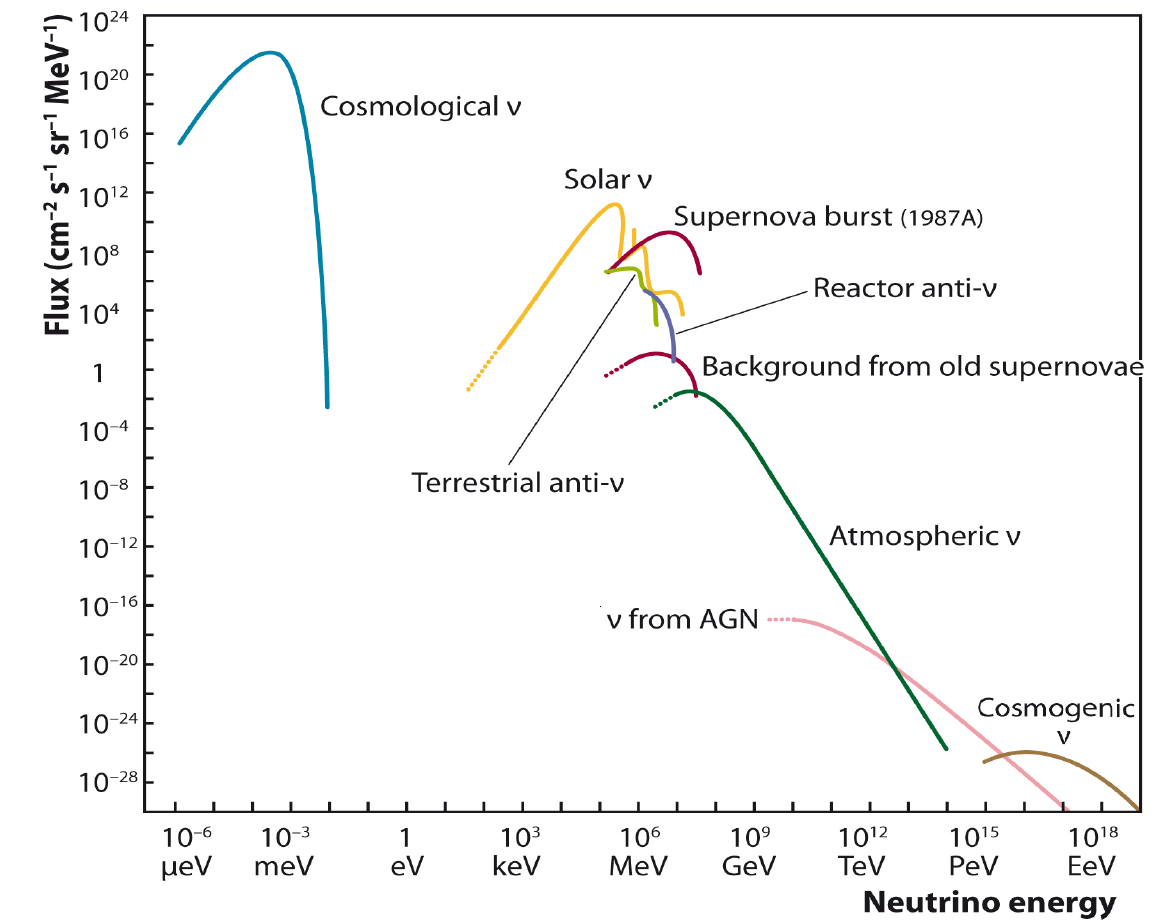,width=9.3cm}
\caption
{Measured and expected fluxes of natural and reactor neutrinos.}
\label{all-nu}
\end{figure}

Already now, neutrino astronomy is reality in the {\it low-energy} sector, where
the detection of neutrinos from the Sun and the supernova SN\,1987A was honoured
by the 2002 Nobel Prize for physics. Figure~\ref{all-nu} shows a compilation of
the spectra of dominant natural and artificial neutrino fluxes. No practicable
idea exists on how to measure the neutrinos of the $1.9\Kelvin$ neutrino
counterpart to the cosmic microwave background. At higher energies, neutrinos
from the Sun, from SN\,1987A, from reactors and from the interior of the Earth
have already been detected, as have so-called ``atmospheric neutrinos'' created
in cosmic ray interactions in the Earth's atmosphere. Still awaiting detection
are high-energy cosmic neutrinos from extraterrestrial sources such as active
galactic nuclei (AGN) or from interactions of ultra-energetic protons with the
cosmic microwave background \cite{Berezinsky-Zatsepin}. These cosmic neutrinos
will hopefully be detected by neutrino telescopes in the next decade, even
though predictions for their fluxes are uncertain by orders of magnitude in many
cases.

The development of high-energy neutrino astronomy is reflected in a series of
previous reviews spanning the period 1995 to 2009 \cite{Gaisser-Halzen-Stanev,
Learned-Mannheim,McDonald-2004,Becker-2007,Chiarusi-Spurio,Anchordoqui-Montaruli}. 
The neutrino telescopes discussed in this review focus on energies beyond a few
$\gev$. First searches for such neutrinos were made in the 1960s in the Kolar
Gold Field mine in India and in the East Rand mine in South Africa (for a review
see \cite{McDonald-2004}). In the 1980s, the spectrum of atmospheric muon
neutrinos was measured with a detector in the Fréjus tunnel between France and
Italy, and a first limit on the diffuse flux of extra-terrestrial $\Tev$
neutrinos was set \cite{Rhode-1996}. Over the following decades, the evolution
of underground neutrino detectors culminated in two experiments with an area of
about $1000\met^2$ each (see Sect.~\ref{sec-det-eff} for a discussion of
effective areas): MACRO in the Gran Sasso Underground Laboratory in Italy and
Super-Kamiokande in the Japanese Kamioka mine. MACRO collected more than
thousand atmospheric neutrinos over six years of data taking. Super-Kamiokande,
with an even larger data sample, is still in operation. The atmospheric neutrino
results from these detectors have demonstrated that neutrinos oscillate between
their flavour states $\nu_\mu$ and $\nu_\tau$, additionally to the $\nu_e$
oscillations observed for solar neutrinos \cite{McDonald-2004}.

The first-generation detectors in water and ice have beaten the largest
underground detectors by a factor of about 30 with respect to their sensitivity
to high-energy neutrinos. The second-stage detectors on the cubic-kilometre
scale will yield another factor of 30. Compared to detectors underground we
therefore enter a ``factor-1000 era''. Arguably, this factor is not a guarantee
for discoveries. On the other hand it rarely happened in astronomy that
improvements of more than an order of magnitude (in sensitivity or in angular or
time resolution) came along without discovering new, unexpected phenomena
\cite{Harwit}. ``Nothing is guaranteed, but history is on our side''
\cite{Halzen-history}: In some years we will know whether we indeed have entered
an era of discovery or not.
 
This review is organised as follows: Section~\ref{sec-sci} summarises the
scientific motivation. Apart from the main topic, neutrino astrophysics, it
includes the indirect search for dark matter, the study of standard and
non-standard neutrino oscillations, the search for exotic particles like
magnetic monopoles, super-symmetric Q-balls or nuclearites and -- last but not
least -- the investigation of environmental effects, be it in deep natural water
or Antarctic ice. The basics of the detection methods are summarised in
Sect.~\ref{sec-det}. In Sect.~\ref{sec-fir} the first-generation neutrino
telescopes are described, in Sect.~\ref{sec-sec} the second-generation projects
on the cubic-kilometre scale. A selection of results obtained with NT200 in Lake
Baikal, ANTARES in the Mediterranean Sea as well as AMANDA and IceCube at the
South Pole is presented in the following Sect.~\ref{sec-phy}. For the highest
energies beyond $100\pev$, even cubic-kilometre detectors are far too small to
detect the feeble neutrino fluxes expected. This is the realm of new
technologies which aim, with a correspondingly high detection threshold, to
monitor volumes of 100 cubic kilometres and beyond. These methods are described
in Sect.~\ref{sec-alt}. The last section finally gives a summary and tries an
outlook to forthcoming developments.

\clearpage
\section{Scientific Background and Motivation}
\label{sec-sci}

The primary motivation to build kilometre-scale neutrino detectors is driven by
the observation of charged cosmic rays. Since long, neutrinos have been supposed
to be a key messenger to identify the sources of cosmic ray acceleration and to
provide a deeper understanding of the associated astrophysical objects. In
Sect.~\ref{sec-sci-neu}, we recall basic information on cosmic rays and
gamma rays which motivate (and constrain) the search for energetic neutrinos. We
sketch the astrophysical source candidates for neutrinos from hadronic
acceleration processes and relate the predicted fluxes to event rates expected
in neutrino telescopes on the cubic-kilometre scale.

The identification of the sources of cosmic rays and, more generally, the
opening of a new observational window to the Universe, is arguably the most
important, but by far not the only purpose of large neutrino telescopes. 
Neutrino detectors are multi-purpose devices addressing also questions of
particle physics and environmental science. Section~\ref{sec-sci-par} gives an
overview on the particle physics issues studied with the help of neutrino
telescopes, and Sect.~\ref{sec-sci-env} sketches the environmental spin-offs.

\subsection{Neutrinos from cosmic accelerators}
\label{sec-sci-neu}

High-energy charged cosmic rays have been known to exist for almost a century,
but their origin is still a mystery. Since charged particles are deflected in
the inter- and extragalactic magnetic fields, their arrival direction at Earth
does not reveal their sources, except at the very highest energies. In contrast,
neutrinos, produced at the acceleration sites or during cosmic ray propagation,
propagate on straight trajectories and point back to their origin.

Several astrophysical object classes have been proposed as potential particle
accelerators. In spite of a vast amount of observational data in all
electromagnetic wavelength regimes -- from radio to gamma -- it is, however,
still unclear, whether the non-thermal processes in these objects are of
electronic or hadronic nature. It appears likely that a final answer to this
question will require the observation of neutrinos (cf.\
Sect.~\ref{sec-sci-neu-pro}).

\subsubsection{Cosmic rays}
\label{sec-sci-neu-cos}

\begin{figure}[ht]
\sidecaption
\epsfig{file=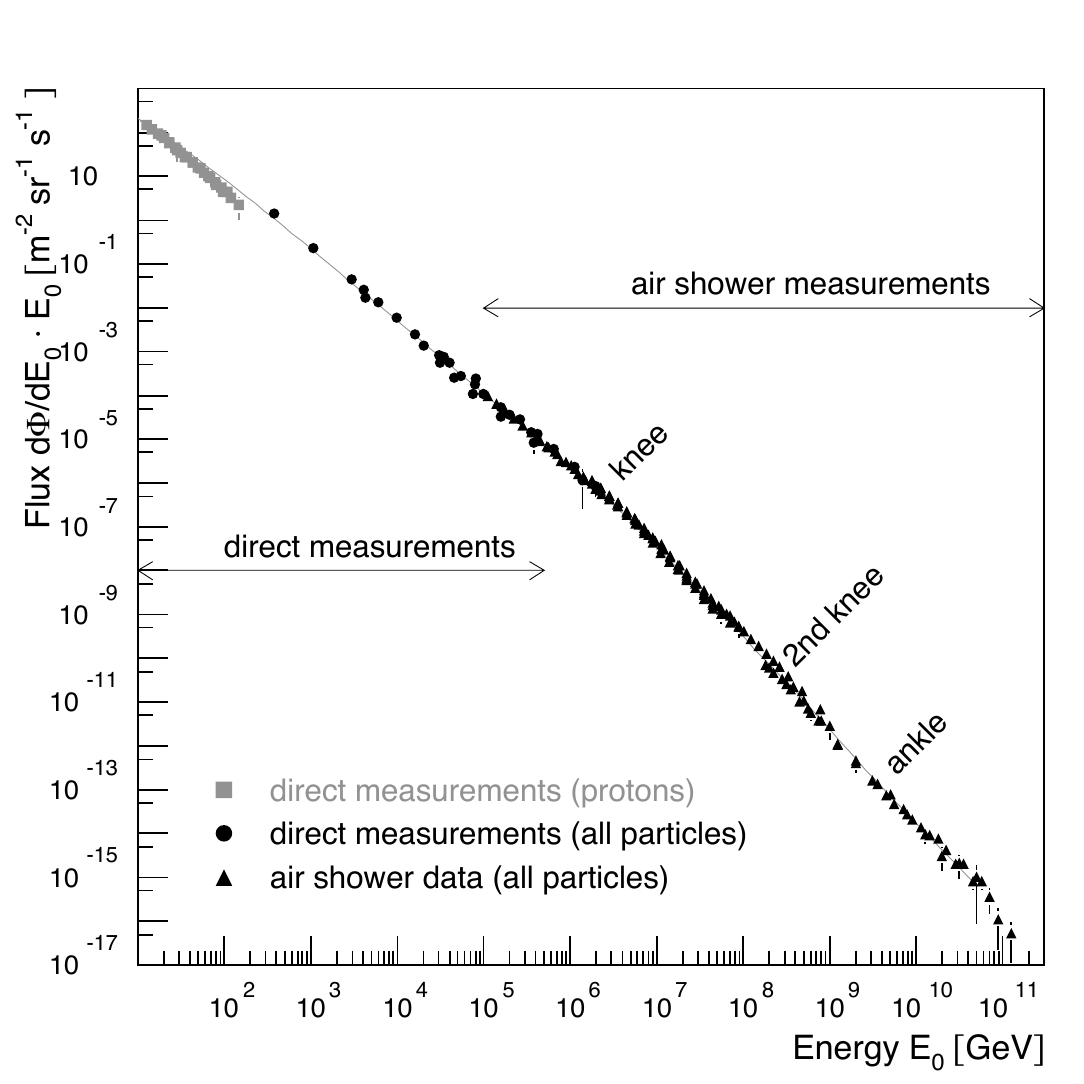,width=10.5cm,trim=5 5 10 25,clip=}
\caption{
Energy spectrum of cosmic rays, illustrating its main features (knee, ankle, and
high-energy cut-off at a few $10^{11}\gev$). Figure taken from
\pcite{Bluemer-etal-2009}.
\label{CR}
}
\end{figure}

Cosmic rays are charged particles -- basically hadrons (protons, light and heavy
nuclei) with only a tiny admixture of electrons -- that impinge on Earth from
outer space, with an energy spectrum extending to energies in excess of
$10^{20}\ev$ ($100\eev$), see Fig.~\ref{CR}.

The energy spectrum follows a broken power law $E^{-\alpha}$. The region where
the decrease with a power index $\alpha=2.7$ steepens to a power index of
$\alpha=3.1$ at about $10^{15}\ev$ ($1\pev$) is called the ``knee''. It is
assumed that cosmic rays up to (and even beyond) the knee are of Galactic
origin. The main accelerator candidates are supernova remnants (SNR): The shock
fronts powered by supernova explosions propagate into the interstellar medium,
and by repeated scattering processes across the shock front particles can gain
energy (``first-order Fermi acceleration''
\cite{Fermi-1949,Fermi-1954,Ginzburg-Ptuskin-1976,Blandford-Eichler-1987,Hillas-2005}).

The spectrum resulting from first-order Fermi acceleration can be shown to
follow roughly an $E^{-2}$ spectrum, with a maximum energy
\begin{equation}
 E_\text{max} \propto Z \cdot v \cdot B \cdot L\;,
 \label{Emax}
\end{equation}
where $Z$ is the charge of the particle, $v$ the velocity of the shock wave, $B$
the magnetic field strength in the acceleration region and $L$ the size of this
region.

With eq.~(\ref{Emax}), the maximum energy to which a SNR can boost particles
turns out to be somewhat below the knee, at about $10^{14}\ev$. However, by
interactions of cosmic rays with the magnetic fields in the acceleration region,
these fields can be amplified, resulting in energies of up to $10^{16}\ev$ (see
\cite{Caprioli-2009} and references therein). The mechanism of converting
kinetic energy of the SNR into energy of accelerated particles is effective over
the first $10^3$--$10^4$\,years of the SNR.

Circumstantial evidence for SNRs as the main sources of cosmic rays can be also
derived from the Galactic energy budget. The energy density of Galactic cosmic
rays is about $1\ev/\Cm^3$ \cite{Gaisser-1990}, corresponding to $\rho_\text{CR}
=10^{-12}\erg/\Cm^3$. The power required to sustain this density is $L =
\rho_\text{CR} \cdot V/t_\text{cont}$, with $t_\text{cont}\approx 10^7$
years \cite{Bluemer-etal-2009} being the average containment time of cosmic rays
in the Galaxy and $V\approx 10^{67}\cm^3$ the volume of the visible part of the
Galaxy. With $L\approx 10^{41}\erg/\Scnd$, the power to maintain the cosmic ray
density turns out to be about 10\% of the power generated by supernovae, which
release on average $10^{51}\erg$ every 20--50\,years, a relation noticed already
in 1934 by Baade and Zwicky \cite{Baade-Zwicky-1934}.

Besides supernova remnants, there are further candidates for Galactic particle
acceleration, most notably pulsars with their extremely high magnetic fields at
the poles, and binary systems with a neutron star or a black hole as one of the
partners. The latter often form relativistic radio jets and are then dubbed
microquasars (for a review, see \cite{Mirabel-2007}). 

Somewhere between $10^{17}$ and $10^{18.5}\ev$, at maximum, known Galactic
source candidates are running out of power and extragalactic sources start
dominating the spectrum. Around $10^{18.5}\ev$ the spectrum flattens again to an
$E^{-2.7}$ shape, a feature named the ``ankle''. Almost all cosmic rays with
energies above the ankle are assumed to be of extragalactic origin. Even if
Galactic sources would accelerate particles beyond the ankle energy, these
particles would right away escape the Galaxy since their gyroradius exceeds the
size of the Galaxy.
 
The cut-off of the cosmic ray spectrum at highest energies, confirmed by recent
Auger measurements \cite{Auger-2011}, is likely due to interactions with the
cosmic microwave background ($p + \gamma_\text{CMB}\to\Delta^+$) and was first
predicted by Greisen, Zatsepin and Kuzmin
\cite{Greisen-1966,Zatsepin-Kuzmin-1966}: the ``GZK cut-off''.

The main candidates for particle acceleration to energies beyond $10^{19}\ev$
are Active Galaxies, Gamma Ray Bursts and starburst galaxies. With
eq.~(\ref{Emax}), the maximum energy to which these objects could accelerate
protons is between $10^{20}$ and $10^{21}\ev$. Similarly as for supernova
remnants and the energy budget of the Galaxy, there is a suggestive coincidence
between the power released by these extragalactic objects and the power required
to sustain an $E^{-2}$ flux up to GZK energies. The latter is
$10^{-7}\ev/\Cm^3$, equivalent to $3\times10^{-19}\erg/\Cm^3$
\cite{Gaisser-1997}. Normalised to the cosmic abundance of AGN and GRB, this
density translates to a required power of about $2\times10^{44}\erg/\Scnd$
(Active Galaxies) and $3\times10^{52}\erg/\Scnd$ (Gamma Ray Bursts) and this
indeed is of a similar order as the corresponding electromagnetic energies
released by these objects.

For a recent review of cosmic rays we refer to \cite{Bluemer-etal-2009}).

\subsubsection{Production of neutrinos}
\label{sec-sci-neu-pro}

It has been recognised for half a century \cite{Reines-1960,Greisen-1960} that
protons from cosmic accelerators would also generate neutrinos, via charged pion
production in collisions with the ambient matter or radiation fields, in
reactions such as:
\begin{eqnarray}
 p + \text{nucleus} 
 &\to& \pi + X\quad(\pi=\pi^\pm,\pi^0)
   \label{pi-generation-1}\\
 p + \gamma 
 &\to& \Delta^+ \to \begin{cases}\pi^0 + p\\ \pi^+ + n \end{cases}\;,
    \label{pi-generation-2} 
\end{eqnarray}
with the subsequent decays $\pi^0\to\gamma\gamma$, $\pi^\pm\to\mu^\pm\nuan_\mu$
and $\mu^+\to e^+\nubar_\mu\nu_e$, $\mu^-\to e^-\nu_\mu\nubar_e$. The resulting
neutrino flavour ratio is approximately $\nu_e:\nu_\mu:\nu_\tau=1:2:0$ at the
sources; neutrino oscillation turns this into a ratio of $\nu_e:\nu_\mu:\nu_\tau
=1:1:1$ upon arrival at Earth (see Sect.~\ref{sec-sci-par-osc}).

The kinematic threshold for process (\ref{pi-generation-2}) is determined by the
photon energies in the radiation field. For ambient photons in the UV region, as
characteristic for many stars and accreting objects, it is in the range of
several $\Pev$. For the cosmic microwave background, it is at about
$10^{19.6}\ev$. If the photon spectrum has a broad spectrum, such as the
power-law for photons generated by synchrotron radiation, the threshold is
``smeared'' to much lower energies. This is particularly important for neutrino
astronomy in the $\Tev$ range.

\begin{figure}[ht]
\sidecaption
\epsfig{file=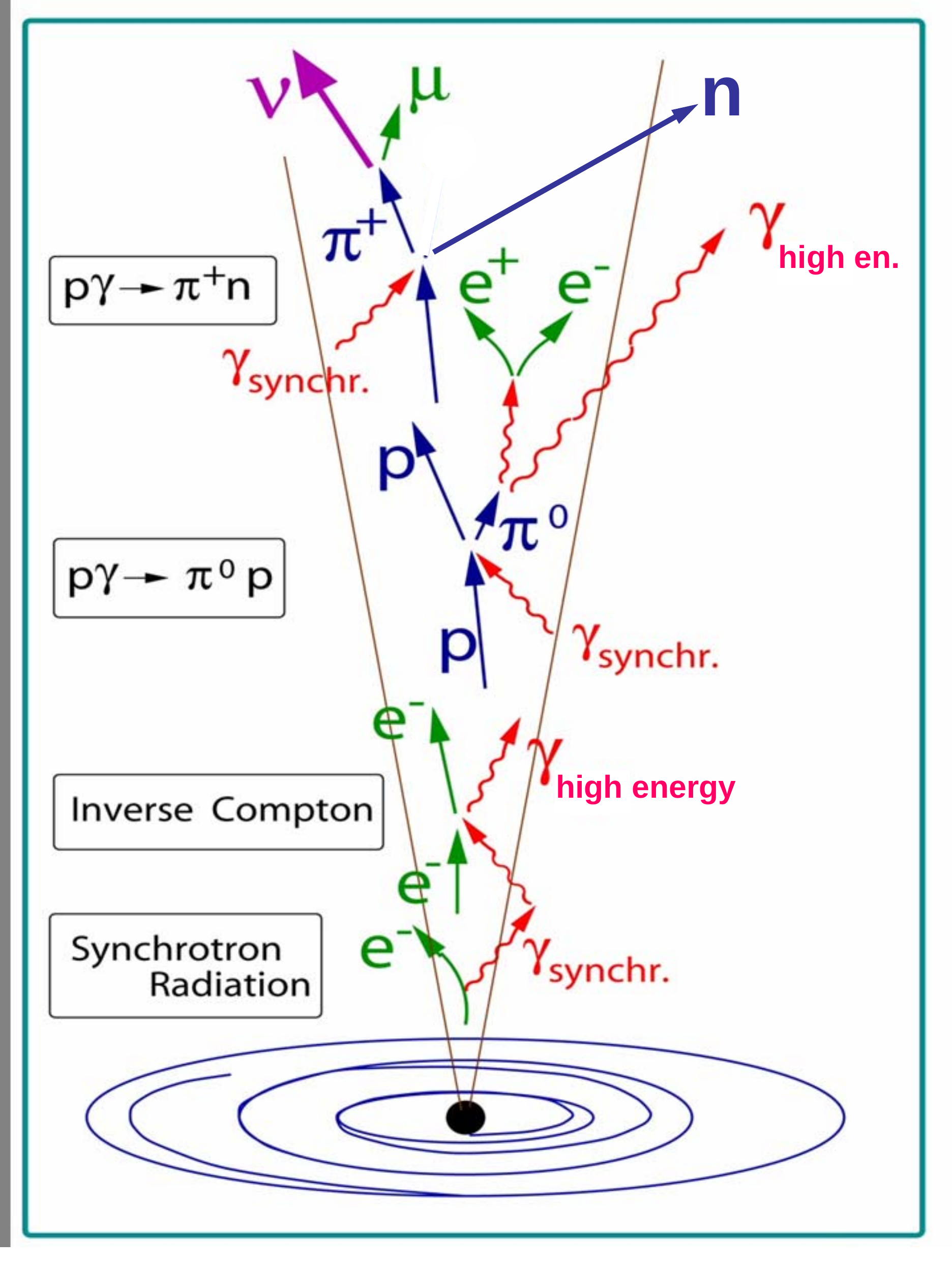,width=5.7cm,trim=12 0 0 0,clip=}
\caption{
Gamma ray and neutrino production in a jet emitted from an Active Galactic
Nucleus, with both hadrons and electrons being accelerated along the jet.
}
\label{AGN-Jet}
\end{figure}

$\Tev$ gamma rays can be produced via the decay of neutral pions, but also by inverse 
Compton scattering:
\begin{equation}
 e^- + \gamma_\text{low energy}\to e^- + \gamma_\text{high energy}\,.
 \label{inverse-compton}
\end{equation}
Actually, most of the measured spectra from $\Tev$ gamma ray sources are
compatible with models based on inverse Compton scattering, and in many cases
with the so-called syn\-chrotron-self Compton (SSC) model where the photon gas is
provided by synchrotron radiation from accelerated electrons (see
Fig.~\ref{AGN-Jet}). Needless to emphasise that pure SSC models are based on
electron, not hadron acceleration and do not directly explain the origin of
cosmic rays. Pure electron acceleration models are called ``leptonic models''. 
In most realistic cases, both electrons and hadrons will be accelerated.

Figure \ref{AGN-Jet} sketches the processes happening in such a combined model. 
The synchrotron radiation from electrons serves as target for Inverse Compton
scattering as well as for proton collisions. Electrons are cooled by synchrotron
emission and may boost synchrotron photons to the $10$--$100\tev$ range but
certainly not to $\Pev$ energies. The observation of $\Pev$ gamma rays would
therefore be a clear proof of hadron acceleration. Unfortunately, the range of
$\Pev$ photons does not exceed the size of our Galaxy, since they are absorbed
by the process $\gamma_\Pev+\gamma_\text{CMB}\to e^+e^-$.

If charged particles are confined by large magnetic fields, only neutral
particles can escape the acceleration region. Apart from gamma rays and
neutrinos, this can be also neutrons (see top right in Fig.~\ref{AGN-Jet}),
provided the source is sufficiently dilute. Neutrons decaying outside the source
would then yield those cosmic ray protons which are detected on Earth and which
can be used to constrain the flux of neutrinos (see
Sect.~\ref{sec-sci-neu-ext}).

\subsubsection{Galactic sources}
\label{sec-sci-neu-gal}

Until the mid-1990s, only one supernova remnant, the Crab nebula, had been
detected in $\Tev$ gamma rays. Therefore, predictions for neutrinos from these
sources were not yet on a firm ground at that time. The Crab belongs to a
special class of SNR, the pulsar wind nebulae, where a central pulsar emits
material into the nebula. The most likely sources of Galactic cosmic rays are
shell-type supernova where this effect is absent or not essential.

\begin{figure}[ht]
\begin{center}
\epsfig{file=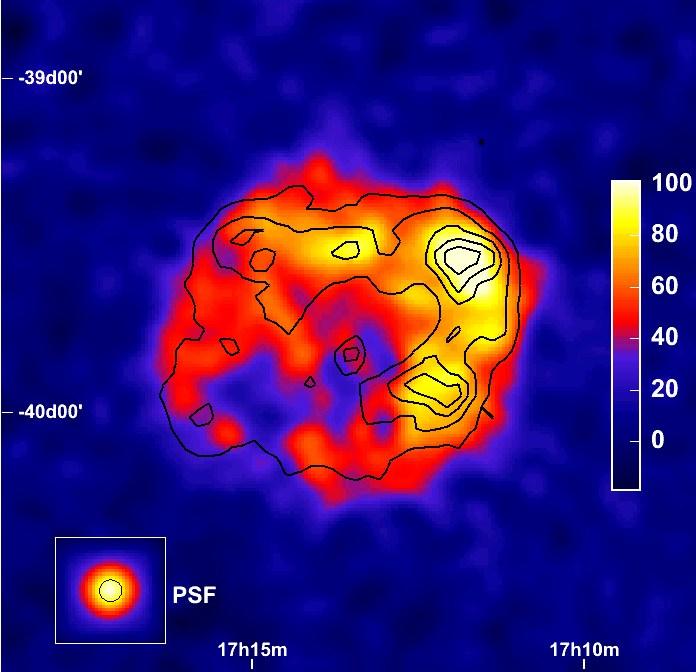,width=0.4\textwidth}
\hspace*{3.mm}
\epsfig{file=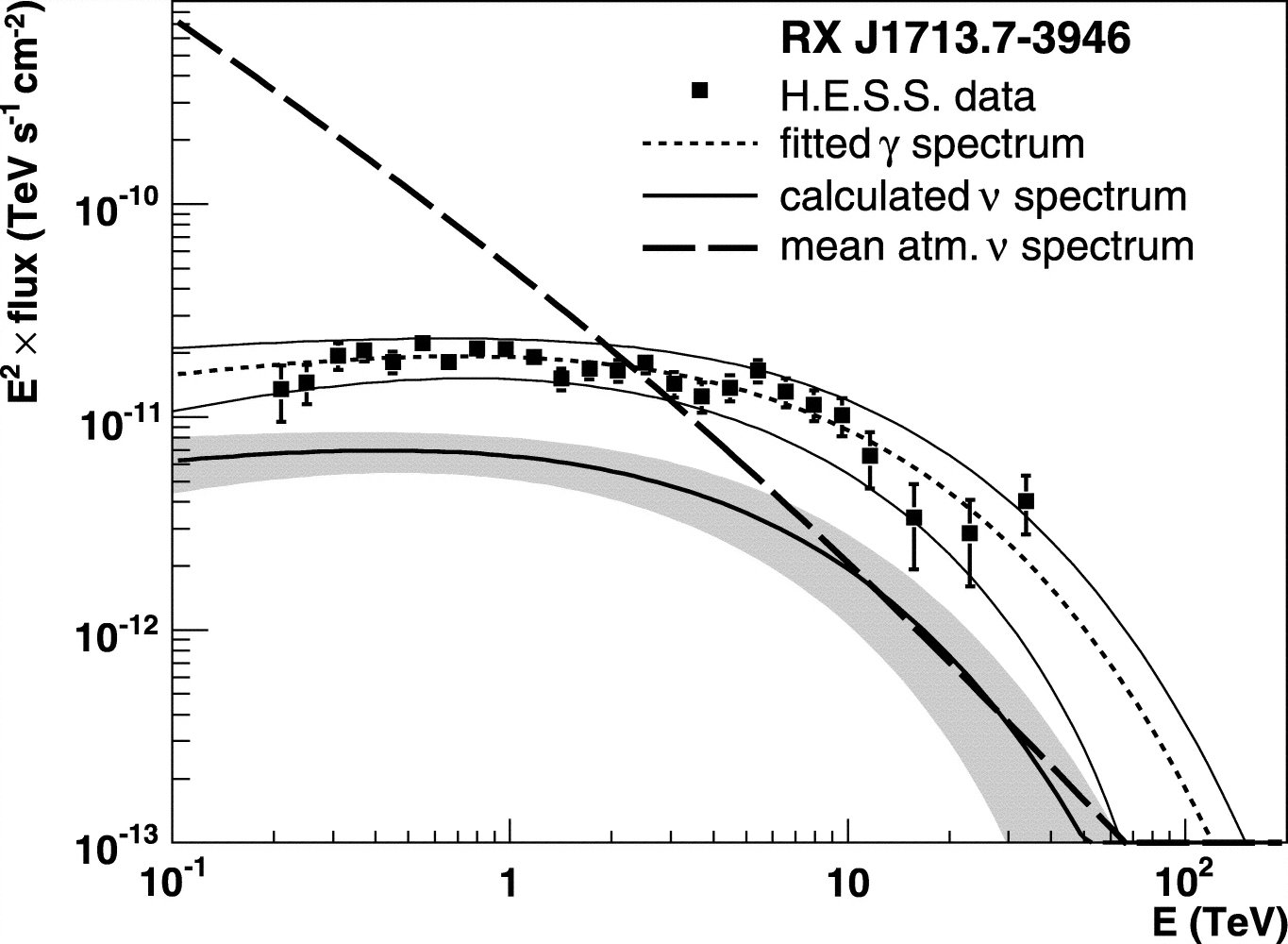,width=0.525\textwidth}
\caption{
Left: Gamma ray image of the supernova remnant RX\,J1713.7-3946, recorded with
the H.E.S.S.\ telescope; superimposed are contours of X-ray surface brightness
recorded by the ASCA satellite (figure taken from \pcite{Berge-2005}). Right:
Measured gamma flux from RX\,J1713.7-3946 and the related neutrino flux
estimated under the assumption that the gamma emission is of purely hadronic
origin (figure taken from \pcite{Kappes-etal-2007}).}
\label{RX-morphology}
\end{center}
\end{figure}

In the mean time the situation has changed dramatically. Imaging Air Cherenkov
Telescopes like H.E.S.S., MAGIC and VERITAS have detected more than hundred
sources of $\Tev$ gamma rays, amongst them about 30\;SNRs (for recent reviews
see \cite{gamma1,gamma2}). From the observed fluxes of gamma rays, estimates or
upper bounds on the flux of neutrinos can be derived, assuming that most of the
observed gamma rays stem from decays of $\pi^0$s generated according to
eqs.~(\ref{pi-generation-1}) and (\ref{pi-generation-2}). Keeping in mind that
high-energy gamma rays may well emerge from inverse Compton scattering,
eq.~(\ref{inverse-compton}), their observation is not a proof that the source
accelerates hadrons rather than only electrons. A certain test can be provided
by detailed information on the $\Mev$--$\Gev$ part of the gamma spectrum and by
information on its high-energy cut-off. For Galactic sources, the morphology of
gamma ray emission can be studied and provides additional information. In this
context, we note the observation of the supernova remnant RX\,J1713.7-3946 with
the H.E.S.S.\ telescope \cite{Aharonian-2004}. The image of this source (see
Fig.~\ref{RX-morphology},\,left) shows an increase of the gamma flux from the
direction of known molecular clouds. The effect can be attributed to protons
accelerated in the SNR and then interacting with the clouds. The spectrum of
gamma rays above some $100\gev$ is well compatible with expectations for the
decay of $\pi^0$s from proton interactions. Recent measurements with the
Fermi-LAT instrument \cite{Fermi-RX}, however, indicate that the spectrum at
lower energies may be better described by leptonic models. This is in accordance
with X-ray line emission around $1\kev$ which should be produced along with pion
decays but has not been observed with the Suzaku satellite detector
\cite{Ellison-2010}. Unambiguous Galactic sites of hadronic acceleration have
thus yet to be identified. Extended gamma ray sources which trace the density of
molecular clouds, such as the SNR Vela (RX\,J0852.0-4622) \cite{Aharonian-2005}
and a large region near the Galactic Centre \cite{Aharonian-2006}, remain good
candidates.

In \cite{Kappes-etal-2007}, the expected neutrino flux from the SNR
RX~J1713.7-3946 is calculated (cf.\ Fig.\ref{RX-morphology}, right). Based on
the assumption that all observed gamma rays stem from $\pi^0$ decays, the
authors calculate the neutrino flux expected from $\pi^\pm$ decays. For five
years of data taking with a cubic-kilometre detector and a threshold at
$1\tev$, the number of events is calculated to be between 7 and 14, over a
background of 21 atmospheric neutrino events in a $2.1^\circ$ search cone
(resulting from the $1.3^\circ$ diameter of the source, multiplied with a
factor of~1.6 to achieve optimal sensitivity). A threshold of $5\tev$ results
in 2.6--6.7 signal events and 8.2 atmospheric neutrino events.  Cutting at
higher energies will eliminate not only the atmospheric background but also the
signal, cutting at lower energies will significantly worsen the
signal-to-background ratio. These estimates suggest that the positive effect of
a detector threshold much below $1\tev$ will be small, at least for steady
sources. It is also clear that neutrinos from this particular source will be
hardly observable if the gamma flux of hadronic origin is sub-dominant; in this
case, however, the quest of identifying the sites of hadronic acceleration
starts afresh, and neutrino observations will be more crucial than ever in
solving this puzzle.

Several candidates for sources with hadron acceleration beyond the knee
(``Pevatrons'') have been identified in the Cygnus region by the Milagro
collaboration \cite{Abdo-2007,Abdo-2008}. The gamma ray spectrum of the
strongest of these sources, MGRO\,J1908+06, is consistent with an $E^{-2}$
behaviour between $500\gev$ and $40\tev$. It does not show evidence for a
cut-off, in accordance with the expectations for Pevatrons, which should emit
gamma rays with energies up to several hundred $\Tev$. In
\cite{Halzen-Kappes-Murchadha-2008}, the associated neutrino fluxes from the six
identified Milagro source candidates have been calculated and the event rates in
the IceCube neutrino telescope estimated. In a simulated neutrino sky-map for 5
years of data taking, two of the sources are discernible ``by eye'' when
applying a lower energy threshold of $40\tev$. Stacking all six sources, an
excess is found with a Poisson probability for being a background fluctuation of
smaller than $10^{-3}$ for lower energy thresholds anywhere between 10 and
$100\tev$. The simulation assumes a gamma-ray cut-off at $300\tev$.

Similarly close to the sensitivity of cubic-kilometre telescopes are the
expected neutrino fluxes from microquasars, as for instance LS\,I\,+61\,303 and
LS\,5039, where the latter could provide a handful of neutrino events per year
\cite{LS5039}.
  
Practically all publications of the last years (see references above and e.g.\
\cite{Lipari-2006,Bednarek-etal-2005,Vissani-2006,NEMO-Aiello-2007,Kappes-etal-2007,
Vissani-2011}) come to the unanimous conclusion that cubic-kilometre detectors
will just ``scrape'' the detection region. The present estimates suggest that
the sensitivity of a cubic-kilometre telescope is ``tantalisingly (and
frustratingly) close'' \cite{Lipari-2006} to the expectations for the brightest
observed Galactic $\Tev$ gamma sources.

\subsubsection{Extragalactic sources}
\label{sec-sci-neu-ext}

There is much less observational guidance to predict neutrino fluxes from
extragalactic than for Galactic sources. The predictions for individual sources
have order-of-magnitude character. The best-motivated extragalactic candidates
for high energy neutrino emission are Active Galactic Nuclei (AGN), Gamma Ray
Bursts (GRB) and galaxy clusters.

\begin{itemize}
\item 
{\it Active Galactic Nuclei} host a super-massive black hole with
$10^6$--$10^9$ solar masses in their extremely bright centre. The black hole
accretes matter and thus transforms huge amounts of gravitational energy into
radiation, typically of the order of $10^{44}\,\text{erg/s}$. This energy can
also be converted to kinetic energy of accelerated particles, see also
Fig.~\ref{AGN-Jet}. Blazars, a particular class of AGN where the jet is aligned
closely to the line of sight, turned out to be strong gamma emitters, with 119
sources at $\Gev$ energies listed in the 2009 Fermi bright gamma source list
\cite{Fermi-2009} and 18 at $\Tev$ energies being reported in
\cite{Aharonian-2008}. Gamma ray emission from blazars is often highly
variable, with the most extreme variation observed by H.E.S.S.\ for the blazar
PKS\,2155-304: An increase by two orders of magnitude within one hour
\cite{PKS2155-304}. Naturally, the observation of neutrino events from such a
source and within such a short time window would be rather significant. Another
interesting outburst has been observed from the blazar 1ES\,1959+650
\cite{Krawczynsky-2004}. This was an ``orphan flare'', where the $\Tev$ emission
was not accompanied by X-ray emission, as it typically would be for SSC models. 
A hadronic model does not require such a correlation between $\Tev$ and X-ray
(synchrotron) emission; therefore orphan flares are interesting environments to
search for neutrino emission.
\item 
{\it Gamma Ray Bursts} are the most cataclysmic phenomena in the Universe,
releasing huge amounts of energy in gamma rays within milliseconds to minutes. 
The favoured explanation for the longer bursts is the collapse of a massive star
into a black hole. The so-called fireball model of GRB assumes that a central
engine ejects large amounts of mass within a short time interval which form
successive plasma shells and have typical Lorentz factors of
$\Gamma=100\rnge1000$. When the outer shells slow down they are hit by the inner
shells and internal shock fronts are piling up. Along these fronts, electrons
and protons are accelerated. Electrons are cooled by synchrotron radiation,
protons can be accelerated up to energies as high as $10^{21}\ev$
\cite{Waxman-Bahcall-1997}. When the shells run into the interstellar medium,
external shocks are built up, leading to afterglow emission visible in X-ray,
optical and radio wavelengths. Neutrino emission has been calculated for three
phases: the precursor phase when the jet is still forming and no electromagnetic
radiation is escaping \cite{Razzaque-2003}; the ``prompt'' phase coinciding with
the burst in gamma rays (see e.g.\ \cite{Waxman-Bahcall-1997}); and the afterglow
phase \cite{Waxman-Bahcall-2000}. We will return to these predictions when
presenting experimental bounds in Sect~\ref{sec-phy}.
\item 
{\it Starburst galaxies} are galaxies undergoing an episode of large-scale star
formation, where the central regions eject a galactic-scale wind driven by the
collective effect of supernova explosions and winds from massive stars. 
Recently, the starburst galaxies NGC253 (southern hemisphere) and M82 (northern
hemisphere) have been detected by the H.E.S.S.\ and VERITAS telescopes
\cite{NGC253,M82}. The gamma ray flux at several hundred $\Gev$ suggests cosmic
ray densities of two to three orders of magnitude above that in our own Galaxy. 
Following \cite{Loeb-Waxman-2006}, the cumulative neutrino flux of all starburst
galaxies is detectable with cubic-kilometre detectors.
\end{itemize}

\begin{figure}[ht]
\sidecaption
\epsfig{file=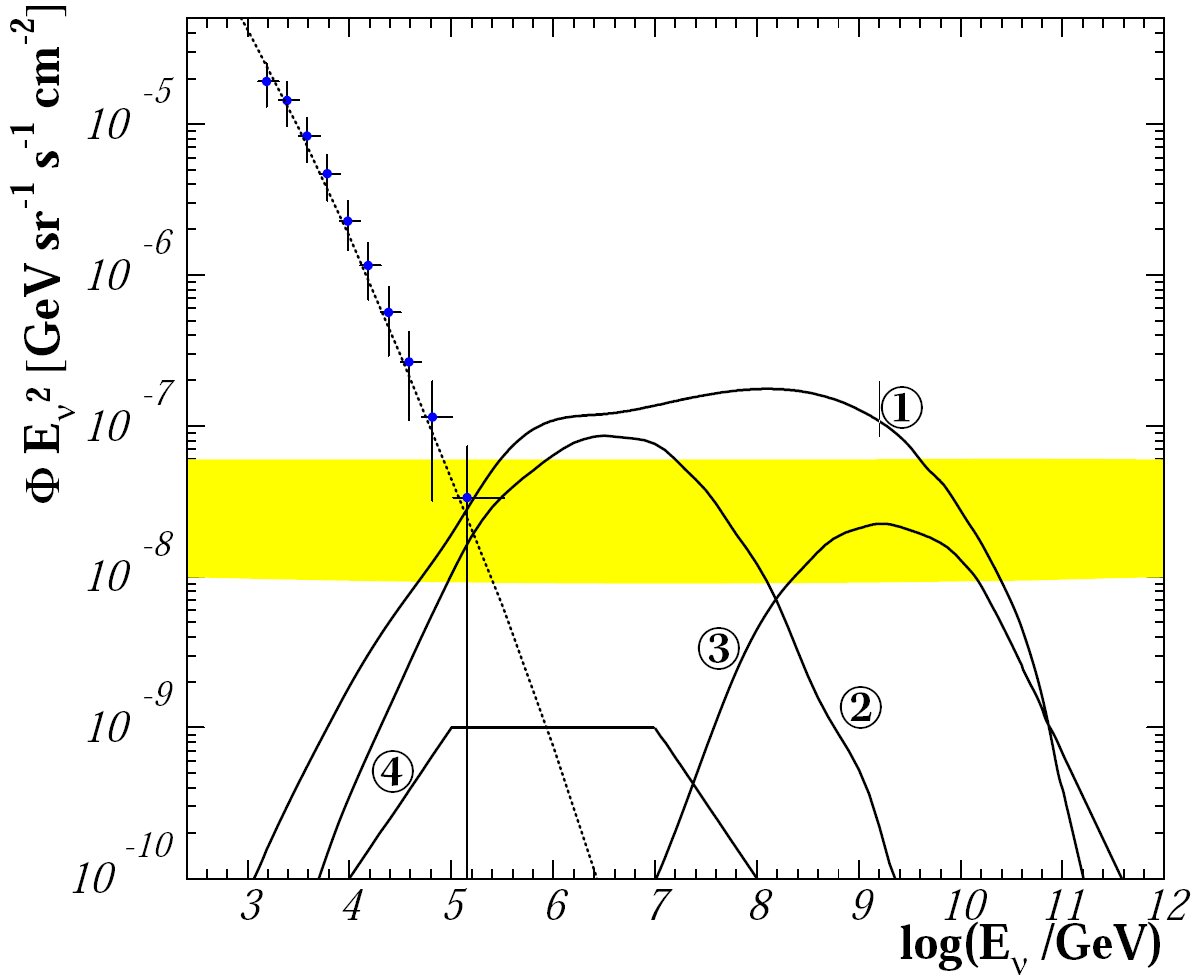,width=8.cm}
\caption{
Spectra of diffuse neutrino fluxes, multiplied with $E^2$. Points: AMANDA
measurements of the atmospheric neutrino flux \pcite{Amanda-atm-2010}, compared
to the prediction from \pcite{Volkova-1980} (dotted line). Horizontal band:
$E^{-2}$ upper bound derived from charged cosmic rays at $E>10^{19}\ev$
\pcite{Waxman-Bahcall-1999}, with the width reflecting the uncertainty of cosmic
evolution parameters. Numbered curves: predictions for the diffuse flux of
neutrinos from all AGN (1,2)
\pcite{Mannheim-Protheroe-Rachen-2001,Stecker-2005}, of cosmogenic neutrinos (3)
\pcite{Berezinsky-Zatsepin} and of neutrinos from GRBs, assuming that they are
the sources of highest energy charged cosmic rays (4)
\pcite{Waxman-Bahcall-1997}. Figure courtesy of J.\,Becker.
\label{fluxes}
}
\end{figure}

Predictions for the integrated flux from all extragalactic sources are based on
the observed gamma and X-rays fluxes or those of charged cosmic rays above
$10^{18}\ev$. A selection of predictions and bounds is shown in
Fig.~\ref{fluxes}. Early normalisations as e.g.\ in \cite{Stecker-Salamon-1996}
assumed neutrinos would be produced in the cores of AGN, accompanied by X-ray
emission, and that most of the observed X-ray background was non-thermal
radiation from a superposition of the fluxes from unresolved AGN. This model as
well as others violated upper bounds derived from the observed cosmic ray
spectrum. Subsequent observations of these AGN showed that most of the X-ray
emission is thermal and therefore cannot be directly related to the production
of relativistic particles. Relying on $\Mev$ gamma ray rather than X-ray
observations, the authors scaled down the original prediction by an order of
magnitude (\cite{Stecker-2005}, cf.\ curve~2 in Fig.~\ref{fluxes}).

Two upper bounds on the diffuse neutrino flux, both derived from charged
cosmic ray fluxes, are frequently used as benchmarks. The first
(``Waxman-Bahcall bound'' \cite{Waxman-Bahcall-1999}) is normalised to the
cosmic ray flux at about $10^{19}\ev$. Assuming a generic $E^{-2}$ spectrum
for all extragalactic sources, the authors obtain a limit of $E^2\Phi_\nu=
1\rnge5\times10^{-8}\flunit$, where $\Phi_\nu$ is the neutrino flux
differential in energy, time, area and solid angle and the uncertainty is
given by different cosmic evolution models (coloured band in
Fig.~\ref{fluxes}). This estimate assumes that the sources are sufficiently
dilute, so that neutrons can escape, decay and provide the observed cosmic
rays to which the estimate is normalised.  If the sources are opaque even for
neutrons, the only remaining estimator is electromagnetic radiation. The decay
of $\pi^0$s (co-produced with charged pions producing neutrinos) yields gamma
rays which develop electromagnetic cascades.  The energy of gammas escaping
the source is mostly in the range $1\mev$ to $100\gev$. The diffuse gamma-ray
background above $30\mev$ was measured by the EGRET satellite
\cite{EGRET-2000} to be $E^2\Phi_\gamma=1.37\times10^{-6}\flunit$ and sets a
bound of similar size to the neutrino flux (not shown in the figure), which
hardly can be circumvented by more sophisticated assumptions on the character
of the sources. Recent Fermi data \cite{Fermi-2010} suggest
\cite{Berezinsky-2011} that this ``gamma bound'' (or ``cascade bound'',
originally derived in \cite{Berezinsky-1975}) is close to the Waxman-Bahcall
bound, and actually has already been superseded by experimental upper limits
from the running underwater/ice telescopes (see below). In a cubic-kilometre
detector, the Waxman-Bahcall flux would lead to 100--500 events per year.

Contrary to Waxman and Bahcall, Mannheim, Protheroe and Rachen (MPR) 
\cite{Mannheim-Protheroe-Rachen-2001} assume that a significant part of
the observed cosmic ray spectrum between $10^{16}\ev$ and $10^{19}\ev$ is due
to extragalactic rather than Galactic sources. Interpreting this spectrum as a
superposition of spectra from many extragalactic source classes, each with a
different cut-off, the neutrino bound considerably weakens towards lower
energies and is about $E^2\cdot\D N/\D E \approx 5\times10^{-7}\flunit$, at a
few $10^{14}\ev$.  Meanwhile, also this bound is only of historical interest,
due to the new results from the Fermi satellite and from neutrino telescopes.

A non-astrophysical contribution to the diffuse neutrino background is assumed
in ``top-down scenarios'', where cosmic rays of highest energy are due to
cascade decays of long-lived super-heavy particles at unification-scale energies
($10^{24}\rnge10^{25}\ev$). Such processes inevitably produce high-energy
neutrinos and photons. An estimate of the resulting neutrino event rate in
IceCube of up to 40 events per year \cite{Barbot-2003} has meanwhile been
superseded by recent, tight limits from Auger on the photon contribution to
energetic cosmic rays \cite{Auger-photons}.
   
The most exciting discovery of neutrino astronomy would be the detection of
point sources rather than just a high-energy excess in diffuse fluxes. However,
experimental limits for diffuse fluxes are setting bounds for expected
point-source fluxes. The argument is as follows: Contributions to the diffuse
flux will come from all the observable universe, up to a distance $c/H_0$,
whereas point sources will be visible at a level of several events per source
only up to a limited distance of a few hundred Mpc, assuming reasonable maximum
luminosities per source. For a homogeneous distribution of extragalactic
sources, one therefore can derive a limit on the number of observable point
sources. In \cite{Lipari-2006} the following assumptions are made: a homogeneous
source density in a Euclidean universe; a ``typical'' (and similar) source
luminosity $L_\text{source}$ for all sources; an $E_\nu^{-2}$ spectrum of the
neutrino fluxes. Given an experimental limit $K_\text{diffuse}$ on the diffuse
neutrino flux and a sensitivity $C_\text{point}$ to point sources, the expected
number of resolvable extragalactic point sources, $N_s$ scales as
\begin{equation}
  N_s\propto\frac{K_\text{diffuse} \cdot \sqrt{L_\text{source}}}{C_\text{point}^{3/2}}\;.
\end{equation}
With the present diffuse flux limit from IceCube and its point source
sensitivity, one obtains $N_s\approx0.2\rnge2$ (in analogy to an estimate made
in \cite{Silvestri-2007} for the AMANDA diffuse limit). This means that, under
the given assumptions, a cubic-kilometre detector would have a remaining chance
to detect extragalactic point sources. Note, however, that a few individual,
very close sources could circumvent the homogeneity assumption and be well
observable. Also, point sources with cut-offs below a few hundred $\Tev$ would
not be covered by the argument above since, in order to obtain the best
sensitivity for diffuse fluxes, a lower energy cut at about $100\tev$ has to be
used \cite{icecube-2004}. In particular this implies that the argument of
\cite{Lipari-2006} does not apply to Galactic sources since they are not
homogeneously distributed and typically are expected to have cut-off energies
below $100\tev$.

\subsection{Particle physics and exotic phenomena}
\label{sec-sci-par}

Beyond charting the high energy universe, neutrino telescopes may reveal first
signatures of new physics beyond the standard model. Opportunities include the
indirect search for dark matter particle candidates, the search for super-heavy
exotic particles like supersymmetric Q-balls or magnetic monopoles, or the
search for deviations from the established neutrino oscillations which may
result from violation of Lorentz invariance. We sketch these opportunities in
the following.

\subsubsection{Indirect search for dark matter}
\label{sec-sci-par-ind}

Favoured dark matter candidates are Weakly Interacting Massive Particles
(WIMPs). They are preferentially discussed in the framework of the minimal
supersymmetric standard model (MSSM), see \cite{DM-Cotta-2009} and references
therein. Specifically, the lightest supersymmetric particle is considered a good
WIMP candidate, with predicted masses in the range from a few $\Gev$ to a few
tens of $\Tev$ \cite{DM-Gilmore-2007}. A more constrained class of theories is
based on minimal supergravity (mSUGRA); see \cite{DM-Olive-2010} for a
comparison of MSSM and mSUGRA. WIMPs can be trapped in the gravitational
potential of celestial bodies such as the Earth or the Sun through elastic
scattering and would accumulate in the cores of these objects. If WIMPs are
Majorana particles (as in supersymmetry), they may subsequently annihilate once
sufficient densities are reached. Neutrinos are produced via decay of the
annihilation products \cite{DM-Jungman-1996}. The mean neutrino energy depends
on the WIMP mass and the annihilation final state. For instance, quark-antiquark
states result in softer, $W^+W^-$ states in harder neutrino spectra.

The capture rate depends on the mass of the celestial body and on local density,
mean velocity and scattering cross section of the WIMPs on nuclei. Slower WIMPs
are captured more efficiently than faster WIMPs. The scattering cross section
can be spin-dependent or spin-independent; the capture rate thus depends on
the spin composition of the nuclei in the celestial body (capture in the Sun,
e.g., is dominated by scattering on hydrogen nuclei with spin $\nicefrac12$ and
is therefore particularly effective for a spin-dependent interaction). The
scattering cross section also depends on the mass ratio of the WIMP and the
scattered nucleus. For instance, the Sun with its lighter nuclei captures light
WIMPs more efficiently, the Earth gains with respect to heavier WIMPs. The
annihilation rate scales with the square of the accumulated WIMP density (which
likely has reached equilibrium for the Sun but less likely for the Earth) and
with the annihilation cross section (which is related to the scattering cross
section through the underlying theoretical model).

WIMP detection via secondary particles from annihilations (in particular gammas
or neutrinos) is called ``indirect detection'', in contrast to ``direct
detection'' of recoil nuclei from elastic WIMP-nucleus scattering. Results from
both methods can be related (see Sect.~\ref{sec-phy}), but one has to keep in
mind that direct event rates are highest for fast WIMPs and scale with the WIMP
density, whereas indirect rates are highest for slow WIMPs and scale with the
square of the WIMP density. In this respect, the two methods are complementary. 
Whereas direct and indirect searches are on similar footing with respect to
spin-independent scattering, indirect searches win when it comes to neutralino
capture via spin-dependent scattering in the Sun \cite{DM-Halzen-2009}. 
Consequently, limits on the spin-dependent scattering cross section from
indirect searches are by nearly two orders of magnitude better than those from
direct searches. For detailed reviews on dark matter searches, we refer to
\cite{DM-Bertone-2005,DM-Bergstroem-2009,DM-Halzen-2009}.

An alternative and in many ways orthogonal approach is the search for WIMP
candidates at accelerators, in particular at the LHC \cite{Ellis-2010}. The
forthcoming years will be particularly interesting in this respect. One has to
keep in mind, however, that a discovery at LHC would not yet pin down the dark
matter properties since it would remain to be shown that the LHC particles
actually (exclusively) form the astrophysical dark matter. In that sense, a full
picture of the very nature of dark matter (if it were supersymmetric particles)
will be obtained only by combining results from all three methods
\cite{DM-Bertone-2005}.

\subsubsection{Super-heavy particles}
\label{sec-sci-par-sup}

Neutrino telescopes can be used to search for several exotic objects on the mass
scale much beyond that of WIMPs which may lead to characteristic signal
patterns. Here, we mention the three most popular ones: magnetic monopoles,
supersymmetric Q-balls and nuclearites. The predicted mass of magnetic monopoles
ranges between $10^4$ and $10^{19}\gev$, nuclearites could have masses from a
few hundred $\Gev$ up to the mass scale of neutron stars, and Q-balls up to
$10^{27}\gev$. The expected signals for these particles are similar; a clear
assignment is therefore excluded for a single event, but may be possible if
angular and light yield distributions can be studied for a sample of several
events.
 
\begin{itemize}
\item 
{\it Magnetic monopoles:} Magnetic monopoles have been introduced by P.\,Dirac
in 1931 in order to explain the quantisation of electric charge
\cite{Dirac-1931}. After decades of unsuccessful monopole searches at
accelerators and in cosmic rays, the efforts gained new momentum when monopoles
turned out to be a consequence of most variants of Grand Unified Theories
\cite{Giacomelli-2008}. A phase transition in the early Universe at
$10^{-34}\scnd$ might have filled the Universe with a significant amount of
monopoles.

Measurements and estimates of cosmic magnetic fields suggest that they could
accelerate magnetic monopoles lighter than $10^{14}\ev$ (sub-GUT-scale) to
relativistic velocities \cite{Wick-2000,Ryu-1998}. The magnetic charge of
monopoles obeys the Dirac quantisation rule $g = n \cdot e/(2\alpha)$, with $n =
1,2,3,\dots$ and $\alpha = 1/137$ ($e$ being the elementary charge). A magnetic
monopole with unit charge $g = 137e/2$ and a velocity above the Cherenkov
threshold in water or ice ($\beta>0.75$) would emit Cherenkov light along its
path. The light intensity would exceed that of a minimally ionising muon by a
factor of 8300, this providing a rather unique signature in a neutrino
telescope.

Typical GUT versions predict monopoles with masses of $10^{16}\gev$ and more. 
These monopoles would have typical virial velocities peaking at
$\beta=10^{-4}\rnge10^{-3}$. They might catalyse baryon decays along their path
which would be detected via the decay particles \cite{Rubakov-1981}. For certain
regions of the parameter space spanned by catalysis cross section and monopole
velocity, the Cherenkov light from the secondary particles would create a
pattern of a slowly propagating light source on a straight trajectory, well
observable in a neutrino telescope.

\item 
{\it Nuclearites:} Nuclearites (``strange quark matter'' or ``strangelets'') are
hypothetical aggregates of $u$, $d$ and $s$ quarks combined with electrons, to
adjust electric neutrality. They might be stable for baryon numbers ranging from
those of usual nuclei up to those of neutron stars ($A\approx10^{57}$)
\cite{Rujula-1984,Bakari-2000}. Nuclearites could have been produced in the
primordial Universe or in certain astrophysical processes like the collision of
neutron stars. They would induce a thermal shock wave along their path through
the detector medium. For virial velocities, the corresponding Planck radiation
could reach $10^3\Kelvin$. A light source so slow and bright would produce a
very specific time pattern similar to that of a slow GUT monopole.

\item 
{\it Supersymmetric Q-balls:} Q-Balls are hypothesised coherent states of
squarks, sleptons and Higgs-fields \cite{Kusenko-1997}. These SUSY soliton
states could be stable for masses $M\gg10^{15}\gev$, the life time for smaller
masses is unknown. Stable Q-Balls could provide a relevant contribution to Dark
Matter. Decays of unstable Q-balls could produce WIMPs, with properties not
compatible with the conventional thermal origin \cite{Bakari-2000}. The
discovery of Q-Balls would have an enormous impact on the understanding of
matter-antimatter symmetry emerging in the early Universe
\cite{Fujii-2002,Dine-2004,Buchmueller-2005}.

Neutral Q-ball objects (SENS, supersymmetric Electrically Neutral Solitons)
could catalyse proton decays along their path, similar to GUT monopoles. 
Electrically charged Q-ball objects (SECS, supersymmetric electrically charged
solitons) would produce light in a similar way as nuclearites.
\end{itemize}

\subsubsection{Neutrino oscillations}
\label{sec-sci-par-osc}

We meanwhile know that neutrinos change their flavours during propagation. There
are two standard mechanisms that contribute to this phenomenon, a mixing between
mass and flavour eigenstates (inducing ``vacuum oscillations''), and
flavour-dependent forward scattering amplitudes in matter (``MSW effect''
\cite{Wolfenstein-1978,Mikheev-Smirnov-1985}). Non-standard oscillation effects
could e.g.\ be associated to a violation of Lorentz invariance. Neutrino
telescopes have the potential to study the high-energy regime of neutrino
oscillations.

\vspace{2mm}
\noindent
\uline{\it Standard oscillations:}
\vspace{2mm}

\begin{figure}[ht]
\sidecaption
\epsfig{file=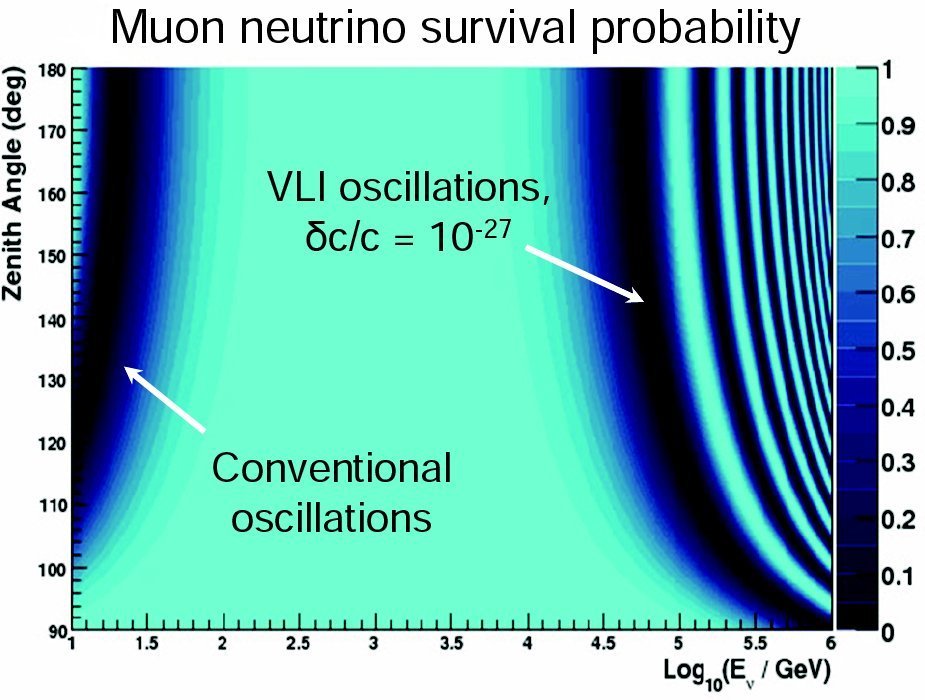,width=8cm, trim=0 5 10 0,clip=}
\caption{
The probability of muon neutrino disappearance as a function of neutrino energy
$E_\nu$ and zenith angle for atmospheric neutrinos. A zenith of $180^\circ$
refers to a vertically-upward going neutrino, i.e.\ propagation along the full
Earth diameter. A zenith angle of $90^\circ$ marks horizontal propagation with a
baseline of a few $100\km$. The pattern at lower energies corresponds to
``standard oscillations'' (without taking into account matter effects), the
patterns at higher energies would be expected for non-standard oscillations due
to the violation of Lorentz invariance (see text).
\label{Oscillations}
}
\end{figure}

Neutrino data from solar, atmospheric, reactor and accelerator experiments
support the concept of neutrino oscillations. The weak flavour eigenstates
$\nu_e, \nu_\mu, \nu_\tau$ are linear combinations of mass
eigenstates $\nu_1, \nu_2, \nu_3$. For the simplified case of
two flavours $\nu_\mu, \nu_\tau$ and two mass eigenstates
$\nu_2, \nu_3$ one has:
\begin{align}
  \nu_\mu  & =   \nu_2 \cos \theta_{23} + \nu_3 \sin \theta_{23} \\
    \nonumber 
  \nu_\tau & = - \nu_2 \sin \theta_{23} + \nu_3 \cos \theta_{23}\;. 
    \label{numunutau} 
\end{align}
If the masses $m_2$ and $m_3$ are different, quantum mechanical time evolution
of an initial $\nu_\mu$ state induces a non-zero transition probability to
$\nu_\tau$. The survival probability for the muon neutrino is
\begin{equation}
 P(\nu_\mu\to\nu_\mu) = 
 1-\sin^2(2\theta_{23})\cdot\sin^2\left(\frac{1.27 \Delta m^2_{23} \cdot L }{E_\nu}\right)\;,
 \label{survival}
\end{equation}
where $L$ (in km) is the distance travelled by the neutrino, $E_{\nu}$ (in
$\Gev$) its energy and $\Delta m^2_{32} = m_3^2 - m_2^2$ (in $\Ev^2$). The
three-flavour case is governed by two independent differences of mass squares
and three mixing angles. The best-fit oscillation parameters derived from
present data are \cite{PDG}:
\begin{align}
\nonumber
\mid\Delta m_{31}^2 \mid & = 2.40  \times10^{-3}\ev^2\approx|\Delta m_{23}^2|
\qquad&
\sin^2 2\theta_{23} & \simeq 1 \\
\nonumber
\Delta m_{21}^2          & = 7.65 \times10^{-5}\ev^2 
\qquad&
\sin^2 \theta_{12}  & = 0.304 \\
&&
\sin^2 \theta_{13}  & < 0.056\quad(3\sigma)\;. 
\label{bestvalues}
\end{align}
If an atmospheric muon neutrino crosses the Earth along its full diameter, the
first minimum of the survival probability occurs around $24\gev$ (see
Fig.~\ref{Oscillations}), the second at $8\gev$. For shorter propagation paths
through the Earth, the minima occur at correspondingly reduced energies and
become unobservable when approaching the horizon.

If their energy threshold is as low as about $10\gev$ (like for IceCube's
DeepCore, see Sect.~\ref{sec-sec}), underwater/ice detectors can approach the
range of conventional oscillations at high energies \cite{Grant-2009},
complementary to underground detectors like Super-Kamiokande which are mostly
sensitive to lower energies.

For Galactic or extragalactic distance scales, kpc to Mpc and beyond, only
average oscillation probabilities are observable since even tiny energy
differences cause large oscillation phase shifts; in addition, the emission
regions may well be larger than oscillation wavelengths. It can be shown that
the mixing parameters listed in eq.~(\ref{bestvalues}) transform a produced
flavour ratio of $\nu_e:\nu_\mu:\nu_\tau=1:2:0$ to an observed one of
$\nu_e:\nu_\mu:\nu_\tau=1:1:1$ \cite{Learned-1995,Beacom-2003}.

\vspace{2mm}
\noindent
\uline{\it Matter effects and neutrino mass hierarchy:}
\vspace{2mm}

One of the remaining puzzles of neutrino physics is the hierarchy of the mass
eigenstates $\nu_1,\nu_2,\nu_3$. We know that $m_2$ is larger than $m_1$, but we
do not know whether $m_3$ is the largest (normal hierarchy) or the smallest mass
(inverted hierarchy). It is expected that eventually double-beta decay
experiments may provide the answer (in case neutrinos are Majorana particles). 
Equally promising are long-baseline neutrino accelerator experiments, provided
$\sin^22\theta_{13}\gtrsim0.001$. Also a 100\,Megaton detector for neutrinos may
give the answer if $\sin^2 2\theta_{13}$ is not too much below 0.1 and the
detection threshold is $5\rnge10\gev$
\cite{Mena-2008,Fernandez-2010,Grant-2009}. This is the energy range where
matter-induced oscillations become important.

For neutrinos below $15\gev$ passing through the centre of the Earth, the MSW
effect slightly enhances the $\nu_\mu$ oscillation probability and suppresses
the $\nubar_\mu$ oscillation probability. For the inverted hierarchy, in
contrast, the oscillation probability for $\nubar_\mu$ is enhanced and for
$\nu_\mu$ suppressed. Water/ice neutrino telescopes cannot distinguish between
neutrinos and antineutrinos (i.e.\ identify the charges of the secondary muons). 
However, in the relevant energy range the interaction cross sections of
neutrinos and anti-neutrinos differ by about a factor of two:
$\sigma(\nu)\approx2\sigma(\nubar)$. This translates into a difference in the
number of observed muon events (about 5\% for $\sin^2 2\theta_{13} = 0.1$). 
Based on the statistical discrimination with respect to the number of events
below $15\gev$, it may therefore be possible to distinguish normal from inverted
hierarchy with DeepCore. Needless to say that firmly establishing such a small
difference requires excellent knowledge of the detector systematics. Given the
complicated optical properties of ice, it remains to be demonstrated whether
systematic uncertainties for these low energies can be kept sufficiently small. 
In this respect, a densely instrumented water detector may be superior.

\vspace{2mm}
\noindent
\uline{\it Non-standard oscillations:}
\vspace{2mm}

Many quantum gravity (QG) models suggest that Lorentz invariance may be violated
and that this effect may be seen in neutrino oscillations
\cite{Coleman-1999,icecube-2009d,Morgan-2007}. Conventional oscillations are due
to different {\it mass} eigenstates. Violation of Lorentz invariance (VLI) may
occur if in addition there are also different {\it velocity} eigenstates, each
with its own limiting velocity, differing from the speed of light. The VLI
magnitude depends on the velocity splitting $\Delta c/c = (c_1 - c_2)/c$. 
Assuming maximum conventional and VLI mixing and the same phase, the muon
survival probability can be written as
\begin{equation}
 \left.P_{\nu_\mu\to\nu_\mu}\right|_\text{max.\ mixing} = 
 1-\sin^2\left(\frac{\Delta m^2L}{4E} + \frac{\Delta c}{c}\frac{LE}{2}\right)\;,
\end{equation}
where the first term comes from standard oscillations (see eq.~(\ref{survival}))
and the second from VLI mixing. Note that the two effects scale differently with
energy, the one with $1/E$, the other with $E$ or, more generally, with $E^n$ if
$\Delta c/c$ is replaced by a generalised VLI term $\Delta \delta = \Delta c/c
\times E^{1-n}$.

Another possible consequence of QG is the evolution of pure neutrino states to
mixed states via interaction with the foamy space-time itself (quantum
decoherence). The resulting effect can be characterised by a set of parameters
$D_1,\dots,D_8$ which in the simplest case are equal and can vary with some
integral power $m$ of energy.

Figure \ref{Oscillations} shows the survival probability as a function of
neutrino energy $E_\nu$ and zenith angle for neutrinos created in the Earth
atmosphere. The pattern at lower energies corresponds to ``standard
oscillations'' (without taking into account matter effects), the pattern at
higher energies would be expected for non-standard oscillations due to the
violation of Lorentz invariance with $\Delta c/c=10^{-27}$.

\subsection{Environmental and marine sciences}
\label{sec-sci-env}

The construction of neutrino telescopes in deep ice or deep water provides
opportunities for long-term, real-time measurements in these environments that
are unique and therefore of utmost interest to a variety of scientific
disciplines beyond astroparticle physics, such as geology and geophysics, marine
biology, oceanography or environmental sciences. Owing to the larger
construction flexibility in water and to the richness of marine phenomena,
corresponding activities are mostly concentrated on marine sites.

Multidisciplinary observatories are associated to all current Mediterranean
deep-sea neutrino telescope projects (see \cite{nemo-sn1,nestor-laertis} and
\cite{Favali-2011} and references therein). For KM3NeT, a dedicated interface to
such instrumentation is integral part of the planning \cite{km3net-tdr}, and a
close cooperation with the {\it European Multidisciplinary Seafloor Observatory
(EMSO)} project \cite{emso-web} has been established. It is expected that
deep-sea data will be taken by a series of instruments that are connected to the
KM3NeT deep-sea cable network; in conjunction, also the neutrino telescope data
themselves will be useful (e.g.\ light measurements for bioluminescence studies,
calibration data for acoustic and water current measurements).

Issues to be addressed with these data are amongst others \cite{km3net-tdr}:
\begin{compactitem}
\item
Investigation of deep-sea phenomena such as internal waves and short time-scale
oscillations in the water column, which are relevant for understanding the
physical processes of the ocean and their effects on the distribution of
suspended geological, chemical and biological materials;
\item 
Real-time tracking of bio-acoustic emissions or vertical migrations of organisms;
\item
Long-term monitoring of the ocean margin environment around Europe;
\item
Studies of oceanic processes and the land--ocean--atmospheric interactions. 
Geological records show that the ecosystem of the Mediterranean Sea amplifies
climatologic variations, making it an ideal test bed for climate studies.
\end{compactitem}
In addition, contributions to hazard warning systems (in particular related to
tsunamis) are conceivable, even though problems related to reliability
standards, data access and usage rights need to be clarified.

In the South Polar ice, the spectrum of multidisciplinary investigations is
limited compared to Ocean detectors, but nevertheless manifold. Light
propagation in deep ice is affected by remnant air bubbles at shallow depths
\cite{Amanda-shallow-1995} and by layered impurities from dust which are due to
climatic effects or volcano eruptions \cite{Amanda-ice-2006} -- see
Fig.~\ref{AMANDA} in Sect.~\ref{sec-fir-ama}. Motivated by these findings, a
project called {\it DeepIce} was proposed in 1999 to the US National Science
Foundation. It was suggested to join efforts of astroparticle physicists,
glaciologists, seismologists, geophysicists and biologists to investigate and
exploit the features of glacial ice at or near the South Pole. The project was
not funded, but many of its ideas have been pursued further and have led to
relevant results, e.g.\ on the propagation of acoustic and radio waves in ice
(see also Sect.~\ref{sec-alt}), on glaciology, climate research and volcanology. 
One legacy of DeepIce is an optical dust-logger, which reads out concentrations
of dust particles and volcanic ash in the 2.5\,km deep water-filled boreholes
with a depth resolution of about $1\mm$ \cite{icecube-geology}. These dust logs,
together with similar results from Greenland ice, have revealed a relationship
between abrupt climate changes and faint volcanic fallout layers and have
permitted for reconstructions of dust and paleowind records of high quality. The
detailed investigation of the IceCube dust logs have provided an unique
characterisation of Polar deep ice and make the South Pole a leading candidate
for a next US deep-ice core project.

Moreover, the thermistors and microinclinometers installed in IceCube allow for
studying the shear strain rate of a large volume of ice in three dimensions as a
function of stress, impurity content and temperatures down to $-35^\circ$C. Last
but not least, we mention an idea on studying biology under extreme conditions
which also evolved from DeepIce: a quantitative model of how micron-sized
microbes could live in liquid veins at triple junctions of ice grains and how
they could be detected \cite{microbiology-2007}.

\clearpage
\section{Detection Principles}
\label{sec-det}

\begin{figure}[ht]
\sidecaption
\epsfig{file=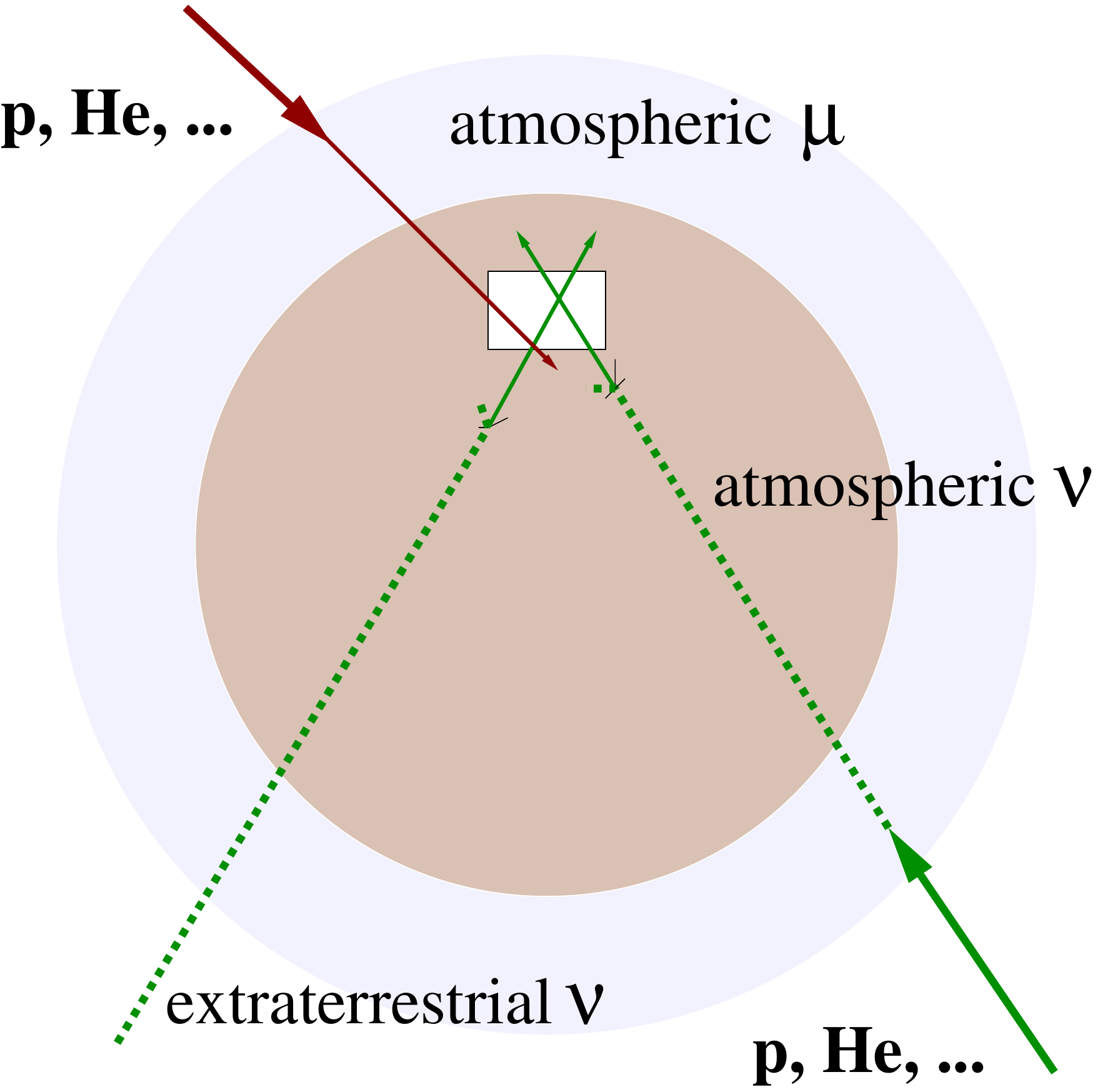,width=6.5cm}
\caption{
Sources of muons in deep underwater/ice detectors. Cosmic nuclei -- protons (p),
$\alpha$ particles (He), etc.\ -- interact in the Earth atmosphere
(light-coloured). Sufficiently energetic muons produced in these interactions
(``atmospheric muons'') can reach the detector (white box) from above. 
Upward-going muons must have been produced in neutrino interactions.
\label{sources}
}
\end{figure}

The classical operation of neutrino telescopes underground, underwater and in
deep ice is recording upward-going muons generated in charged current muon
neutrino interactions. The upward signature guarantees the neutrino origin of
the muon since no other known particle can traverse the Earth. Neutrino
telescopes need to be situated at more than a kilometre depth in order to
suppress downward-moving muons which may be misreconstructed as upward-moving
ones (Fig.~\ref{sources}). Apart from these, only one irreducible background to
extraterrestric neutrinos remains: neutrinos generated by cosmic ray
interactions in the Earth's atmosphere (``atmospheric neutrinos''). This
background cannot be reduced by going deeper. On the other hand, it provides a
standard calibration source and a reliable proof of principle.

Underwater/ice neutrino telescopes consist of an array of photomultiplier tubes
(PMTs) housed in transparent pressure spheres which are spread over a large
volume in oceans, lakes or glacial ice. The PMTs record arrival time and
amplitude, sometimes even the full waveform, of Cherenkov light emitted by muons
or particle cascades. In most designs the spheres are attached to strings which
-- in the case of water detectors -- are moored at the ground and held
vertically by buoys. The typical PMT spacing along a string is $10\rnge20\met$,
and the distance between adjacent strings $60\rnge150\met$. The spacing is thus
by orders of magnitude larger than in an underground detector like
Super-Kamiokande. This allows for covering large volumes, but makes the detector
practically blind with respect to events with energies below about $10\gev$.

\subsection {Neutrino interactions}
\label{sec-det-int}

At neutrino energies $E_\nu$ above some $10\gev$, as relevant for this review,
charged-current (anti)neutrino-nucleon reactions are dominated by deep-inelastic
scattering, $\nuan_\ell N\to\ell^\mp X$ (charged current, CC) or $\nuan_\ell
N\to\nuan_\ell X$ (neutral current, NC), with $\ell=e,\mu,\tau$. The
leading-order differential cross section for the most important process,
$\nu_\mu$ CC reactions, is
\begin{equation}
 \frac{\D^2\sigma_{\nuan N\to\mu^\mp X}}{\D x\D y}
 =\frac{2G_F^2M_NE_\nu}{\pi}\cdot\left(\frac{M_W^2}{Q^2+M_W^2}\right)^2\cdot x
  \begin{cases}d(x,Q^2)+(1-y)^2\ubar(x,Q^2)&\text{for $\nu$}\\
               (1-y)^2u(x,Q^2)+\dbar(x,Q^2)&\text{for $\nubar$}
  \end{cases}
  \label{eq-csdis}
\end{equation}
where $x$ and $y$ are the Bjorken scaling variables ($x$ being the fraction of
the nucleon momentum carried by the struck quark, $y$ the fraction of the
incoming neutrino energy transferred to the hadronic system $X$),
$Q^2\approx2xyE_\nu M_N$ is the negative four-momentum transfer squared, $G_F$
is the Fermi constant and $M_N$ and $M_W$ are the nucleon and the $W$ mass,
respectively (see \cite{Katz-2000} for more details). The parton distributions
$d(x,Q^2)$ and $u(x,Q^2)$ represent the sums of all $d$-type and $u$-type quark
flavours, respectively; likewise $\dbar(x,Q^2)$ and $\ubar(x,Q^2)$ are the
corresponding antiquark distributions. These functions have been measured in
fixed-target experiments and at HERA (see
\cite{Forte-2010} and references therein).

\begin{figure}[ht]
\center{
\epsfig{file=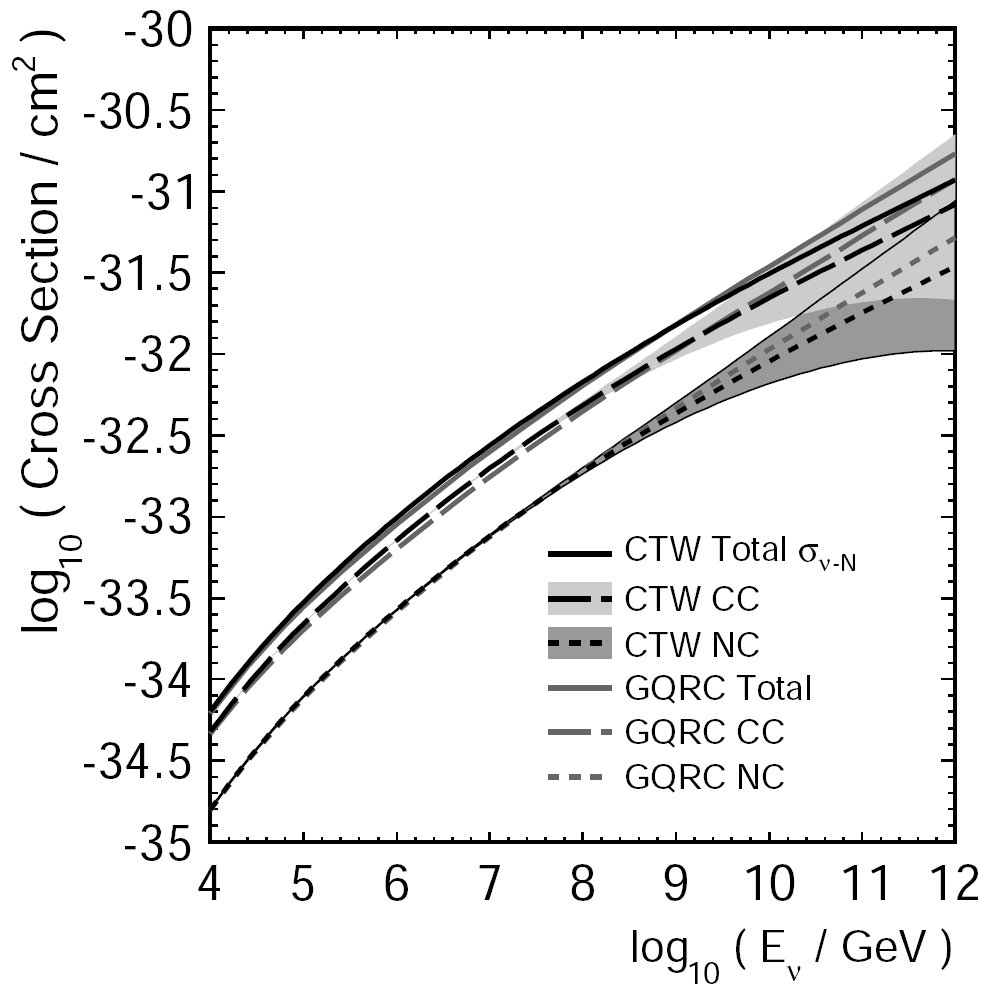,width=0.3\textwidth}
\hspace*{1.5mm}
\epsfig{file=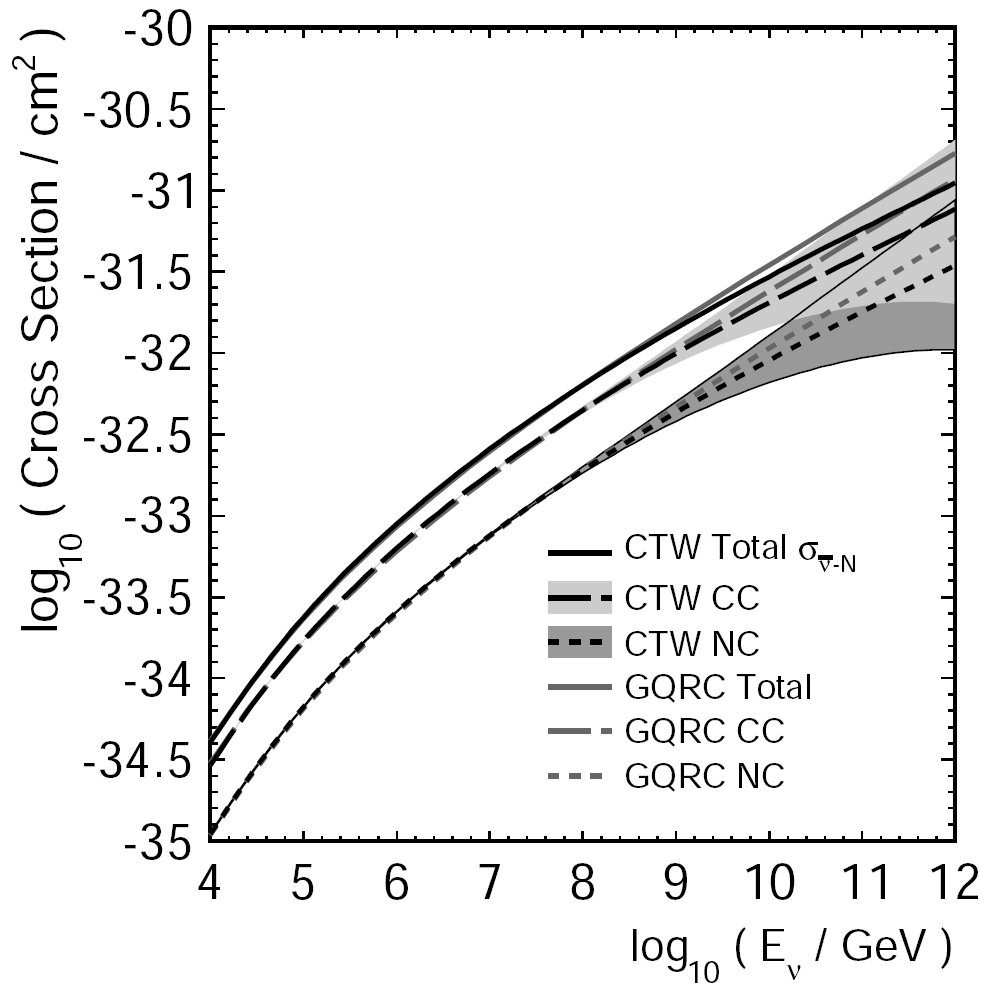,width=0.3\textwidth}
\hspace*{1.5mm}
\epsfig{file=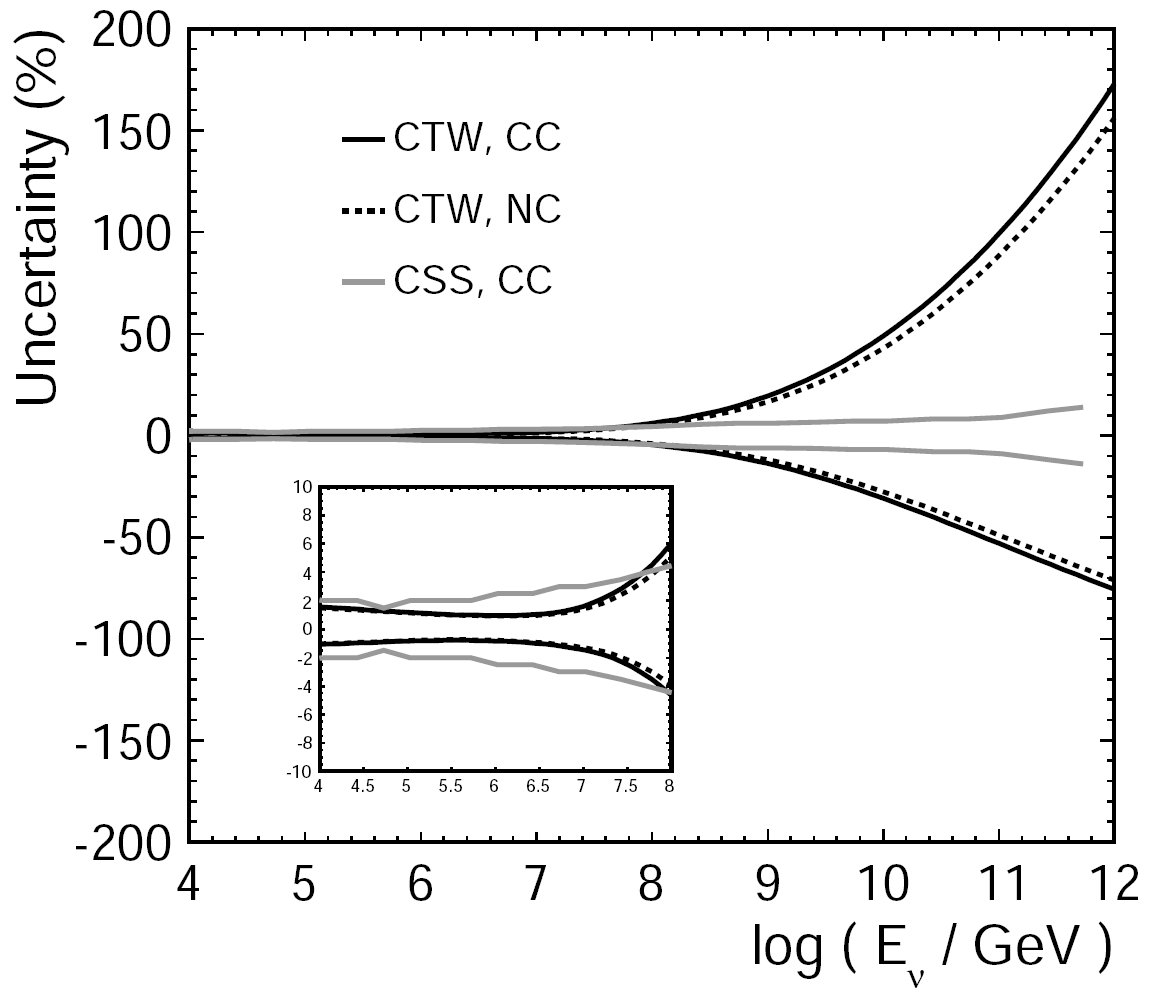,width=0.35\textwidth}
\caption{
Total cross sections for neutrino (left) and antineutrino (middle) scattering on
isoscalar nucleons. Shown are results of two different analyses, both for
charged current (CC) and neutral current (NC) scattering. In the right panel,
the relative uncertainties in both analyses are indicated. Figures taken from
\pcite{Connolly-2011} (CTW); shown are also results from \pcite{Gandhi-1998}
(GQRC) and \pcite{Cooper-Sarkar-2008} (CSS).
\label{fig-ncs}
}}
\end{figure}

Integrating eq.~(\ref{eq-csdis}) over $x$ and $y$ yields the total cross section
$\sigma_{\nuan N\to\mu^\mp X}$ relevant for neutrino telescope observations. For
$E_\nu\lesssim 10\tev$ we have $\langle Q^2\rangle\ll M_W^2$; in this regime the
cross section is linear in $E_\nu$ to a good approximation, with a value of
$\sigma\approx 10^{-35}\cm^2$ at $E_\nu=1\tev$. For higher energies, the
propagator term and the QCD-induced $Q^2$ dependence of the parton distributions
result in a slower rise, roughly proportional to $E_\nu^{0.4}$. The cross
sections resulting from a recent analysis \cite{Connolly-2011} for CC and NC
reactions and their uncertainties are shown in Fig.~\ref{fig-ncs}. Note that the
cross section is per nucleon and is different for neutrinos and antineutrinos;
for isoscalar targets, the cross sections roughly have ratio $\nu:\nubar=2:1$,
whereas they are approximately equal for proton targets. The uncertainties at
large neutrino energies are mainly caused by the lack of experimental
constraints on the parton distributions at $x\lesssim10^{-5}$.

The final state lepton follows the initial neutrino direction with a mean square
root mismatch angle $\theta$ decreasing with the square root of the neutrino
energy \cite{Gaisser-1990}:
\begin{equation}
 \sqrt{\langle\theta^2\rangle} \approx \frac{1.5^\circ}{\sqrt{E_\nu\,[\Tev]}}\;.
 \label{eq-mnangle}
\end{equation}
In \cite{antares-proposal}, the parameterisation $\langle\theta\rangle\approx
0.7^\circ/(E_\nu\,[\Tev])^{0.6}$ is given for the average angle. On the one
hand, this principally allows for source tracing with CC muon neutrino
reactions, but on the other hand represents an intrinsic kinematic limit to the
ultimate angular resolution, which is slightly worse than for high-energy
gamma-ray astronomy and orders of magnitude worse than for conventional
astronomy.

For a CC reaction of a muon neutrino with energy $E_\nu$, we define
$P_{\nu\to\mu}(E_\nu,E_\mu^\text{min})$ to be the probability to produce a muon
which reaches the detector with an energy exceeding the minimum detectable
energy, $E_\mu^\text{min}$. This probability depends on the cross section
$\D\sigma_{\nu N\to\mu X}(E_\nu,E_\mu)/\D E_\mu$ and the effective muon range
$R_\text{eff}$, which is defined as the range after which the muon energy has
decreased to $E_\mu^\text{min}$ \cite{Gaisser-1990}:
\begin{equation}
 P_{\nu\to\mu}(E_\nu,E_\mu^\text{min}\eql1\gev)
 =\frac{\rho}{M_N} \intl^{E_\nu}_{E_\mu^\text{min}}\D E_\mu 
  \frac{\D\sigma_{\nu N\to\mu X}(E_\nu,E_\mu)}{\D E_\mu}
  \cdot R_\text{eff}(E_\mu^\text{min},E_\mu)
\end{equation}
with $\rho$ being the density of the target material
and $M_N$ the nucleon mass. For water and $E_\mu^\text{min}=1\gev$ the following
approximation holds \cite{Gaisser-Halzen-Stanev}:
\begin{eqnarray}
 P_{\nu\to\mu}(E_\nu,E_\mu^\text{min})
 =\begin{cases}
  1.3\times10^{-6} \cdot (E_\nu/1\tev)^{2.2}&\text{for $E_\nu<1\tev$} \\
  1.3\times10^{-6} \cdot (E_\nu/1\tev)^{0.8}&\text{for $E_\nu>1\tev$}\;.
  \end{cases}
\end{eqnarray}
This implies that a neutrino telescope can detect a muon neutrino with $1\tev$
energy with a probability of about $10^{-6}$ if the telescope is on the
neutrino's path.

\begin{figure}[ht]
\sidecaption
\epsfig{file=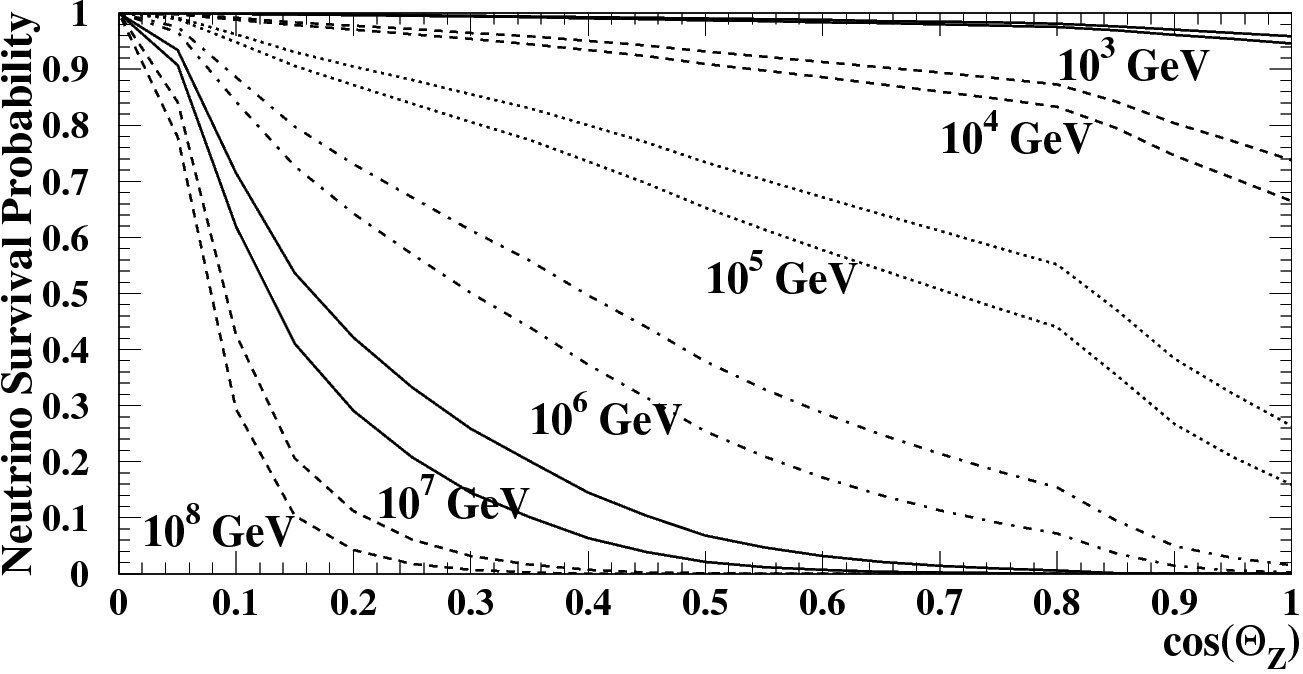,width=9.cm}
\caption{
Transmission probability through the Earth for neutrinos of different energies,
as a function of the zenith angle. For each energy, the upper line is for CC
interactions and the lower one for CC and NC interactions. Figure taken from
\pcite{Albuquerque-2002}.
\label{Transmission}
}
\end{figure}

For a neutrino flux $\Phi_\nu(E_\nu,\vartheta)=[\D^4 N_\nu/(\D E_\nu\,\D t\,\D
A\,\D\Omega)]\cdot\Delta\Omega$ arriving from a solid angle region
$\Delta\Omega$ at zenith $\vartheta$, the number of events recorded by a
detector with area $A$ within a time $T$ is given by
\begin{equation}
 \frac{N_\mu(E_\mu^\text{min},\vartheta)}{AT}
 =\intl^{E_\nu}_{E_\mu^\text{min}}\D E_\nu\, \Phi_\nu(E_\nu,\vartheta) 
  \cdot P_{\nu\to\mu}(E_\nu,E_\mu^\text{min}) 
  \cdot\exp\left[-\frac{\sigma_{\nuan N\to\mu^\mp X}(E_\nu)\cdot Z(\vartheta)}{M_N}\right]\;.
\end{equation}
Here $Z(\vartheta)$ is the matter column density in the Earth crossed by the
neutrino, in $\gramm/\Cm^2$. For sub-$\Tev$ energies, absorption in the Earth is
negligible and the exponential term is close to unity; for larger neutrino
energies, absorption becomes relevant (see Fig.\,\ref{Transmission}).

\begin{figure}[ht]
\sidecaption
\epsfig{file=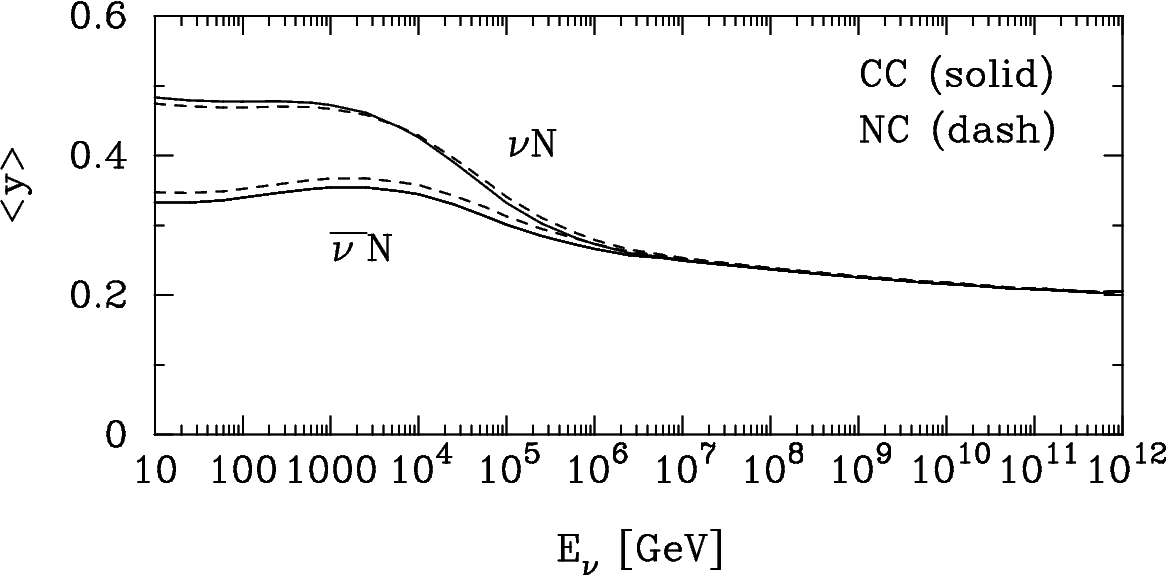,width=9.cm}
\caption{
Average $y$ as a function of neutrino energy, for CC (solid lines) and NC
(dashed) reactions. Figure taken from \pcite{Gandhi-1996}.
\label{y-vs-Enu}
}
\end{figure}

The fraction of the neutrino energy carried by the final-state lepton is given
by $1-y$. For lower energies, where $\langle Q^2\rangle\ll M_W^2$, integration
of eq.~(\ref{eq-csdis}) yields $ \nicefrac14<\langle y\rangle<\nicefrac12$,
depending on the relative contribution of quarks and antiquarks. For larger
$E_\nu$, the dependence of the propagator term on $Q^2\approx2xyM_NE_\nu$
reduces the mean $y$ (see Fig.~\ref{y-vs-Enu}). The muon in $\nu_\mu$ CC
reactions on average thus carries the major fraction of the neutrino energy,
whereas the hadronic cascade has lower energy.

\subsection {Cherenkov light}
\label{sec-det-che}

Charged particles moving faster than the speed of light in a medium with index
of refraction $n$, i.e.\ $v>c/n$, emit Cherenkov light under the characteristic
angle
\begin{equation}
  \cos\theta_c=\frac1{\beta\cdot n},
  \label{eq-cherangle}
\end{equation}
where $\beta = v/c$ and $n$ depends on the frequency $\nu$ of the emitted
photons, $n = n(\nu)$. The spectral distribution of Cherenkov photons per path
length of an emitting particle with charge $\pm ze$ is given by
\begin{equation}
 \frac{\D N}{\D x\D\lambda}
 =\frac{2\pi\cdot z^2\alpha}{\lambda^2}
  \cdot\left(1-\frac1{\beta^2 \cdot n^2}\right)
 \label{eq-cherspec}
\end{equation}
with $\alpha$ being the fine structure constant. The total amount of released
energy per particle path length is obtained by multiplying
eq.~(\ref{eq-cherspec}) with $E_\gamma=h\nu$ and integrating over $\nu$:
\begin{equation}
 -\left(\frac{\D E}{\D x}\right)_c
 =\frac{2\pi z^2\alpha h}{c}
  \intl_{\beta\cdot n(\nu)\ge1}\nu\left(1-\frac1{\beta^2 \cdot n^2(\nu)}\right)\,\D\nu\;.
\end{equation}
For $\lambda=550\nm$ the index of refraction for water is $n\approx1.33$,
yielding about $400\ev/\cm$, or approximately 200 Cherenkov photons per cm in
the transparency window of water, i.e.\ for wavelengths
$400\nm\le\lambda\le700\nm$. The average Cherenkov angle of these photons is
$\theta_c\approx43^\circ$.

\subsection{Light propagation and detection} 
\label{sec-det-lig}

The propagation of light in water is governed by absorption and scattering,
which both depend on the wavelength $\lambda$. In the first case the photon is
lost, in the second case it changes its direction. Scattering effectively delays
the propagation of photons between the points of emission and detection. The
parameters generally chosen to describe these phenomena are:
\begin{enumerate}
\item 
The absorption length $L_a(\lambda)$ -- or the absorption coefficient 
$a(\lambda) = 1/L_a$ -- describes the exponential decrease of the number 
$N$ of non-absorbed photons as a function of photon path length $r$, 
$N=N_0\cdot\exp(-r/L_a)$.
\item 
The scattering length $L_b(\lambda)$ and scattering coefficient
$b(\lambda)=1/L_b$, defined in analogy to $L_a(\lambda)$ and $a(\lambda)$.
\item 
The scattering function $\chi(\theta, \lambda)$, i.e.\ the 
distribution in scattering angle $\theta$. 
\item 
Instead of the ``geometrical'' scattering length $L_b(\lambda)$ one frequently
uses the effective scattering length
$L_\text{eff}=L_b/(1-\langle\cos\theta\rangle)$, where $\langle\cos\theta
\rangle$ is the mean cosine of the scattering angle. $L_\text{eff}$
``normalises'' scattering lengths for different distributions
$\chi(\theta,\lambda)$ of the scattering angle to the extreme case
$\langle\cos\theta\rangle=0$, i.e.\ $L_\text{eff}$ can be interpreted as
isotropisation length. For $\langle\cos\theta\rangle\approx0.8\rnge0.95$, as for
all media considered here, photon delay effects in media with the same
$L_\text{eff}$ are approximately the same.
\end{enumerate}

In Table~\ref{tab-optprop}, typical parameter values for Lake Baikal, oceans and
the Antarctic ice are summarised (see \cite{Spiering-Handbook-2010} for
references). All values are given for the wavelength where they are maximal.

\begin{table}[ht]
\begin{center}
\begin{tabular}{|l|c|c|}
\hline
Site & $L_a\;[\Met]$ & $L_\text{eff}\;[\Met]$\\
\hline
Lake Baikal, 1 km depth & 22 & 150--400 \\
& & (seasonal variations) \\
\hline
Ocean, $>1.5\km$ depth 
& 40--70 (depends      & 200--400 (depends \\[-1.mm] 
& on site and season)  &  on site and season) \\
\hline
Polar ice, $1.5\rnge2.0\km$ depth 
& $\approx95$ (average)& $\approx20$ (average) \\
\hline
Polar ice, $2.2\rnge2.5\km$ depth 
& $\approx150$ (average)   &  $\approx40$ (average) \\
\hline
\end{tabular}
\caption{
Absorption length and effective scattering length for deep lake and ocean water
and for deep ice.
\label{tab-optprop}
}
\end{center}
\end{table}

Scattering and absorption in water and ice are measured with artificial light
sources. The scattering coefficient, to a large part due to particulate
matter, changes with wavelength less pronouncedly than the absorption
coefficient (see Fig.\,\ref{SP-Ice} for ice). In water, the depth dependence
over the vertical dimensions of a neutrino telescope is small, but parameters
may vary with time, due to transient water inflows loaded with bio-matter or
dust, or due to seasonal changes in water parameters. They must therefore be
permanently monitored. In glacial ice at the South Pole, the situation is
different. The parameters are constant in time but strongly depend on depth
(see below).

Strong absorption leads to reduced photon collection, strong scattering
deteriorates the time information which is essential for the precise
reconstruction of tracks and showers (see below).

Efficient and accurate detection and reconstruction of neutrino events requires
light detection with a sensitivity at the single-photon level and a measurement
of the arrival time with nanosecond precision. Furthermore, the integrated
photo-sensitive area, multiplied with the single-photon detection efficiency, is
the prime parameter determining the sensitivity of the neutrino telescope and is
to be maximised. Currently, PMTs \cite{UChicago-PMTs,Hamamatsu-PMTs} are the
only devices matching these constraints at an affordable price. Relevant
parameters are the photocathode size (i.e.\ the diameter of the tube); the
quantum efficiency (probability that an incoming photon causes emission of a
photo-electron) and its spectral dependence; the collection efficiency
(probability that a photo-electron causes a measurable signal); the transit time
spread (jitter of the delay between photon interaction and output signal); the
gain (electric amplification, i.e.\ average number of electrons in the signal of
one photo-electron).

\begin{figure}[ht]
\center{
\epsfig{file=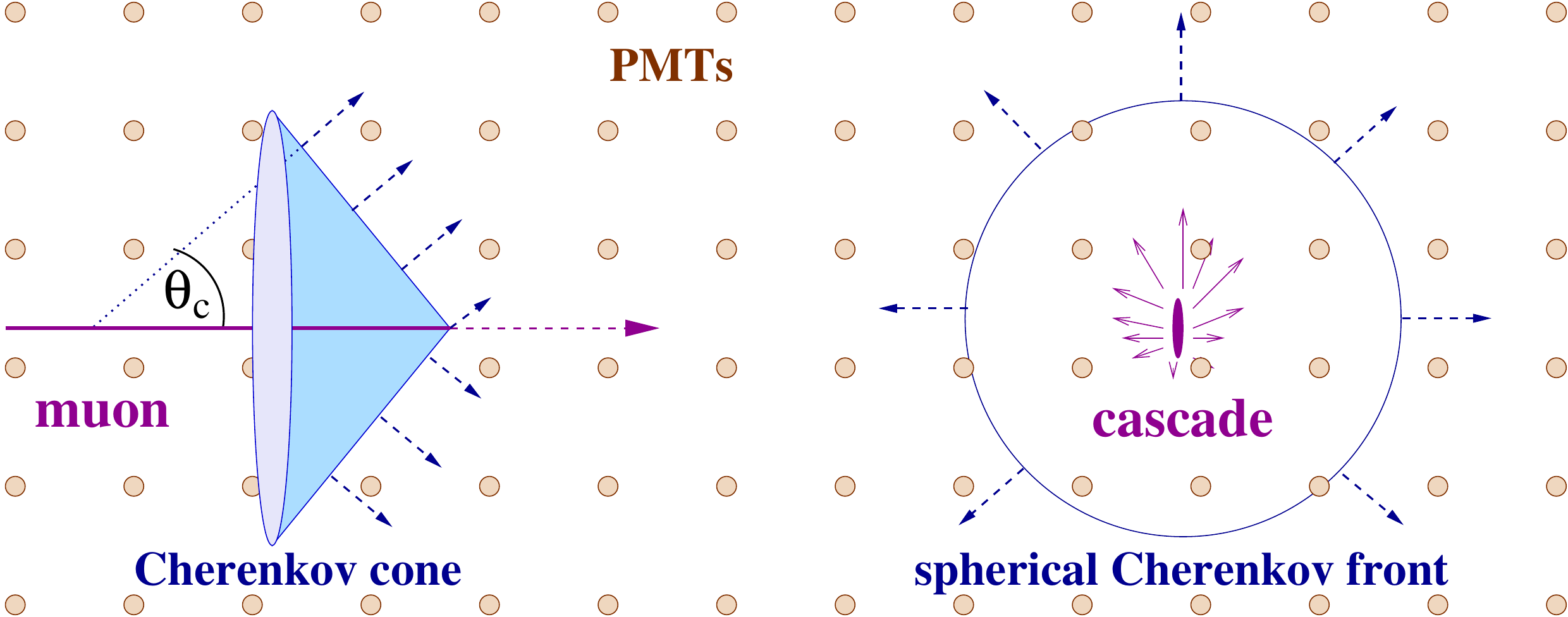,width=12cm}
\caption{
Detection principles for muon tracks (left) and cascades (right) in
underwater/ice detectors. Note that the Cherenkov light emission by cascades is
peaked at the Cherenkov angle $\theta_c$ with respect to the cascade axis but
has a wide distribution covering the full solid angle.}
\label{DMT}
}
\end{figure}

\begin{figure}[ht]
\epsfig{file=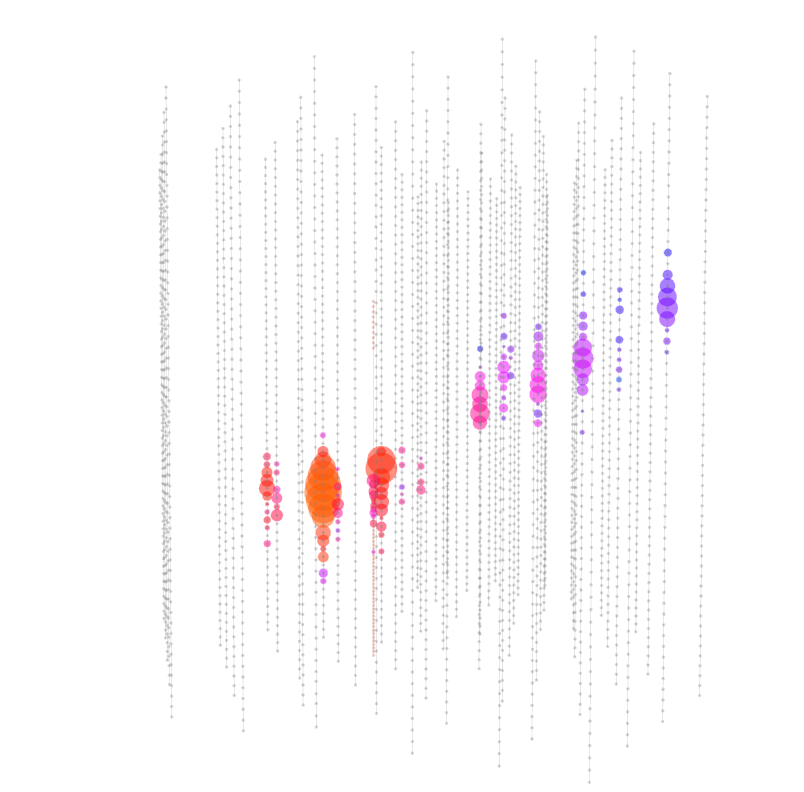,trim=50 0 20 0,clip=,width=0.48\textwidth}
\hfill
\epsfig{file=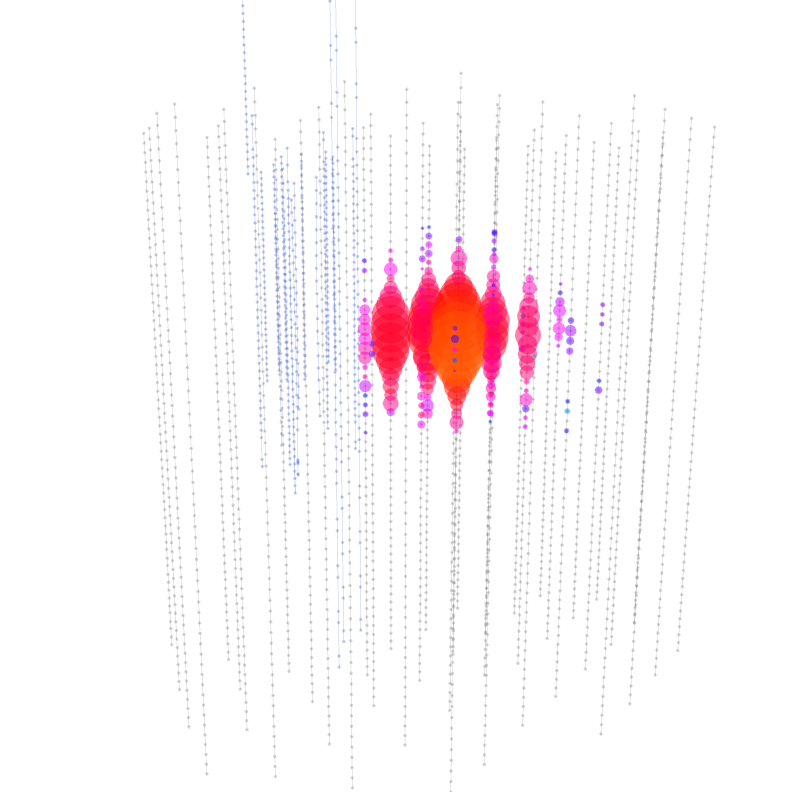,trim=50 0 20 0,clip=,width=0.49\textwidth}
\caption{
Event displays of a muon (left) and a cascade (right) event recorded in IceCube. 
The coloured circles indicate the photomultipliers hit; their sizes correspond
to the numbers of photons detected. The colour code visualises the hit times,
with red hits earliest and violet ones latest. The reconstructed zenith angle
and energy of the muon are $117^\circ$ and $20\tev$, respectively; the
reconstructed cascade energy is $175\tev$.}
\label{icecube-event}
\end{figure}

\subsection{Detection of muon tracks and cascades}
\label{sec-det-det}

Figure~\ref{DMT} sketches the two basic detection modes of underwater/ice
neutrino telescopes. In the following, $N$ denotes the target nucleon and $X$
the hadronic final state. CC muon neutrino interactions, $\nuan_\mu N\to\mu^\mp
X$ produce a muon track (left; an IceCube event display of a $10\tev$ muon is
shown in Fig.~\ref{icecube-event}), whereas other neutrino reaction types cause
hadronic and/or electromagnetic cascades (right). This is in particular true for
NC reactions ($\nuan N\to\nuan X$, hadronic cascade) or CC reactions of electron
neutrinos ($\nuan_eN\to e^\mp X$, overlapping hadronic and electromagnetic
cascades). CC tau neutrino interactions, $\nuan_\tau N\to\tau^\mp X$, can have
either signature, depending on the $\tau$ decay mode. In most astrophysical
models, neutrinos are produced with a flavour ratio
$\nu_e:\nu_\mu:\nu_\tau\approx1:2:0$. Over cosmic distances, oscillations turn
this ratio to $\nu_e:\nu_\mu:\nu_\tau\approx1:1:1$ (cf.\
Sect.~\ref{sec-sci-par-osc}), which means that almost $\nicefrac23$ of the
charged current interactions initiate cascades only.

\subsubsection{Muon tracks}
\label{sec-det-det-muo}

The most precise measurement of the neutrino direction is achieved for charged
current muon (anti)neutrino interactions; this channel is therefore central to
all investigations of astrophysical neutrino sources. The muon direction, and
also its energy, are reconstructed from the Cherenkov cone accompanying
upward-going muons. The upward signature guarantees the neutrino origin of the
muon since no other particle can cross the Earth. The effective target volume
considerably exceeds the actual detector volume due to the large range of muons
(about $0.96\times10^5\gramm/\Cm^2$ ($1.04\times10^5\gramm/\Cm^2$) at $300\gev$ and 
$14\times10^5\gramm/\Cm^2$ ($12\times10^5\gramm/\Cm^2$) at $100\tev$ in
water/ice (standard rock), logarithmically rising towards larger energies
\cite{Groom-2001}; $10^5\gramm/\Cm^2$ corresponds to $1\km$ in pure water).

The muon loses energy mainly via ionisation, pair production, bremsstrahlung and
photo-nuclear reactions. The energy loss can be parameterised by
\cite{Gaisser-1990,Groom-2001}
\begin{equation}
 -\frac{\D E_\mu}{\D x} = a+b\cdot E_\mu\;,
\end{equation}
where $a$ and $b$ are weakly energy dependent. For water, the ionisation loss is
roughly given by $a=2\mev/\Cm$. The energy loss from pair production,
bremsstrahlung and photo-nuclear reactions becomes dominant at energies beyond
$1\tev$, where it is approximately linear in $E_\mu$; for water and
$E_\mu=100\tev$ it is given by $b_\text{tot}=b_\text{pair}+b_\text{brems}+
b_\text{phonuc}=(1.7 + 1.2 + 0.6)\times10^{-6}\gramm\cm^{-2} =
3.5\times10^{-6}\gramm\cm^{-2}$ \cite{Groom-2001}. Note that all energy loss
mechanisms producing relativistic charged particles -- in particular $e^\pm$
from pair production and conversion of bremsstrahlung photons -- also lead to
additional Cherenkov light.

For the reconstruction of a muon track, its Cherenkov photons need to be
measured at least at five positions. In reality, because of timing inaccuracies
due to scattering (mostly in ice) and to light background from radioactive
decays (deep sea) and bioluminescence (lakes and deep sea), a reliable
reconstruction requires typically 10 or more PMT signals.

Underwater/ice telescopes are optimised for the detection of muon tracks at
energies of the order $\Tev$ or above, for the following reasons:
\begin{enumerate}
\item
The flux of neutrinos from cosmic accelerators is expected to be harder
(generic shape $E_\nu^{-2}$) than that of atmospheric neutrinos
($E_\nu^{-3.7}$), yielding a better signal-to-background ratio at higher
energies.
\item
Neutrino cross section and muon range increase with energy. The larger the muon
range, the larger the effective detector volume (even though this argument must
be used with care since a muon that has lost most of its energy when reaching
the detector may be discarded due to energy-dependent selection cuts).
\item
The mean angle between muon and neutrino decreases with energy (see
eq.~(\ref{eq-mnangle})), resulting in better source tracing and
signal-to-background ratio at high energy.
\item
For energies above a $\Tev$, the light emission increases linearly with $E_\mu$,
providing an estimate for the muon energy with an accuracy of $\sigma(\log
E_\mu)\approx0.3$. By unfolding procedures, a muon energy spectrum can be
translated into a neutrino energy spectrum.
\end{enumerate}

Muons which have been generated by cosmic ray interactions in the atmosphere
above the detector and punch through the water or ice down to the detector
outnumber neutrino-induced upward-moving muons by several orders of magnitude
(about $10^6$ at $1\km$ depth and $10^4$ at $4\km$ depth) and have to be removed
by careful up/down discrimination.

At energies above a few hundred $\Tev$, where the Earth becomes opaque even to
neutrinos, neutrino-generated muons arrive preferentially from directions close
to the horizon, at $\Eev$ energies essentially only from the upper hemisphere
(Fig.\,\ref{Transmission}). The high energy deposition of muons from
$\Pev$--$\Eev$ extraterrestrial neutrinos provides a handle to distinguish them
from downward-going atmospheric muons (that have a spectrum decreasing much more
steeply with energy).

\subsubsection{Cascades}
\label{sec-det-det-cas}

Charged current interactions of electron and, in most cases, tau neutrinos and
all neutral current interactions do not lead to high energy muons but to
electromagnetic or hadronic cascades. Their length increases only with the
logarithm of the cascade energy and is of the order $5\rnge20\met$. Cascade
events therefore typically need to be contained in the instrumented volume of
the detector to be reconstructible (``contained events'').

With their small length and a diameter of the order of $10\rnge20\cm$, cascades
may be considered as quasi point-like compared to the spacing of the PMTs. The
effective volume for cascade detection is close to the instrumented volume of
the detector. For first-generation neutrino telescopes it is therefore much
smaller than that for muon detection. However, for kilometre-scale detectors and
moderate energies both can be of the same order of magnitude. The total amount
of light is proportional to the cascade energy and thus provides a direct
measurement of $E$ for contained cascades, in contrast to muons, for which only
$\D E/\D x$ can be measured. Therefore, in charged current $\nu_e$ interactions,
the neutrino energy can be determined with an accuracy of about $30\%$ (note,
however, that this is not true for NC events; also, in $\nu_\tau$ CC events the
$\nu_\tau$ from the decay of the final-state $\tau$ carries away a sizeable
fraction of ``invisible'' energy). Whereas the energy resolution is much better
than for muons, the directional accuracy is worse since the lever arm for
fitting the direction is small (for actual values, see below). The irreducible
background from atmospheric electron neutrinos is significantly smaller than
that of atmospheric muon neutrinos for the muon channel. All this makes the
cascade channel particularly interesting for searches for a diffuse, high-energy
excess of extraterrestrial over atmospheric neutrinos.

\subsection{Effective area and sensitivity}
\label{sec-det-eff}

The minimum energy of a muon to be recorded depends on the impact point and the
impact angle, resulting in a smearing for $E_\mu^\text{min}$. Even muons outside
the geometrical area may fulfil the trigger condition. On the other hand,
background suppression usually requires tight cuts during the analysis, so that
muons having triggered the detector might be rejected at a later stage. 
Therefore, the geometrical area $A$ is replaced by the effective muon area
$A^\mu_\text{eff}$ which is calculated from Monte Carlo simulations as follows:
$N_\text{gen}$ muons are generated perpendicular to an area $A_\text{gen}$
significantly larger than the detector cross section. With $N_\text{passed}$
muons accepted after all cuts, the effective area is then given by
\begin{equation}
 A^\mu_\text{eff}=A_\text{gen} \cdot \frac{N_\text{passed}}{N_\text{generated}}\;.
\end{equation}
The effective area depends on $\theta$ and $\phi$, i.e.\ the muon direction with
respect to the detector. Due to the increasing range and light emission,
$A^\mu_\text{eff}$ increases with muon energy. Note that $A^\mu_\text{eff}$ also
depends on the distance between the point where the muons are generated and the
detector.

\begin{figure}[ht]
\sidecaption
\epsfig{file=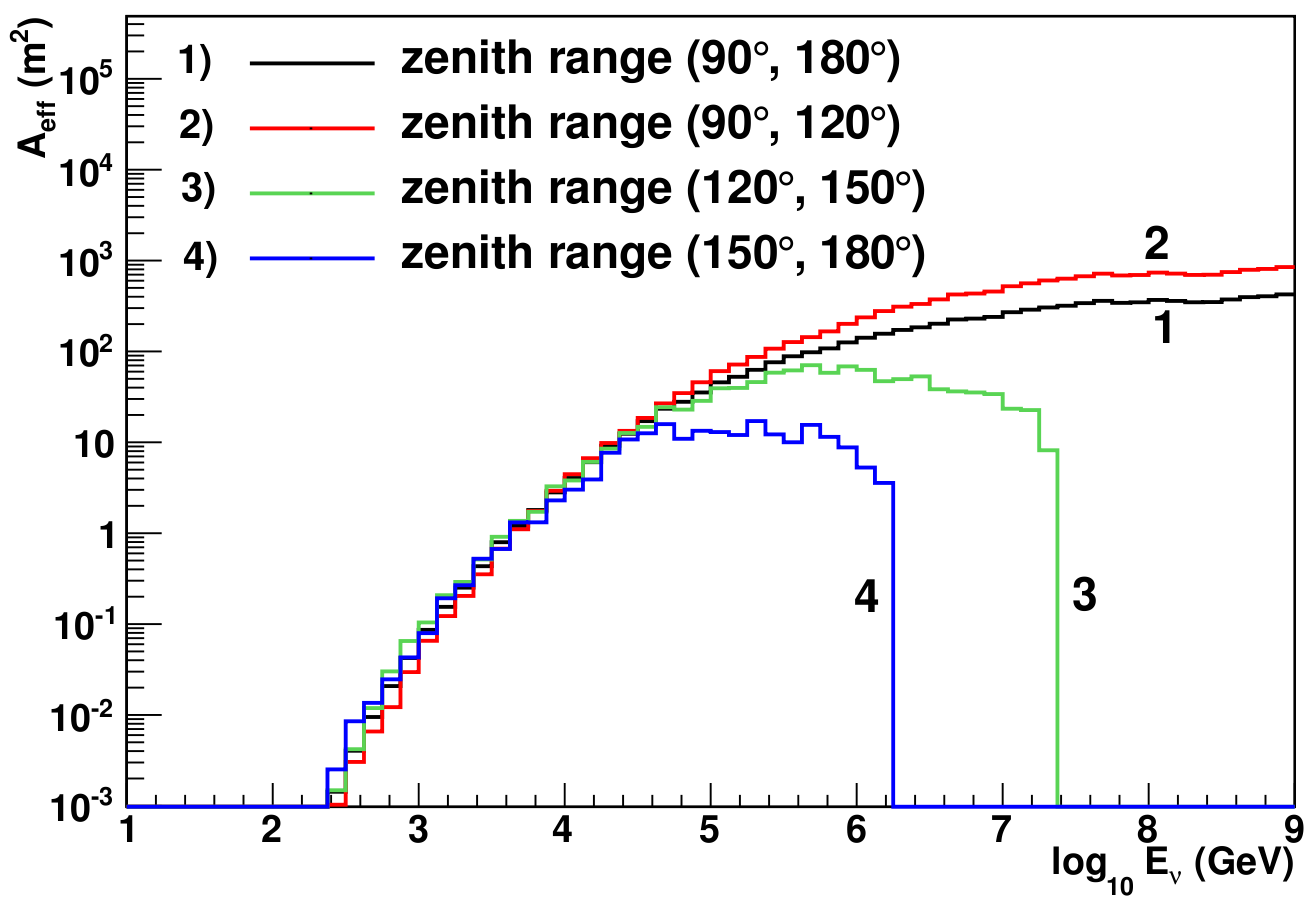,width=10.cm}
\caption{
Effective neutrino area of the 40-string configuration of IceCube as a function
of energy, averaged over different zenith angle ranges. At energies beyond
$10^4\gev$, neutrino absorption n the Earth becomes important. Figure taken from
\pcite{icecube-2011f}.}
\label{Effar}
\end{figure}

The concept of an effective area can also be applied to neutrinos, taking into
account the neutrino nucleon cross section, the muon range and the absorption in
the Earth according to Sect.~\ref{sec-det-int}. Figure~\ref{Effar} shows the
effective neutrino area of IceCube as a function of energy and zenith angle. The
strong increase with energy is due to the increasing neutrino cross section and
to the increasing muon range. The decrease at high energies for large zenith
angles is due to neutrino absorption in the Earth.

Due to the small neutrino interaction cross section, the effective neutrino area
is many orders of magnitude smaller than the muon effective area. Even
cubic-kilometre neutrino telescopes reach only neutrino areas between a few
square metres and a few hundred square metres, depending on energy. This has to
be compared to several ten thousand square metres typical for air Cherenkov
telescopes which detect gamma-ray initiated air showers. A ratio 1:1000
($10\met^2:10000\met^2$) may appear desperately small. However, one has to take
into account that Cherenkov gamma telescopes can only observe one source at a
time, and that their observations are restricted to Moon-less, clear nights. 
Neutrino telescopes observe a full hemisphere, 24 hours per day. Therefore,
cubic-kilometre detectors reach a flux sensitivity similar to that reached by
first-generation Cherenkov gamma telescopes like Whipple and HEGRA
\cite{Whipple-1989,Hegra-1996} for $\Tev$ gamma rays, i.e.\
$\Phi(\gre1\tev)\approx10^{-12}\cm^{-2}\scnd^{-1}$.

\subsection{Reconstruction} 
\label{sec-det-rec}

In this section, some relevant aspects of event reconstruction are discussed for
the case of muon tracks \cite{amanda-2004b,Antares-fastreco,Baikal-mureco-1999}. 
For cascades, see \cite{Amanda-2004c,Antares-casc-2006,Baikal-showerreco-2009}. 
The reconstruction procedure for a muon track consists of several consecutive
steps which are typically:
\begin{enumerate}
\item  
Rejection of noise hits;
\item  
Simple pre-fit procedures providing a first-guess estimate for the following
iterative maximum-likelihood reconstruction;
\item  
Maximum-likelihood reconstruction;
\item  
Quality cuts in order to reduce background contaminations and to enrich the
sample with signal events. This step is strongly dependent on details of the
actual analysis -- diffuse fluxes at high energies, searches for steady point
sources, searches for transient sources etc.
\end{enumerate}

\begin{figure}[ht]
\sidecaption
\epsfig{file=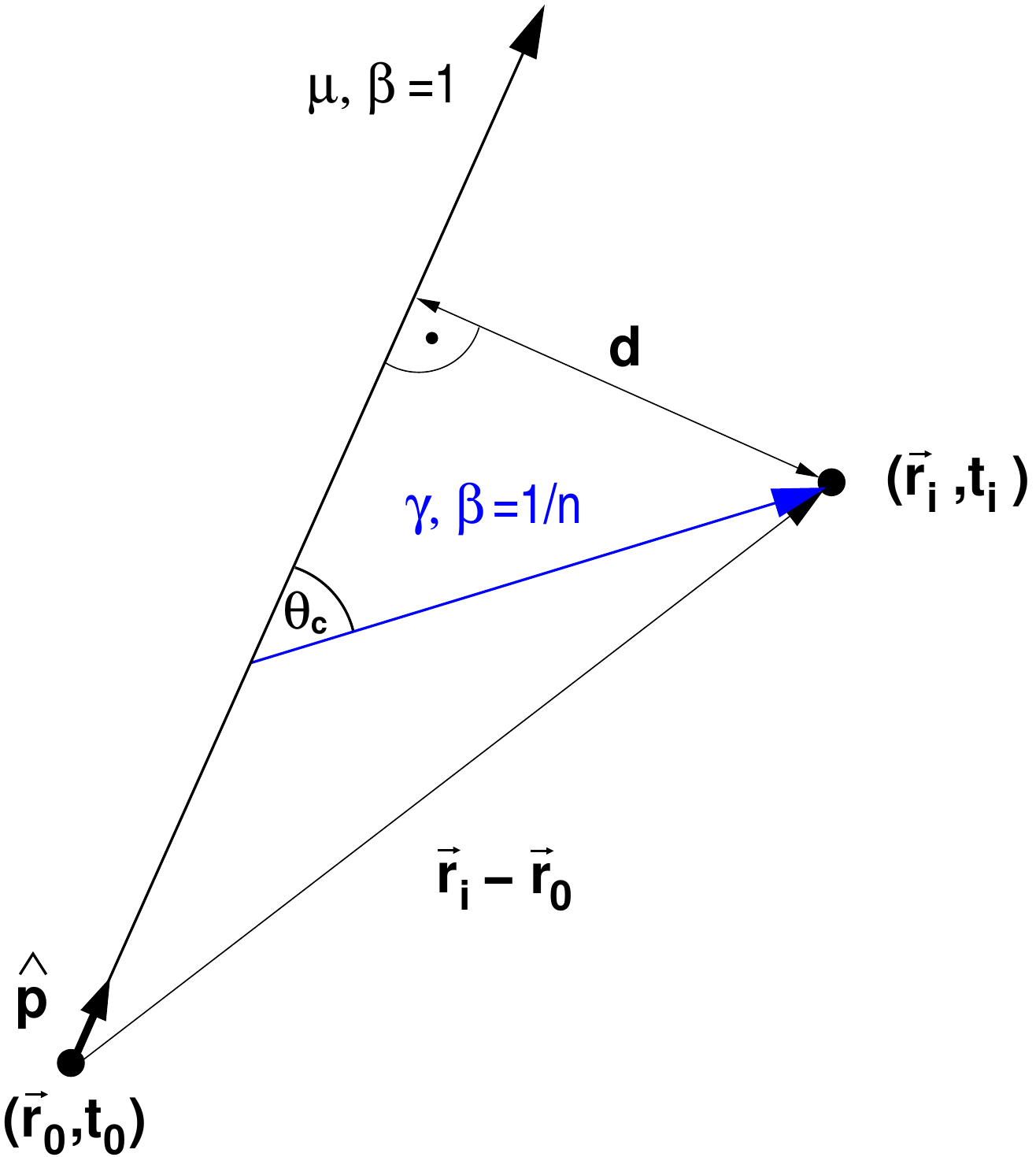,width=7.cm}
\caption
{Geometry of muon reconstruction: The muon, propagating with $\beta=1$ in very
good approximation, radiates a Cherenkov photon under the angle $\theta_c$ (with
$\cos\theta_c=1/n$) that propagates with $\beta=1/n$ on a straight line and hits
the photon sensor at position $\vec r_i$. Under these assumptions the hit time
can be calculated to be $t_i=t_\text{geo}$ (see eq.~(\ref{eq-hittime})). The
time difference $t_i-t_0$ thus relates the muon track parameters (position $\vec
r_0$, direction $\hat p$) to the position of the photon hit.}
\label{mureco}
\end{figure}

An infinitely long muon track can be described by an arbitrary point $\vec r_0$
on the track which is passed by the muon at time $t_0$, with a direction $\hat
p$ and energy $E_0$ (see Fig.~\ref{mureco}). Since the muon in very good
approximation propagates with the vacuum speed of light, $c$, photons emitted
under the Cherenkov angle $\theta_c$ (see eq.~(\ref{eq-cherangle})) and
propagating on a straight path (``direct photons'') are expected to arrive at
PMT\;$i$ located at $\vec r_i$ at a time
\begin{equation}
 t_\text{geo} = t_0 + \frac{\hat p\cdot(\vec r_i-\vec r_0)+ d\cdot\tan\theta_c}{c}\;,
 \label{eq-hittime}
\end{equation}
where $d$ is the closest distance between PMT\;$i$ and the track. The time
residual $t_\text{res}$ is given by the difference between the measured hit time
$t_\text{hit}$ and the hit time expected for a direct photon, $t_\text{geo}$:
\begin{equation}
 t_\text{res} = t_\text{hit} - t_\text{geo}\;.
\end{equation}
 
\begin{figure}[ht]
\sidecaption
\epsfig{file=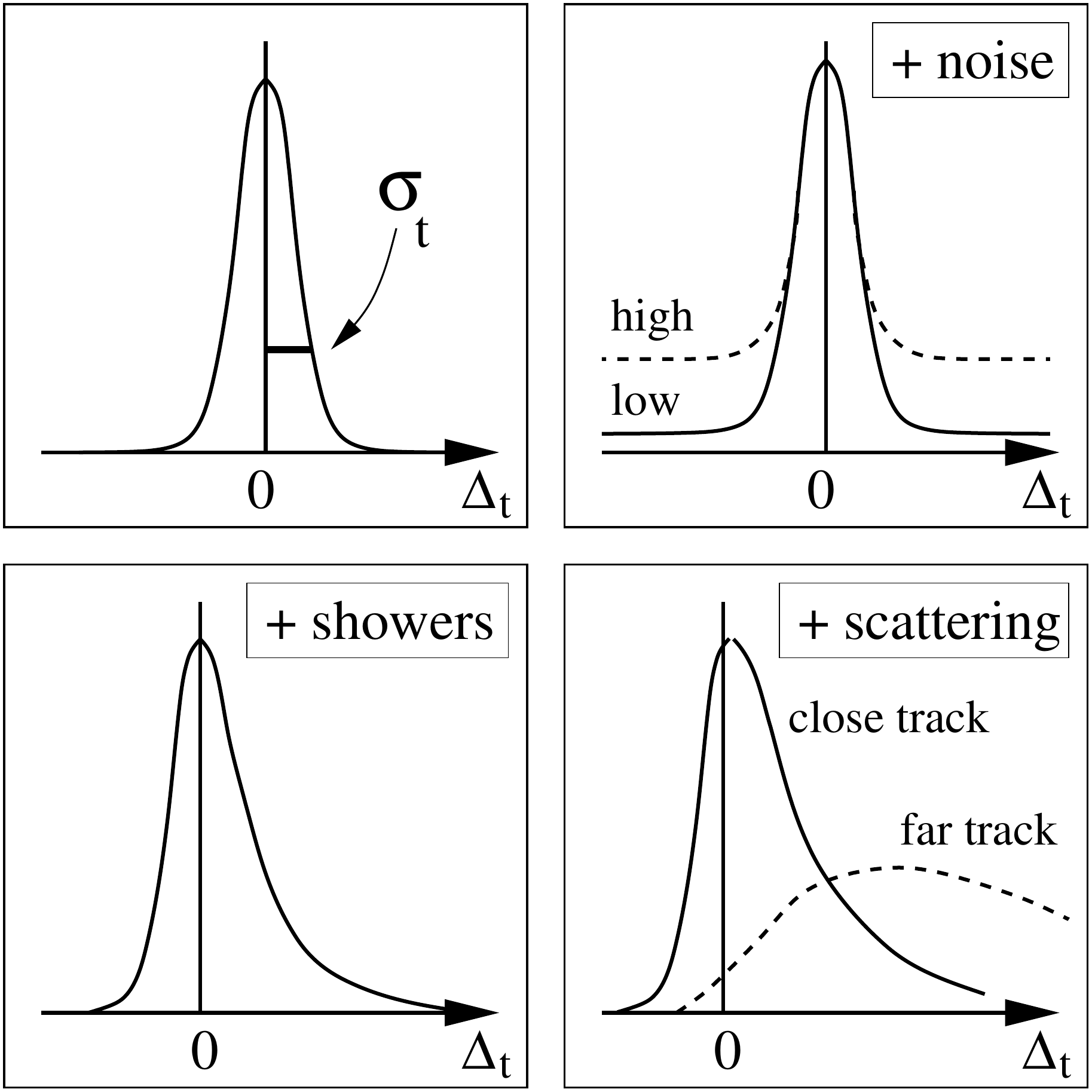,width=7.cm}
\caption
{Schematic distributions of arrival times for different cases (see text).}
\label{Schedis}
\end{figure}

Schematic distributions for time residuals are given in Fig.~\ref{Schedis}. An
unavoidable symmetric contribution around $\Delta_t=0$ in the range of a
nanosecond comes from the PMT/electronics time jitter, $\sigma_t$. An admixture
of noise hits to the true hits from a muon track adds a flat pedestal
contribution like shown in top right of Fig.~\ref{Schedis}. Electromagnetic
cascades along the track, initiated by bremsstrahlung and pair production, lead
to a tail towards larger (and only larger) time residuals (bottom left). 
Scattering of photons can lead to an even stronger delay of the arrival time
(bottom right). These residuals must be properly implemented in the probability
density function for the arrival times used in the maximum-likelihood procedure.

The simplest likelihood function is based exclusively on the measured arrival 
times. It is the product of all $N_\text{hit}$ probability density functions 
$p_i$ to observe, for a given value of track parameters $\{a\}$, photons 
at times $t_i$ at the location of the PMTs hit: 
\begin{equation}
 L_\text{time} = \prod\limits^{N_\text{hit}}_{i=1}  
                 p(t_{\text{res},i}\left|\{a\})\right.\;.     
\end{equation}

More complicated likelihood functions include the probability of hit PMTs to be
hit and of non-hit PMTs not to be hit, or of the respective amplitudes. Instead
of referring only to the arrival time of the first photon for a given track
hypothesis and the amplitude for a given energy hypothesis, one may also refer
to the full waveform from multiple photons hitting the PMT. For efficient
background suppression, the likelihood may also incorporate information about
the zenith angular dependence of background and signal (Bayesian probability). 
The reconstruction procedure finds the best track hypothesis by maximising the
likelihood.

\clearpage
\section{First-Generation Neutrino Telescopes}
\label{sec-fir}

An explorative phase of more than two decades was required to solve the
technical problems of instrumenting target masses of the order of 10\,Megatons
in deep ice, fresh or sea water and to overcome the difficulties which these
hostile environments pose. Nevertheless, neutrino telescopes of this size have
been constructed and successfully operated in all three media. In this Section,
the technical setup of these installations is discussed, also with reference to
the history of major steps leading to their construction.

\subsection{DUMAND}
\label{sec-fir-dum}

The history of underwater neutrino telescopes starts with a project which
eventually failed but left an incredibly rich legacy of ideas and technical
principles: The DUMAND project. DUMAND stands for Deep Underwater Muon and
Neutrino Detector. Its early history is excellently covered in a "Personal
history of the DUMAND project" by A.\,Roberts \cite{DUMAND-Roberts}. At the 1973
International Cosmic Ray Conference (ICRC), a small group of physicists
conceived a deep-water detector to clarify puzzles in muon depth-intensity
curves. The puzzles faded away, but it was obvious that such a detector could
also work for neutrinos. The year 1975 saw the first of a -- meanwhile legendary
-- series of DUMAND Workshops. Soon, the decision was taken to deploy the
detector 30\,km off the coast of Big Island, Hawaii, at a depth of 4.8\,km.

\begin{figure}[ht]
\sidecaption
\epsfig{file=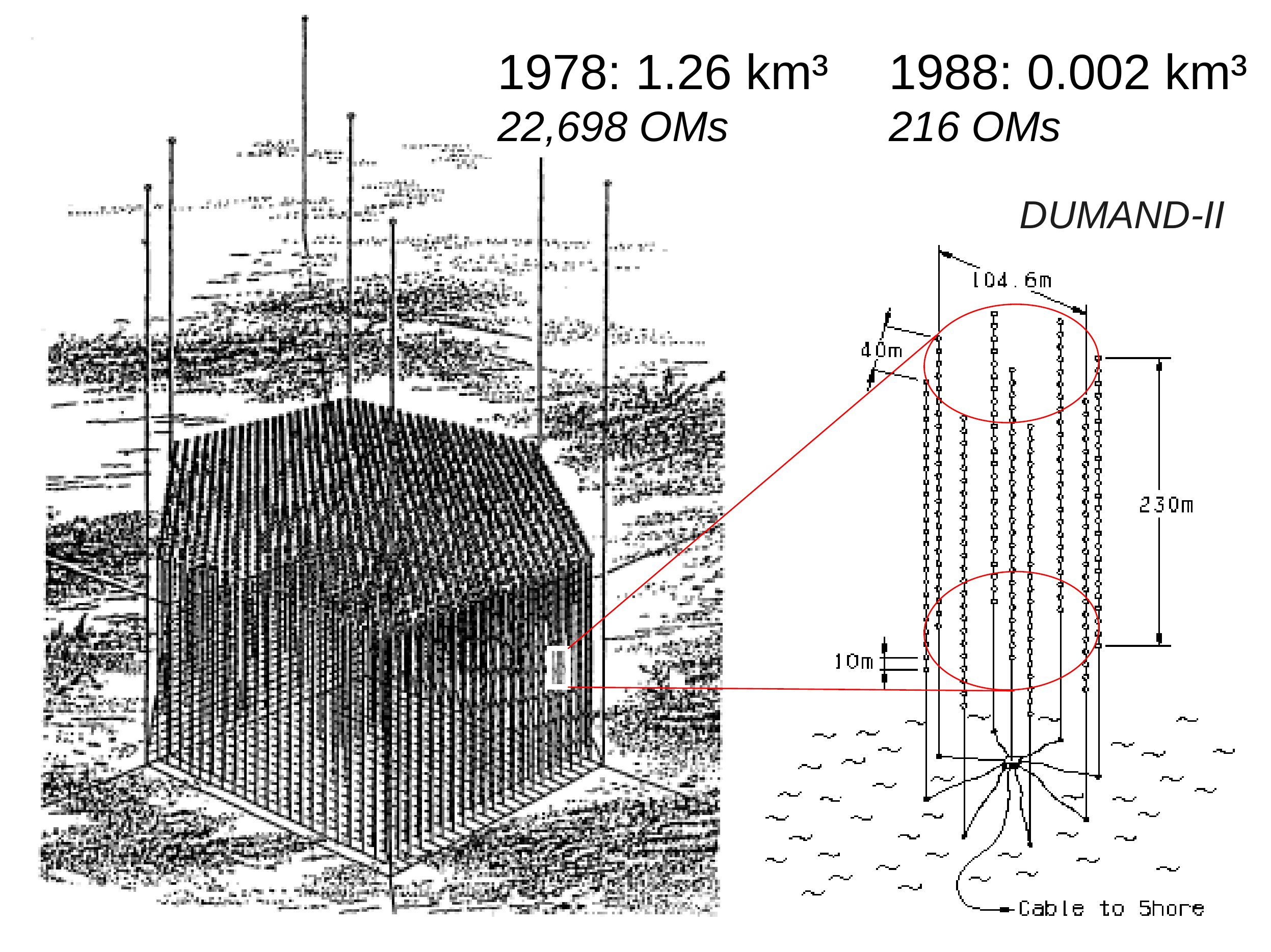,width=12.cm}
\caption{
The originally conceived DUMAND cubic-kilometre detector and the 1988 plan for
the first-generation underwater neutrino telescope DUMAND-II.
\label{DUMAND}
}
\end{figure}

The original idea to construct a cubic-kilometre detector (1978) with more than
20\,000 large-size photomultipliers (see Fig.~\ref{DUMAND}) was quickly
abandoned due to technical and financial reasons. A half-sized configuration
(1980) met the same fate, as did a much smaller array with 756 phototubes
(1982). The latter design was comparable in size to the meanwhile decommissioned
AMANDA detector at the South Pole (see Sect.~\ref{sec-fir-ama}) and the
operating ANTARES telescope in the Mediterranean Sea, close to Toulon (see
Sect.~\ref{sec-fir-ant}). What finally emerged as a technical project was a
216-phototube version, dubbed DUMAND-II or ``The Octagon'' (eight strings at the
corners of an octagon and one in the centre), 100\,m in diameter and 230\,m in
height \cite{DUMAND-Project}.

Signals on at least three strings are required for full spatial reconstruction
of a muon trajectory. Consequently, the sequential deployment was planned to
start with TRIAD, a sub-array of three full strings. Unfortunately, pressure
housings of the first string developed leaks during deployment in 1993 and soon
the communication to shore failed. The ``3-string race'' between DUMAND, the
Baikal project and AMANDA was eventually won by the Baikal collaboration in
1993, whereas the DUMAND project was terminated in 1995.

Ironically, the most sustaining physics result from the DUMAND project has been
obtained with a 7-phototube test string deployed for only some hours from a ship
\cite{DUMAND-Babson}. It gives the measured muon intensity as a function of
depth and thus returned to the initial idea of the 1973 ICRC.
 
\subsection{The Baikal neutrino telescope}
\label{sec-fir-bai}

The Baikal Neutrino Telescope is installed in the Southern part of Lake Baikal
\cite{Baikal-Web,Baikal-principle-Belolaptikov-1997}. The distance to shore is
3.6\,km, the depth of the lake at this location is 1366\,m, the active part of
the detector is located at a depth of about 1.1 km.

First site surveys started in 1980. In 1984, a first stationary string was
deployed and muons recorded \cite{Baikal-1984}. This was followed by another
stationary string in 1986 which was optimised for the detection of magnetic
monopoles catalysing proton decay \cite{Baikal-1986,Baikal-1990}.

\begin{figure}[ht]
\sidecaption
\epsfig{file=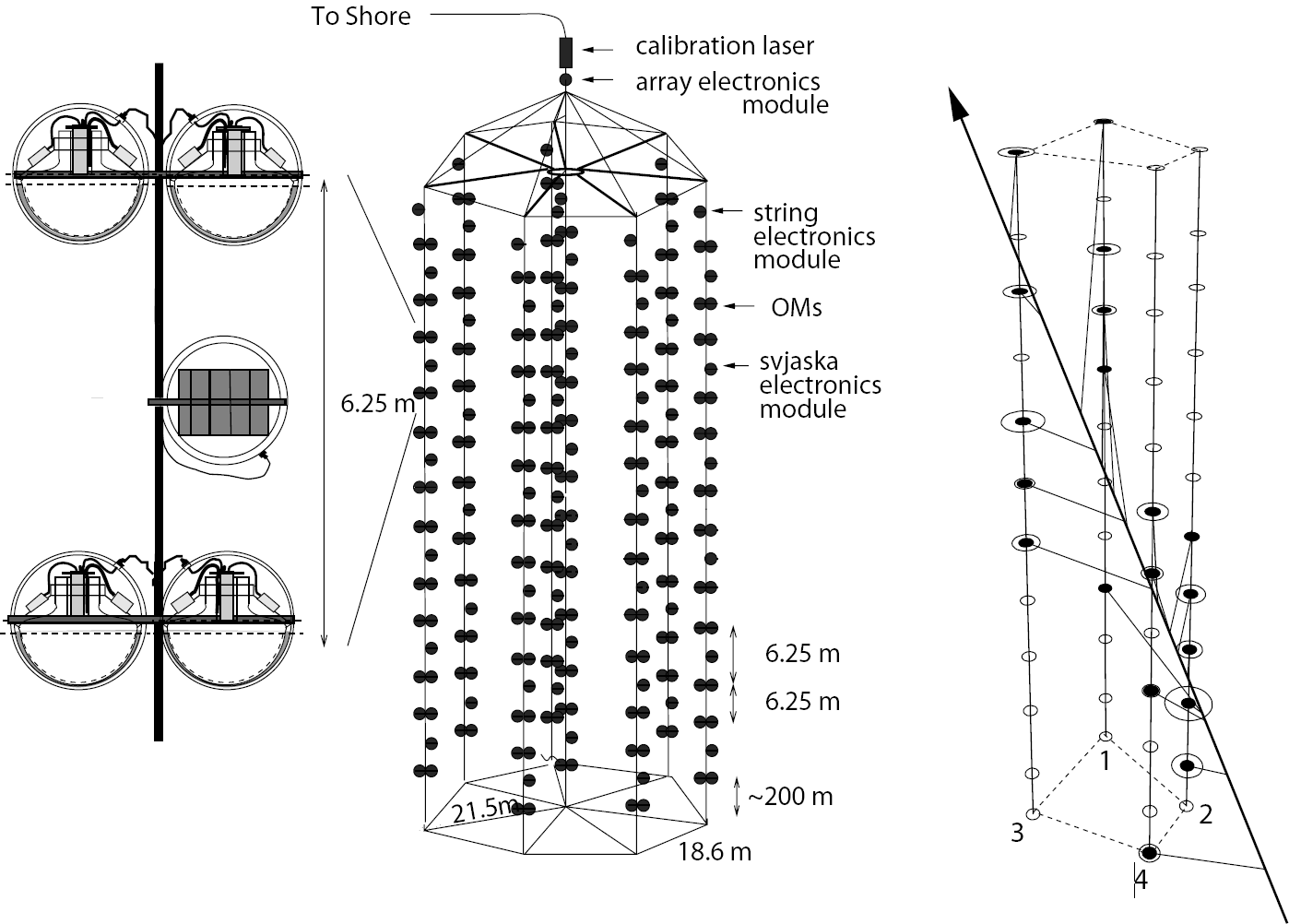,width=11.3cm}
\caption{
Left: The Baikal Neutrino Telescope NT200. Right: One of the first upward moving
muons from a neutrino interaction recorded with the 4-string stage of the
detector in 1996 \pcite{Baikal-atm-Balkanov-1999}. The Cherenkov light from the
muon is recorded by 19 channels.}
\label{Baikal}
\end{figure}

A new period began with the development of the QUASAR photodetector (see below),
which replaced the former 15\,cm flat photomultipliers, and the design of the
array NT200 (see Fig.~\ref{Baikal}, left). The BAIKAL collaboration was not only
the first, in 1993, to deploy three strings (as necessary for full spatial
reconstruction of muon trajectories), but also reported the first atmospheric
neutrino detected underwater (Fig.~\ref{Baikal}, right).

The central part of the Baikal configuration is NT200, an array of 192 optical
modules which was completed in April 1998 and has been taking data since then. 
The optical modules are carried by eight strings which are attached to an
umbrella-like frame. The strings are anchored by weights at the lake floor and
held in a vertical position by buoys at various depths. The configuration spans
72\,m in height and 43\,m in diameter. The detector is deployed (or hauled up
for maintenance) within a period of about 6~weeks in February to April, when the
lake is covered with a thick ice layer providing a stable working platform. It
is connected to shore by several copper cables on the lake floor which allow for
operation over the full year.

The optical modules are glass spheres equipped with QUASAR-370 phototubes; they
are grouped pair-wise along a string. In order to suppress accidental hits from
dark noise (about $30\kHz$) and bioluminescence (typically 50\,kHz but
seasonally raising up to hundreds of kHz), the two photomultipliers of a pair
are switched in coincidence, defining a {\it channel}, with typically only
$0.1\kHz$ noise rate. The basic cell of NT200 consists of a {\it svjaska}
(Russian for ``bundle''), comprising two optical module pairs and an electronics
module for time and amplitude conversion and slow control functions
(Fig.\,\ref{Baikal},\,left). A majority trigger is formed if $\ge4$ channels are
fired within a time window of 500\,ns (this is about twice the time a
relativistic particle needs to cross the NT200 array). Trigger and inter-string
synchronisation electronics are housed in an array electronics module at the top
of the umbrella frame. This module is less than 100\,m away from the optical
modules, allowing for nanosecond synchronisation over copper cable.

\begin{figure}[ht]
\epsfig{file=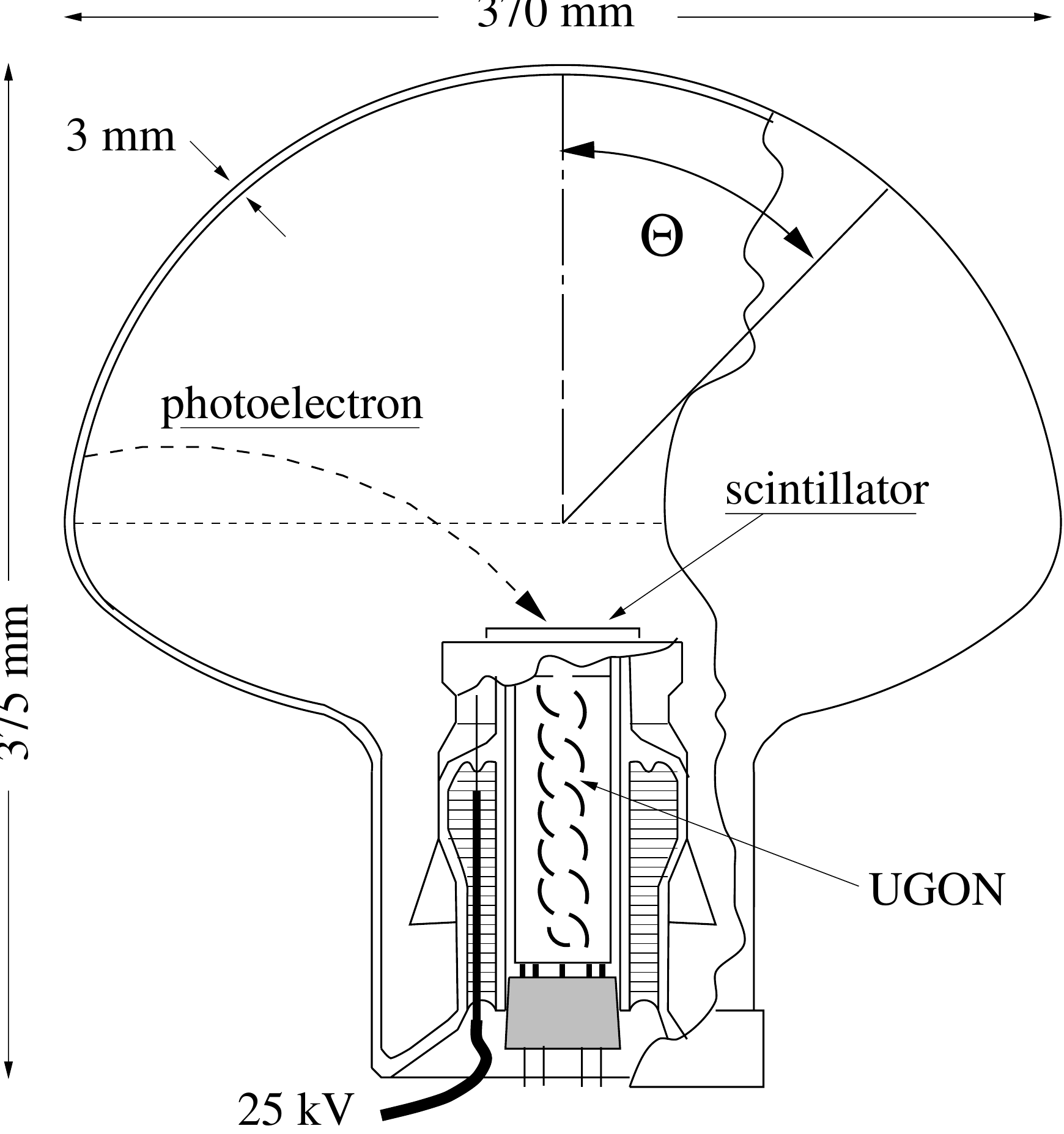,width=6cm}
\hfill
\epsfig{file=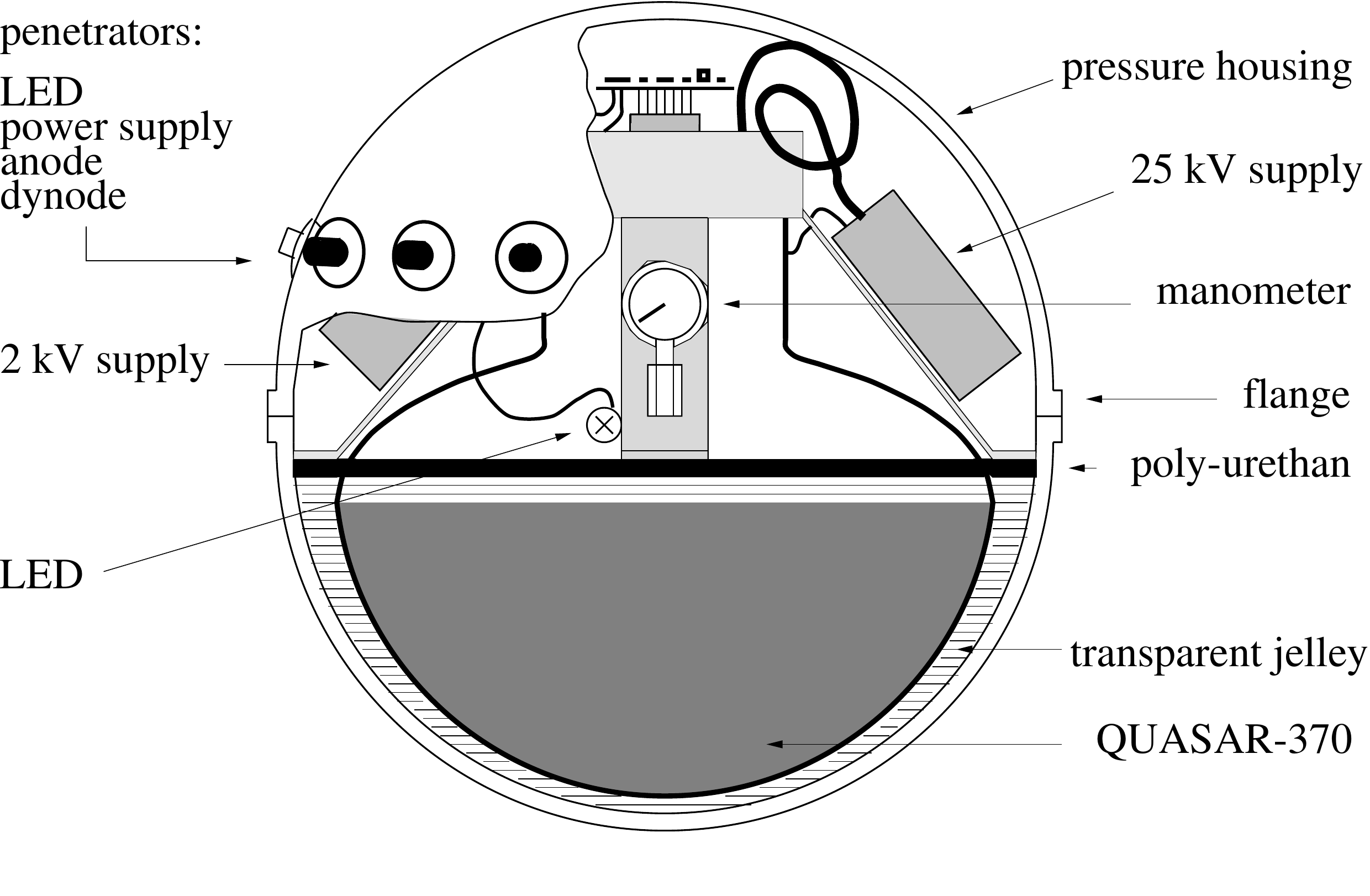,width=9cm}
\caption{
Left: The QUASAR-370 phototube. Right: A full Baikal optical module.
\label{Quasar}
}
\end{figure}

Figure \ref{Quasar} shows the phototube and the full optical module 
\cite{Baikal-OM-Bagduev-1999}. The QUASAR-370 is a hybrid device.
Photoelectrons from a large hemispherical cathode (K$_2$CsSb) with $>2\pi$
viewing angle are accelerated by 25\,kV to a fast, high-gain scintillator which
is placed near the centre of the glass bulb. The light from the scintillator is
read out by a small conventional photomultiplier (type UGON). One photo-electron
from the hemispherical photocathode yields typically 20 photoelectrons in the
conventional photomultiplier. This high multiplication factor results in an
excellent single electron resolution of 70\%. Furthermore, the QUASAR-370 is
characterised by a small time jitter (2\,ns) and a small sensitivity to the
Earth's magnetic field. 

The small spacing of modules leads to a comparably low energy threshold of about
$15\gev$ for muon detection. About 400 upward muon events were collected over
5~years. Still, NT200 could compete with the much larger AMANDA for a while by
searching for high energy cascades {\it below} NT200, surveying a volume about
ten times as large as NT200 itself \cite{Baikal-diff-Aynutdinov-2006}. In order
to improve pattern recognition for these studies, NT200 was fenced in 2005--2007
with three sparsely instrumented outer strings (6 optical module pairs per
string). This configuration is named NT200+ \cite{Baikal-NT200+}.

\subsection{AMANDA}
\label{sec-fir-ama}

AMANDA (Antarctic Muon And Neutrino Detection Array) used the 3\,km thick ice
layer at the South Pole as target and detection medium
\cite{amanda-2000,amanda-2001}. AMANDA was located some hundred metres away from
the Amundsen-Scott station. Holes of 60\,cm diameter are drilled with
pressurised hot water; strings with optical modules are deployed in the molten
water which subsequently refreezes. Installation operations at the South Pole
are performed in the Antarctic summer, November to February. For the rest of the
time, two operators (of a winter-over crew of 25--40 persons in total) maintain
the detector, connected to the outside world via satellite communication.

\begin{figure}[ht]
\sidecaption
\epsfig{file=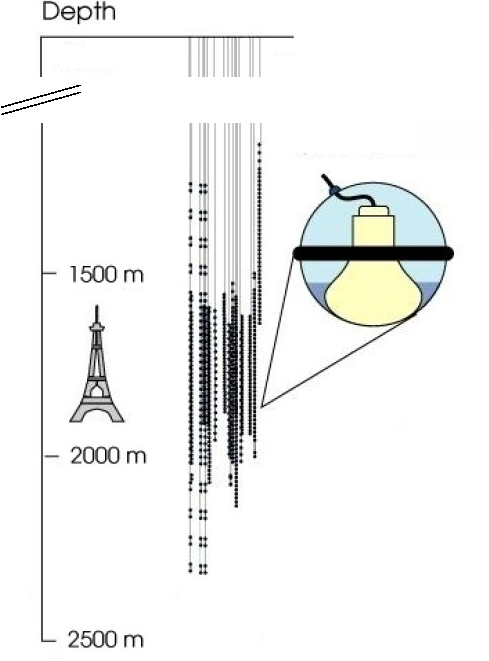,width=7.cm}
\caption{
The AMANDA configuration. Three of the 19 strings have been sparsely
equipped towards larger and smaller depth in order to explore ice properties,
one string got stuck during deployment at too shallow depth and was not used in
analyses. The Eiffel tower is shown to scale for size comparison.
\label{AMANDA}
}
\end{figure}

Figure~\ref{AMANDA} shows the AMANDA configuration. A first test array with 80
optical modules at four strings (not shown) was deployed in 1993/94, at depths
between 800 and 1000\,m \cite{Amanda-shallow-1995}. The effective scattering
length $L_\text{eff}$ was found to be extremely small, between 40\,cm at 830\,m
depth and 80\,cm at 970\,m. The scattering is due to air bubbles trapped in the
ice and makes track reconstruction impossible. The tendency of the scattering to
decrease with depth, as well as results from ice core analyses at other places
in Antarctica, suggested that the bubbles should disappear below 1300\,m. This
expectation was confirmed with a second 4-string array which was deployed in
1995/96. The remaining scattering, averaged over 1500--2000\,m depth,
corresponds to $L_\text{eff}\approx20\met$ and is assumed to be due to dust. 
This is still considerably worse than for water but sufficient for track
reconstruction \cite{Amanda-ice-2006,amanda-2004b}. The array was upgraded
stepwise until January 2000 and eventually comprised 19 strings with a total of
677 optical modules, most of them at depths between 1500 and 2000\,m.

\begin{figure}[ht]
\epsfig{file=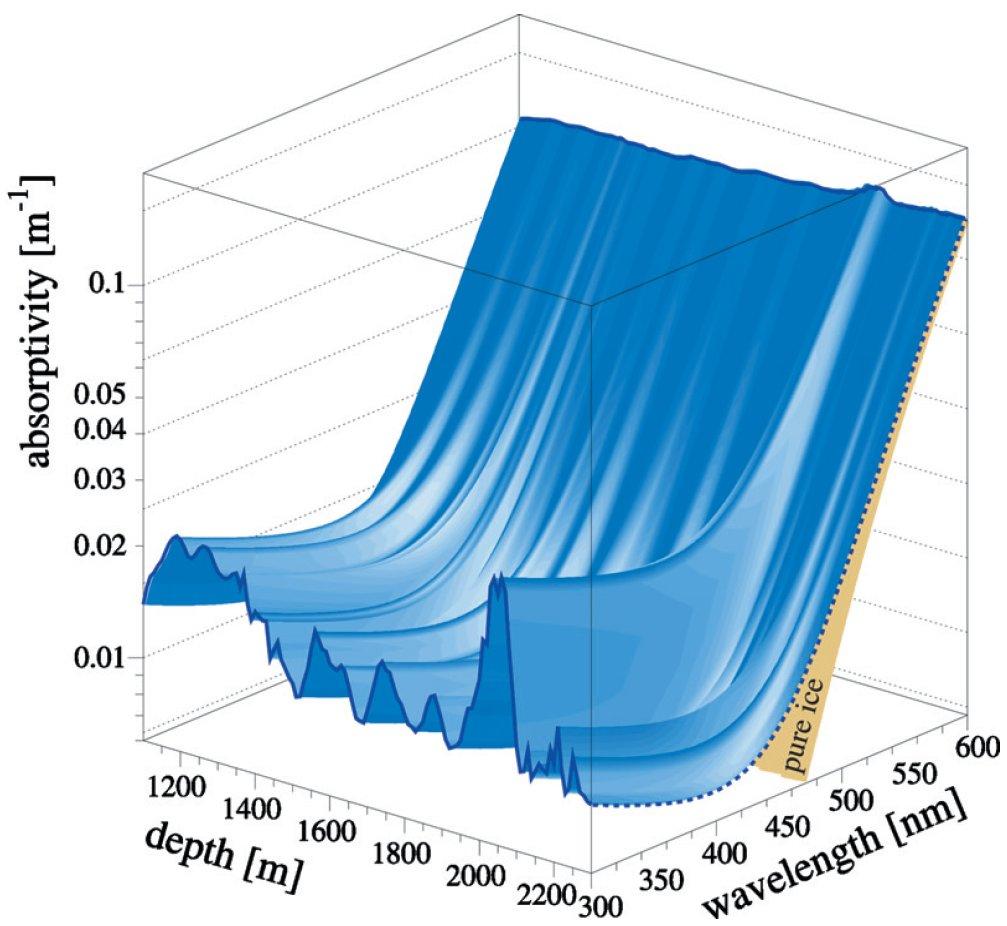,width=7.5cm}
\hfill
\epsfig{file=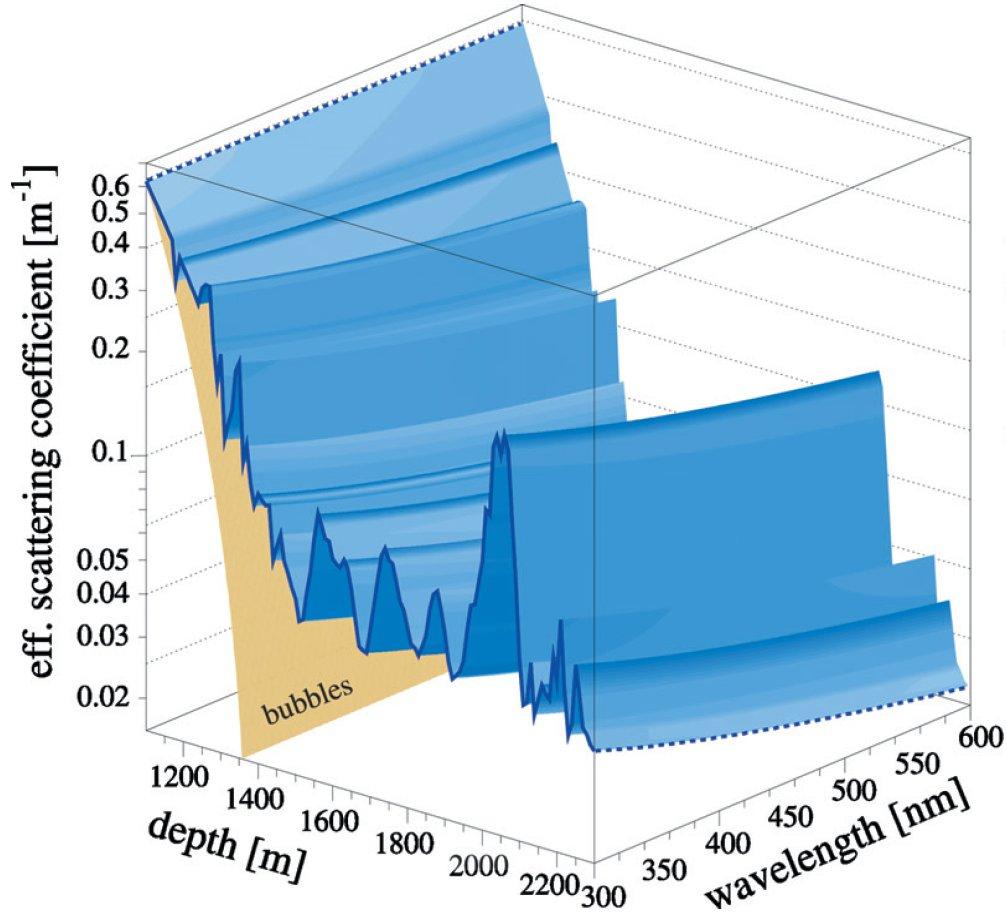,width=7.5cm}
\caption{
Absorption coefficient (left) and scattering coefficient (right) in the South
Polar ice as functions of depth and wavelength.
\label{SP-Ice}
}
\end{figure}

In Fig.~\ref{SP-Ice}, absorption and scattering coefficients are shown as
functions of depth and wavelength \cite{Amanda-ice-2006}. The variations with
depth are due to bubbles at shallow depth leading to very strong scattering and,
at larger depths, to dust and other material transported to Antarctica during
varying climate epochs. The quality of the ice improves substantially below a
major dust layer at a depth of about $2000\rnge2100\met$, with a scattering
length about twice as large as for the region above 2000\,m. The depth
dependence of the optical properties complicates the analysis of the
experimental data. Furthermore, the large delays in photon propagation due to
the strong scattering cause a worse angular resolution of deep-ice detectors
compared to water (see below). On the other hand, the large absorption length,
with a cut-off below 300\,nm instead of 350--400\,nm in water, results in better
photon collection.

The relatively short distance between optical modules and surface electronics
allowed for transporting the analogue signals of the photomultipliers to surface
over 2\,km of cable instead of digitising them in situ. This requires a large
output signal of the photomultiplier, a specification met by the 8-inch R5912-2
from Hamamatsu with 14 dynodes and a gain of $10^9$. The first ten strings used
copper cables for both high-voltage supply and signal transmission; for the last
9 strings the anode signal was used to drive a LED and the light signal was
transmitted to surface via optical fibre. The time resolution achieved was
5--7\,ns; given the strong smearing of photon arrival times due to light
scattering in ice, this jitter appeared to be acceptable. An event was defined
by a majority trigger formed in the surface counting house, requesting $\ge8$
hits within a sliding window of $2\musec$.

Time calibration of the AMANDA array was performed with a YAG laser at surface,
sending short pulses via optical fibres to each optical module. This laser
system was also used to measure the delay of optical pulses propagating between
strings and so to determine the optical ice properties as well as the
inter-string distances. A nitrogen laser (337\,nm), halogen lamps (350 and
380\,nm) and LED beacons (450\,nm) located in deep ice yielded further
information about the ice properties across a large range of wavelengths. The
measured time delays were fitted and the resulting parameterisations implemented
in the probability density functions for the reconstruction procedure (see
Sect.~\ref{sec-det-rec}).

A big advantage compared to underwater detectors is the small photomultiplier
noise rate, about 1\,kHz in an 8-inch tube, compared to 20--40\,kHz due to
K$^{40}$ decays and bioluminescence in lakes and oceans. The contamination of
hit patterns from particle interactions with noise hits is thus small and makes
hit selection much easier than in water.

The angular resolution of AMANDA for muon tracks is $2^\circ\rnge2.5^\circ$,
with a lower energy threshold around $50\gev$. Although better than for Lake
Baikal ($3^\circ\rnge4^\circ$), it is much worse than for ANTARES ($<0.5^\circ$,
see below). This is the result of the strong light scattering which deteriorates
the original information contained in the Cherenkov cone. The effect is even
worse for cascades, where the angular resolution achieved with present
algorithms is only $25^\circ$ (compared to $5^\circ\rnge8^\circ$ in water
\cite{Antares-casc-2006}).

\subsection{ANTARES}
\label{sec-fir-ant}

Starting with the decline of the DUMAND project (see Sect.~\ref{sec-fir-dum}),
the participating European groups started to explore options for a deep-see
neutrino telescope in the Mediterranean Sea. The NESTOR (see
Sect.~\ref{sec-fir-nes}), ANTARES and NEMO (see Sect.~\ref{sec-fir-nem})
projects were initiated, in the temporal order indicated. The NESTOR and NEMO
Collaborations have performed technical R\&D work and deployed prototype
installations; they are now, together with the ANTARES collaboration, pursuing
further work in the KM3NeT framework (see Sect.~\ref{sec-sec-kmt}). ANTARES has
succeeded in installing and is operating the first working deep-sea neutrino
telescope, thus providing the proof of feasibility for such devices.

\begin{figure}[ht]
\sidecaption
\epsfig{file=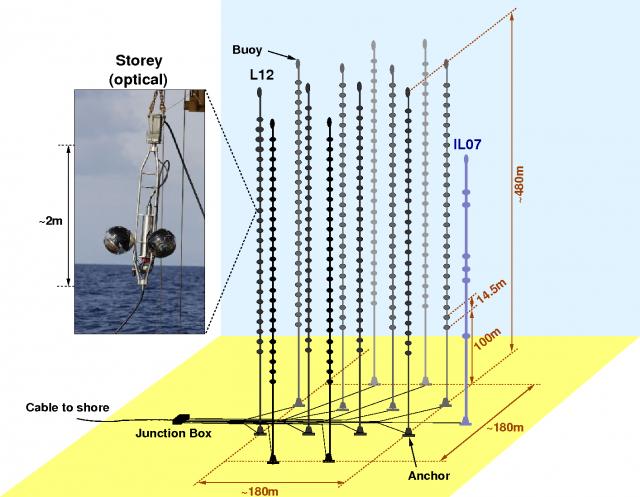,width=12.cm}
\caption{
Schematic of the ANTARES detector. Indicated are the 12 strings and the
instrumentation line in its 2007 configuration (IL07). Shown as an inset is the
photograph of a storey carrying 3 photomultipliers.}
\label{ANTARES}
\end{figure}

The ANTARES (Astronomy with a Neutrino Telescope and Abyss environmental
RESearch) \cite{antares-web} proposal \cite{antares-proposal} was presented in
1999. It was based on the operation of a demonstrator string
\cite{antares-demo1,antares-demo2} as well as on the results of extensive site
exploration campaigns in the region off Toulon at the French Mediterranean
coast, indicating that the optical background \cite{antares-light} as well as
sedimentation and biofouling \cite{antares-sedi} are acceptable at that site. 
The ANTARES design encompasses 12 strings, each carrying 25~``storeys'' equipped
with three optical modules, an electronics container and calibration devices
where necessary. A further string, the ``instrumentation line'', carries devices
for environmental monitoring. The strings are anchored to the sea floor with a
dead weight and kept upright by a submerged buoy at their top. The cable
connecting the storeys encompasses copper leads and optical fibres and at the
same time has to sustain the mechanical tension; this combination was found to
be conceptually challenging. At the sea floor, electro-optical cables connect the
strings to a junction box, from which the main cable goes to shore. The strings
are deployed from the surface and subsequently connected to the junction box
employing a submersible. The inter-string distances are about 70\,m, the
vertical distance between adjacent storeys is $14.5\met$. The depth at the
ANTARES site is 2475\,m. The schematic setup is shown in Fig.~\ref{ANTARES}, a
detailed technical description can be found in \cite{antares-detector}.

The ANTARES construction started in 2002 with the deployment of the main cable
and the junction box. In 2002/2003, a preproduction string was deployed and
operated for some months. Several technical problems were identified that
required further studies, design modifications and the operation of a mechanical
test string \cite{antares-line0}. The detector in its final configuration was
eventually installed in 2006--2008 and has been operational since then, with a
break of a few months in 2009 due to a failure of the main cable that required
repair.

\begin{figure}[ht]
\sidecaption
\epsfig{file=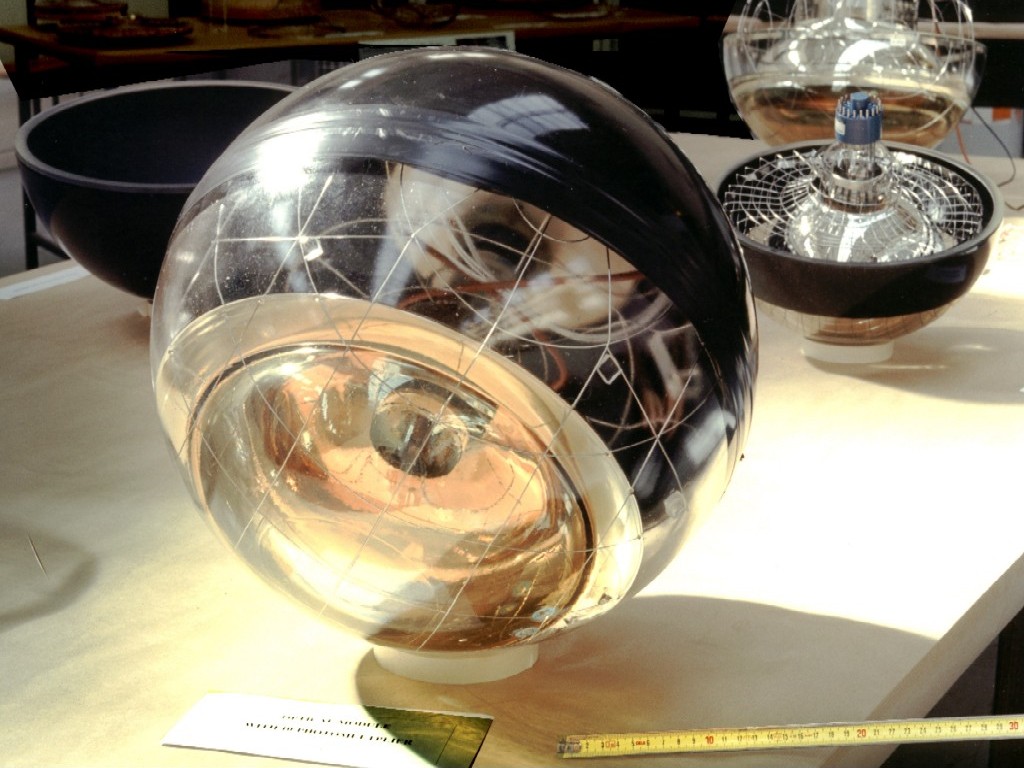,width=7.cm}
\caption{
Photograph of an ANTARES optical module. Its upper hemisphere is painted black to
reduce background light.}
\label{ANTARES-OM}
\end{figure}

The ANTARES optical module \cite{antares-om} consists of a 17-inch glass sphere
housing a hemispherical 10-inch photomultiplier (Hamamatsu R7081-20), connected
to the glass surface by optical gel, and its high-voltage base. A mu-metal cage
shields the photomultiplier against the Earth magnetic field (see
Fig.~\ref{ANTARES-OM}). A single cable with copper leads is used for voltage
supply and for transporting the analogue photomultiplier signals to the
digitisation electronics in the electronics container of the storey; dispersion
effects are negligible due to the short cable length of about a metre. The
optical modules are oriented such that the photomultipliers look downward at an
angle of $45^\circ$.

The photomultipliers provide an intrinsic timing resolution (transit time
spread) of about $\sigma_\text{TTS}=1.3\ns$ \cite{antares-pmt}, thus allowing
for precise measurements of the arrival times of Cherenkov photons. Owing also
to the minute light scattering in deep-sea water \cite{antares-water} and
accurate timing and optical module position calibration (see below), muon
trajectories can be reconstructed with a precision of about $0.2^\circ$ for muon
energies exceeding $1\tev$ \cite{antares-status-brunner}. Degradations are
expected for down-going muons (where the direct Cherenkov light is partly
outside the photomultiplier acceptance and the fraction of scattered light is
therefore increased) and due to background light from K$^{40}$ decays and
bioluminescence. The single photon rate from K$^{40}$ decays is about $30\kHz$
per photomultiplier, the bioluminescence rate exhibits a slowly varying baseline
of typically a few $10\kHz$ to a few $100\kHz$, and second-scale ``bursts'' in
the MHz region. Efficient data taking is possible up to baseline rates of
$200\kHz$.

The ANTARES data acquisition \cite{antares-daq} follows the
``all-data-to-shore'' concept, i.e.\ all photomultiplier signals exceeding an
adjustable threshold (default 0.3\,photo-electrons) are read out, thus avoiding
any inter-storey or inter-string trigger processing off-shore. The signals are
digitised and time-stamped by the storey electronics and then sent to shore via
optical fibres. The resulting data rate is dominated by background hits and
exceeds by far the capacity of data mass storage. Data filter software, running
on an on-shore farm of PCs, selects event candidates based on multi-hit
coincidences and/or multi-photo-electron hits in single photomultipliers. The
selected data are stored for offline analysis.

The two main calibration tasks are the synchronisation of the signal time
measurements at the individual photomultipliers and the position and orientation
monitoring of the optical modules that move with their strings in the sea
current. For the timing calibration \cite{antares-timecal}, light signal running
times through the optical fibres from shore to each storey and back are used, as
well as pulsed light emissions by laser and LED beacons \cite{antares-beacon}. 
An accuracy of about $0.5\ns$ is achieved, complying with the requirements. The
position and orientation calibration \cite{antares-pos1,antares-pos2} uses data
from compasses and tiltmeters on each storey and from an acoustic system
measuring running times of acoustic pulses between transmitters on the sea floor
and receivers (hydrophones) on 5 storeys per string. Also here, the precision of
roughly 10\,cm, translating into a timing uncertainty of $0.5\ns$, is within
specifications.

ANTARES is equipped with an acoustic detection system, AMADEUS
\cite{antares-acou}, for feasibility studies towards acoustic neutrino
detection. For a more detailed discussion see Sect.~\ref{sec-alt-aco}.

\subsection{NEMO}
\label{sec-fir-nem}

The NEMO (NEutrino Mediterranean Observatory) project \cite{nemo-web} is pursued
by Italian groups and started 1998 with the objective to investigate the
feasibility of a cubic-kilometre-scale deep-sea detector and to identify and
explore a suitable site.

\begin{figure}[ht]
\sidecaption
\epsfig{file=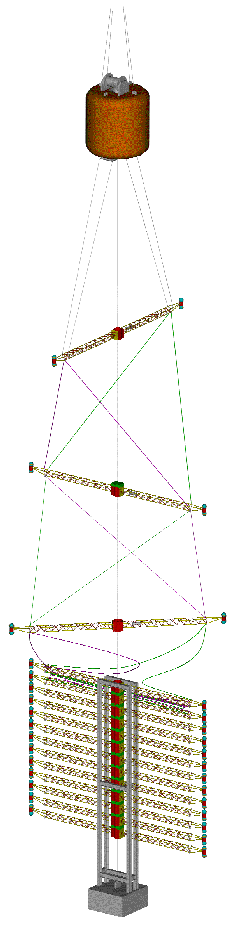,width=2.cm}
\caption{
Sketch of the principle of a flexible tower during the unfurling process. Note
that exact configuration and the packaging of the bars does not correspond to
the current design framework.}
\label{nemo-tower}
\end{figure}

Technical solutions have been devised, investigated and optimised for an
easy-to-deploy, stable and cost-effective detector
\cite{nemo-status-2004,nemo-status-2006,nemo-status-2009}. The supposedly most
important new concept developed in NEMO is that of ``flexible towers''
constructed of horizontal bars of a length of up to 15\,m, interconnected by
ropes forming a tetrahedral structure, so that adjacent bars are positioned
orthogonal to each other (see Fig.~\ref{nemo-tower}). The optical modules are
fixed to the bars. This design has several advantages, in particular (i) that a
tower can be folded together and deployed to the sea floor as a compact object
that is subsequently unfurled; (ii) the separation of the mechanical tension
(carried by the ropes) from the electro-optical backbone cable; (iii) the
provision of a 3-dimensional arrangement of photomultipliers per tower, allowing
for local reconstruction of muon directions.

Concurrently with the technical developments, a suitable site at a depth of
3.5\,km, about 100\,km off Capo Passero at the South-Eastern coast of Sicily has
been identified and investigated during various campaigns (see \cite{km3net-tdr}
and references therein).

During the first prototyping phase (``NEMO Phase-1''), a cable to a test site
near Catania at a depth of 2\,km was installed and equipped with a junction box. 
In 2007, a ``mini-tower'' with 4 bars was deployed, connected and operated for
several weeks. Although the data taking period was limited to a few months due
to technical problems, the mini-tower provided the proof of concept for the
technologies and most of the components employed. The flux of atmospheric muons
was determined in good agreement with the expectations from simulation
\cite{nemo-atmuons}.

The Phase-2 \cite{nemo-latest} setup includes shore infrastructure at Capo
Passero and a 100\,km long cable to the site at 3.5\,km depth; both are
meanwhile in place. A remotely operated vehicle (ROV) is available for the
deep-sea operations. A mechanical test tower of limited size was successfully
deployed and unfurled in early 2010. The plans to deploy a full-size prototype
tower will be pursued in the KM3NeT framework.

During the Phase-1 activities, an acoustic setup, O$\nu$DE (Ocean noise
Detection Experiment), was operated for almost two years and yielded important
data on the acoustic deep-sea background \cite{nemo-acou} and also on sperm
whales \cite{nemo-whales}.
 
\subsection{NESTOR} 
\label{sec-fir-nes}

NESTOR \cite{nestor-web} started out as a Greece-centred, originally
international collaboration around the beginning of the 1990s. A series of
workshops was held in 1991--1993 at the homonymous institute in Pylos at the
West coast of the Peloponnesus and a detector concept presented in 1994
\cite{nestor-1994}. The basic idea is to attach the optical modules to rigid
hexagonal structures (``floors'') of 30\,m diameter, of which 12 are stacked to
a tower with 20--30\,m inter-floor distance. The deployment would be from the
sea top, the connection to the main cable or already-deployed components being
made in the sequence recovery--connection--redeployment, thus avoiding the use
of deep-sea submersibles.

Several suitable site options off the coast of Pylos have been identified. They
are between 3000\,m and 5200\,m deep and have distances to shore of 30--50\,km. 
The NESTOR collaboration has performed numerous studies on the site
characteristics, most recently on the water optical properties (see
\cite{km3net-tdr} and references therein).

\begin{figure}[hbt]
\sidecaption
\epsfig{file=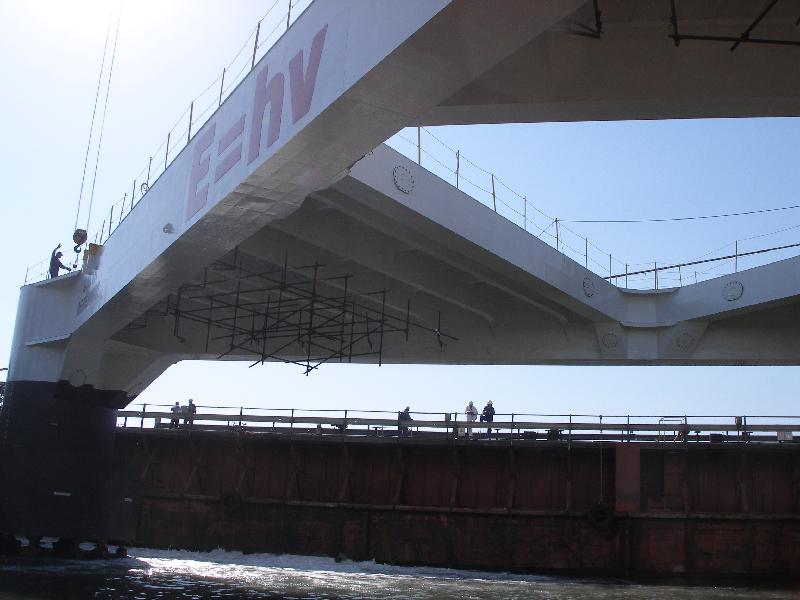,width=7.cm}
\caption{
Photograph of the Delta Berenike platform in the final phase of its
construction.}
\label{Delta-Berenike}
\end{figure}

After a long phase of technical development a cable was installed to a site at
4\,km depth. A single, reduced-size floor with 12 optical modules was deployed,
connected and operated for more than a month; its operation had to be terminated
due to a failure of the cable to shore. The data recorded sufficed to validate
the detector functionality within specifications \cite{nestor-test} and to
measure the atmospheric muon flux, which was found to agree with the
expectations \cite{nestor-atmuons}.

To facilitate the deployment of the floors, NESTOR has developed and constructed
a triangular working platform for sea operations, the {\it Delta Berenike} (see
Fig.~\ref{Delta-Berenike}). The platform with side lengths of about 50\,m is
carried by three cylindrical pontoons at its apices and has a triangular
aperture for deployment operations.

\clearpage
\section{Second-Generation Neutrino Telescope Projects}
\label{sec-sec}

There are three projects which reach, or even exceed, the size originally
conceived by the DUMAND pioneers: IceCube at the South Pole, in its basic
configuration, has been completed in December 2010; KM3NeT (km$^3$ Neutrino
Telescope) in the Mediterranean Sea and GVD (Gigaton Volume Detector) in Lake
Baikal are in their preparatory/prototyping phase. IceCube instruments $1\km^3$
of ice. GVD is also planned to cover $1\km^3$, but with a much higher energy
threshold than IceCube. KM3NeT envisages an instrumented volume of several
km$^3$. KM3NeT and GVD could be completed by 2017 but will start data taking
earlier with intermediate configurations. It is expected that these detectors
will form a global network or even join to a global neutrino observatory.

\subsection{IceCube}
\label{sec-sec-ice}

\begin{figure}[ht]
\sidecaption
\epsfig{file=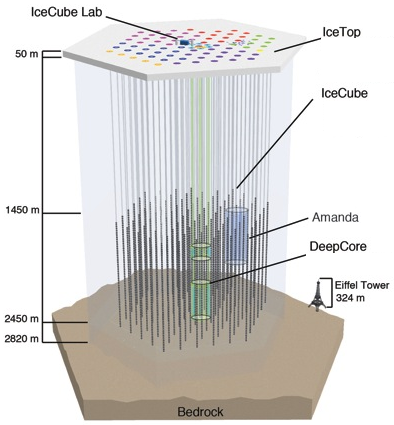,width=9.cm}
\caption{
Schematic view of the IceCube neutrino observatory. Also shown is the location
of AMANDA (cf.\ Sect.~\ref{sec-fir-ama}) and of DeepCore, a nested low-threshold
array (cf.\ Sect.~\ref{sec-sec-ice-dee}). At the surface, the air shower array
IceTop (cf.\ Sect.~\ref{sec-sec-ice-top}) and the IceCube counting house are
indicated. The Eiffel tower is shown to scale for a size comparison.
\label{IceCube}
}
\end{figure}

IceCube \cite{icecube-web,icecube-sensi-2004} is the successor of AMANDA. It
consists of 5160 digital optical modules (DOMs) installed on 86 strings at
depths of 1450 to 2450\,m in the Antarctic ice \cite{IceCube-PDD-2001}. 
320~further DOMs are installed in IceTop, an array of detector stations on the
ice surface directly above the strings (see Fig.~\ref{IceCube}). AMANDA,
initially running as a low-energy sub-detector of IceCube, was decommissioned in
2008 and replaced by DeepCore, a high-density sub-array of six strings at large
depths (i.e.\ in the best ice layer) at the centre of IceCube.

For IceCube construction, the thermal power of the hot-water drill factory has
been upgraded to 5\,MW, compared to 2\,MW for AMANDA, thus reducing the average
time to drill a 2450\, m deep hole with a diameter of 60\,cm to $35$\,hours. The
subsequent installation of a string required typically 12 hours. Installation of
IceCube started in January 2005 with the first string and was completed with the
deployment of the $86^\text{th}$ string at Dec.~18, 2010.

As the components are not accessible after refreezing of the holes, the IceCube
architecture avoids single point failures in the ice. A string carries 60~DOMs,
with 30 twisted copper pair cables providing power and communication. 
Neighbouring DOMs share the same wire pair and are thus connected to enable fast
local coincidence triggering in the ice.

\begin{figure}[ht]
\sidecaption
\epsfig{file=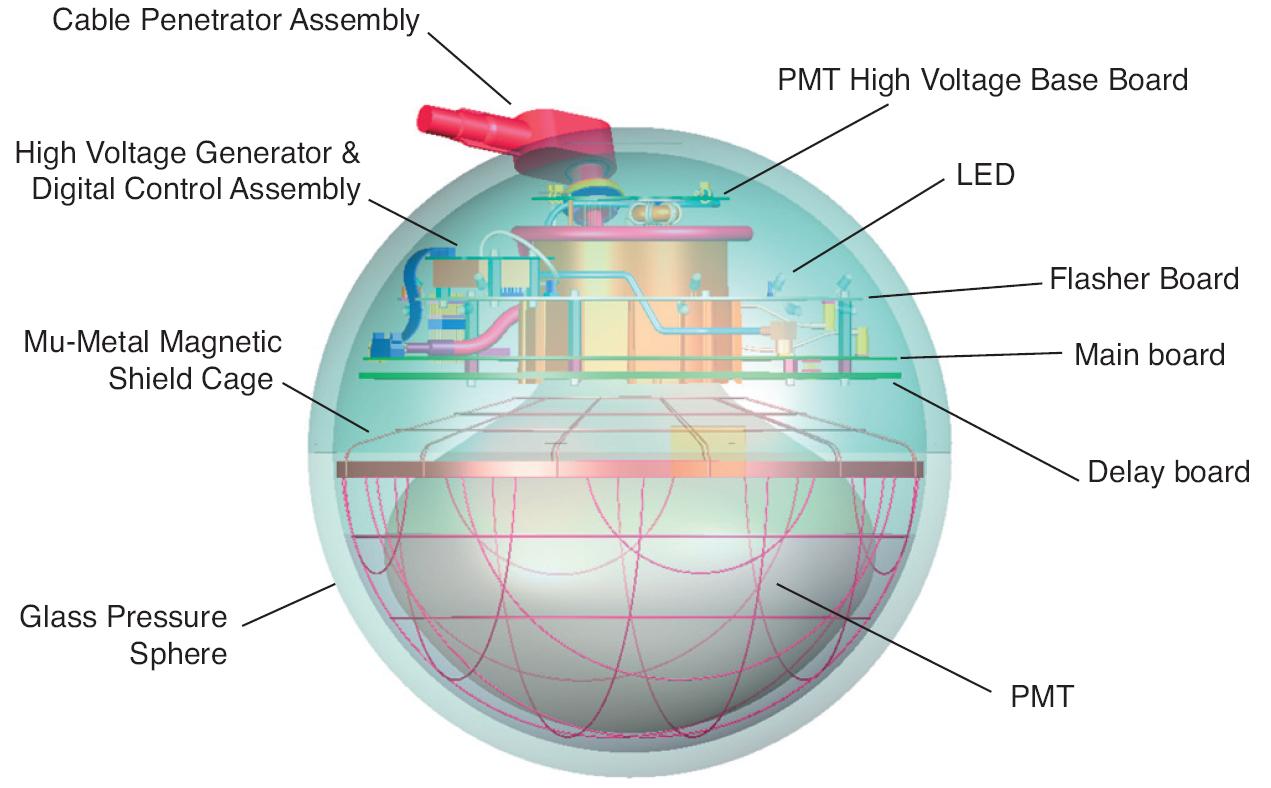,width=10.cm}
\caption{
Schematic view of an IceCube Digital Optical Module (DOM). The analogue
photomultiplier signals are digitised on the main board. For details see
\protect\cite{icecube-daq-2009}.}
\label{DOM}
\end{figure}

A schematic view of a DOM is shown in Fig.~\ref{DOM}. A 10-inch photomultiplier
(Hamamatsu R7081-02) is embedded in a 13-inch glass sphere. A mu-metal grid
reduces the influence of the Earth magnetic field. The programmable high voltage
is generated inside the DOM. The average photomultiplier gain is set to $10^7$. 
Signals are sent to the main board where they are digitised by a fast analogue
transient waveform recorder (ATWD, 3.3\,ns sampling) and by a FADC (25\,ns
sampling). The photomultiplier signal is amplified by 3 different gains to
extend the dynamic range of the ATWD to 16\,bits. The resulting linear dynamic
range is 400~photo-electrons in 15\,ns; the dynamic range integrated over
$2\musec$ is about 5000~photo-electrons \cite{icecube-daq-2009}. The digital
section of the main board is based on a field-programmable gate array (FPGA),
which communicates with the surface electronics and allows for uploading new
program code. LEDs on a ``flasher board'' emit calibration pulses at a
wavelength of 405\,nm which can be adjusted in intensity over a wide range up to
$10^{11}$ photons.

All digitised photomultiplier pulses are sent to the surface. In order to
compress data for isolated hits which are mostly noise pulses, the full waveform
is only sent for pulses appearing in local (neighbour or next-to-neighbour)
coincidences on a string. All DOMs have precise quartz oscillators providing
local clock signals, which are synchronised every few seconds to a central GPS
clock. The time resolution is about 2\,ns. The noise rate for DOMs in the deep
ice is about $540\Hz$ and is reduced to ca.\ $280\Hz$ if an artificial deadtime
of $50\musec$ is applied to discard after-pulses (only for the supernova burst
trigger, see Sect.~\ref{sec-sec-ice-snb}). These very low noise rates are
essential for the detection of the low-energy neutrino emission associated with
a supernova collapse (see Sect.~\ref{sec-sec-ice-snb}).

At the surface, 8 custom PCI cards per string provide power, communication and
time calibration. Subsequent processors sort and buffer hits until the array
trigger and event builder process is completed. This architecture allows for a
deadtime-free operation. The raw data rate of the full array is about
800\,GB/day, which are written to tape and processed online on a computer farm
to extract interesting event classes, like up-going muon candidates, high-energy
events, IceTop/IceCube coincidences, cascade events, events from the direction
of the Moon or events in coincidence with Gamma Ray Bursts (GRBs). The filtered
data stream (about $70$\,GB/day) is transmitted to the Northern hemisphere via
satellite.

The muon angular resolution achieved with present reconstruction algorithms is
about $1^\circ$ for $1\tev$ muons. It is expected that improved reconstruction
algorithms using the full waveform will lead to a resolution below $0.5^\circ$
for energies above $10\tev$. The very clear ice below a depth of 2100\,m has a
particular potential for improved resolution. This will be even more important
for the angular reconstruction of cascades, for which the presently achieved
angular resolution is only $30^\circ$ -- much worse than for water, mainly due
to the strong light scattering in ice.

\subsubsection{DeepCore}
\label{sec-sec-ice-dee}

\begin{figure}[ht]
\sidecaption
\epsfig{file=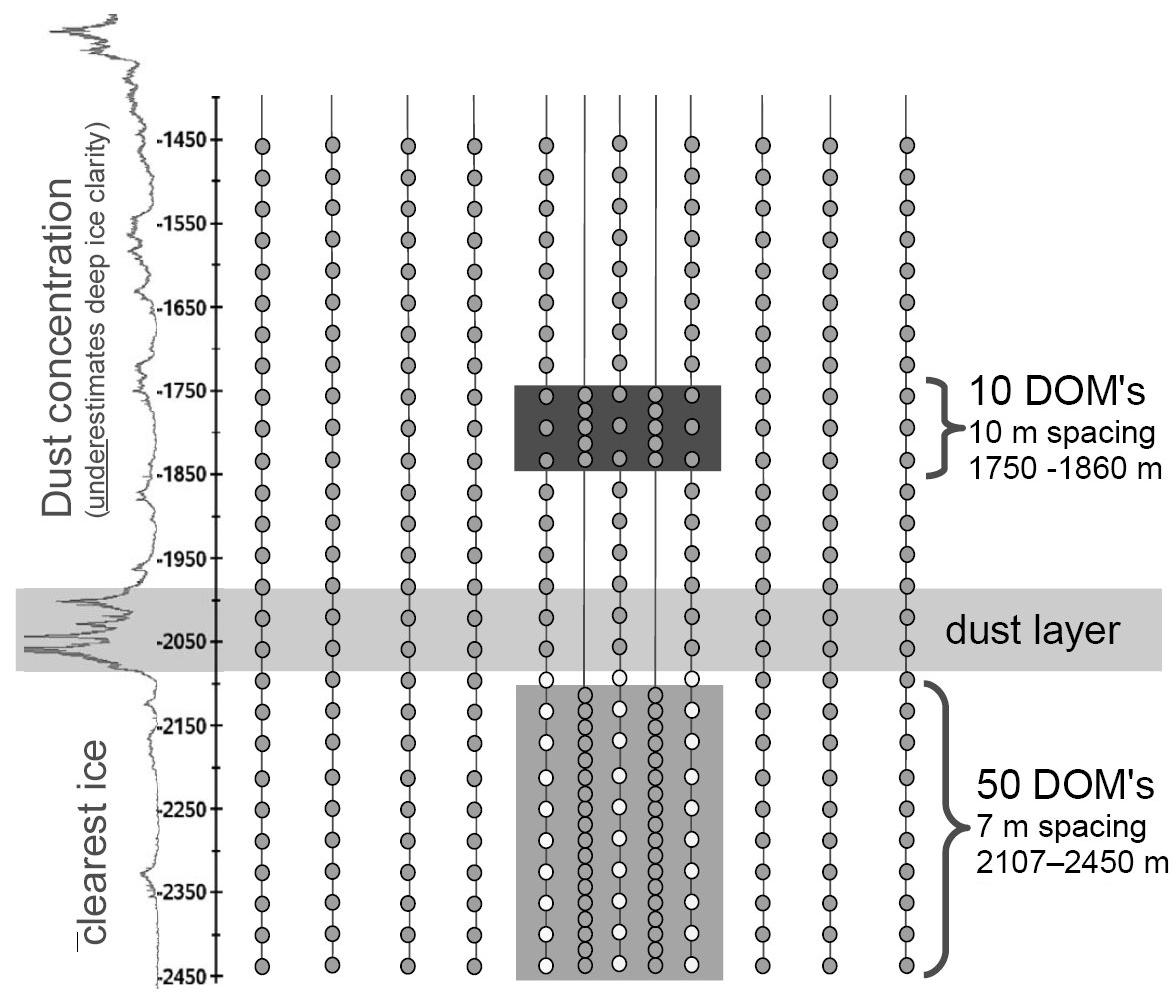,width=8cm}
\caption{
Layout of the DeepCore sub-detector. Shown are the positions of the DeepCore
DOMs; for a 3-dimensional representation see Fig.~\ref{IceCube}. The depth
profile of the ice transparency is indicated on the left.}
\label{DeepCore}
\end{figure}

The geometry of DeepCore is sketched in Fig.~\ref{DeepCore}. DeepCore consists
of 7 central standard IceCube strings plus additional 6 special strings. Ten of
the DOMs of these strings are arranged at 1750--1850\,m depth, above the dust
layer with bad optical transparency. They are used as veto-detector for the
deeper component. The deep component comprises 50~DOMs per string and is
installed in the very clear ice at depths between 2100 and 2450\,m, where the
effective scattering length ranges up to 50\,m and the absorption length to
230\,m. The six additional DeepCore strings are equipped with photomultipliers
with an enhanced quantum efficiency (Hamamatsu R7081-MOD).

DeepCore has a factor of about six better sensitivity in photon collection than
IceCube, due to the smaller spacing between strings (72 instead of 125\,m) and
OMs along a string (7 instead of 17\,m); the better ice quality; and the 30\%
higher quantum efficiency of the new photomultipliers. Together with the veto
provided by IceCube, this results in an expected threshold of less than
$10\gev$. This opens a new window for oscillation physics not tested by other
experiments and allows for probing dark matter models not covered by direct
searches. The veto will also allow for identification of neutrinos from above if
they interact within DeepCore. This enlarges the field of view of IceCube to the
full sky \cite{Wiebusch-2009}.

\subsubsection{IceTop}
\label{sec-sec-ice-top}

IceCube is the only large neutrino telescope which can be permanently operated
together with a surface air shower array, IceTop \cite{IceCube-IceTop-2008}. 
IceTop consists of tanks filled with ice, each instrumented with 2~DOMs. With
IceTop, the energy spectrum of air showers can be measured up to primary
particle energies around $10^{18}\ev$. The combination of IceTop information
(reflecting dominantly the electron component of the air shower) and IceCube
information (muons from the hadronic component) restricts the mass range of the
primary particle. The comparison of air shower directions measured with IceTop
and directions of muons from these showers in IceCube establishes a tool for
angular calibration of IceCube (absolute pointing and angular resolution).

\subsubsection{IceCube as a supernova burst detector}
\label{sec-sec-ice-snb}

Last but not least, IceCube can be operated in a mode that is only possible in
ice: The detection of burst neutrinos from supernovae. The low dark-count rate
of the photomultipliers ($280\Hz$, see above) allows for detection of the
feeble increase of the summed count rates of all photomultipliers during several
seconds, which would be produced by millions of interactions of few-$\Mev$
neutrinos from a supernova burst \cite{amanda-sn-2002,icecube-sn-2009}. IceCube
records the counting rate of all photomultipliers in millisecond steps. A
supernova in the centre of the Galaxy would be detected with extremely high
significance and the onset of the pulse could be measured in unprecedented
detail. Even a SN\,1987A-type supernova in the Large Magellanic Cloud would
provide a recognisable signal and be sufficient to provide a trigger to the
SuperNova Early Warning System, SNEWS \cite{SNEWS-2004}.

\subsection{KM3NeT}
\label{sec-sec-kmt}

In recognition of the fact that at least a cubic-kilometre sized detector will
be necessary to really observe abundant astrophysical high-energy neutrino
sources, the High Energy Neutrino Astronomy Panel (HENAP) of the
PaNAGIC\footnote{Particle and Nuclear Astrophysics and Gravitation International
Committee} Committee of {IUPAP}\footnote{International Union of Pure and Applied
Physics} has concluded in its 2002 report \cite{henap-2002} that ``a
km$^3$-scale detector in the Northern hemisphere should be built to complement
the IceCube detector being constructed at the South Pole''.

Following this recommendation, the Mediterranean neutrino telescopes groups --
together with deep-sea technology and marine science groups -- have formed the
KM3NeT collaboration to prepare, construct and operate such a device. In
2006--2009, a Design Study, supported with 9\,M\Euro by the EU, was conducted;
its major achievements are a Conceptual Design Report (CDR) \cite{km3net-cdr}
and a Technical Design Report (TDR) \cite{km3net-tdr}. A further EU project, the
KM3NeT Preparatory Phase (2008--2012) provides resources and a framework for
work directed towards solving the funding, governance, legal and strategic
questions and also for engineering activities. KM3NeT will be a deep-sea
research infrastructure hosting a neutrino telescope, but also providing
continuous, long-tern access to deep-sea measurements to a variety of science
communities, such as marine biologists, oceanographers, geophysicists and
environmental scientists. In recognition of its high scientific potential,
KM3NeT has been included in the priority project list of the European Strategy
Forum on Research Infrastructures, ESFRI \cite{esfri-2006,esfri-2008}.

Based on the experience and expertise of the first-generation projects, a
variety of new, cost-effective design solutions for the neutrino telescope have
been elaborated. The original goal of reducing the capital investment for a
cubic kilometre of instrumented sea water to 200\,M\Euro has been outperformed
by a factor of at least 4. Whereas the TDR still presented a set of design
options, in particular for the mechanical structure of the strings/towers
(``detection units (DUs)'') and the optical modules, further investigations have
meanwhile led to convergence on a specific design, which will be described in
the following. Since final prototyping and deployment tests are still to come,
the discarded solutions partly serve as backup options.

For the mechanical DU structure, a solution along the NEMO design (see
Sect.~\ref{sec-fir-nem}) has been chosen. The bars will be 6\,m long and be
equipped with one optical module (see below) at each end; a schematic view is
given in Fig.~\ref{km3net-du}. Oil-filled hoses in equipressure with the
deep-sea water will be used as vertical cables for electrical leads and optical
fibres. Two such backbone cables, wound loosely around the ropes, will be used
per DU, with break-outs for two copper leads and one optical fibre at each
optical module. The DUs will be stacked in a cubicle pile for transport and
deployment, fitting into a standard shipping container. The deployment will
proceed as discussed in Sect.~\ref{sec-fir-nem}.

\begin{figure}[ht]
\epsfig{file=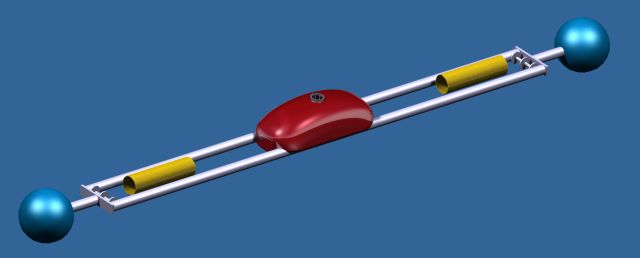,width=8.4cm}
\hfill
\epsfig{file=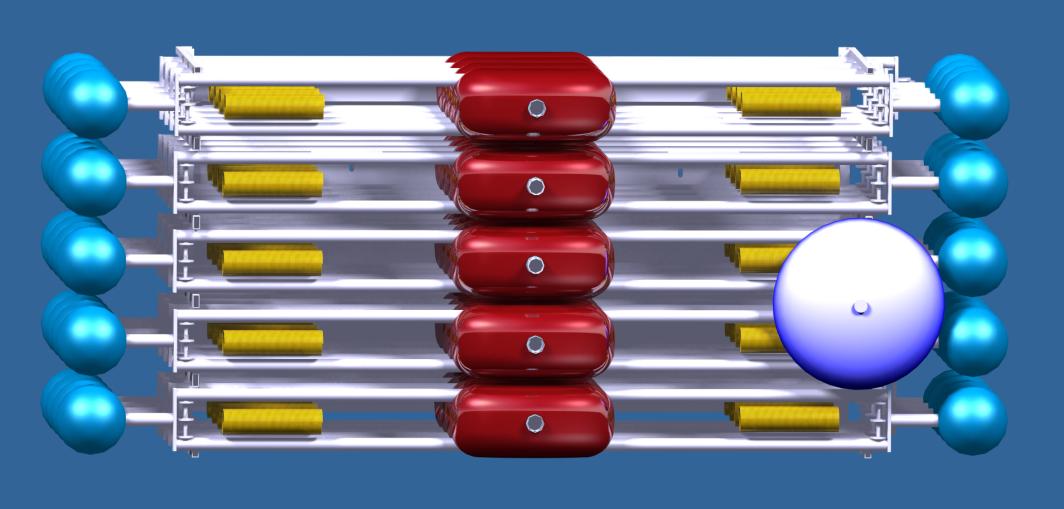,width=7.1cm}
\caption{
Left: Schematic drawing of a bar of the KM3NeT DU. The overall length is
slightly below 6\,m. The red object in the centre is a buoy made of syntactic
foam, the yellow cylinders house the ropes and vertical backbone cables before
unfurling. Right: A full DU stacked for transport and deployment (top view). The
round bluish object is the top buoy.}
\label{km3net-du}
\end{figure}

\begin{figure}[t]
\centerline{
\epsfig{file=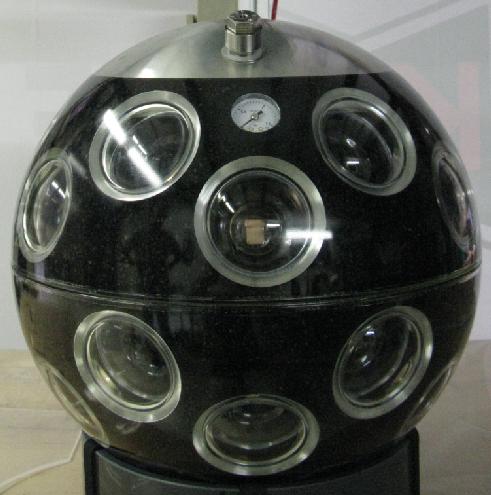,width=6.5cm}
\hspace*{5.mm}
\epsfig{file=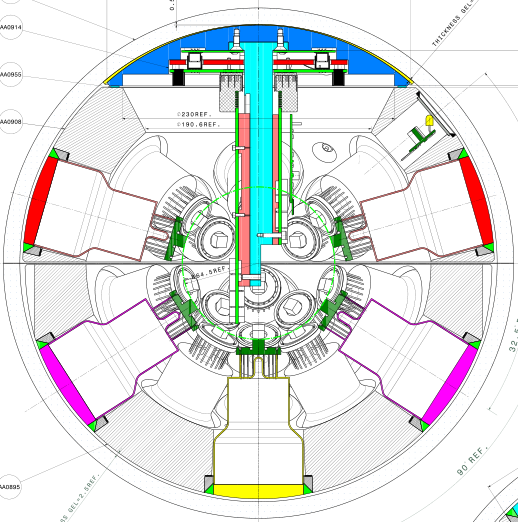,width=6.5cm}
}
\caption{
The KM3NeT digital optical module (DOM). Left: Photograph of a prototype; right:
technical drawing, with the mushroom-shape heat conductor and the electronics
components.}
\label{km3net-om}
\end{figure}

The digital optical module (DOM, Fig.~\ref{km3net-om}) will be a 17-inch glass
sphere, equipped with 31 3-inch photomultipliers, their high-voltage bases and
the digitisation electronics. High-voltage bases with a power consumption as low
as $140\,\text{mW}$ for a complete optical module have been designed for this
application. The photomultipliers are oriented from vertically downwards to
about $45^\circ$ upwards. They are supported by a foam structure and fixed to
the glass sphere by optical gel. An aluminium structure is used to conduct the
heat to the glass and to provide support for the electronic boards. Major
advantages of this design are:
\begin{itemize}
\item
The overall photocathode area exceeds that of a 10-inch photomultiplier by more
than a factor of three; a further increase is possible by extending the light
collection area using reflective rings \cite{Kavatsyuk-2009}. The number of
penetrators and separate electronics containers, which are expensive and
failure-prone, is thus reduced to a minimum.
\item
Since the photomultipliers are read out individually, a very good one-vs.-two
photo-electron resolution is obtained, which is essential for efficient online
data filtering.
\item
Some directional sensitivity is provided.
\end{itemize}

The KM3NeT data acquisition will follow the all-data-to-shore concept (see
Sect.~\ref{sec-fir-ant}). For each photomultiplier, the time intervals during
which the analogue output signal exceeds an adjustable threshold will be
digitised and sent to shore. This time-over-threshold information allows for
precise photo-electron counting for small signals and provides a logarithmic
measure of the amplitude for large signals, i.e.\ effectively an infinite
dynamic range. The technical implementation can either be achieved using a
custom-designed ASIC \cite{km3net-tdr} or a fast time-to-digital converter (TDC)
on a field programmable gate array (FPGA). The latter option is new and might
significantly simplify the system. The digitised data are transported to shore
via optical point-to-point connections using the Gigabit ethernet protocol and
dense wavelength division multiplexing. The system is driven by on-shore lasers,
which allows for easy maintenance and also for sending clock signals to the DOMs
by the same laser beam that carries the photomultiplier signals on its way back
to shore.

Time and position calibration will be based on the same principles as in ANTARES
(see Sect.~\ref{sec-fir-ant}). For the acoustic receivers on the DUs, the
stand-alone hydrophones (that require extra penetrators) will presumably be
replaced by piezo elements glued to the inner glass surface of the DOMs.

The full KM3NeT neutrino telescope will consist of about 300 DUs. Their
geometrical layout on the sea-floor (the ``footprint'') is still subject to
optimisation. For the sensitivity studies in the TDR two equal-size homogeneous
blocks of hexagonal layout were assumed, with an inter-DU distance of 180\,m and
a vertical distance between adjacent bars of 40\,m. The partition into two
blocks takes into account the facts that deployment and maintenance of the
deep-sea cable network becomes increasingly difficult for large homogeneous
setups, and that at least two main cables to shore will be needed for the full
detector.

Since the exact footprint is not yet known, the deep-sea cable network topology
is still open. It will either be star-like with a primary and a set of secondary
junction boxes or consist of a cable ring surrounding the detector, with several
primary junction boxes connected directly to the DUs. In the cable network,
single points of failure for the full or at least substantial parts of the
detector are unavoidable, making recovery and maintenance operations necessary. 
In contrast, no maintenance is planned for the DUs themselves.

\begin{figure}[ht]
\sidecaption
\epsfig{file=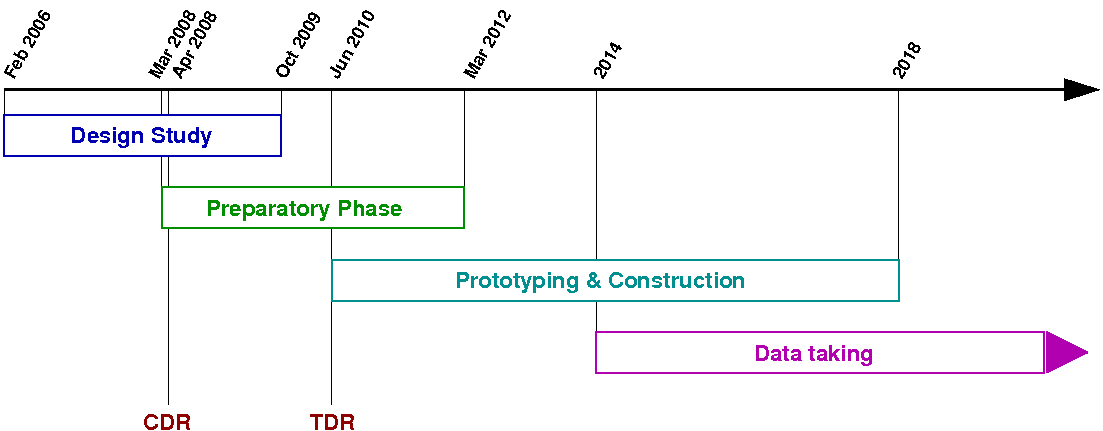,width=11.cm}
\caption{
Time-line towards KM3NeT construction and operation.}
\label{km3net-timeline}
\end{figure}

The time-line of the further steps towards construction and operation of KM3NeT
is shown in Fig.~\ref{km3net-timeline}. The three sites of ANTARES, NEMO and
NESTOR are under consideration for KM3NeT. The site question has scientific
aspects that have been investigated in detail (depth, water transparency,
bioluminescence etc., see \cite{km3net-tdr} and references therein);
additionally, the site choice is strongly linked to the availability of
regional funding sources and is therefore closely interwoven with a political
decision on KM3NeT. It is currently hoped that a site decision can be made by
late 2011 or early 2012. Subsequently, the final technical design will be worked
out and presented in a detailed proposal. Assuming that funding, legal and
administrative issues are sorted out by then, it will be possible to launch
production at that point. Data taking will start as soon as the first DUs are
operational. From a very early stage of its construction on, the data from the
KM3NeT neutrino telescope will exceed data from first-generation
Northern-hemisphere neutrino telescopes in quality and statistics and thus
provide an exciting discovery potential.

\subsection{GVD in Lake Baikal}
\label{sec-sec-gvd}

\begin{figure}[ht]
\sidecaption
\epsfig{file=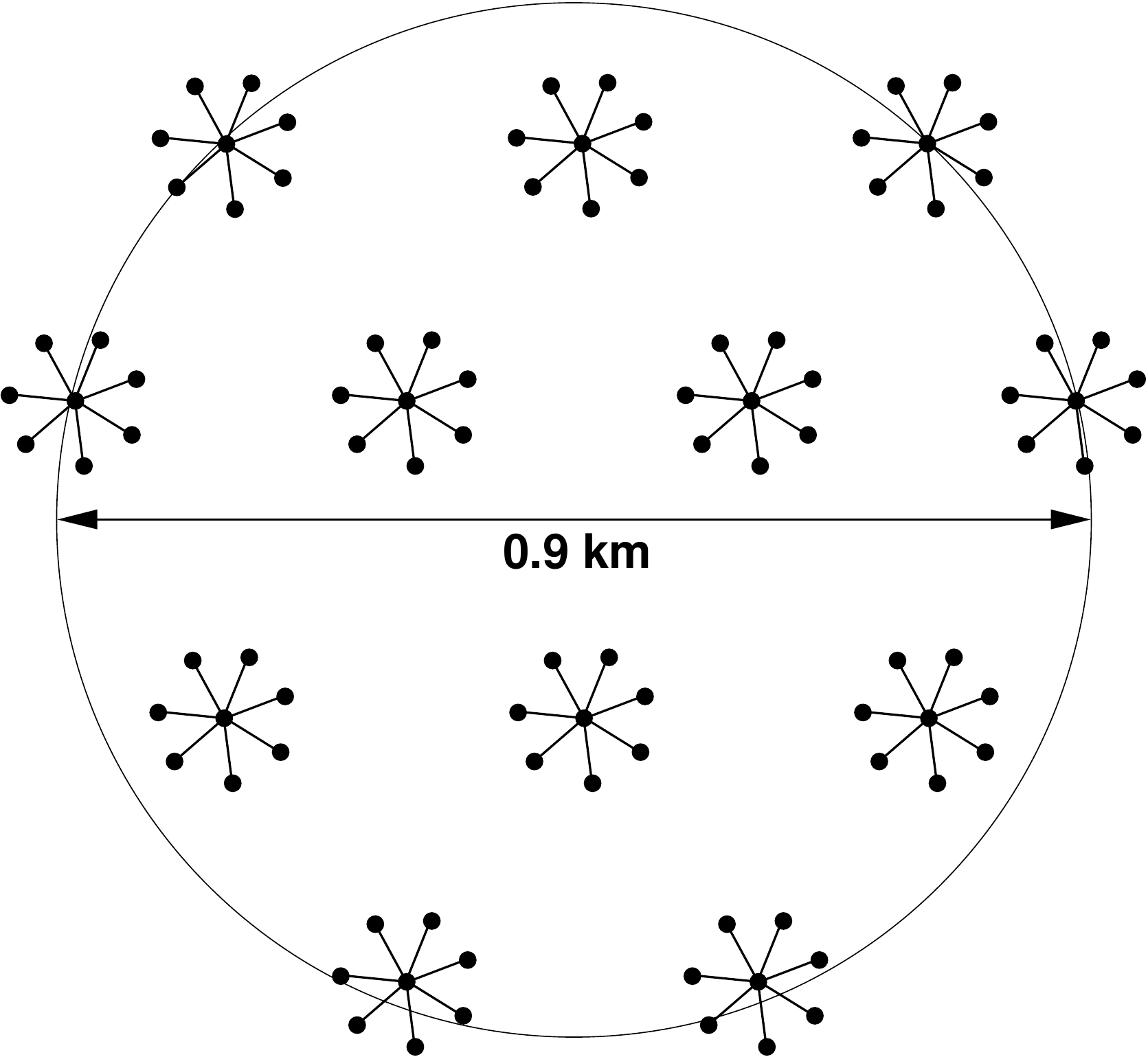,width=5.5cm}
\kern3.mm
\epsfig{file=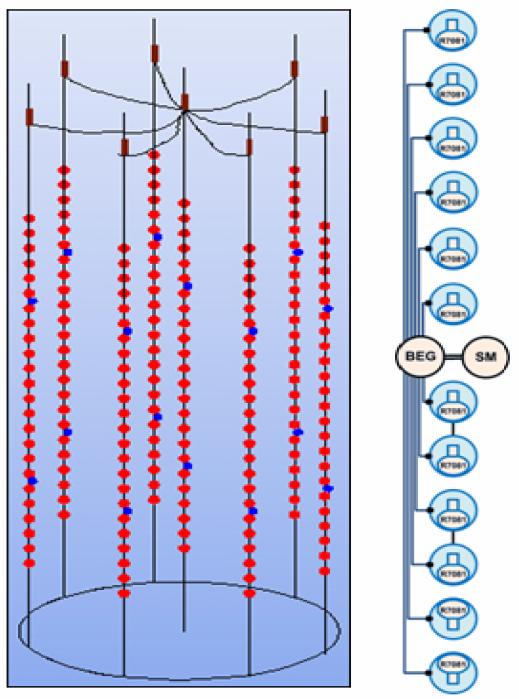,width=5.5cm}
\caption{
The Baikal Gigaton Volume Detector (GVD). Left: Arrangement of the 12 clusters;
middle: One cluster; right: Schematic view of a string section.}
\label{GVD}
\end{figure}

The Baikal Collaboration plans the stepwise installation of a kilometre-scale
array in Lake Baikal, the Gigaton Volume Detector, GVD
\cite{Baikal-GVD-Aynutdinov-2007}. It will consist of strings which are grouped
in clusters of eight (see Fig.~\ref{GVD}). This results in a relatively flexible
structure, which allows for rearranging the clusters and meets best the
deployment conditions from the ice. Each string carries 24 optical modules
spaced uniformly from 900\,m down to about 1250\,m depth
\cite{Baikal-GVD-Avrorin-2010}. The modules will house 10-inch
photomultipliers (likely Hamamatsu R7081-HQE) with a peak quantum efficiency of
about 35\%.

The optical modules on a string are grouped into two sections, each consisting
of 12 OMs, a service module (SM) and the electronics unit (BEG) with its
200\,MHz FADCs (see Fig.~\ref{GVD} right). Analogue signals from the optical
modules are transmitted to the BEG through coaxial cables. A trigger is formed
by a coincidence of any neighbouring optical modules. Digitised waveforms for
each triggered channel are transmitted via Ethernet from the BEG to the central
underwater micro-PC of the cluster. The cluster DAQ provides inter-section time
synchronisation, on-line data selection, and communication to shore through an
optical cable. Prototype strings have been operated in 2009 and 2010. They have
demonstrated a time accuracy of about 2\,ns. In April 2011, a prototype cluster
with three mini-strings and all key elements of DAQ electronics and the
communication system was deployed.

Simulations have been performed for 96 strings in 12 clusters and a total of
2304 OMs. A compromise between large volume for cascade detection and
reasonable efficiency for muons was found for an instrumented height of
345\,m, a cluster diameter of 120\,m and a vertical spacing between optical
modules of about 15\,m \cite{Baikal-GVD-Avrorin-2010}. At trigger level, the
effective detection area for muons with energies above $3\tev$ is
$0.2\rnge0.5\km^2$; the effective detection volume for cascades above $50\tev$
is $0.3\rnge0.8\km^3$. We note that cuts for background suppression will
reduce these values significantly, in particular at lower energies. The
directional accuracies are $0.5\rnge1.0^\circ$ for muons and $3\rnge7^\circ$
for cascades.

A threshold for muons of about $3\rnge10\tev$ appears to be rather high when
compared to IceCube and KM3NeT. On the other hand, the optimum energy cut to
obtain the best signal-to-noise ratio (extraterrestrial versus atmospheric
neutrinos) for the weakest detectable sources is at a few $\Tev$ for point
sources, and in the $100\tev$ range for diffuse fluxes
\cite{Gabici-2008}. Therefore, a sparse detector configuration such as GVD may
offer a favourable physics/cost ratio for neutrino signals extending to the
$100\tev$ region and beyond, as e.g.\ expected for Gamma Ray Bursts and AGN
jets.

\clearpage
\section{Physics Results and Perspectives}
\label{sec-phy}

After two decades of data taking with neutrino telescopes, a lot has been
learned -- despite the fact that not a single high-energy neutrino of cosmic
origin has been clearly identified as yet. In this section, the results as of
early 2011 are presented for atmospheric neutrinos (Sect.~\ref{sec-phy-atm}),
for cosmic neutrinos including those from supernova bursts
(Sect.~\ref{sec-phy-cos}), for searches for Dark Matter and other exotic
particles (Sect.~\ref{sec-phy-dar}) and from cosmic-ray studies with
IceCube/IceTop (Sect.~\ref{sec-phy-ray})

\subsection{Atmospheric neutrinos}
\label{sec-phy-atm}

Atmospheric neutrinos and muons are produced in cosmic-ray interactions in the
atmosphere. Up to energies of about $100\tev$, their flux is dominated by pion
and kaon decays. The corresponding neutrinos are referred to as ``conventional''
atmospheric neutrinos. The spectrum follows approximately an $E^{-3.7}$ shape. 
It can be calculated with uncertainties of about 25\% in the energy range
$100\gev\rnge1\tev$ (see \cite{Barr-2006} and references therein, a widely used
analytical calculation is given in \cite{Volkova-1980}). At higher energies,
``prompt'' atmospheric neutrinos from the decay of charm and bottom particles
take over. These particles decay before having a chance for further
interactions, and the resulting neutrinos therefore closely follow the primary
cosmic ray power law spectrum, i.e.\ an $E^{-2.7}$ shape.

\begin{figure}[ht]
\sidecaption
\epsfig{file=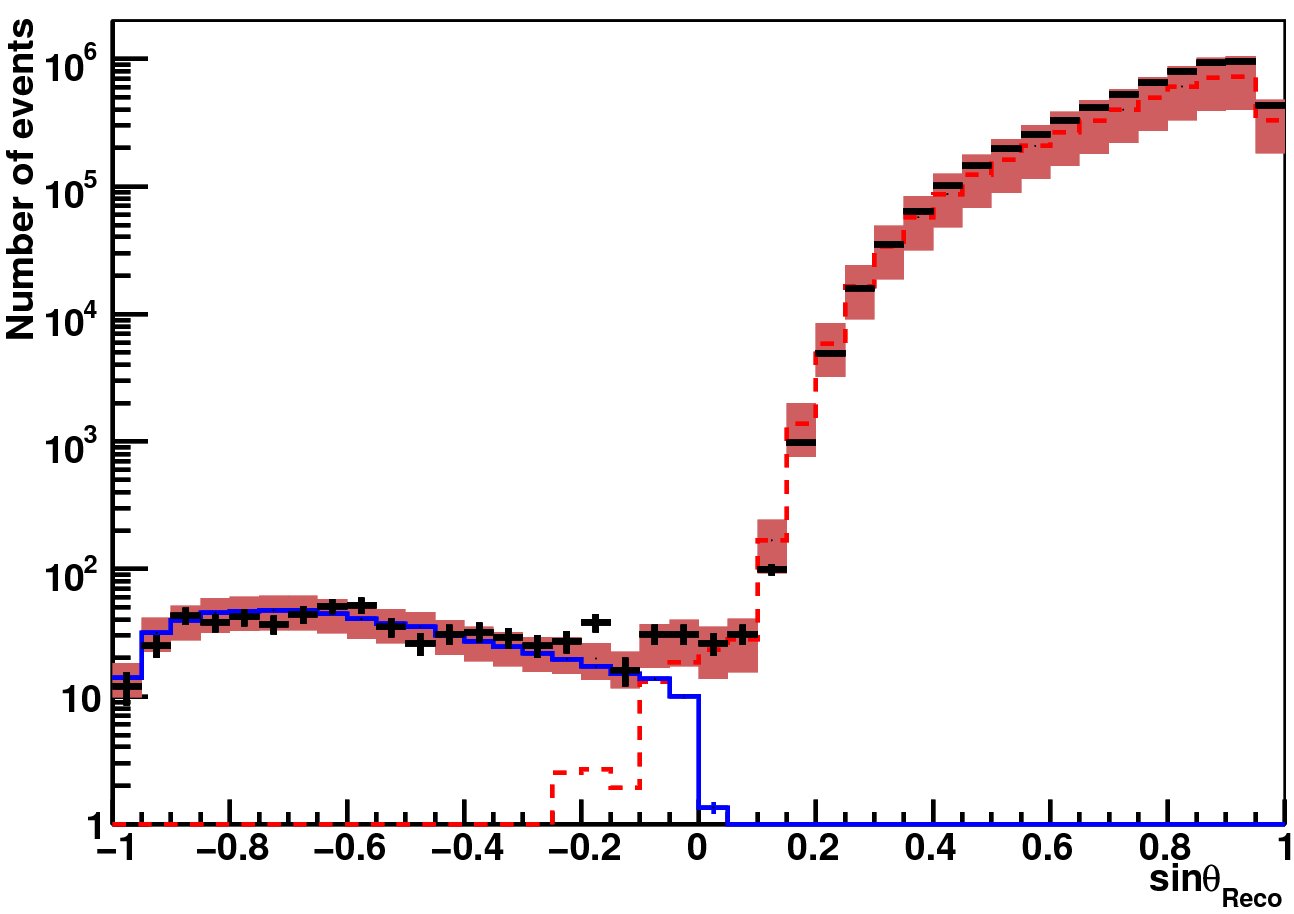,width=9.cm}
\caption{
Number of reconstructed muons in the 2008 ANTARES data, as a function of the
reconstructed zenith angle $\theta_\text{Reco}$ (black error bars). Also
indicated are the simulation results for atmospheric muons (red dashed), and
muons induced by atmospheric neutrinos (blue). The shaded band indicates the
systematic uncertainties. Figure taken from \pcite{Antares-fastreco}.
}
\label{atm-ang}
\end{figure}

An almost background-free separation of neutrino-induced upward-going muons from
the huge background of downward-going muons is the central requirement for an
underwater or under-ice telescope (see Sect.~\ref{sec-det-det-muo}). The
first-generation experiments (Baikal, ANTARES and AMANDA) have quickly mastered
this challenge, even more so IceCube. Figure~\ref{atm-ang} shows the rate of
muons as a function of the zenith angle $\theta$ as measured with ANTARES. Below
the horizon ($\theta<0$) the rate is well described by the expectation for
atmospheric neutrinos, above the horizon by that for atmospheric muons.

\begin{figure}[ht]
\sidecaption
\epsfig{file=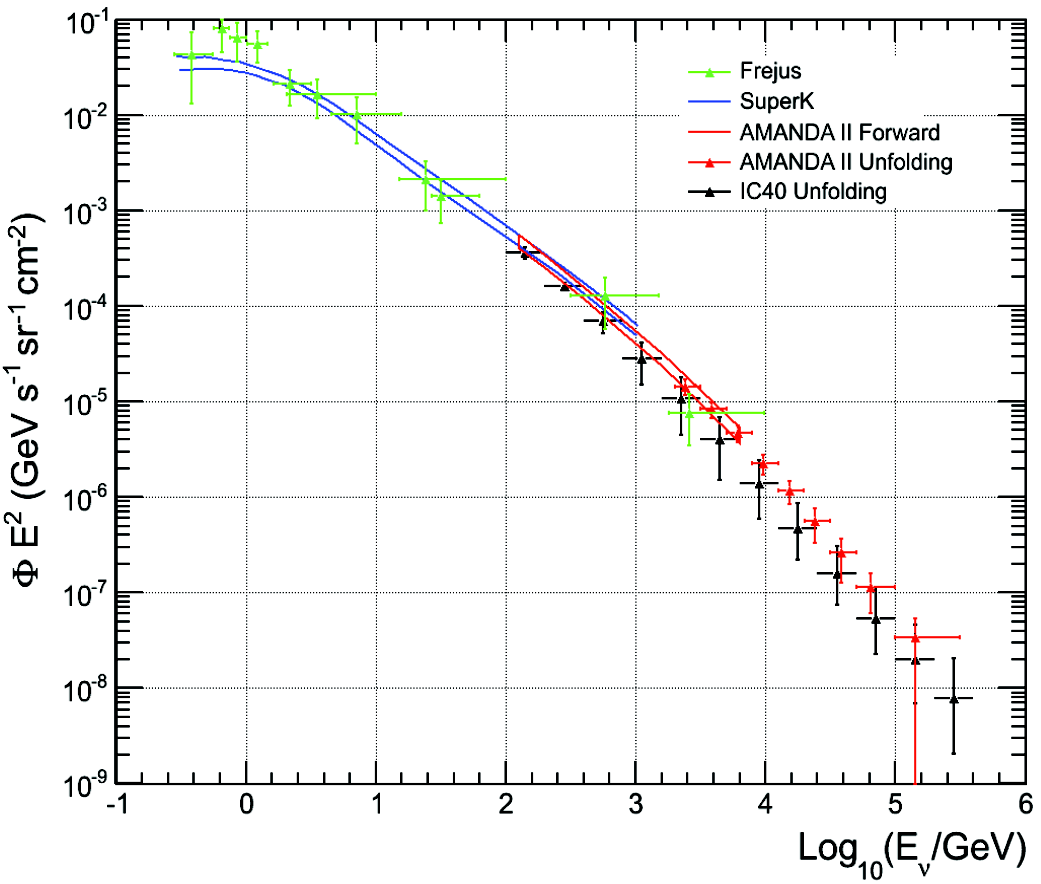,width=9.cm}
\caption{
Energy spectrum of atmospheric neutrinos. Green triangles: Fréjus
\pcite{Daum-1995}; blue band: Super-Kamiokande \pcite{SK-atmnu}; red band:
AMANDA forward folding analysis \pcite{icecube-2009d}; red triangles: AMANDA
unfolding analysis \pcite{icecube-2010d}; black triangles: IceCube-40 unfolding
analysis \pcite{icecube-2011c}.}
\label{atm-energy}
\end{figure}

The energy spectrum of atmospheric neutrinos is inferred from muon neutrino
charged current events since they offer the best event statistics and, for this
measurement, the least background. Measuring the energy spectrum is difficult
since the weak dependence of the Cherenkov light yield (i.e.\ of $\D E_\mu/\D
x$) on $E_\mu$ needs to be exploited; additionally, one has to take into account
that most muons lose an unknown fraction of their energy before reaching the
detector. The neutrino energy spectrum is therefore determined with the help of
sophisticated deconvolution procedures, inducing significant point-to-point
correlations. Two underground experiments have published atmospheric neutrino
energy spectra up to a few $\Tev$: The Fréjus experiment \cite{Daum-1995} and
Super-Kamiokande \cite{SK-atmnu}. AMANDA and IceCube have extended this energy
range by two orders of magnitude, up to 200 and $400\tev$, respectively. Figure
\ref{atm-energy} shows the spectra as published by these experiments. The data
are well compatible with the predictions for conventional atmospheric neutrinos. 
In particular, no excess at high energies is observed as yet; improved data
statistics from IceCube, however, will soon allow to test flux models for prompt
neutrinos from the decay of charm and bottom hadrons, which would show up as a
shoulder at some $100\tev$. In addition, constraints on the neutrino flux in
this energy range and beyond are used to place upper limits on the flux of
extra-terrestrial sources with a hard spectrum like $E_\nu^{-2}$ (see
Sect.~\ref{sec-phy-cos}).

\begin{figure}[ht]
\epsfig{file=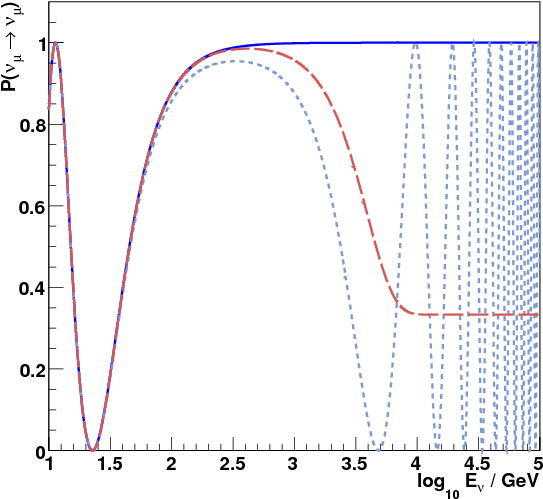,width=7.8cm}
\hfill
\epsfig{file=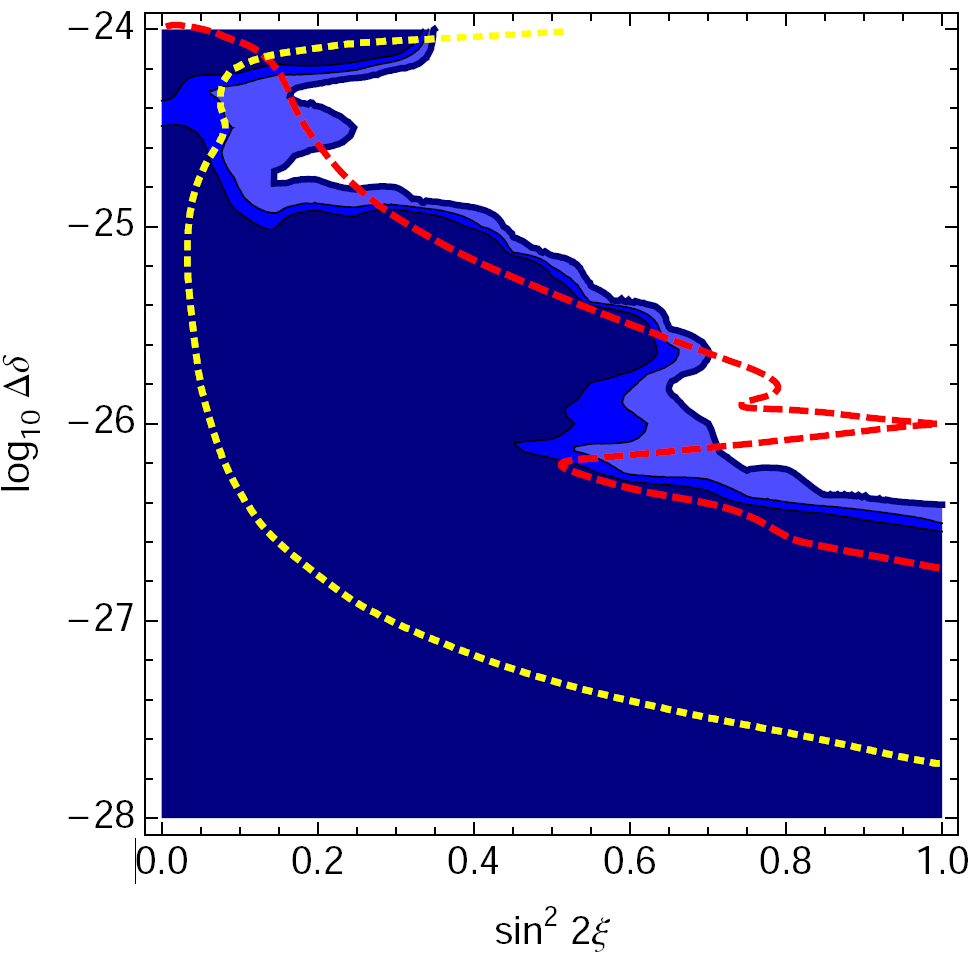,width=7.4cm}
\caption{
Left: $\nu_\mu$ survival probability as a function of energy for a baseline of
the Earth diameter and for conventional oscillations (solid line), VLI (dotted
line) with $n=1$, $\Delta\delta = 10^{-26}$ and maximal mixing, and quantum
decoherence (dashed line) with $n=2$ and $D_i^*= 10^{-30}\gev^{-1}$. See
\pcite{icecube-2009d} for analysis methods taking into account the limited
energy resolution. Right: allowed regions at 90\%, 95\% and 99\% confidence
levels (from darkest to lightest) for VLI-induced oscillation effects ($n=1$)
from AMANDA data, together with the 90\%-contour from a combined
Super-Kamiokande and K2K analysis (dashed line) and the projected IceCube
10-year 90\% sensitivity (dotted line). Figures are taken from
\pcite{icecube-2009d}.}
\label{QG}
\end{figure}

Atmospheric neutrinos also provide a tool to investigate neutrino oscillations. 
Standard oscillation lengths scale with $E_\nu$. For distances of the order of
the Earth diameter the first oscillation minimum is at $E_\nu\simeq24\gev$ (see
Fig.~\ref{QG}). Violation of Lorentz Invariance (VLI), as suggested by certain
quantum gravity theories, also leads to oscillation effects, with oscillation
lengths that scale with $1/E_\nu$ in the simplest case, $n=1$ (see
Sect.~\ref{sec-sci}). The left panel of Fig.~\ref{QG} shows the survival
probability as a function of neutrino energy for a baseline of the Earth
diameter and assuming maximal mixing for all three cases (conventional
oscillations, VLI oscillations and quantum decoherence, with the assumptions
given in the figure caption). Large new neutrino telescopes are an ideal tool to
look for non-standard oscillations, due to their ability to collect large
statistics of high-energy neutrinos having travelled over distances of several
thousand kilometres. From the non-observation of a deficit at high energies and
the angular dependence, limits on the relevant parameters can be derived. For
$n=1$ the 90\% CL upper limits from AMANDA are $2.8\times10^{-27}$ for
$\Delta\delta$ and $1.2\times10^{-27}$ for $D_i$ (see Sect.~\ref{sec-sci}). This
is similar to limits obtained from the much smaller underground experiments. For
$n=2$ the limits are $2.7 \times 10^{-31}\gev^{-1}$ for $\Delta\delta$ and
$1.3\times 10^{-31}\gev^{-1}$ for $D_i^*=D_i/E^{n-1}$
\cite{icecube-2009d}. IceCube promises another factor of ten in sensitivity over
the forthcoming years.

Another exotic signature of VLI would be directional dependences of neutrino
interactions or oscillations. Accelerator experiments like MINOS and K2K have
searched for a sidereal modulation of their interactions rate (i.e.\ for a
dependence on the orientation of their beam axes with respect to the Sun) and
did not find any effect. A search for a sidereal effect of rates of atmospheric
neutrinos in IceCube led to limits on the relevant parameters which improve
MINOS/K2K limits by factors of $3\rnge1000$, due to the longer baseline and the
higher energy (see \cite{icecube-siderial} and references therein).

\subsection{Cosmic neutrinos}
\label{sec-phy-cos}

High-energy cosmic neutrinos may either be identified as accumulation of events
pointing to a particular celestial direction (``point sources'') or as extended
diffuse emission, ranging from a few degrees (as for nearby supernova remnants)
to fully diffuse, expectedly isotropic neutrino flux; in both cases, the signal
needs to be distinguishable from the intrinsic background of atmospheric
neutrinos.

Point-source searches use the directional and energy information to reduce this
background. Cosmic neutrinos from a given source would cluster around the source
direction, with a point spread function determined by the angle between muon and
neutrino and by the detector angular resolution for muons, both depending on
energy. A further handle in point source searches comes from the fact that
generic extraterrestrial sources have a harder spectrum than atmospheric
neutrinos. For variable sources the time distribution may be used as additional
criterion, in particular if independent information is available (as e.g.\ gamma
observations in the case of GRB signals).

Searches for diffuse fluxes can only use the measured energy as criterion for
separating cosmic and atmospheric neutrinos. They thus critically depend on
a detailed understanding of the detector response as a function of
energy. Moreover, the high-energy tail of atmospheric neutrinos is dominated by
prompt neutrinos, whose flux has larger uncertainties than that of conventional
atmospheric neutrinos. 

In the following, we will first summarise the results on diffuse fluxes obtained
so far, then those for steady point sources, and finally methods and results for
variable sources.

\subsubsection{Searches for diffuse cosmic neutrino fluxes}
\label{sec-phy-cos-dif}

A diffuse neutrino signal may reveal itself as an excess of
\begin{enumerate}
\item 
high energy upward moving muons from $\Tev\rnge\Pev$ muon neutrinos;
\item 
high energy contained cascades;
\item 
extremely high energy events emerging from downward moving $\Pev\rnge\Eev$
neutrinos of all flavours.
\end{enumerate}
The first case corresponds to the standard signature of muon neutrinos
interacting via charged current reactions. Due to neutrino oscillations, the
expected ratio of neutrino flavours at Earth is $\nu_e:\nu_\mu:\nu_\tau=1:1:1$
(see Sect.~\ref{sec-sci-par-osc}). Since only charged current reactions of
$\nu_\mu$ and, in about 17\% of all cases, of $\nu_\tau$ produce high-energy
final-state muons, whereas all other reaction channels appear as cascades, the
relevance of detecting cascade events is obvious. The reduced directional
accuracy (much worse for cascades than for muons) is not very important for
diffuse fluxes.

For first-generation detectors, the ratio between the ``fiducial'' volume
(within which cascades can be reliably identified) and the effective volume for
muon detection is small. Therefore the cascade channel does not necessarily
provide superior sensitivity for diffuse fluxes. This is particularly true for
detectors in ice where light scattering makes the identification of cascades
challenging. The situation is more favourable in water, where even events beyond
the geometrical volume of the detector may be detected and reconstructed. Even
without precise reconstruction of distant cascades, however, the non-observation
of large signals from outside the geometrical volume can be used to derive upper
limits on the diffuse flux. This approach has been pursued by the Baikal
collaboration, who -- in spite of the small volume of the NT200 detector --
could compete with the much larger AMANDA detector over many years.

For detectors on the cubic-kilometre scale, with a much smaller
surface-to-volume ratio, the situation is more favourable for the cascade
channel. The relevance of contained cascades for diffuse flux measurements is
enhanced by the fact that their energy deposit in the detector volume -- and
thus their Cherenkov light output -- correlates much more strongly with the
neutrino energy than that of muons. Furthermore, the outer part of the
instrumented volume can be used as a veto against events mimicking cascades.

\begin{figure}[ht]
\begin{center}
\epsfig{file=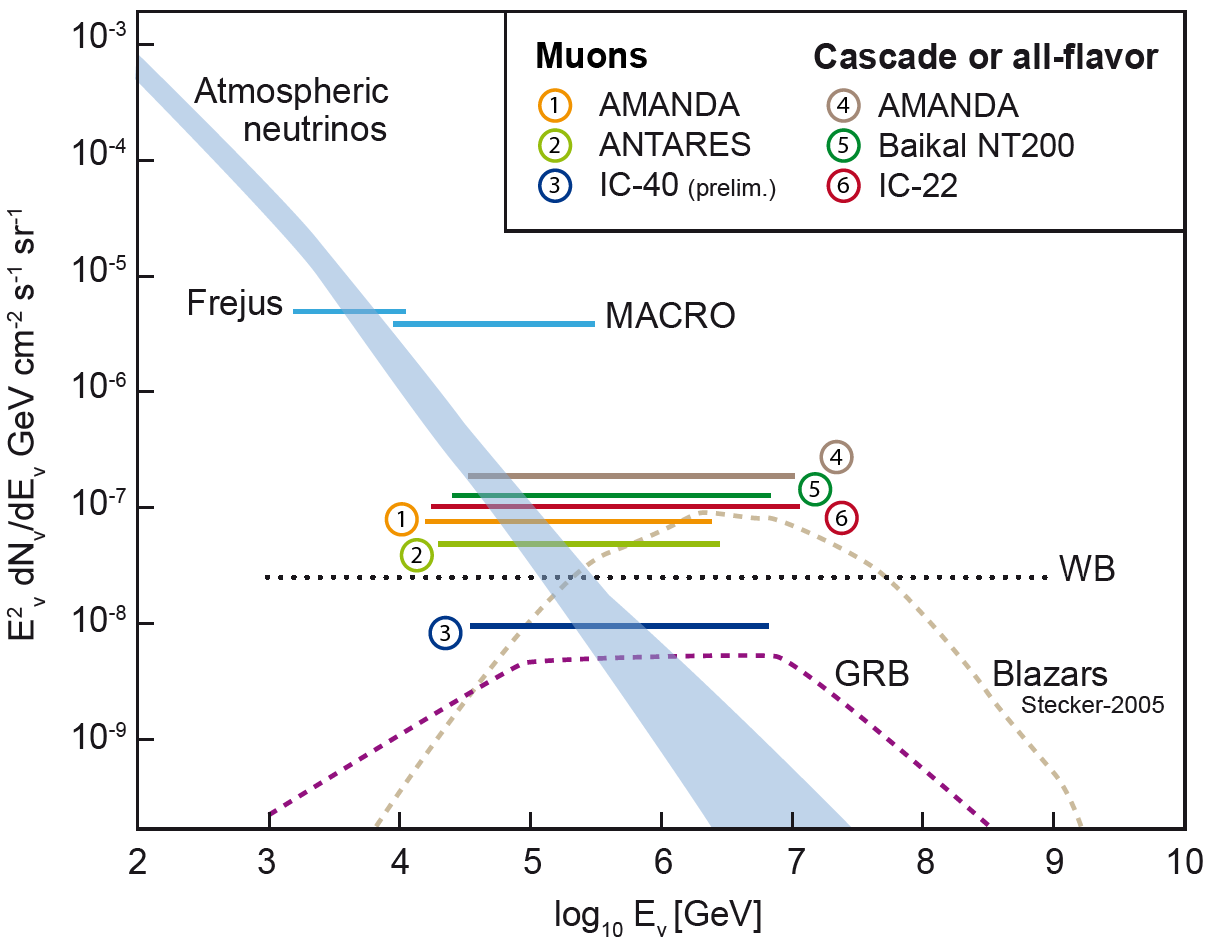,width=10.5cm}
\caption{
90\% C.L.\ integral upper limits on the diffuse flux of extraterrestrial
neutrinos. The horizontal lines extend over the energy range which would cover
90\% of the detected events from an $E^{-2}$ source (5\% would be below and 5\%
above the range). All model predictions have been normalised to one flavour,
i.e.\ all of the all-flavour limits have been divided by 3. The coloured band
indicates the measured flux of atmospheric neutrinos (see also
Fig.~\ref{atm-energy}), the broadening at higher energies reflects the
uncertainties for prompt neutrinos. The limits on muon neutrinos are from 807
days AMANDA \pcite{icecube-2007b}, 334 days ANTARES \pcite{Antares-diffuse}, and
375 days IceCube-40 (preliminary). Cascade/all flavour limits are from 807 days
AMANDA \pcite{icecube-2011b}, 1038 days Baikal-NT200
\pcite{Baikal-diff-Aynutdinov-2006,Baikal-diff-2009}, and 257 days IceCube-22
\pcite{IC-22-cascades}. The Fréjus and MACRO limits have been published in
\pcite{Rhode-1996} and \pcite{MACRO-diffuse}, respectively. Also indicated
is the Waxman-Bahcall (WB) bound \pcite{Waxman-Bahcall-1999}, see 
Sect.~\ref{sec-sci-neu-ext}.}
\label{UHE-diffuse}
\end{center}
\end{figure}

No significant excesses over atmospheric neutrinos or other kinds of background
has been observed so far, resulting in upper limits on the diffuse flux of
extraterrestrial high energy neutrinos. Figure~\ref{UHE-diffuse} summarises the
limits obtained in the $\Tev\rnge\Pev$ region. For each experiment and each
method only the best limit is shown. Remarkably, from the first limit derived
from the underground experiment Fréjus (1996) to the 2010 IceCube-40 limit, a
factor of 500 improvement has been achieved. Several models as e.g.\ the blazar
model of Stecker \cite{Stecker-2005} shown in the figure can be excluded. A
further factor of 10 improvement is expected over the next 2--3 years, using the
full IceCube detector and combining muon and cascade information. The expected
sensitivity is more than an order of magnitude below the Waxman-Bahcall bound,
and prompt atmospheric neutrinos will be detectable for all but the lowest
predictions \cite{Kowalski-2005}. Also, a test of the Waxman-Bahcall prediction
for the diffuse flux of neutrinos from Gamma Ray Bursts
\cite{Waxman-Bahcall-1997} will be in reach. We note, however, that this model
is much more easily tested with point source methods for transient, triggered
sources (see below).

\begin{figure}[ht]
\begin{center}
\epsfig{file=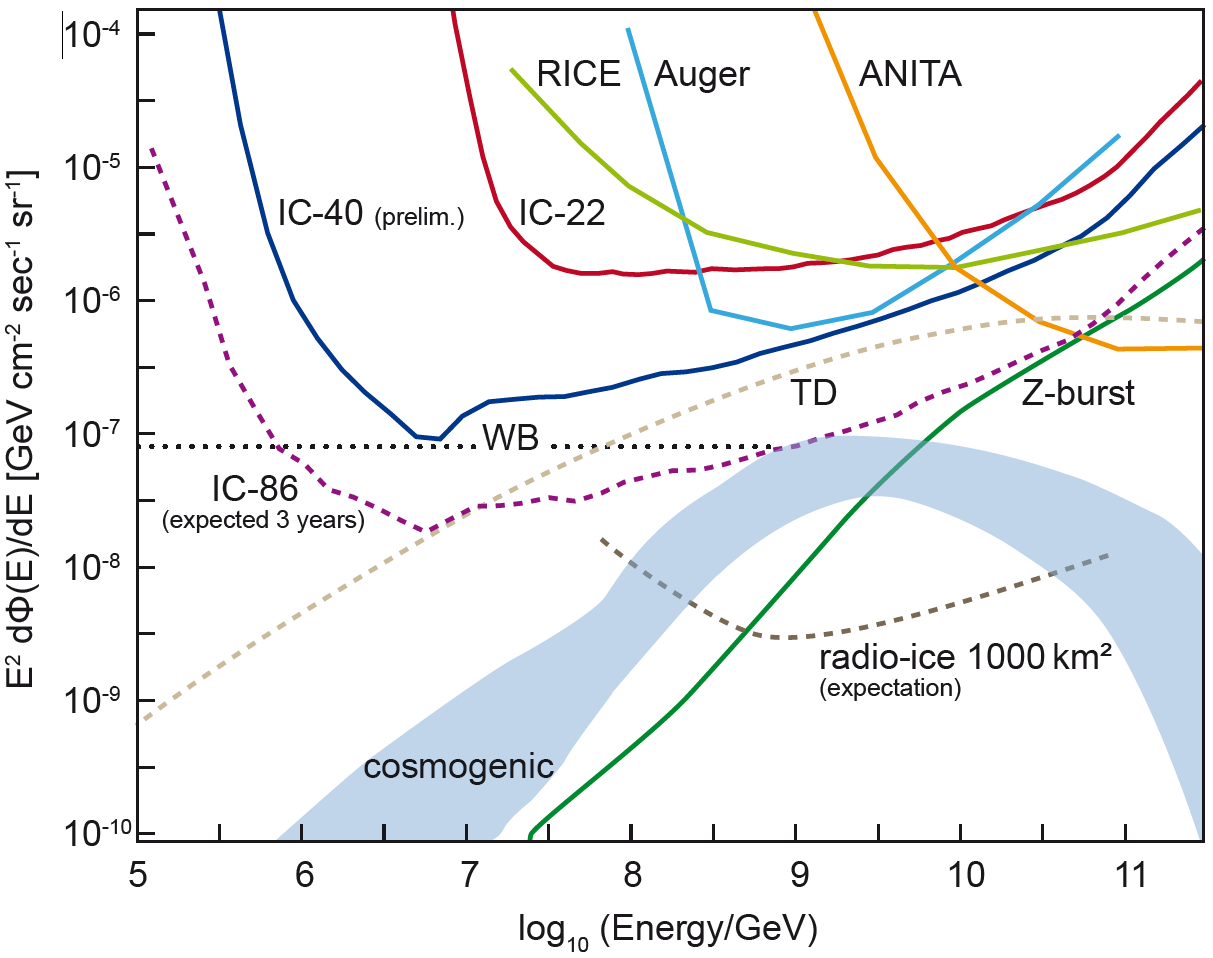,width=10.5cm}
\caption{
All-flavour 90\% C.L.\ differential upper limits on the flux of extraterrestrial
neutrinos in the $\Pev\rnge\Eev$ region. Limits are from the under-ice radio
array RICE \pcite{RICE-2006}, the air shower detector Auger \pcite{Auger-2010},
the radio balloon experiment ANITA \pcite{ANITA-2009}, IceCube-22
\pcite{icecube-2010d}, and IceCube-40 (preliminary) \pcite{icecube-EHE2011}. 
Also given are the expectations for 3 years of operation of the full IceCube
detector and for a $1000\km^2$ shallow radio detector at the South Pole. The
coloured band corresponds to different predictions for cosmogenic neutrinos from
GZK processes. References for the GZK scenarios, the Z-burst prediction and the
SUSY top-down scenario (TD) can be found in \pcite{icecube-EHE2011}. For
recent predictions we refer to \pcite{Berezinsky-2011,Kotera-2011,Gelmini-2011}.}
\label{EHE-diffuse}
\end{center}
\end{figure}

Neutrinos in the multi-$\Pev$ to $\Eev$ region must come from above or from
close to the horizon to be detected since the Earth is not transparent for
neutrinos of such energies (see Fig.~\ref{Transmission} in
Sect.~\ref{sec-det-int}). The IceCube collaboration has performed analyses which
are tailored to these neutrinos of extreme energies and yield limits
\cite{icecube-2010d,icecube-EHE2011} that meanwhile are equal or better than
those obtained from radio or air shower detectors (see Sect.~\ref{sec-alt}). 
Figure~\ref{EHE-diffuse} summarises the differential limits in the
$\Pev\rnge\Eev$ range obtained by IceCube and by radio and air shower methods.

The figure demonstrates that IceCube can merely detect cosmogenic neutrinos from
cosmic ray interactions with the cosmic microwave background radiation. Also
Auger (with a multi-year exposure) and ANITA (with another flight) will have
little chances to detect this ``guaranteed source''. Only dedicated detectors
like e.g.\ next-generation radio or acoustic detectors or maybe a very large,
sparsely instrumented configuration of KM3NeT will provide the required
sensitivities. Arguably, other sources than cosmogenic neutrinos may populate
the highest-energy region, making explorative studies in this energy range
valuable. The alternative detection methods (radio, acoustic, air shower
detection) are addressed in Sect.~\ref{sec-alt}.

\subsubsection{Searches for steady neutrino point sources}
\label{sec-phy-cos-ste}

In initial searches for point sources
\cite{MACROpoint-2001,SuperK-point-2009,icecube-2009a}, the sky was subdivided
into bins of a size which was optimised to the detector resolution and the
expected signal energy spectrum (typically $E^{-2}$). In these ``binned
searches'', the signal would appear as an excess over atmospheric neutrinos in a
certain bin. In order not to lose sources through signal sharing between
adjacent bins, the search had to be repeated with shifted bins, resulting in
trial factors which effectively reduced the sensitivity. Present searches, in
contrast, use likelihood functions which account for the smearing of the signal
with a given point spread function. The probability of an event originating from
a given source is calculated from a 2-dimensional probability density (typically
a Gaussian), with the width defined by the uncertainty of the event direction. 
This ``unbinned method'' turned out to be up to 40\% more sensitive than the
binned \cite{Braun-2008}. Using an energy estimator to distinguish hard
extraterrestrial from soft atmospheric neutrino spectra further enhances the
sensitivity.

\begin{figure}[ht]
\epsfig{file=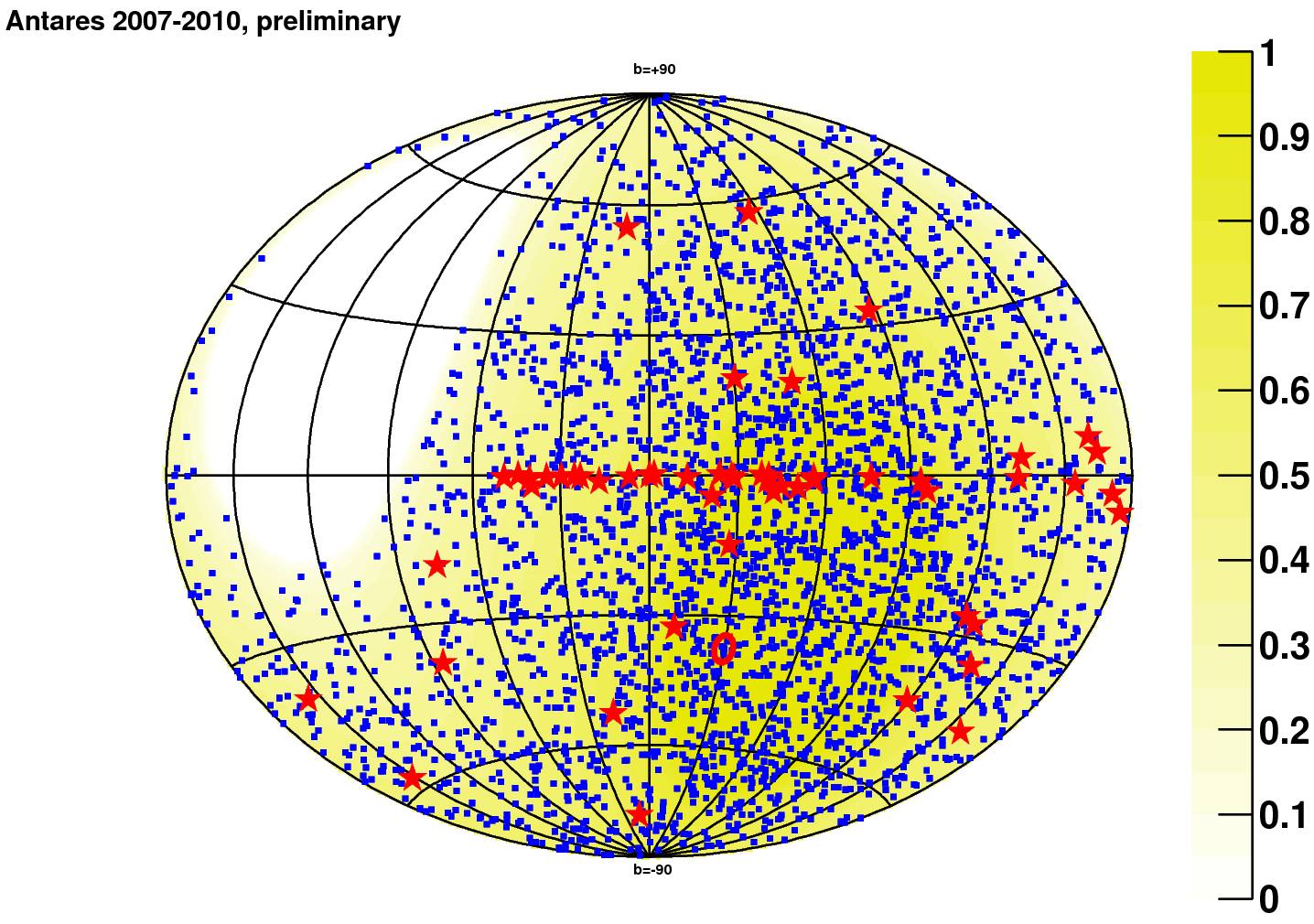,width=9.1cm}
\hfill
\epsfig{file=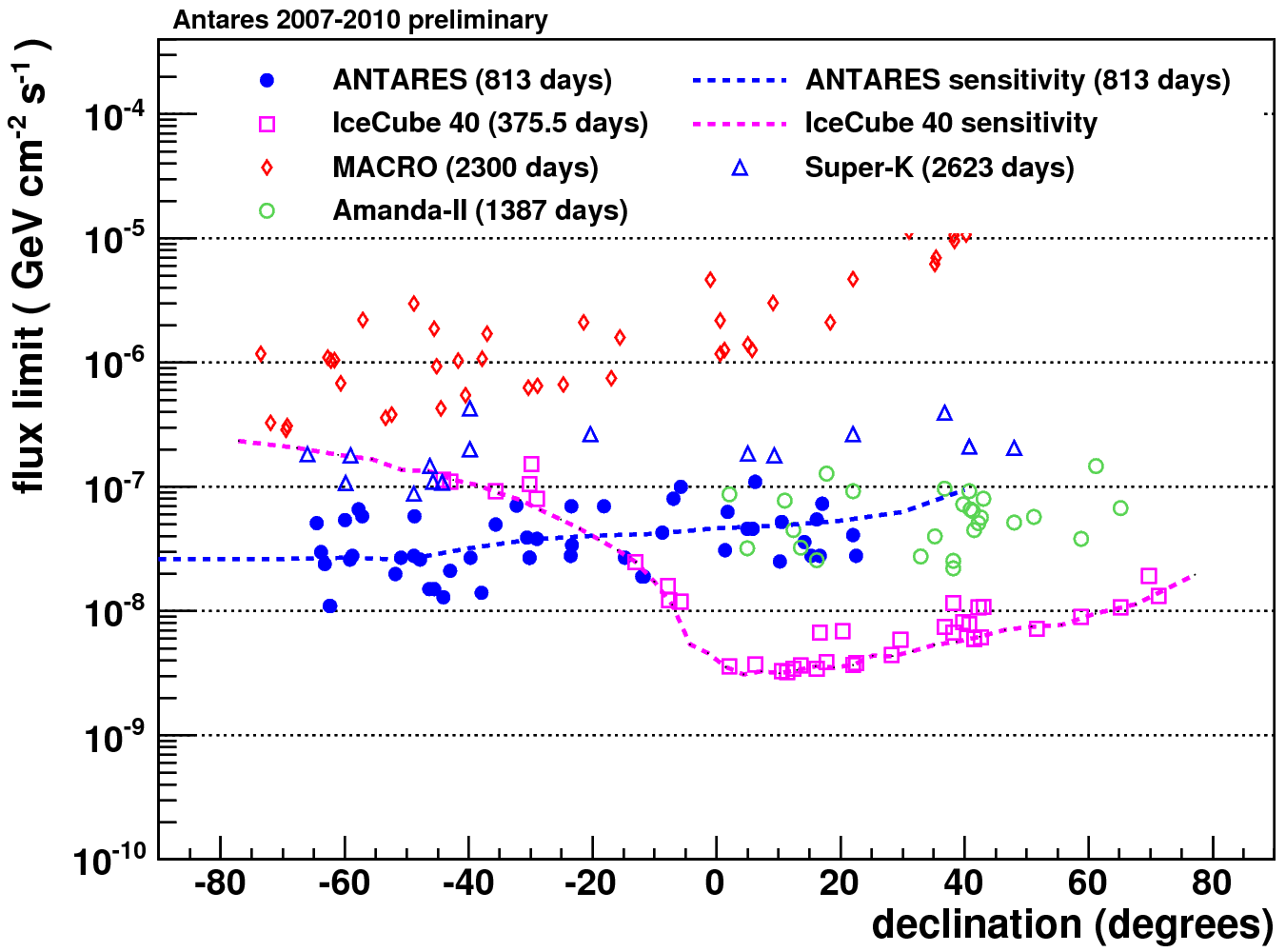,width=6.9cm}
\caption{
Left: Equatorial skymap of neutrino-induced muon events from 813 days of ANTARES
data from 2007--2010 (preliminary). The background colour scale indicates the
sky visibility in percent of the time. The most significant accumulation of
events, marked with a red circle, is compatible with the background expectation. 
Right: Flux limits for the 51 candidate sources marked in the skymap, compared
to the expected average ANTARES sensitivity (blue dotted line) and to
corresponding results from MACRO \pcite{MACROpoint-2001}, Super-Kamiokande
\pcite{SuperK-point-2009}, AMANDA \pcite{icecube-2009a} and IceCube
\pcite{IC-40-point}.}
\label{skymap-Antares}
\end{figure}

The upper limit on a $E^{-2}$ neutrino flux from any point source in the
Northern sky measured with AMANDA over seven years is
$E^2\phi<5.2\times10^{-8}\flunit$ (averaged over the Northern hemisphere, using
a binned method) \cite{icecube-2009a}. Based on 813 days of data taking, ANTARES
has recently released a sky map and preliminary values \cite{Antares-pointlims}
for the flux limits of a set of preselected potential point sources (see
Fig.~\ref{skymap-Antares}), which improve the Southern sky limits from 14 years
of Super-Kamiokande data.

The real big leap, however, is being made with IceCube. First point source
results were published for the 22-string configuration operated in 2007
\cite{IceCube-2009e}. In \cite{icecube-2009g} this search was extended up to
$45^\circ$ above the horizon. The background from down-going atmospheric muons
was suppressed by hard cuts selecting only few thousand of the billions of
background events, including most of the $\Pev\rnge\Eev$ neutrinos. In that way,
a reasonable sensitivity to the high-energy tail of hard neutrino spectra (e.g.\
$E^{-2}$) was achieved. Naturally, the IceCube sensitivity to a corresponding
source in the Southern hemisphere is worse than for Northern sources since the
analysis relies exclusively on the tiny high-energy tail of the neutrino flux. 
Note, however, that IceCube would be almost blind to Southern sources with an
energy cut-off in the $\Pev$ range or below (some remaining sensitivity is
expected from DeepCore, which, using the rest of IceCube as a veto layer, could
identify neutrinos which interact within the DeepCore volume with $4\pi$
acceptance). For unbroken $E^{-2}$ spectra, a cubic-kilometre detector at the
South Pole can compete with a Northern first-generation detector like ANTARES up
to a declination of $45^\circ$. This, on the other hand, means that there is a
broad declination region where the combination of IceCube and ANTARES data will
give a better sensitivity than IceCube or ANTARES alone. Such combined analyses
are presently underway.
 
\begin{figure}[ht]
\sidecaption
\epsfig{file=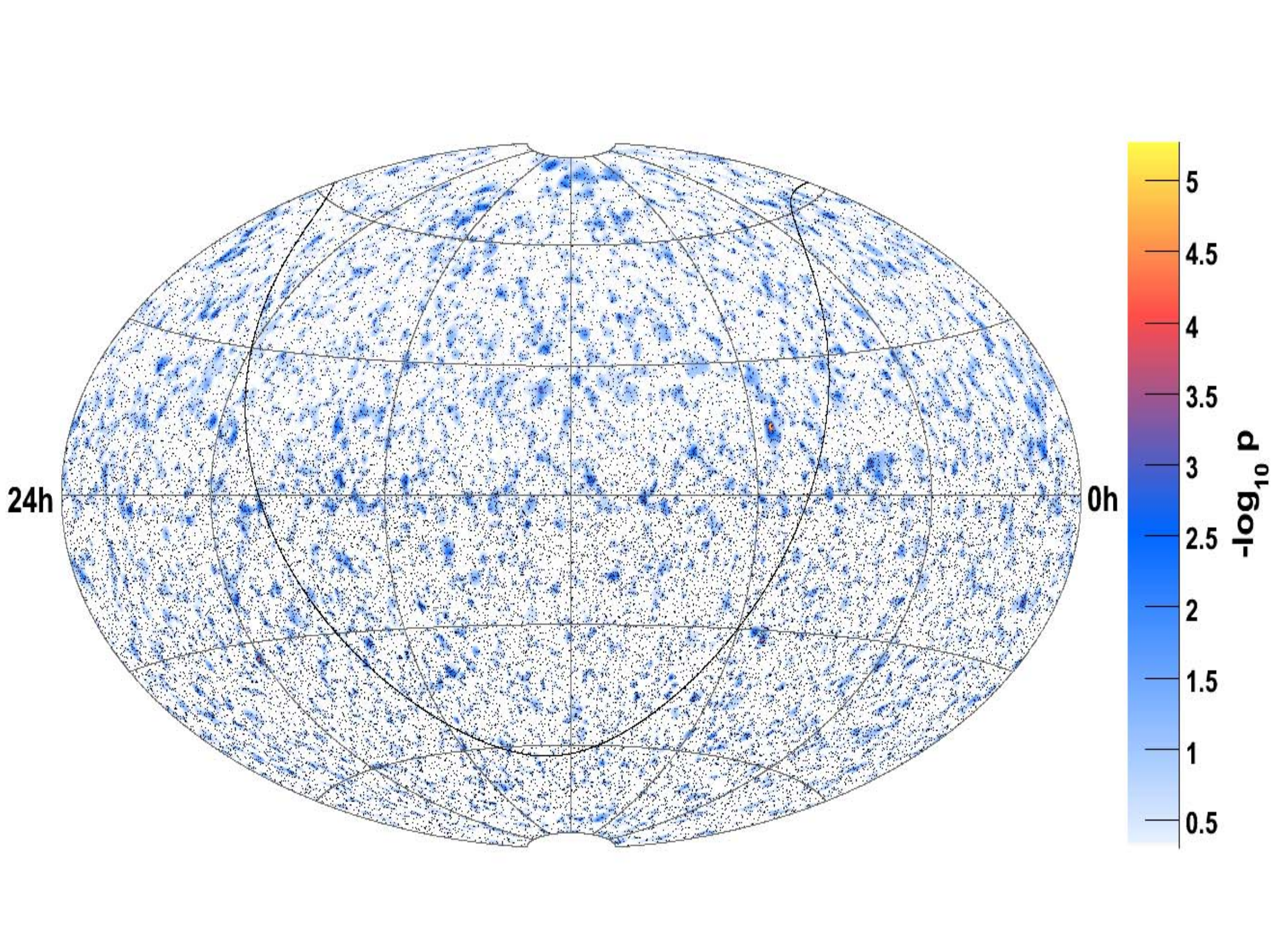,width=10.5cm,trim=0 50 0 70}
\caption{
Equatorial skymap of pre-trial significances of the all-sky point source search
with IceCube-40 \pcite{IC-40-point}. Each dot represents one neutrino event, the
colour scale indicates the significance of event accumulations. The Galactic
plane is shown as black curve.}
\label{skymap-IceCube}
\end{figure}

\begin{figure}[ht]
\sidecaption
\epsfig{file=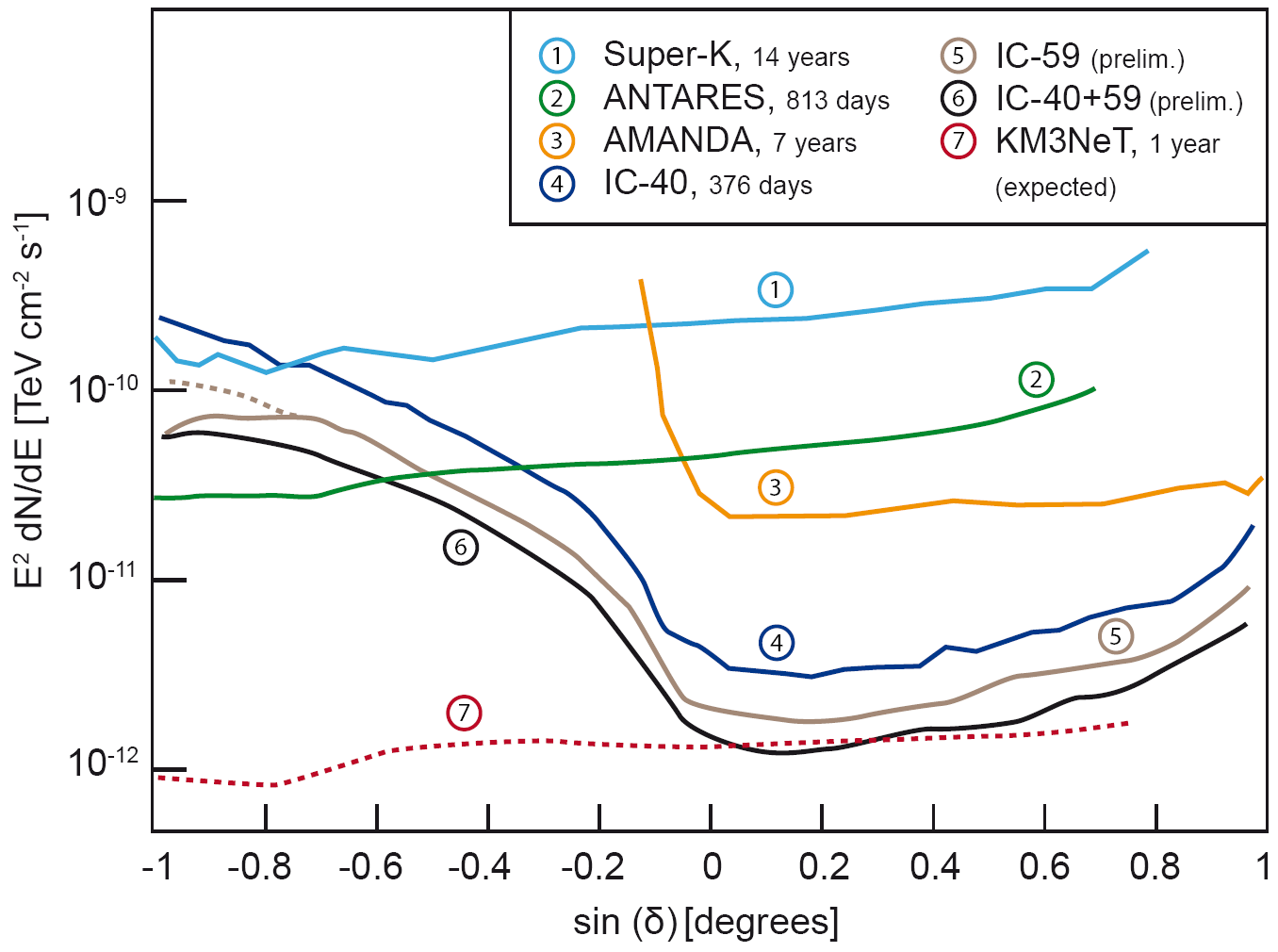,width=9.cm}
\caption
{Point source neutrino flux sensitivities (median expected limits at 90\%
C.L.) from various experiments: Super-Kamiokande \cite{SuperK-point-2009},
AMANDA \cite{icecube-2009a}, IceCube-40 \cite{IC-40-point}, IceCube-59, the
sum of IceCube-40 and IceCube-59 (preliminary results) and ANTARES
\cite{Antares-pointlims}. Also shown is the expected sensitivity from one year
of KM3NeT data \cite{km3net-tdr}.}
\label{point-limits}
\end{figure}

Figure \ref{skymap-IceCube} shows the full-sky map derived from IceCube-40 data
\cite{IC-40-point} taken between April 2008 and May 2009. The data sample
contains 14\,121 events from the Northern sky, mostly muons from atmospheric
neutrinos, and 22\,779 events from the Southern sky, mostly high energy
atmospheric muons. With this sample, five searches were performed: (i) a scan of
the entire sky for point sources; (ii) an analysis for 39 predefined potential
sources (reducing the huge trial factor inherent to a full-sky search); (iii)
the remaining three analyses stacked source candidates of different
astrophysical nature and then searched for an excess. These classes were (iii-a)
16 sources of $\Tev$ gamma rays; (iii-b) 127 starburst galaxies and (iii-c) 5
nearby galaxy clusters. The predefined sources included, among others, the Crab
Nebula, the Geminga pulsar, the star cluster Cygnus OB2 and the active galaxies
Markarian 421, Markarian 501, 3C273 and M87. The results of the predefined
point-source search (ii) are shown in Fig.~\ref{skymap-Antares} (right). All
search results are consistent with the null hypothesis; the ``hottest spot''
from the all-sky scan is found at a right ascension of $\text{RA}=113.75^\circ$
and a declination of $\text{DEC}=15.15^\circ$ and has a chance value of 18\%. 
Figure~\ref{point-limits} compiles the limits from previous experiments, from
the different IceCube stages and the expected sensitivity of one year of KM3NeT
data. We note that the combined data of IceCube-40 and IceCube-59 surpass the
mark of $1\km^3\times1\,$year and thus exceed 1 year worth of data from the full
IceCube detector. Very soon, a factor of 1000 improvement of the sensitivity to
point sources will have been reached when compared to the very first AMANDA
point source paper from 2000 \cite{AMANDAB10-2000}.

\begin{figure}[ht]
\sidecaption
\epsfig{file=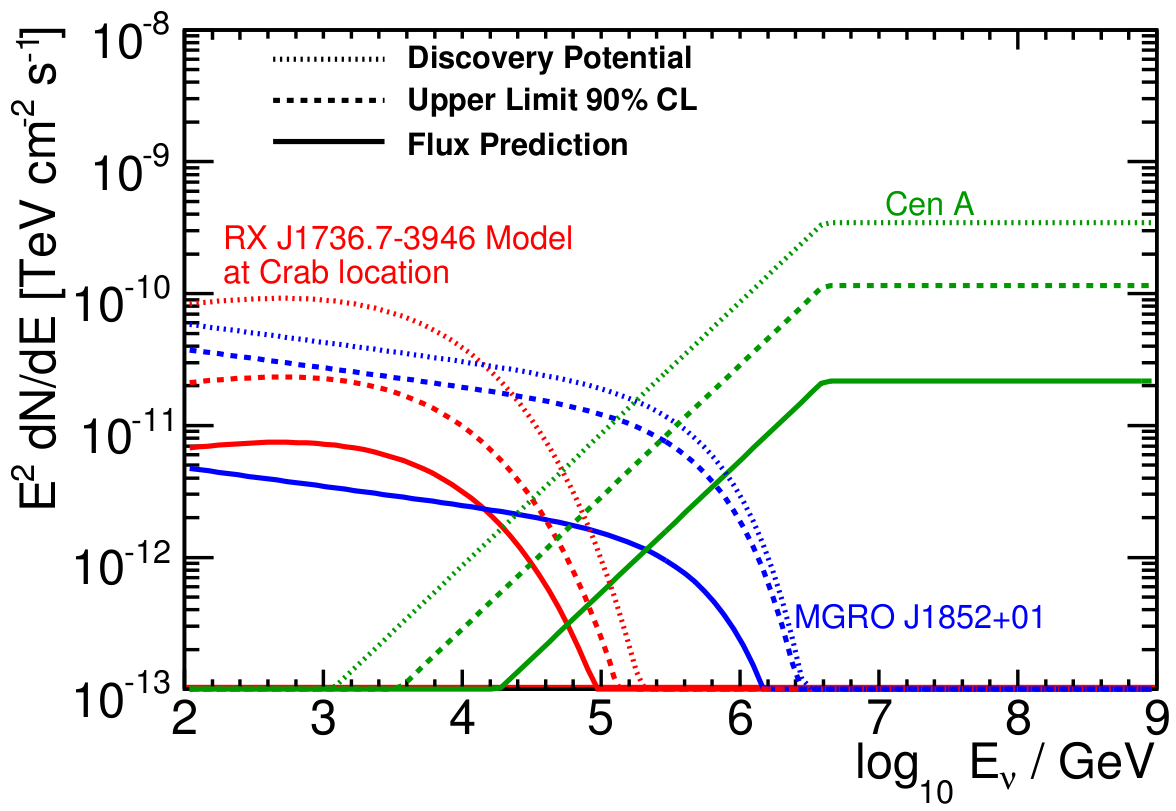,width=10.cm}
\caption{
Differential flux for three theoretical models of neutrino point sources,
confronted with the 90\% C.L.\ upper limit and the $5\sigma$ discovery potential
from IceCube-40 (see text for explanations). Figure taken from
\pcite{IC-40-point}.}
\label{ps-models-and-limits}
\end{figure}

Where do we stand with respect to predictions for known sources? 
Figure~\ref{ps-models-and-limits} gives a ``taste'' of the answer. It shows the
differential fluxes for three theoretical models and confronts them with the
90\% C.L.\ upper limit and the $5\sigma$ discovery potential from IceCube-40. 
Its shows the $\nu_\mu + \nubar_\mu$ predictions for the supernova remnant
RX-J1713.7-3946 \cite{Morlino-2009} which was moved from its real position at
the Southern sky the to the location of the Crab Nebula at the Northern sky, for
MGRO J182+01 \cite{Halzen-Kappes-Murchadha-2008}, a gamma-ray source observed by
the Milagro experiment, and for the active galaxy Centaurus A \cite{Koers-2008}. 
Note that recent measurements from the Fermi satellite \cite{Fermi-RX} seem to
disfavour a dominant hadronic origin of the gamma rays observed from
RX-J1713.7-3946 -- the underlying assumption for the shown prediction (see also
Sect.~\ref{sec-sci-neu-gal}). The conclusion is that optimistic model
predictions are about one order of magnitude below present IceCube-40 limits. 
This does not rule out a discovery with a few years of full IceCube data. On the
other hand, it seems that with IceCube -- after having made the gigantic leap of
a factor 1000 in sensitivity improvement -- we are just scraping the discovery
region. The main discovery potential therefore may remain for source phenomena
which are not covered by the models addressed in
Fig.~\ref{ps-models-and-limits}, e.g.\ ``dark'' sources without significant
high-energy gamma emission, or variable sources discussed in
Sect.~\ref{sec-phy-cos-var}. The implications for a Northern-hemisphere
detector, however, are obvious: Its sensitivity must substantially exceed that
of IceCube!

\subsubsection{Searches for variable neutrino sources}
\label{sec-phy-cos-var}

Many astrophysical sources are known to have a variable flux at different
wavelengths. Examples for such flaring sources are Active Galactic Nuclei, Soft
Gamma Ray Repeaters, and Gamma Ray Bursts. Binary systems often show a periodic
behaviour, as pulsars do. Neutrino searches for steady point sources integrate
over time and continuously accumulate atmospheric neutrinos. That makes them
less sensitive to the detection of flares when compared to searches over smaller
time windows. Actually, compared to searches integrating over a full year,
time-dependent searches can achieve $5\sigma$ discoveries with two (five) times
less signal events for flares of 1 day (1 second) duration
\cite{Braun-2009,Alba-2009,Gora-2011} -- even if the there is no information on
the flaring state from electromagnetic observations (so-called ``untriggered''
searches). In case this information is available and used (``triggered
searches''), the gain can even be larger, in particular for Gamma Ray Bursts
signalled by satellite detectors.

IceCube has performed an untriggered search for flares from selected sources
using IceCube-40 data (April~2008 to May~2009), and a triggered search for
sources monitored by Fermi-LAT, SWIFT and Imaging Cherenkov Telescopes using
IceCube-22 and IceCube-40 data (May~2007 to May~2009)
\cite{IceCube-2011-time-dependent}. These analyses cover time intervals from
$40\musec$ to a year. No significant signal evidence was found.

In this context it may be worth mentioning that of seven flares which were
selected for the triggered search with IceCube-22, five did not show a related
neutrino signal in the selected time window, while one event was observed for
each of the other two sources. One of these sources is the Active Galaxy
1ES1959+650, which has a ``neutrino history'': In an analysis of AMANDA data
taken from 2000-2003, five events where recorded from the direction of
1ES1959+650. Interestingly, three of these came within 66~days in 2002. Two of
the three neutrinos were coinciding within about a day with gamma-ray flares
observed by the gamma-ray telescopes HEGRA and Whipple
\cite{Markus-thesis,Bernardini-2005}. Excitingly, one of these 2 flares was not
accompanied by an X-ray flare; such ``orphan flares'' would be expected for
hadronic outbursts where the X-ray flux from synchrotron radiation of the
electron plasma is absent. This result was quickly followed by two theoretical
papers, one claiming that the corresponding neutrino flux would not fit any
reasonable assumption on the energetics of the source \cite{Reimer-2005}, the
other claiming that scenarios yielding such fluxes were conceivable
\cite{Halzen-Hooper-2005}. Since the analysis was not a fully blind analysis, it
turned out to be impossible to determine chance probabilities for this event,
and actually the result was never published in a journal. However, it initiated
consideration to send alerts to gamma-ray telescopes in case time-clustered
events from a certain direction would appear. These ``Target-of-Opportunity''
alert programs are described below.

A special analysis \cite{icecube-2011d} was performed for SN\,2008D, a
core-collapse supernova at a distance of 27\,Mpc discovered by the SWIFT
satellite on Jan.~9, 2008. Core-collapse supernovae might emit mildly
relativistic jets (Lorentz factor $\Gamma=3\rnge10$) in which neutrinos could be
produced by proton-proton collisions. Assuming that the jet pointed to the
Earth, the non-observation of coinciding neutrinos with IceCube was used to
constrain the total energy of the released energy and $\Gamma$. Actually,
according to current models, the full IceCube detector could detect up to 100
events for a core-collapse supernova at 10\,Mpc distance.

\begin{figure}[ht]
\begin{center}
\epsfig{file=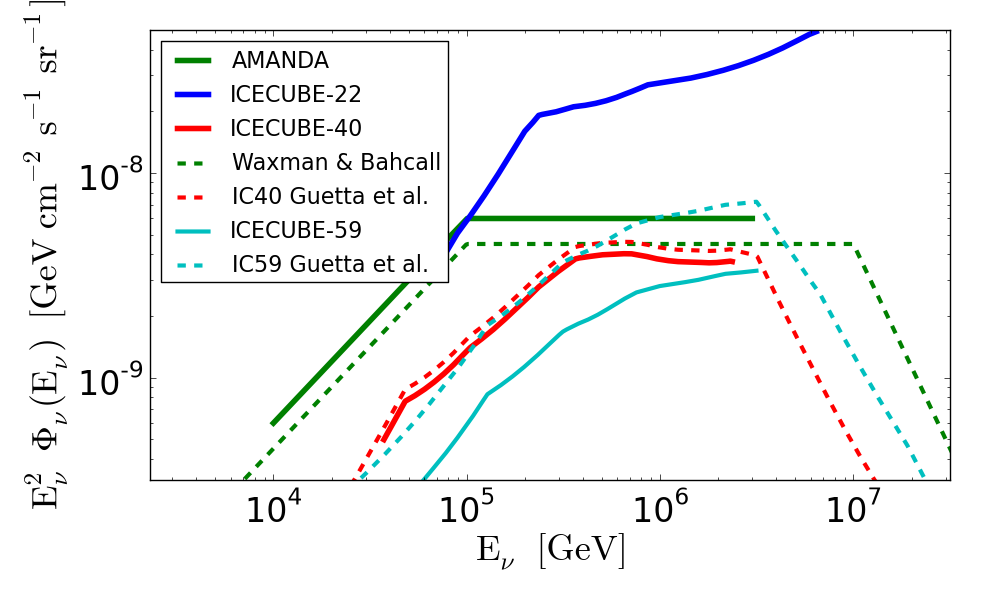,width=11.cm}
\caption{
90\%C.L.\ upper limits on the neutrino fluxes from GRBs set by AMANDA,
IceCube-22, IceCube-40 and IceCube-59 (preliminary). The IceCube limits are
calculated with respect to the flux expected from the model of Guetta et al.\
\cite{Guetta-2004}, the AMANDA limit with respect to the Waxman-Bahcall flux
\cite{Waxman-Bahcall-1997} which assumes an average shape of the GRB spectra. 
The fluence limits for single GRBs have been translated to a limit on the
diffuse neutrino flux from GRBs, multiplied with $E^2$. The corresponding model
predictions are indicated as dashed lines. The strikingly worse limit of
IceCube-22 as compared to AMANDA and later IceCube configurations is due to
small statistics and one observed conicidence. Figure adapted from
\cite{icecube-2011h}.}
\label{GRB-limits}
\end{center}
\end{figure}

One of the most promising sources of high energy neutrinos are Gamma Ray Bursts
(GRBs). As mentioned in Sect.~\ref{sec-sci-neu-ext}, neutrino emission can be
modeled for three GRB phases: the precursor phase when the jet is still forming
and no electromagnetic radiation is escaping \cite{Razzaque-2003}; the prompt
phase coinciding with the burst in gamma rays (see e.g.\
\cite{Waxman-Bahcall-1997,Guetta-2004}); and the afterglow phase
\cite{Waxman-Bahcall-2000}. Direction, time, duration and gamma spectrum of GRBs
are provided by satellite observations. Early searches for neutrino events
coinciding with GRBs have been performed with Super-Kamiokande, Baikal and
AMANDA. The AMANDA analysis \cite{icecube-2008a} looked for neutrinos coinciding
with any of 408 well-located GRBs recorded in the Northern hemisphere between
1997 and 2003. No neutrinos were found during or immediately prior to the GRBs. 
An upper limit on the diffuse flux from all GRBs was derived, which was still a
factor of 1.5 above the flux predicted in \cite{Waxman-Bahcall-1997} for the
prompt phase, but already two times below the precursor model published in
\cite{Razzaque-2003}. In a recent analysis, neutrinos recorded with IceCube-40
have been analysed with respect to coincidences with any of 117
Northern-hemisphere GRBs recorded in the time interval April 2008 to May 2009
\cite{icecube-2011e}; preliminary results for IceCube-59 are also already
available \cite{icecube-2011h}. Differently to the AMANDA analysis, the expected
neutrino spectra have been calculated according the the observed gamma spectra
from each individual GRB and following the prompt phase model of
\cite{Guetta-2004}. Again no excess of neutrinos close in time and direction is
observed, resulting in a limit below the predictions of
\cite{Waxman-Bahcall-1997,Guetta-2004}, see Fig.~\ref{GRB-limits}. Even though
this does not rule out the general picture of the GRB fireball model of
\cite{Waxman-Bahcall-1997} which assumes that cosmic rays of highest energy
essentially all emerge from GRB, the amount of energy of the accelerated protons
transfered to pions is obviously smaller than assumed in these models. The
transfer parameter is uncertain by a factor of 3 or more; this and other
uncertainties in the model parameters leave room for a possible confirmation of
these models with future IceCube data.

\subsubsection{Alert programs}
\label{sec-phy-cos-ale}

Neutrino telescopes monitor essentially a full hemisphere. In contrast, most
gamma-ray, X-ray and optical observatories can observe only a small fraction of
the sky at any given moment. Therefore only a small subset of neutrino data can
be used for correlation studies with these data. The ability to identify such
correlations can be improved by optical follow-up (OFU) programs
\cite{Kowalski-Mohr-2007,Antares-OFU-2008,Antares-OFU-2011}. In these programs,
the observation of doublets or triplets of neutrino events from the same
direction (or one exceptionally energetic single neutrino event) triggers a
small network of automated 1--2 metre telescopes to point in that direction. If
the neutrinos are emitted from jets in core-collapse supernovae in other
galaxies or from GRBs, the optical telescopes could identify the rising light
emission from the supernova or the GRB afterglow. As was shown in
\cite{Kowalski-Mohr-2007}, OFUs can improve the sensitivity to neutrinos from
supernovae and GRBs by a factor of 2--3. Both IceCube and Antares are running
OFU programs, triggering the optical telescopes TAROT and ROTSE (for details see
\cite{Antares-OFU-2011}. No coincidences have been reported so far, although the
IceCube analysis provided some interesting (but not significant) doublets. 
Presently, these programs are extended to include the X-ray satellite SWIFT.

Another follow-up program in IceCube is known as Neutrino Target of Opportunity
(NToO) \cite{NToO-2007}. It was motivated by the coincidence in 2002 between two
neutrinos from the direction of the AGN 1ES1959+650 and gamma ray flares from
the same source (see Sect.~\ref{sec-phy-cos-var}). In this case, the problem is
complicated by the comparatively long duration of AGN flares (hours to weeks
instead of seconds to minutes for supernovae or GRBs). Doublets within such long
time intervals appear too often from {\it any} direction of the sky to be useful
for such a program. Therefore only selected sources, which are known to show
flaring behaviour, are used for the NToO. In case a ``cluster in time'' of
events from the direction of one of these sources has accumulated, a trigger is
sent to the MAGIC gamma-ray telescope in La Palma. As soon as possible MAGIC is
then pointed to that direction.

A further spectacular alert program is the ``standard'' IceCube supernova alert. 
As mentioned in \ref{sec-sec-ice-snb}, the low dark noise rate of the
photomultipliers allows for a mode of operation which is best suited for ice:
the detection of burst neutrinos via the feeble increase of the summed count
rates of all photomultipliers. This increase would be produced by millions of
neutrino interactions at energies up to some tens of $\Mev$ within several
seconds \cite{amanda-sn-2002,icecube-SN-2010}. The tiny amount of light from one
of these interactions would usually fire only the closest photomultiplier, so
that no event-by-event reconstruction is possible. With this method, a supernova
in the centre of the Galaxy would be detected with extremely high confidence,
and the onset of the pulse could be measured in unprecedented detail since
IceCube records the counting rate in millisecond steps. Even a SN\,1987A-type
supernova in the Large Magellanic Cloud would be identifiable.

Since neutrinos leave a supernova a few hours before light is emitted, neutrino
signals can be used to issue an early alert to optical astronomers. Actually,
IceCube alerts are fed into the SuperNova Early Warning System, SNEWS
\cite{SNEWS-2004}. Currently, the detectors Super-Kamiokande (Japan), LVD and
Borexino (Italy) as well as IceCube contribute to SNEWS, with a number of other
neutrino detectors and gravitational wave detectors planning to join in the near
future.

The average supernova rate for a galaxy like ours is estimated to be 2--5 per
century (see \cite{Raffelt-2007} for an overview). Since 1987 no supernovae have
been observed in our Galaxy or its neighbourhood. There might have been,
however, Galactic supernovae obscured by matter between them and us which
consequently could have been visible only in neutrinos. The non-observation of
low-energy neutrino signals by underground neutrino detectors like
Super-Kamiokande and by AMANDA/IceCube sets a 90\% C.L.\ upper limit of about
eight per century on the average rate of such bursts (we have rescaled the 2007
limit obtained in \cite{Raffelt-2007} for 25 years of non-observation to
meanwhile 29 years).

\subsection{Dark matter and other exotic particles}
\label{sec-phy-dar}

In addition to neutrinos from astrophysical objects, as discussed above,
neutrino telescopes are also sensitive to potential neutrino fluxes from dark
matter annihilations (indirect dark matter searches) and to hypothesised exotic
particles (see Sect.~\ref{sec-sci-par}). The status of the corresponding
experimental search results and the future expectations are discussed in this
section.

\subsubsection{Dark matter}
\label{sec-phy-dar-dar}

As explained in Sect.~\ref{sec-sci-par}, dark matter could accumulate in the
Galactic centre, in the Sun or in the centre of the Earth and eventually release
neutrinos via self-annihilation processes. These neutrinos would be detectable
as an excess over the irreducible background of atmospheric neutrinos. Several
underground and underwater/ice experiments (Baksan, MACRO, Super-Kamiokande,
Baikal, AMANDA, IceCube (see \cite{amanda-2006} and references therein) as well
as ANTARES \cite{Lim-2009,Motz-2009,Lim-2011}) have searched for an excess of
neutrinos from the centre of the Earth or the Sun, IceCube also for signals from
the Galactic halo and the Galactic centre \cite{Rott-2009,Heros-2011} -- all
without having identified any significant signal. Alternative indirect searches
using gamma rays \cite{Conrad-2011} or other particle messengers
\cite{Porter-2011} have not found clear dark matter signals either.

\begin{figure}[ht]
\sidecaption
\raisebox{6.5cm}{\begin{minipage}[t]{0.6\textwidth}
\begin{center}
\epsfig{file=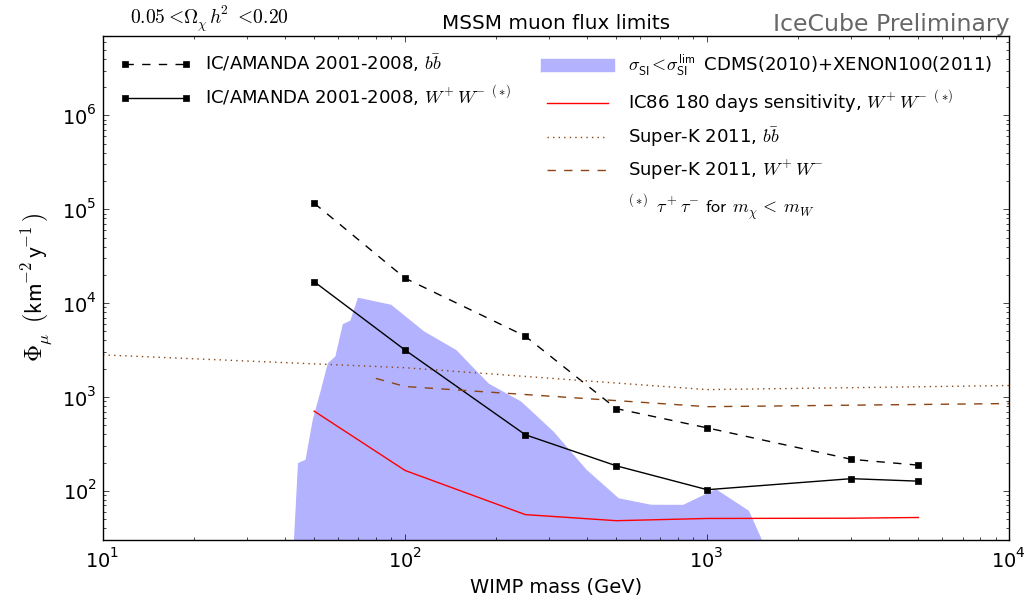,width=\textwidth,height=6.5cm}
\epsfig{file=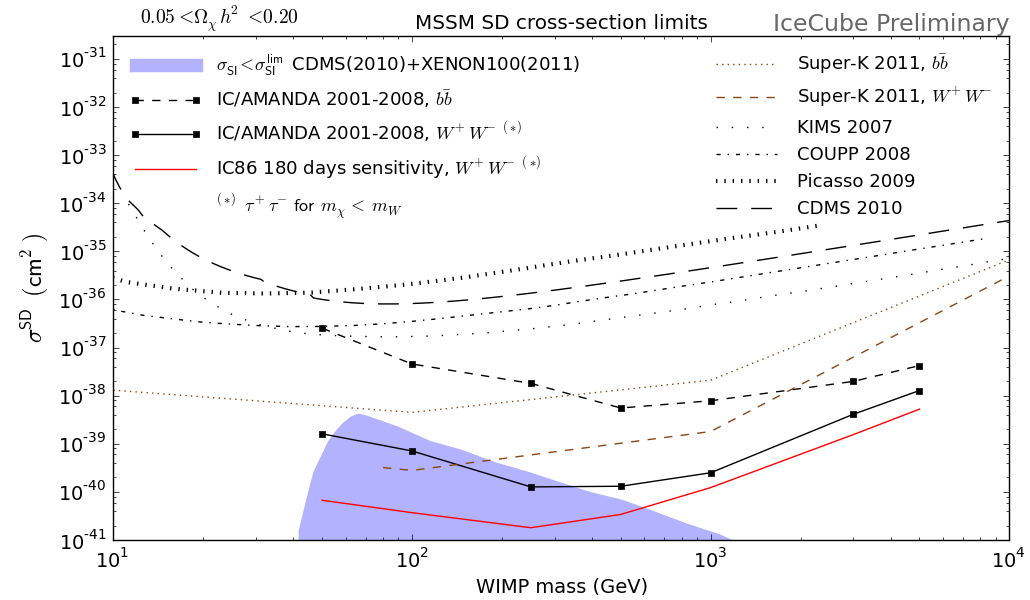,width=\textwidth,height=6.5cm}
\end{center}
\end{minipage}}
\caption{
90\% C.L.\ exclusion limits on the muon flux induced by neutrinos from
neutralino annihilations in the Sun (top) and on the spin-dependent
neutralino--proton cross section (bottom) from indirect dark matter searches in
the neutrino channel, as functions of the WIMP mass. The shaded areas indicate
the MSSM parameter regions not excluded by direct searches (see text). Indicated
are the limits obtained with AMANDA, IceCube-22 and IceCube-40 (squares)
assuming soft ($b\bbar$) and hard ($W^+W^-$) neutrino spectra from WIMP
annihilations; the expected sensitivities of IceCube with DeepCore (red lines);
and exclusion limits obtained by other experiments. Limits have been rescaled to
a common muon threshold of $1\gev$. See \pcite{Heros-2011} for references to the
various experimental data; the Super-Kamiokande results are taken from
\pcite{Kappl-2011}.}
\label{DM-limits}
\end{figure}

The most sensitive search for neutrino-induced muons from the direction of the
Sun has been performed with IceCube-22, using data taken between March and
September 2007 \cite{Heros-2011}. In these months, the Sun is below the horizon
at the South Pole. Simulations have been run for several neutralino masses each
with two annihilation channels, a hard channel (annihilation into $W^+W^-$), and
a soft channel (annihilation into $b\bbar$). Figure~\ref{DM-limits}\,(left)
shows the limits on the flux of muons induced by neutrinos produced in MSSM
neutralino annihilations in the centre of the Sun as a function of the
neutralino mass $m_\chi$. Since the average energy of the muons would increase
with neutralino mass and hardness of the decay channel, and since the muon
detection probability increases with energy, the limits are tighter for harder
decay spectra and for larger $m_\chi$. The green shaded area represents flux
predictions from presently allowed combinations of the MSSM parameters based on
direct search experiments. The grey shaded area represents the parameter space
that would still be allowed if direct searches would improve their sensitivity
by a factor of 1000.

Since the muon flux is proportional to the capture rate of neutralinos in the
Sun, the muon flux limit can be converted into a limit on the neutralino-proton
scattering cross section, in particular its spin-dependent part,
$\sigma^{SD}_{\chi+p}$ (the Sun is mainly a proton system). In
\cite{Heros-2011}, this conversion has been performed by assuming equilibrium
between capture and annihilation rates. A conservative result is obtained by
further assuming that the spin-independent cross section vanishes. The result of
this analysis is given in Fig.~\ref{DM-limits}\,(right) and compared to other
limits from indirect searches by Super-Kamiokande and IceCube-22 as well as the
best current limits from direct searches. The figure illustrates the potential
of indirect searches to explore the parameter space of spin-dependent neutralino
interactions with a sensitivity exceeding by far that of direct searches.

Using the same method, data from AMANDA and IceCube have been used to set limits
on other dark matter candidates, like the lightest Kaluza-Klein particle
(hypothesised in theories with extra dimensions, with an expected mass range
similar to that for neutralinos) and Simpzillas (super-heavy relic particles
with masses in the range of $10^4\rnge10^{18}\gev$). We refer to
\cite{icecube-2010f,Heros-2011} for details.

\begin{figure}[ht]
\sidecaption
\epsfig{file=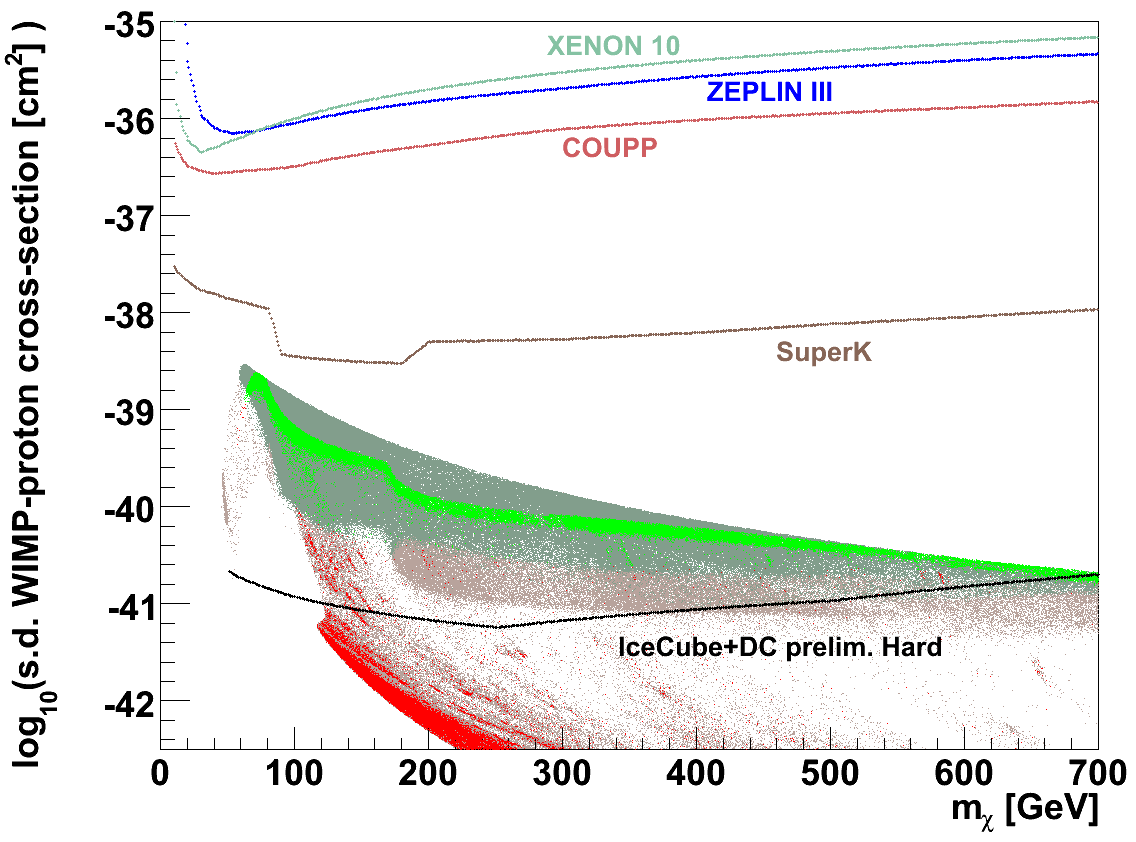,width=9.cm}
\caption{
Spin-dependent WIMP-proton cross section as a function of the neutralino mass. 
The coloured points correspond to mSUGRA parameters consistent with constraints
from cosmic microwave measurements. The models within (beyond) the discovery
reach of KM3NeT are coloured green (red). Limits of selected other experiments
are also indicated. The KM3NeT and IceCube limits are for 10 years of data
taking.
\label{DM-msugra}
}
\end{figure}

An alternative approach was pursued to estimate the KM3NeT sensitivity to
neutrinos from dark matter annihilations. The mSUGRA parameter space was
scanned; for each parameter set the resulting neutrino spectrum was calculated
and the resulting event rate in KM3NeT determined, taking properly into account
neutrino propagation through the Sun. The resulting limits for the
spin-dependent cross section, equivalent to Fig.~\ref{DM-limits}\,(right), are
shown in Fig.~\ref{DM-msugra} (see \cite{km3net-tdr} for further explanations
and references to the experimental data). KM3NeT in the currently planned
configuration (see Sect.~\ref{sec-sec-kmt}), with its comparably high detection
threshold, is less sensitive to indirect dark matter signals than IceCube with
DeepCore.

\subsubsection{Magnetic monopoles, Q-balls and nuclearites}
\label{sec-phy-dar-mag}

\begin{figure}[ht]
\sidecaption
\epsfig{file=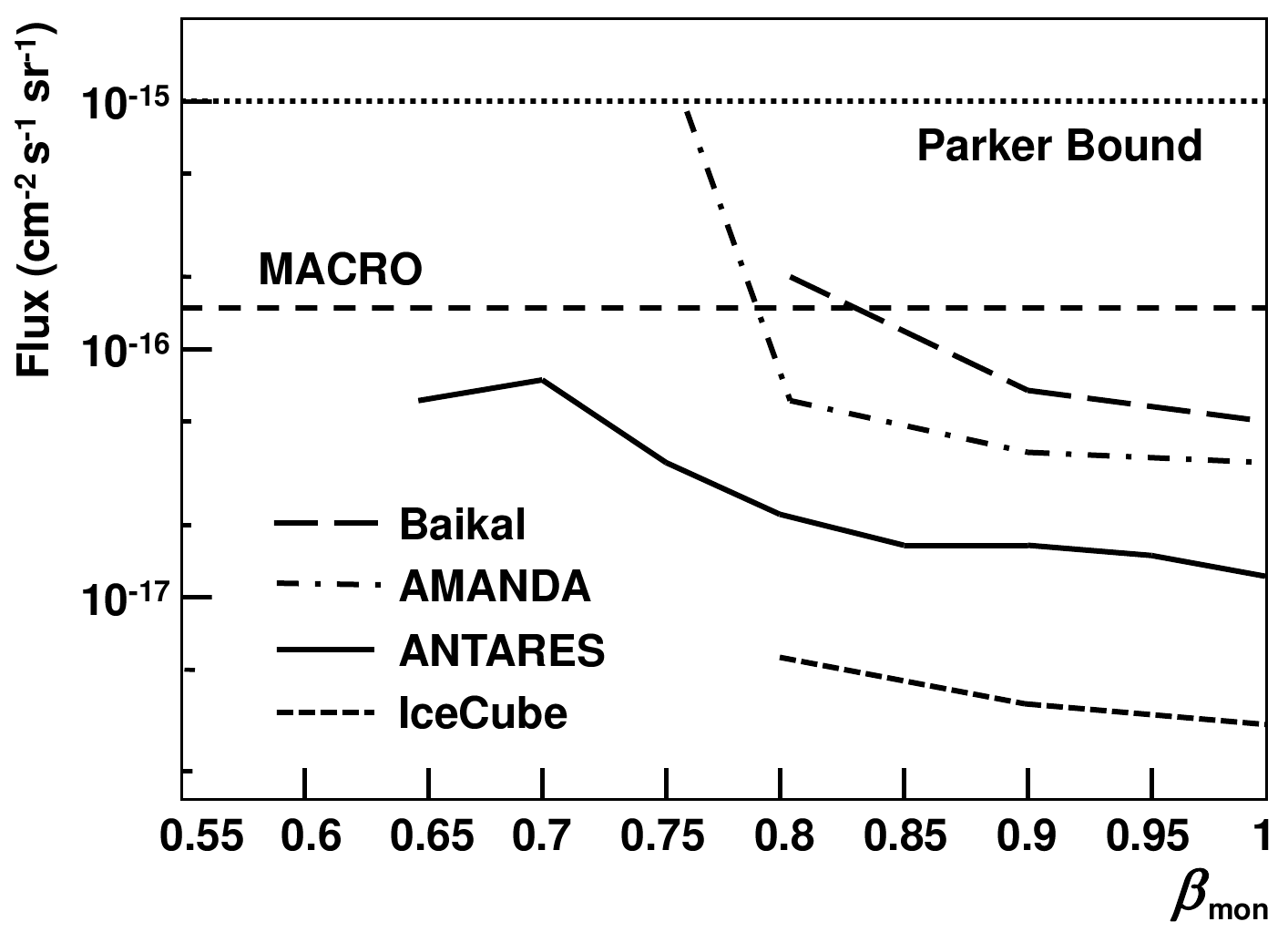,width=8.0cm}
\caption{
90\% C.L.\ upper limits on the flux of relativistic magnetic monopoles as a
function of their velocity, $\beta_\text{mon}=v_\text{mon}/c$. The ANTARES and
IceCube results are preliminary.
\label{Monopoles}
}
\end{figure}

In water and ice detectors, relativistic magnetic monopoles can be identified
via their strong light emission, 8300 times more intense than that of a
minimally ionising muon. Even below the Cherenkov threshold (velocity
$v_c\approx0.75\,c$), down to $v\approx0.5\,c$, the light emission is large due
to accompanying delta electrons. Figure \ref{Monopoles} shows the monopole flux
limits obtained by Baikal, AMANDA and ANTARES, where only in the ANTARES
analysis light emission below $v_\text{mon}=0.75\,c$ was addressed. The Baikal
flux limit of $4.7\times10^{-17}\cm^{-2}\scnd^{-1}\sr^{-1}$ for upward-moving
monopoles with $\beta\approx1$ is based on 1040 live days of NT200 (1998-2002)
\cite{Baimon}; that of AMANDA of $2.8\times10^{-17}\cm^{-2}\scnd^{-1}\sr^{-1}$
for upward-going and $3\times10^{-16}\cm^{-2}\scnd^{-1}\sr^{-1}$ for
downward-going monopoles on the data taken with AMANDA in a single year (2000)
\cite{icecube-2010e}; the preliminary ANTARES limit for upward-moving monopoles
with $\beta_\text{mon}\approx 1$ is at $1\times10^{-17}\cm^{-2}\scnd^{-1}\sr^{-1}$
\cite{antares-monopoles}; the preliminary IceCube limit derived from one year of
data taking with the 22-string configuration is about
$3\times10^{-18}\cm^{-2}\scnd^{-1}\sr^{-1}$ \cite{icecube-mon-2011}. IceCube in
its full configuration will be yet another factor 3--4 more sensitive and thus
advance to more than three orders of magnitude below the Parker bound
\cite{Turner-1982}, which is set by the condition that magnetic monopoles must
not destroy the Galactic magnetic fields. Note that due to the absence of
predictions for magnetic monopole fluxes a non-observation cannot exclude
certain theoretical models (as it is the case for dark matter searches).

Slow GUT monopoles, Q-balls and nuclearites (see Sect.~\ref{sec-sci-par}) would
heavily ionise the matter they traverse, or even catalyse proton decays along
their path. Such particles with $\beta$ in the range $10^{-5}\rnge10^{-2}$ have
been searched for with underground and underwater/ice detectors; in absence of
any signals, upper limits on their fluxes of the order of
$10^{-16}\cm^{-2}\scnd^{-1}\sr^{-1}$ are derived
\cite{Giacomelli-2007,baikal-1,baikal-2,Arvid-Thesis}. Note that limits based on
the catalysis of proton decay are conditional since they depend on an assumed
catalysis cross section. For instance, the limit of the Baikal experiment on GUT
monopoles is $2[50]\times10^{-16}\cm^{-2}\scnd^{-1}\sr^{-1}$ for catalysis cross
sections of $10^{-28}[10^{-30}]\cm^2$ and $\beta=10^{-4}$. Again, IceCube data
will improve this sensitivity by about two orders of magnitude. Needless to
mention that intense light emission of a slow object would be a spectacular
signature which would make the first clear observation of such an event a real
sensation -- regardless of its concrete interpretation as GUT monopole, Q-ball
or nuclearite.

\subsection{Cosmic ray physics}
\label{sec-phy-ray}

In detectors underwater/ice and underground, cosmic rays are recorded via the
punch-through muons generated in air showers above the detector. Ideally, the
deep detector is complemented by a surface air shower array for such
measurements. In this case, the air shower array essentially responds to the
total energy contained in the electromagnetic shower component, whereas the deep
detector responds to the hadronic shower component from which the muons
originate. Combination of both pieces of information allows for disentangling
energy spectra and mass composition of the primary particles on a statistical
basis. This approach has been pioneered with underground/air shower hybrid
detectors like Baksan/Carpet (Russia) or MACRO/EAS-Top (Italy), followed by
AMANDA/SPASE (South Pole Air Shower Experiment) and recently IceCube/IceTop
(both at South Pole). The experimental results indicate that the average mass of
the primary particles increases with energy in the knee region around
$10^{15}\ev$, see \cite{SPASE} and references therein. The combination
IceCube/IceTop covers an enormously enlarged solid angle and an area of a full
square kilometre, and yields data of unprecedented quality. With its huge
statistics it will extend the energy range up to $10^{18}\ev$, where the
transition from Galactic to extragalactic cosmic rays is expected.

Surface detectors can also be used to estimate the angular resolution of the
deep detector, by comparing the reconstructed directions of air showers and the
corresponding muon. Comparing data from SPASE and AMANDA, the muon angular
resolution of AMANDA was measured to be $2^\circ\rnge2.5^\circ$
\cite{SPASE-calib}. The angular resolution of IceCube was determined via the
shadow of the moon with respect to cosmic rays. This shadow can be measured via
down going muons with a high significance within a few weeks, and confirms that
angular resolution and absolute pointing of IceCube are about $1^\circ$ or
better \cite{IceCube-Moon}.

\begin{figure}[ht]
\sidecaption
\epsfig{file=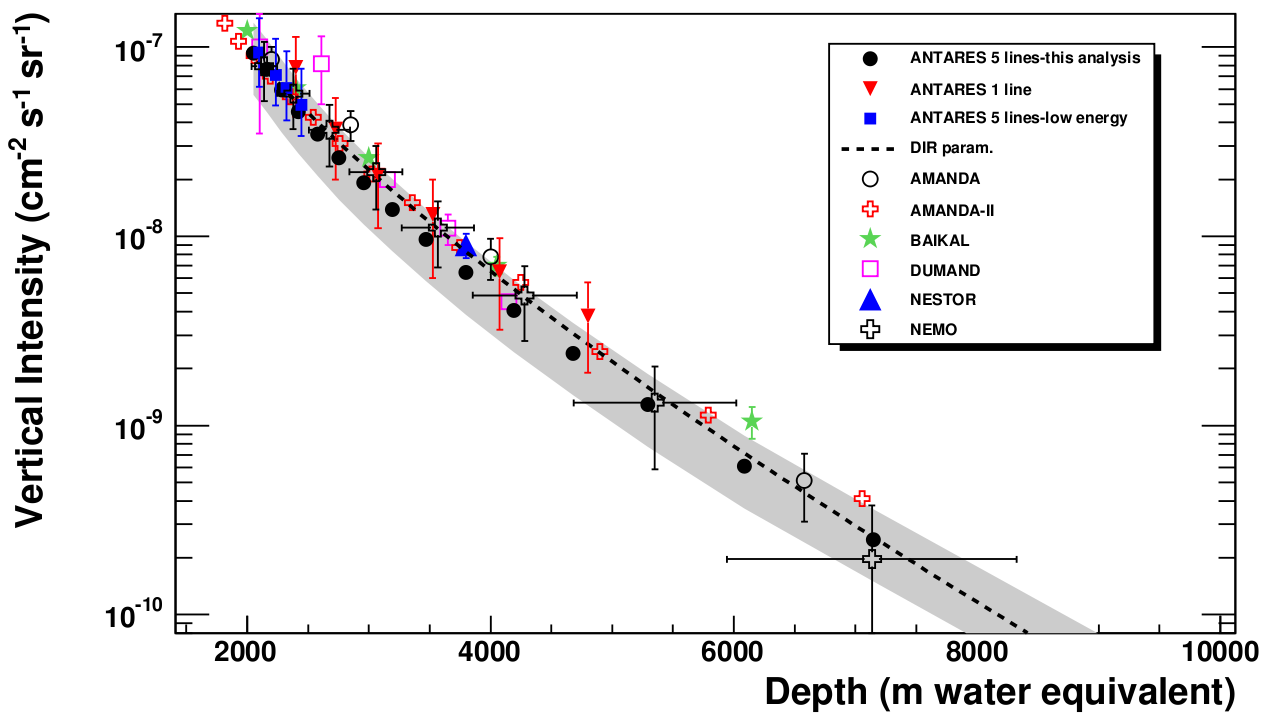,width=9.8cm}
\caption{
Depth-intensity relation for muons measured by various detectors underground and
underwater/ice. The shaded band indicates the systematic uncertainty of the
ANTARES analysis, the dashed curve is the expectation derived from
\pcite{Bugaev-1998}. Note that single experiments can cover extended depth
ranges by investigating muons incident under different zenith angles. Figure
taken from \pcite{antares-muon-zenith-2010}.
\label{depth-intensity}
}
\end{figure}

One of the initial motivations for deep-water detectors was to clarify puzzles
in the depth-intensity relations that had emerged from underground muon
measurements (see Sect.~\ref{sec-fir-dum}). Due to the energy dependence of the
muon range in matter, these measurements are sensitive to the energy spectrum of
muons from cosmic-ray-induced air showers. Figure~\ref{depth-intensity}
\cite{antares-muon-zenith-2010} shows a compilation of recent data, dominated by
muons with energies below $100\tev$; the measured muon flux excellently follows
a model prediction \cite{Bugaev-1998}, thus demonstrating that this aspect of
cosmic ray physics is well under control in the energy range covered.

Over the last years, however, a new cosmic ray puzzle has emerged: Several
detectors located at the Northern hemisphere have found tiny anisotropies in the
arrival direction of cosmic rays in the range of several $\Tev$ to several
hundreds of $\Tev$ (Milagro \cite{anisotropy-Milagro}, Tibet air shower array
\cite{anisotropy-Tibet}, ARGO/YBJ \cite{anisotropy-Argo}) and of down-going
atmospheric muons (Super-Kamiokande \cite{anisotropy-SK}). Using recorded data
of several $10^{10}$ down-going muons with energies above a few $\Tev$, IceCube
has detected per-mille anisotropies on all angular scales down to about
$15^\circ$. A multipole analysis yields the strongest power for dipole ($l=1$)
and quadrupole ($l=2$) contributions. The dipole does not point to the direction
suggested by the Compton-Getting effect (anisotropy induced by the movement of
the Earth \cite{Compton-Getting}) if the cosmic ray plasma were at rest with
respect to the Galactic centre. This indicates that Galactic cosmic rays
co-rotate with the local Galactic magnetic field. Also smaller structures on
scales between $15^\circ$ and $30^\circ$ are visible. The relative amplitude of
the smaller structures is about a factor of five weaker than that of the
dipole/quadrupole structure. The minima and maxima on all scales are
reproducible in all hitherto analysed data samples, from IceCube-22
\cite{icecube-2010b} to IceCube-59 \cite{icecube-2011g}.

\begin{figure}[ht]
\sidecaption
\epsfig{file=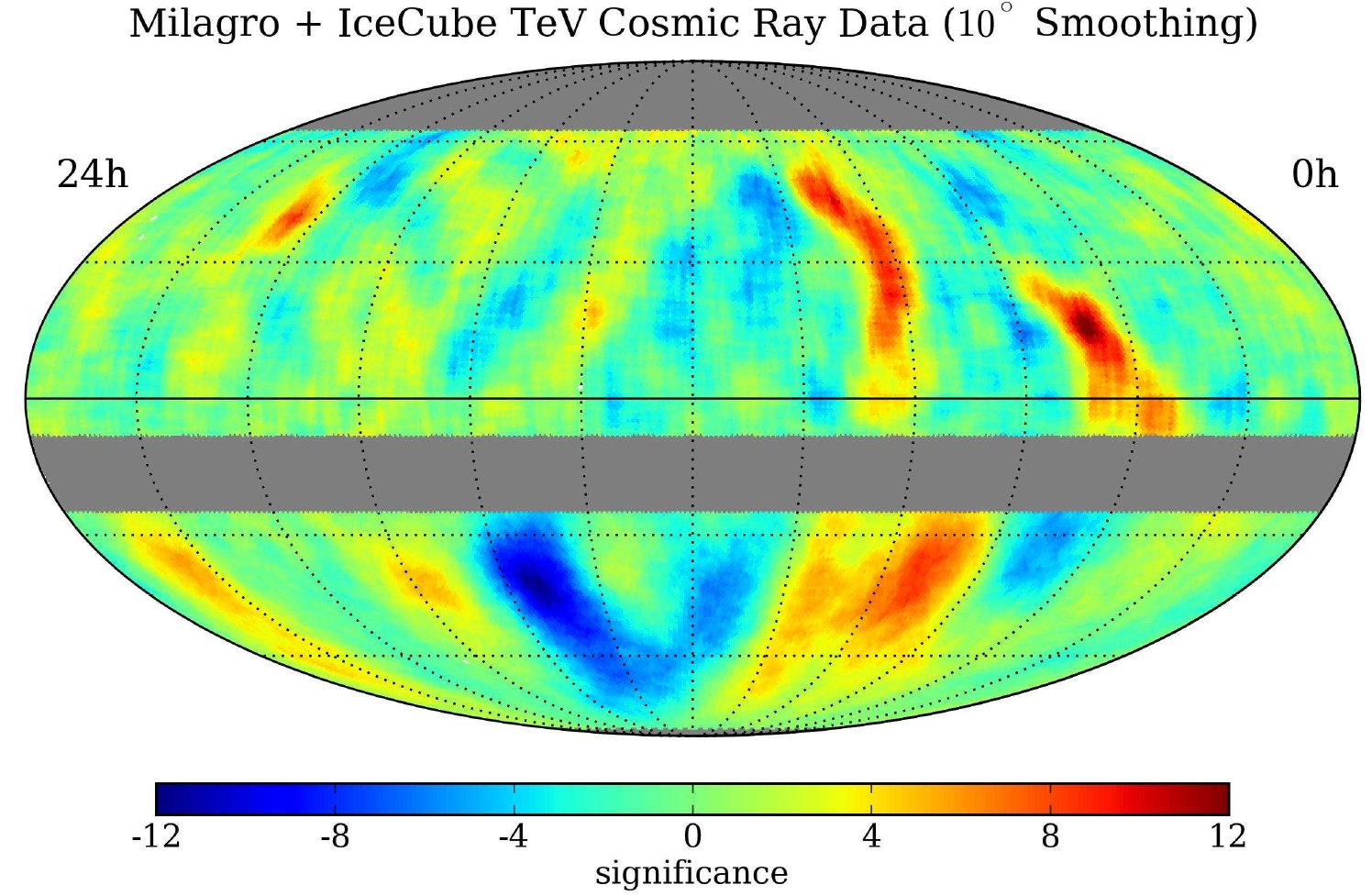,width=10.5cm}
\caption{
Significance skymap of the medium-scale anisotropies of atmospheric muon arrival
directions in IceCube, combined with the Milagro cosmic-ray sky-map for the
Northern hemisphere. The colour code indicates the significance of deviations
from the average in $\sigma$. Note that structures at very small and very large
scales have been filtered out (see text). Figure taken from \pcite{icecube-2011g}.
\label{anisotropies}
}
\end{figure}

Figure \ref{anisotropies} shows the combined Milagro/IceCube significance
sky-map after applying a band-pass filter to remove small-scale fluctuations 
($10^\circ$ smoothing) and the dipole and quadrupole terms.

The Milagro data contain $2.2\times10^{11}$ air showers with a median energy of
about $1\tev$, the overall IceCube data contain $3.4\times10^{10}$ muons with a
median energy of the primary particle of about $20\tev$. The origin of the
medium scale anisotropies is still unknown. Possible explanations include a
magnetic ``nozzle'' to a nearby supernova remnant, with a scatter-free transport
along a field connected to the source \cite{Drury-2008}; at lower energies and
small scales, the heliotail may play a role \cite{Lazari-2010}. What ever the
origin of the anisotropies is: These structures observed with IceCube may
contain key information on nearby sources and/or cosmic ray propagation. To
understand the anisotropies is an interesting challenge that will be
re-addressed with the increasing statistics expected in the coming years.

\clearpage
\section{Alternative Detection Principles for Extreme Energies}
\label{sec-alt}

The alternative detection technologies described in this section are tailored to
signals which propagate with kilometre-scale attenuation. Consequently, they
allow for the observation of much larger volumes than those achievable for
optical neutrino telescopes. Detectors on the 100 to 1000 $\Km^3$ scale are
necessary, for instance, to record more than just a few cosmogenic neutrinos, in
the typical energy range of $100\pev$ to $10\eev$. This section is intended to
give an outlook to future experimental opportunities. Some results of initial
projects are included in Sect.~\ref{sec-phy}, see in particular
Fig.~\ref{EHE-diffuse}.

\subsection{Detection via air showers}
\label{sec-alt-air}

\begin{figure}[ht]
\sidecaption
\epsfig{file=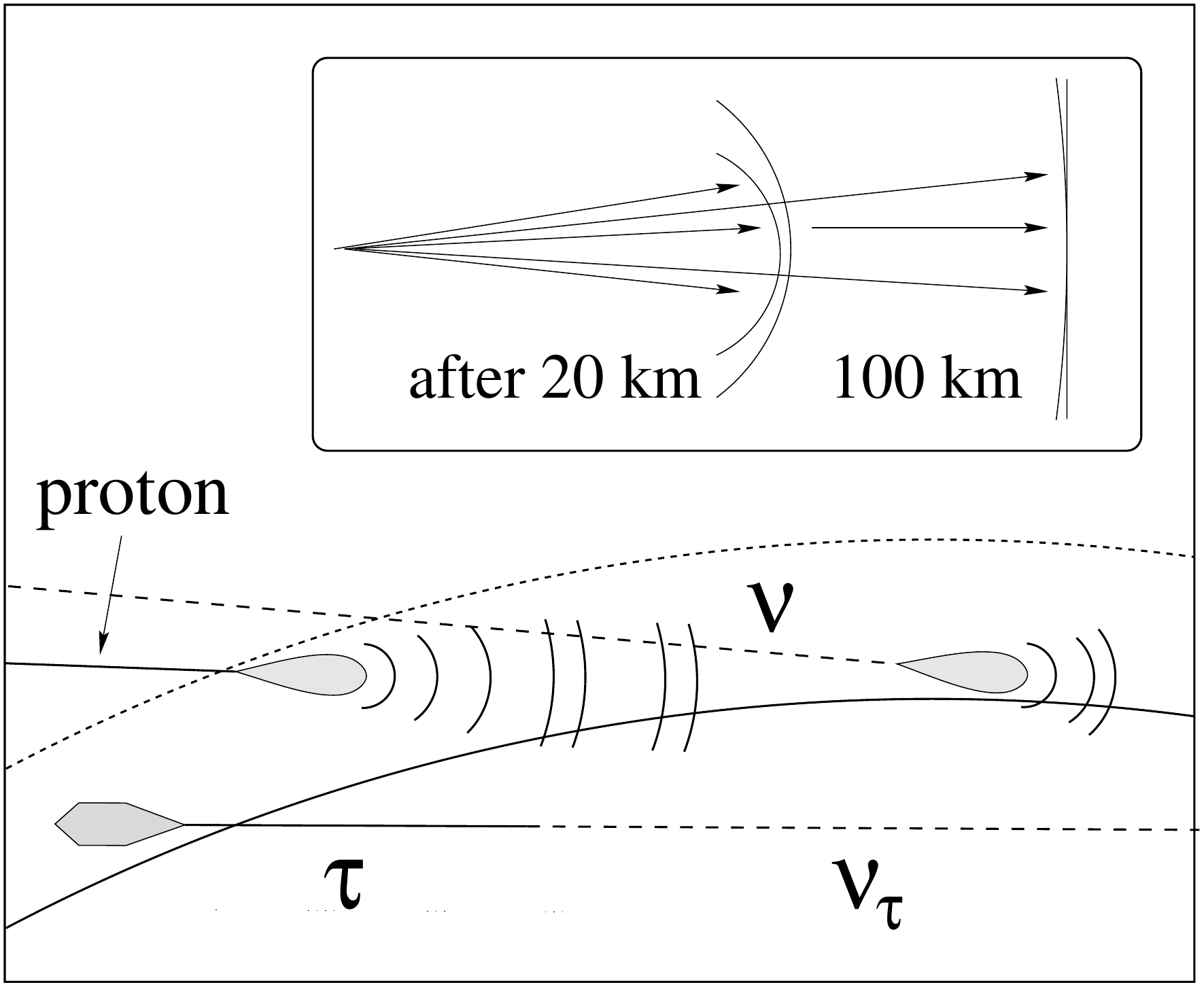,width=9.cm}
\caption{
Detection of particles or fluorescence light emitted by horizontal or
upwards-directed air showers from neutrino interactions.}
\label{horizontal}
\end{figure}

At energies above $10^{17}\ev$, large air shower arrays like the Pierre Auger
Observatory (PAO) in Argentina \cite{Auger-Web} or the Telescope Array in Utah,
USA \cite{TA-2003} are seeking for horizontal air showers induced by neutrino
interactions deep in the atmosphere (showers caused by charged cosmic ray
interactions start much higher up in the atmosphere). Figure~\ref{horizontal}
explains the detection principle. PAO consists of an array of water tanks
covering an area of $3000\km^2$ that record the Cherenkov light of charged
air-shower particles crossing the tanks. The array is combined with telescopes
recording the atmospheric fluorescence light from air showers. The optimum
sensitivity window for this method is at $1\rnge100\eev$, the effective target
mass is up to 20\,Gigatons. An even better sensitivity might be obtained for tau
neutrinos, $\nu_\tau$, scratching the Earth and interacting close to the array
\cite{Fargion-2002,Fargion-2004}. The charged $\tau$ lepton produced in
charged-current interaction can escape the rock around the array (in contrast to
electrons) and mostly decays into hadrons (branching ratio ca.\ 65\%) after a
short path length (in contrast to muons). If this decay happens in the field of
view of the fluorescence telescopes, the decay cascade can be recorded. Provided
the experimental pattern allows for a clear identification, the acceptance for
this kind of signals can be large. For the optimal energy scale of $1\eev$, the
present $\nu_\tau$ limit for an $E^{-2}$ tau neutrino flux is about
$E^2\phi<10^{-7}\flunit$ \cite{Auger-nutau}.

Space-based observation of extended air showers is an approach to even further
increase the target mass for highest-energy cosmic-ray and neutrino detection,
at energies beyond $10^{19}\ev$. The ``Extreme Universe Space Observatory
(EUSO)'', proposed around the year 2000, is a wide-field camera observing the
atmosphere from an orbit at several 100\,km height and registering the
fluorescence light from extended air showers and, if circumstances allow, also
the reflection at Earth surface of the Cherenkov light emitted in shower
direction \cite{EUSO-2002}. After several years of technical development and a
long phase of uncertainty concerning the space carrier, plans are now to install
the device -- meanwhile renamed to JEM-EUSO \cite{JEM-EUSO-Web} -- on the
Japanese Experiment Module (JEM) of the International Space Station, with the
launch expected around 2015. JEM-EUSO will observe an atmospheric target volume
with a mass of more than one Teraton and will thus exceed the PAO sensitivity by
two orders of magnitude for energies above some $10^{19}\ev$
\cite{JEM-EUSO-2010}. Similarly as for PAO, neutrino-induced air showers can be
separated from those from hadrons or gammas by the depth of the interaction
point in the atmosphere. The JEM-EUSO physics opportunities in the neutrino
channel have been studied in general \cite{Fargion-2002b,JEM-EUSO-2010b} and
specifically for cosmogenic neutrinos \cite{Kotera-2010} and GRB neutrinos
\cite{JEM-EUSO-2009}. It is not obvious that JEM-EUSO will detect neutrinos from
known astrophysical or cosmogenic sources, but its measurements will explore
possible neutrino fluxes in a hitherto inaccessible energy region.

\subsection{Radio detection}
\label{sec-alt-rad}

Electromagnetic cascades, e.g.\ generated by high energy neutrino interactions
in ice or salt, emit coherent Cherenkov radiation at radio frequencies. The
effect was predicted in 1962 \cite{radio-Askaryan-1962} and confirmed by
measurements at accelerators \cite{radio-Saltzberg-2001,radio-Gorham-2007}. 
Electrons from the material traversed are swept into the developing shower,
which thus acquires an electric net charge. This charge propagates like a
relativistic pancake of about $1\cm$ thickness and $10\cm$ diameter. For
wavelengths exceeding the cascade diameter, coherent emission of electromagnetic
radiation (Cherenkov radiation and synchrotron radiation caused by the Earth
magnetic field) occurs. The signal amplitude increases with the square of the
net charge in the cascade, i.e.\ it is proportional to $E_\nu^2$, thus making
the method particularly attractive for high-energy cascades. The resulting
bipolar pulse is in the radio frequency band and has a width of $1\rnge2\ns$. In
ice, attenuation lengths of more than a kilometre are observed for radio
signals, depending on the frequency band and the ice temperature, implying that
for energies above a few $10\pev$ radio detection becomes competitive or
superior to optical detection (with its attenuation length of the order 100\,m)
\cite{Price-1996}.

The prototype ``Radio Ice Cerenkov Experiment (RICE)'' was operated at the South
Pole, with 20 receivers and emitters buried at depths between 120 and 300\,m. 
From the non-observation of very large pulses, limits on the diffuse flux of
neutrinos with $E>100\pev$ and on the flux of relativistic magnetic monopoles
have been derived \cite{radio-RICE-2006}.

\begin{figure}[ht]
\begin{center}
\epsfig{file=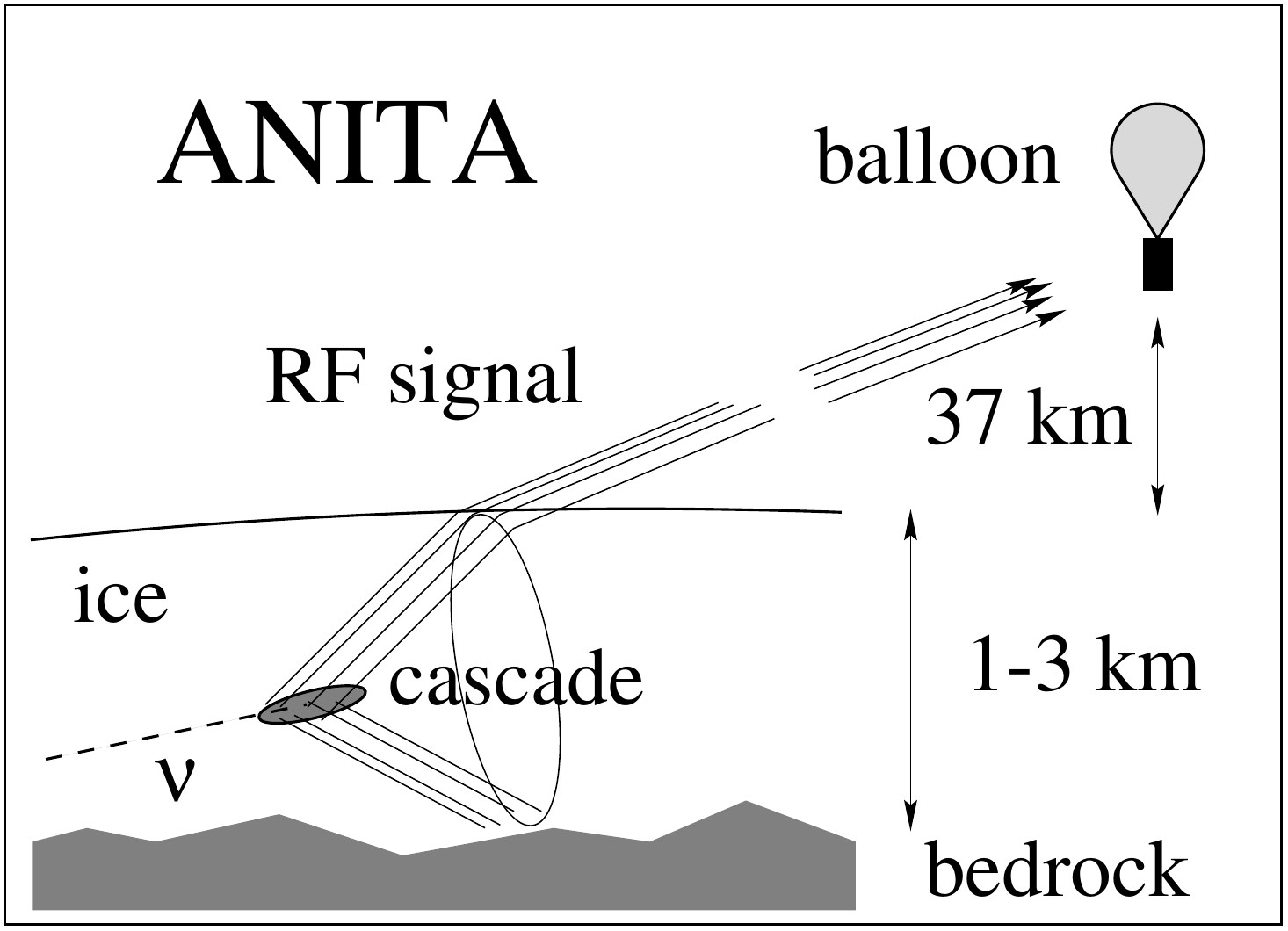,width=8.4cm}
\hfill
\epsfig{file=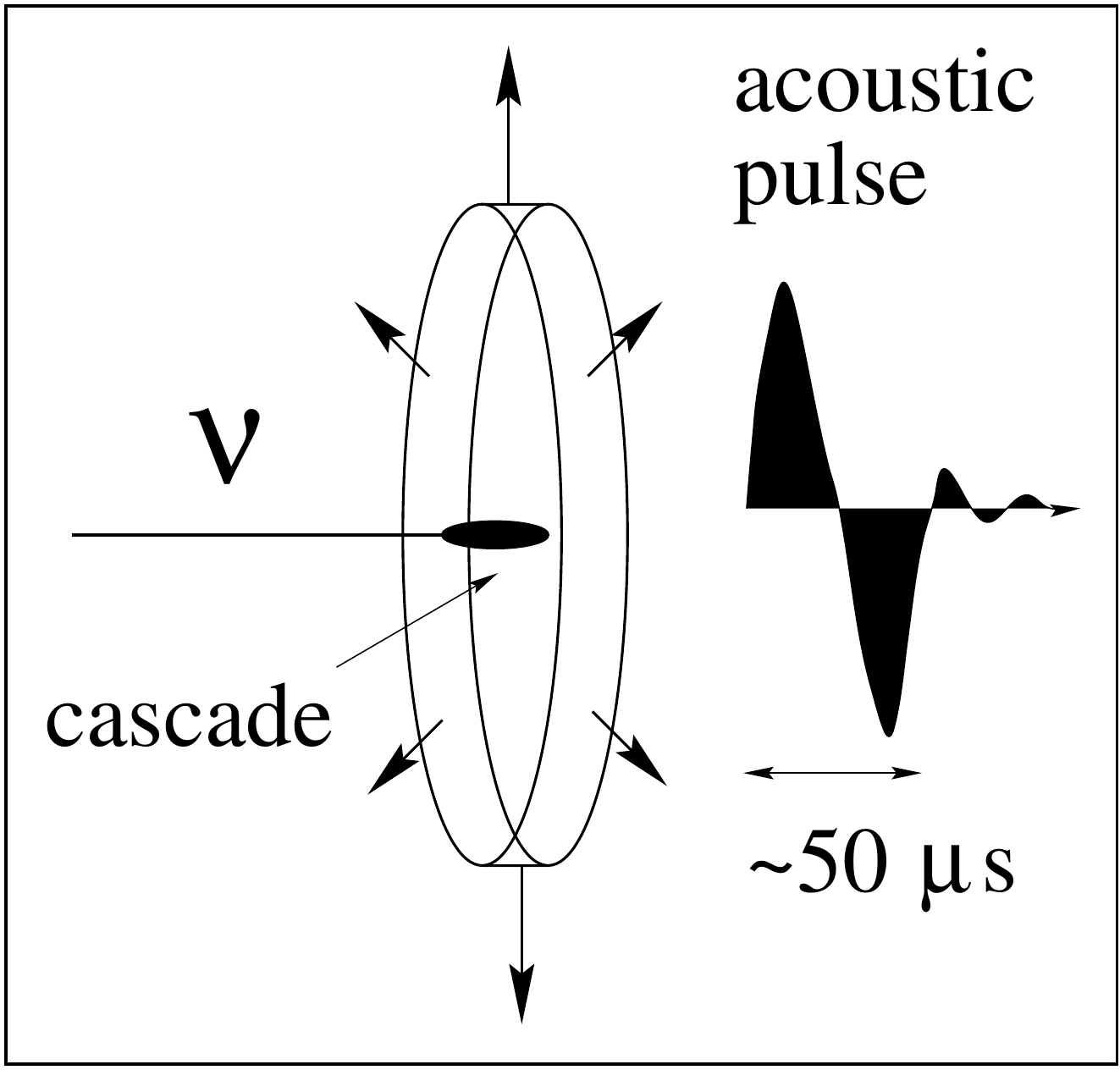,width=6.4cm}
\caption{
Principles of radio and acoustic detection of neutrinos. Left: ANITA balloon
experiment. Right: Acoustic emission of a particle cascade.}
\label{ANITA}
\end{center}
\end{figure}

The ``Antarctic Impulsive Transient Array (ANITA)'' \cite{radio-ANITA-2006} is
an array of radio antennas which has been flown at a balloon on an Antarctic
circumpolar path in 2006 and 2008/09 (see Fig.~\ref{ANITA}, left). From 35\,km
altitude it searched for radio pulses from neutrino interactions in the thick
ice cover and monitored, with a threshold in the $\Eev$ range, a volume of the
order of $10^6$ Gigatons. The resulting neutrino flux limits \cite{ANITA-2009} are 
presented in Fig.~\ref{EHE-diffuse} in Sect.~\ref{sec-phy-cos-dif}.

Future plans for radio detection in ice foresee large arrays of antennas on the
surface of the antarctic ice shelf (ARIANNA \cite{radio-ARIANNA-2007}) or in the
South Polar ice close to the surface (ARA \cite{radio-ARA-2011}).

Even more exotic is the search for radio emission from extremely-high energy
cascades induced by neutrinos or cosmic rays skimming the moon surface. An
example is the ``Goldstone Ultra-high Energy Neutrino Experiment (GLUE)'' which
used two NASA antennas and reached a maximum sensitivity at several
$\Zev=1000\eev$ \cite{radio-GLUE-2004}. With the same method, the NuMoon
experiment at the Westerbork Radio Telescope was searching for extremely
energetic neutrinos \cite{radio-NUMOON-2008}. Similar activities are under
preparation in the context of the LOFAR experiment \cite{LOFAR-2010}.

\subsection{Acoustic detection}
\label{sec-alt-aco}

Production of pressure waves by charged particles depositing energy in
liquids or solid media was predicted in 1957 \cite{acoustic-Askaryan-1957} and
experimentally proven with high-intensity proton beams two decades later
\cite{acoustic-Learned-1979}. In the case of a particle cascade, its entire
energy is deposited into the medium, mostly through ionisation, and converted to
heat on a time scale that is very short compared to the typical time scales
relevant for generation and propagation of acoustic pulses. The effect is a fast
expansion, generating a bipolar acoustic pulse with a width of a few ten
microseconds in water or ice (see Fig.~\ref{ANITA}, right), corresponding to a peak
signal power at 20\,kHz. Transversely to the pencil-like cascade, the acoustic
pulse propagates into the medium within a disk-shaped volume with a thickness
corresponding to the cascade length of about 10\,m. Exploiting this method would
require to detect the acoustic pulses on the background of the ambient and
intrinsic noise with a sparsely instrumented detector. This implies a very high
detection threshold, in the $\Eev$ range at best.

Acoustic detection is an option both for ice and sea water. For ice, the signal
itself is expected to be higher and ambient noise to be lower than in sea water. 
A test array, SPATS (South Pole Acoustic Test Setup), has been deployed at the
South Pole in order to determine the depth dependence of the speed of sound
\cite{acoustic-SPATS-2009}, the attenuation length of acoustic signals
\cite{acoustic-SPATS-2010} and the ambient noise \cite{acoustic-SPATS-2011}. The
results of the latter two measurements are slightly discouraging as the
attenuation length turns out to be about 300\,m, an order of magnitude smaller
than expected, and the noise level is roughly the same as in the deep sea at a
calm sea state. As a variation of the ice approach, even the use of permafrost
as medium has been discussed \cite{acoustic-Nahnhauer-2008}.

Test of acoustic detection in sea water have been performed close to Sicily
(O$\nu$DE setup, see Sect.~\ref{sec-fir-nem}), close to Scotland and in Lake
Baikal (see \cite{ARENA-Zeuthen,ARENA-Scotland,ARENA-Rome} for overviews). 
Another project, SAUND, has been using a very large but extremely sparsely
instrumented hydrophone array of the US Navy, close to the Bahamas
\cite{acoustic-Lehtinen-2002,acoustic-Vandenbroucke-2005,acoustic-SAUND-2006}. 
The array of hydrophones covers an area of $250\km^2$, has good sensitivity at
$1\rnge500\kHz$ and can trigger on events above $100\eev$ with a tolerable
background rate.

\begin{figure}[ht]
\sidecaption
\epsfig{file=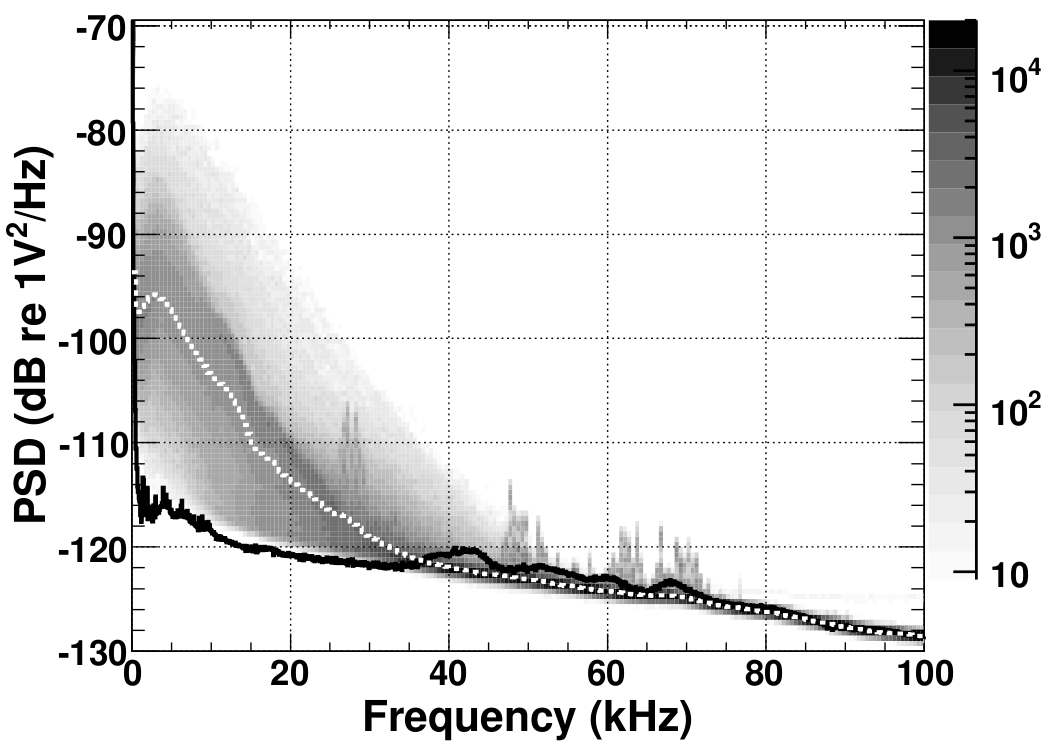,width=10.cm}
\caption{
Power spectral density (PSD) of the ambient noise recorded in AMADEUS. Shown in
shades of grey is the occurrence rate in arbitrary units, where dark colours
indicate higher rates. The white dotted line is the median value of the in-situ
PSD and the black solid line indicates the noise level recorded in the
laboratory prior to deployment.}
\label{amadeus-noise}
\end{figure}

An extended test configuration with various hydrophones, named AMADEUS, has been
deployed together with the ANTARES detector \cite{acoustic-AMADEUS}. The sensors
are arranged in local clusters (size scale one metre), with inter-cluster
distances between 12.5\,m and 340\,m. First studies focussed on the ambient
noise levels (see Fig~\ref{amadeus-noise}) and their dependence on sea state and
precipitation, with the conclusion that under calm conditions the noise in the
relevant frequency range is below typical acoustic neutrino signals. In
addition, the 3-dimensional arrangement of hydrophones allows for locating
transient signals and investigating the rate of background signals of the
expected bipolar shape that come from an appropriately defined fiducial volume. 
For the first time in the deep sea, the intrinsic background to acoustic
neutrino signals can thus be studied down to an amplitude level of a few
10\,mbar, corresponding to detection thresholds at the $\Eev$ scale.

\clearpage
\section{Summary and Outlook}
\label{sec-sum}

Five decades after first proposals, and three decades after first practical
attempts to build high-energy neutrino telescopes, we may be close to a turning
point. IceCube, the first cubic-kilometre neutrino telescope, has started data
taking in its final configuration, and preparations for counterparts on the
Northern hemisphere are advanced.

The strong case for high-energy neutrino astronomy has remained unchanged over
all the time, but the requirements on the necessary sensitivity have tightened
continuously. The detection of first high-energy neutrino sources from outer
space lays still ahead and may, with some optimism, be expected in the next few
years. Galactic ``Pevatrons'' as those detected by Milagro are in reach after a
few years of IceCube data taking if the corresponding predictions are correct. 
Models assigning the most energetic cosmic rays to Gamma Ray Bursts will be
tested within a couple of years. However, clear observations are all but
guaranteed.

Discoveries or non-discoveries of IceCube will have a strong impact on the
future of the field and possibly mark a ``moment of truth''. Clear evidence for
neutrino sources would pave the straight way for KM3NeT in the Mediterranean
Sea, with possible modifications to the present design (high energies vs.\ low
energies) according to the IceCube results. KM3NeT would then complement IceCube
in its field of view; for results not depending on the hemisphere (e.g.\ diffuse
fluxes, exotic particles) it could confirm and extend the IceCube findings using
a detector with different systematic uncertainties, and in addition it would
explore the the central parts of the Galactic plane with superior sensitivity.

Missing or marginal evidence for sources from IceCube may lead to different
developments. One option would be to continue the venue of detectors which
explore the energy range characteristic for Gamma Ray Bursts and Active Galactic
Nuclei. This would require configurations with significantly larger spacing than
the present KM3NeT design, resulting in a penalty at energies below a few tens
of $\Tev$, and in sacrificing many Galactic source candidates. Abundance and
characteristics of extragalactic sources are not expected to differ much between
the Northern and Southern skies. A factor of three or so in sensitivity compared
to IceCube would therefore only provide a limited additional discovery window
for extragalactic sources. Since firm flux predictions are absent, the resulting
``gambling'' would be justified only if indeed a substantially larger
sensitivity increase with respect to IceCube could be reached.
     
Predictions for Galactic sources are much more firm, provided we assume
dominantly hadronic emission. In contrast to extragalactic sources, we have good
reasons to assume that we may be close to the discovery region, so that a
sensitivity increase by a factor 3--5, with a telescope observing the Southern
hemisphere, indeed counts. Therefore a second option would be to focus on
Galactic sources like Supernova Remnants and tailor the array to these sources,
i.e.\ keeping high sensitivity down to about a TeV. The uncertainty here comes
from the question to what extent the observed gamma rays are indeed due to
hadronic production.

The third option would be an even larger leap in size than for the first option. 
It would address energies above $100\pev$ with the help of new technologies like
radio or acoustic detection and envisage $100\rnge1000$ cubic kilometres of
instrumented volume. This option might still have sensitivity to neutrinos from
AGN jets but would also well cover the energy range of cosmogenic neutrinos. In
contrast to optical detectors, new-technology detectors are still in the R\&D
stage and also have no natural calibration source like atmospheric neutrinos for
optical detectors.

The fourth option would define, at least for the time being, an end to the
search for neutrinos from cosmic accelerators. The field would focus on optical
neutrino detection with dense detectors optimised for investigating oscillations of
accelerator neutrinos (Mediterranean Sea) and atmospheric neutrinos, or, even
more pretentious, to study supernova bursts beyond our own Galaxy or even proton
decay.

For the moment, however, we don't see a reason to be pessimistic. We have made a
factor-of-thousand step in sensitivity compared to a dozen years ago. This is
far more than the traditional factor of ten which so often led to the discovery
of new phenomena. For instance, looking back to our own field, the prospects for
discovery had not been estimated overly high before launching the first X-ray
rocket in 1962, or before detecting the Crab in $\Tev$ gamma rays in 1989. 
History told a different story, as we know by today.

We have a good chance to open a new window to the Universe, but we don't know it
for sure -- and that, at the end, may be the most tempting and challenging
situation that we possibly can imagine!

\vspace*{3.cm}

\noindent
{\bf Acknowledgements:} The authors wish to thank their colleagues in the
ANTARES, Baikal, IceCube and KM3NeT collaborations for valuable input and in
particular Jürgen Brunner and Alexander Kappes for carefully reading the
manuscript and for many important remarks and corrections.

\clearpage
\section*{References}
\label{sec-ref}
\addcontentsline{toc}{section}{References}

\providecommand{\etal}{et al.\xspace}
\providecommand{\coll}{Coll.\xspace}
\providecommand{\Lodz}{{\Lslash\'od\'z}}
{
\raggedright
\bibliographystyle{./styles/elsarticle-num}
\bibliography{./NT_Review}

\begin{thebibliography}{100}
\expandafter\ifx\csname url\endcsname\relax
  \def\url#1{\texttt{#1}}\fi
\expandafter\ifx\csname urlprefix\endcsname\relax\def\urlprefix{URL }\fi
\expandafter\ifx\csname href\endcsname\relax
  \def\href#1#2{#2} \def\path#1{#1}\fi

\bibitem{Greisen-1960}
K.\,Greisen, {\it Cosmic ray showers}, Ann.\ Rev.\ Nucl.\ Part.\ Sci. 10 (1960)
  63.

\bibitem{Reines-1960}
F.\,Reines, {\it Neutrino interactions}, Ann.\ Rev.\ Nucl.\ Part.\ Sci. 10
  (1960) 1.

\bibitem{Markov-1960}
M.A.\,Markov, {\it On high energy neutrino physics}, in: Proc.~10th ICHEP,
  Rochester, 1960, p. 578.

\bibitem{Berezinsky-Zatsepin}
V.S.\,Berezinsky and G.T.\,Zatsepin, {\it Cosmic rays at ultra high energies
  (neutrino?)}, Phys.\ Lett. B\,28 (1969) 423.

\bibitem{Gaisser-Halzen-Stanev}
T.K.\,Gaisser, F.\,Halzen and T.\,Stanev, {\it Particle astrophysics with
  high-energy neutrinos}, Phys.\ Rep. 258 (1995) 173.
\newblock \href {http://arxiv.org/abs/hep-ph/9410384} {\path{
  arXiv:hep-ph/9410384}}.

\bibitem{Learned-Mannheim}
J.G.\,Learned and K.\,Mannheim, {\it High-energy neutrino astrophysics}, Ann.\
  Rev.\ Nucl.\ Part.\ Sci. 50 (2000) 679.

\bibitem{McDonald-2004}
A.B.\,McDonald \etal, {\it Astrophysical neutrino telescopes}, Rev.\ Sci.\
  Instrum. 75 (2004) 293.
\newblock \href {http://arxiv.org/abs/astro-ph/0311343} {\path{
  arXiv:astro-ph/0311343}}.

\bibitem{Becker-2007}
J.K.\,Becker, {\it High-energy neutrinos in the context of multimessenger
  physics}, Phys.\ Rep. 458 (2008) 173.
\newblock \href {http://arxiv.org/abs/0710.1557 [astro-ph]} {\path{
  arXiv:0710.1557 [astro-ph]}}.

\bibitem{Chiarusi-Spurio}
T.\,Chiarusi and M.\,Spurio, {\it High-energy astrophysics with neutrino
  telescopes}, Eur.\ Phys.\ J. C\,65 (2010) 649.
\newblock \href {http://arxiv.org/abs/0906.2634 [astro-ph.HE]} {\path{
  arXiv:0906.2634 [astro-ph.HE]}}.

\bibitem{Anchordoqui-Montaruli}
L.A.\,Anchordoqui and T.\,Montaruli, {\it In search for extraterrestrial high
  energy neutrinos}, Ann.\ Rev.\ Nucl.\ Part.\ Sci. 60 (2010) 129.
\newblock \href {http://arxiv.org/abs/0912.1035 [astro-ph.HE]} {\path{
  arXiv:0912.1035 [astro-ph.HE]}}.

\bibitem{Rhode-1996}
Fréjus Coll., W.\,Rhode \etal, {\it Limits on the flux of very high-energetic
  neutrinos with the {Fréjus} detector}, Astropart.\ Phys. 4 (1996) 217.

\bibitem{Harwit}
M.\,Harwit, {\it Cosmic discovery}, Basic Books Inc., New York, 1981.

\bibitem{Halzen-history}
F.\,Halzen.
\newblock {\it Ice fishing for neutrinos} [online] (1995).
\newblock Available from:
  \url{http://icecube.berkeley.edu/amanda/ice-fishing.html}.

\bibitem{Bluemer-etal-2009}
J.\,Blümer, R.\,Engel and J.R.\,Hörandel, {\it Cosmic rays from the knee to the
  highest energies}, Prog.\ Part.\ Nucl.\ Phys. 63 (2009) 293.

\bibitem{Fermi-1949}
E.\,Fermi, {\it On the origin of the cosmic radiation}, Phys.\ Rev. 75 (1949)
  1169.

\bibitem{Fermi-1954}
E.\,Fermi, {\it Galactic magnetic fields and the origin of cosmic radiation},
  Astrophys.\ J. 119 (1954) 1.

\bibitem{Ginzburg-Ptuskin-1976}
V.L.\,Ginzburg and V.S.\,Ptuskin, {\it On the orgin of cosmic rays: {Some}
  problems in high-energy astrophysics}, Rev.\ Mod.\ Phys. 48 (1976) 161.

\bibitem{Blandford-Eichler-1987}
R.\,Blandford and D.\,Eichler, {\it Particle acceleration at astrophysical
  shocks: {A} theory of cosmic ray orgin}, Phys.\ Rep. 154 (1987) 1.

\bibitem{Hillas-2005}
A.M.\,Hillas, {\it Can diffuse shock acceleration in supernova remnants account
  for high-energy galactic cosmic rays?}, J.\ Phys. G\,31 (2005) R95.

\bibitem{Caprioli-2009}
D.\,Caprioli, P.\,Blasi, E.\,Amato and M.\,Vietri, {\it Dynamical feedback of
  self-generated magnetic fields in cosmic rays modified shocks}, Mon.\ Not.\
  Roy.\ Astron.\ Soc. 395 (2009) 895.

\bibitem{Gaisser-1990}
T.\,Gaisser, {\it Cosmic rays and particle physics}, Cambridge University
  Press, Cambridge, UK, 1991.

\bibitem{Baade-Zwicky-1934}
W.\,Baade and F.\,Zwicky, {\it Remarks on super-novae and cosmic rays}, Phys.\
  Rev. 46 (1934) 76.

\bibitem{Mirabel-2007}
I.F.\,Mirabel, {\it Gamma-ray binaries}, Astrophys.\ Space Sci. 309 (2007) 267.
\newblock \href {http://arxiv.org/abs/astro-ph/0610707} {\path{
  arXiv:astro-ph/0610707}}.

\bibitem{Auger-2011}
Pierre Auger Coll., P.\,Abreu \etal, {\it The {Pierre Auger Observatory I: The}
  cosmic ray energy spectrum and related measurements}, contribution to 32nd
  Int.\ Cosmic Ray Conf., Beijing, China (2011).
\newblock \href {http://arxiv.org/abs/1107.4809} {\path{ arXiv:1107.4809}}.

\bibitem{Greisen-1966}
K.\,Greisen, {\it End to the cosmic ray spectrum?}, Phys.\ Rev.\ Lett. 16
  (1966) 748.

\bibitem{Zatsepin-Kuzmin-1966}
G.T.\,Zatsepin and V.A.\,Kuzmin, {\it Upper limit of the spectrum of cosmic
  rays}, JETP Lett. 4 (1966) 78.

\bibitem{Gaisser-1997}
T.K.\,Gaisser, {\it Neutrino astronomy: Physics goals, detector parameters},
  {OECD} Megascience Forum, Taormina, Italy (1997).
\newblock \href {http://arxiv.org/abs/astro-ph/9707283} {\path{
  arXiv:astro-ph/9707283}}.

\bibitem{gamma1}
R.A.\,Ong, {\it The status of {VHE} gamma-ray astronomy}, rapporteur Talk at
  29th Int.\ Cosmic Ray Conf., Pune, India (2005).
\newblock \href {http://arxiv.org/abs/astro-ph/0605191} {\path{
  arXiv:astro-ph/0605191}}.

\bibitem{gamma2}
J.\,Hinton, {\it Ground based gamma-ray astronomy with {Cherenkov Telescopes}},
  New J.\ Phys. 11 (2009) 055005.
\newblock \href {http://arxiv.org/abs/0803.1609 [astro-ph]} {\path{
  arXiv:0803.1609 [astro-ph]}}.

\bibitem{Aharonian-2004}
H.E.S.S.\ Coll., F.A.\,Aharonian \etal, {\it High-energy particle acceleration
  in the shell of a supernova remnant}, Nature 432 (2004) 75.

\bibitem{Fermi-RX}
Fermi LAT Coll., A.A.\,Abdo \etal, {\it Observations of the young supernova
  remnant {RX\,J1713.7-3946} with the {Fermi Large Area Telescope}},
  Astrophys.\ J. 734 (2011) 28.
\newblock \href {http://arxiv.org/abs/1103.5727 [astro-ph.HE]} {\path{
  arXiv:1103.5727 [astro-ph.HE]}}.

\bibitem{Ellison-2010}
D.C.\,Ellison, D.J.\,Patnaude, P.\,Slane and J.\,Raymond, {\it Efficient cosmic
  ray acceleration, hydrodynamics, and self-consistent thermal {X-ray} emission
  applied to {SNR RX\,J1713.7-3946}}, Astrophys.\ J. 712 (2010) 287.
\newblock \href {http://arxiv.org/abs/1001.1932 [astro-ph.HE]} {\path{
  arXiv:1001.1932 [astro-ph.HE]}}.

\bibitem{Aharonian-2005}
H.E.S.S.\ Coll., F.A.\,Aharonian \etal, {\it Detection of {$\Tev$} $\gamma$-ray
  emission from the shell-type supernova remnant {RX\,J0852.0-4622} with
  {HESS}}, Astron.\,\&\,Astrophys. L7 (2005) 437.
\newblock \href {http://arxiv.org/abs/astro-ph/0505380} {\path{
  arXiv:astro-ph/0505380}}.

\bibitem{Aharonian-2006}
H.E.S.S.\ Coll., F.A.\,Aharonian \etal, {\it Discovery of very-high-energy
  gamma-rays from the {Galactic Centre} ridge}, Nature 439 (2006) 695.

\bibitem{Berge-2005}
D.\,Berge, M.\,Lemoine-Goumard and M.\,de Naurois for the H.E.S.S.\ Coll., {\it
  Observations of {SNR RX\,J1713.7-3946} with {H.E.S.S.}}, AIP Conf.\ Proc. 745
  (2005) 223, contribution to 3rd Int.\ Symp.\ High-Energy Gamma-Ray Astron.,
  Heidelberg, 2004.

\bibitem{Kappes-etal-2007}
A.\,Kappes \etal, {\it Potential neutrino signals from {Galactic} gamma-ray
  sources}, Astrophys.\ J. 656 (2007) 870.
\newblock \href {http://arxiv.org/abs/astro-ph/0607286} {\path{
  arXiv:astro-ph/0607286}}.

\bibitem{Abdo-2007}
A.A.\,Abdo \etal, {\it Discovery of {$\Tev$} gamma-ray emission from the
  {Cygnus} region of the {Galaxy}}, Astrophys.\ J. 658 (2007) L33.
\newblock \href {http://arxiv.org/abs/astro-ph/0611691} {\path{
  arXiv:astro-ph/0611691}}.

\bibitem{Abdo-2008}
A.A.\,Abdo \etal, {\it A measurement of the spatial distribution of diffuse
  {$\Tev$} gamma ray emission from the {Galactic Plane} with {Milagro}},
  Astrophys.\ J. 688 (2008) 1078.
\newblock \href {http://arxiv.org/abs/0805.0417 [astro-ph]} {\path{
  arXiv:0805.0417 [astro-ph]}}.

\bibitem{Halzen-Kappes-Murchadha-2008}
F.\,Halzen, A.\,Kappes and A.\,O'Murchadha, {\it Prospects for identifying the
  sources of the {Galactic} cosmic rays with {IceCube}}, Phys.\ Rev. D\,78
  (2008) 063004.
\newblock \href {http://arxiv.org/abs/0803.0314 [astro-ph]} {\path{
  arXiv:0803.0314 [astro-ph]}}.

\bibitem{LS5039}
F.A.\,Aharonian \etal, {\it Microquasar {LS\,5039}: {A $\Tev$} gamma-ray
  emitter and a potential {$\Tev$} neutrino source}, J.\ Phys.\ Conf.\ Ser. 39
  (2007) 408.
\newblock \href {http://arxiv.org/abs/astro-ph/0508658} {\path{
  arXiv:astro-ph/0508658}}.

\bibitem{Lipari-2006}
P.\,Lipari, {\it Perspectives of high energy neutrino astronomy}, Nucl.\ Inst.\
  Meth. A\,567 (2006) 405.
\newblock \href {http://arxiv.org/abs/astro-ph/0605535} {\path{
  arXiv:astro-ph/0605535}}.

\bibitem{Bednarek-etal-2005}
W.\,Bednarek, T.F.\,Burgio and T.\,Montaruli, {\it Galactic discrete sources of
  high energy neutrinos}, New Astron.\ Rev. 49 (2005) 1.
\newblock \href {http://arxiv.org/abs/astro-ph/0404534} {\path{
  arXiv:astro-ph/0404534}}.

\bibitem{Vissani-2006}
F.\,Vissani, {\it Neutrinos from galactic sources of cosmic rays with known
  gamma-ray spectra}, Astropart.\ Phys. 26 (2006) 310.
\newblock \href {http://arxiv.org/abs/astro-ph/0607249} {\path{
  arXiv:astro-ph/0607249}}.

\bibitem{NEMO-Aiello-2007}
NEMO Coll., S.\,Aiello \etal, {\it Sensitivity of an underwater {{\v Cerenkov}}
  km$^3$ telescope to {$\Tev$} neutrinos from {Galactic Microquasars}},
  Astropart.\ Phys. 28 (2007) 1.
\newblock \href {http://arxiv.org/abs/astro-ph/0608053} {\path{
  arXiv:astro-ph/0608053}}.

\bibitem{Vissani-2011}
F.\,Vissani, F.\,Aharonian and N.\,Sahakyan, {\it On the detectability of
  high-energy {Galactic} neutrino sources}, Astropart.\ Phys. 34 (2011) 778.
\newblock \href {http://arxiv.org/abs/1101.4842 [astro-ph.HE]} {\path{
  arXiv:1101.4842 [astro-ph.HE]}}.

\bibitem{Fermi-2009}
Fermi LAT Coll., A.A.\,Abdo \etal, {\it {Fermi Large Area Telescope} bright
  gamma-ray source list}, Astrophys.\ J.\ Suppl. 183 (2009) 46.
\newblock \href {http://arxiv.org/abs/0902.1340 [astro-ph]} {\path{
  arXiv:0902.1340 [astro-ph]}}.

\bibitem{Aharonian-2008}
F.A.\,Aharonian, J.\,Buckley, T.\,Kifune and G.\,Sinnis, {\it High energy
  astrophysics with ground-based gamma ray detectors}, Rep.\ Prog.\ Phys. 71
  (2008) 096901.

\bibitem{PKS2155-304}
W.\,Benbow \etal, for the H.E.S.S.\ Coll., {\it A spectcular {VHE} gamma-ray
  outburst from {PKS\,2155-304} in 2006}, in: R.~Caballero \etal (Ed.),
  Proc.~30th Int.\ Cosmic Ray Conf., Merida, Mexico, Vol.~3, 2007, p. 1081.
\newblock \href {http://arxiv.org/abs/0709.4608 [astro-ph]} {\path{
  arXiv:0709.4608 [astro-ph]}}.

\bibitem{Krawczynsky-2004}
H.\,Krawczynski \etal, {\it Multiwavelength observations of strong flares from
  the {$\Tev$}-blazar {1ES\,1959+650}}, Astrophys.\ J. 601 (2004) 151.
\newblock \href {http://arxiv.org/abs/astro-ph/0310158} {\path{
  arXiv:astro-ph/0310158}}.

\bibitem{Waxman-Bahcall-1997}
E.\,Waxman and J.N.\,Bahcall, {\it High energy neutrinos from cosmological
  gamma-ray burst fireballs}, Phys.\ Rev.\ Lett. 78 (1997) 2292.
\newblock \href {http://arxiv.org/abs/astro-ph/9701231} {\path{
  arXiv:astro-ph/9701231}}.

\bibitem{Razzaque-2003}
K.\,Razzaque, P.\,Meszaros and E.\,Waxman, {\it Neutrino tomography of
  gamma-ray bursts and massive stellar collapses}, Phys.\ Rev. D\,68 (2003)
  083001.
\newblock \href {http://arxiv.org/abs/astro-ph/0303505} {\path{
  arXiv:astro-ph/0303505}}.

\bibitem{Waxman-Bahcall-2000}
E.\,Waxman and J.N.\,Bahcall, {\it Neutrino afterglow from gamma-ray bursts:
  $\sim 10^{18}\ev$}, Astrophys.\ J. 541 (2000) 707.
\newblock \href {http://arxiv.org/abs/hep-ph/9909286} {\path{
  arXiv:hep-ph/9909286}}.

\bibitem{NGC253}
H.E.S.S.\ Coll., F.\,Acero \etal, {\it Detection of gamma rays from a starburst
  galaxy}, Science 326 (2009) 1080.
\newblock \href {http://arxiv.org/abs/0909.4651 [astro-ph.HE]} {\path{
  arXiv:0909.4651 [astro-ph.HE]}}.

\bibitem{M82}
N.\,Karlsson for the VERITAS Coll., {\it Discovery of {VHE} gamma-ray emission
  from the starburst galaxy {M82}}, presented at FERMI symposium, Washington
  D.C., USA (2009).
\newblock \href {http://arxiv.org/abs/0912.3807 [astro-ph.HE]} {\path{
  arXiv:0912.3807 [astro-ph.HE]}}.

\bibitem{Loeb-Waxman-2006}
A.\,Loeb and E.\,Waxman, {\it The cumulative background of high energy
  neutrinos from starburst galaxies}, J.\ Cosm.\ Astropart.\ Phys. 0605 (2006)
  003.
\newblock \href {http://arxiv.org/abs/astro-ph/0601695} {\path{
  arXiv:astro-ph/0601695}}.

\bibitem{Stecker-Salamon-1996}
F.W.\,Stecker and M.H.\,Salamon, {\it High-energy neutrinos from quasars},
  Space Sci.\ Rev. 75 (1996) 341.
\newblock \href {http://arxiv.org/abs/astro-ph/9501064} {\path{
  arXiv:astro-ph/9501064}}.

\bibitem{Stecker-2005}
F.W.\,Stecker, {\it A note on high energy neutrinos from {AGN} cores}, Phys.\
  Rev. D\,72 (2005) 107301.
\newblock \href {http://arxiv.org/abs/astro-ph/0510537} {\path{
  arXiv:astro-ph/0510537}}.

\bibitem{Amanda-atm-2010}
IceCube Coll., R.\,Abbasi \etal, {\it The energy spectrum of atmospheric
  neutrinos between 2 and {$200\tev$} with the {AMANDA-II} detector},
  Astropart.\ Phys. 34 (2010) 48.
\newblock \href {http://arxiv.org/abs/1004.2357 [astro-ph.HE]} {\path{
  arXiv:1004.2357 [astro-ph.HE]}}.

\bibitem{Volkova-1980}
L.V.\,Volkova, {\it Energy spectra and angular distributions of atmospheric
  neutrinos}, Sov.\ J.\ Nucl.\ Phys. 31 (1980) 784.

\bibitem{Waxman-Bahcall-1999}
E.\,Waxman and J.\,Bahcall,, {\it High-energy neutrinos from astrophysical
  sources: {An} upper bound}, Phys.\ Rev. D\,59 (1999) 023002.
\newblock \href {http://arxiv.org/abs/hep-ph/9807282} {\path{
  arXiv:hep-ph/9807282}}.

\bibitem{Mannheim-Protheroe-Rachen-2001}
K.\,Mannheim, R.J.\,Protheroe and J.P.\,Rachen, {\it On the cosmic ray bound
  for models of extragalactic neutrino production}, Phys.\ Rev. D\,63 (2001)
  023003.
\newblock \href {http://arxiv.org/abs/astro-ph/9812398} {\path{
  arXiv:astro-ph/9812398}}.

\bibitem{EGRET-2000}
EGRET Coll., P.\,Sreekumar \etal, {\it {EGRET} observations of the
  extragalactic gamma-ray emission}, Astrophys.\ J. 494 (1998) 523.
\newblock \href {http://arxiv.org/abs/astro-ph/9709257} {\path{
  arXiv:astro-ph/9709257}}.

\bibitem{Fermi-2010}
Fermi-LAT Coll.\, A.\,Abdo \etal, {\it The spectrum of the isotropic diffuse
  gamma-ray emission derived from first-year {Fermi Large Area Telescope}
  data}, Phys.\ Rev.\ Lett. 104 (2010) 101101.
\newblock \href {http://arxiv.org/abs/1002.3603 [astro-ph.HE]} {\path{
  arXiv:1002.3603 [astro-ph.HE]}}.

\bibitem{Berezinsky-2011}
V.\,Berezinsky, A.\,Gazizov, M.\,Kachelriess and S\,.Ostapchenko, {\it
  Restricting {UHECRs} and cosmogenic neutrinos with {Fermi-LAT}}, Phys.\ Lett.
  B\,695 (2011) 13.
\newblock \href {http://arxiv.org/abs/1003.1496 [astro-ph.HE]} {\path{
  arXiv:1003.1496 [astro-ph.HE]}}.

\bibitem{Berezinsky-1975}
V.S.\,Berezinsky and A.Y.\,Smirnov, {\it Cosmic neutrinos of ultra-high
  energies and detection possibility}, Astrophys.\ Space Sci. 32 (1975) 461.

\bibitem{Barbot-2003}
C.\,Barbot, M.\,Drees, F.\,Halzen and D.\,Hooper, {\it Neutrinos associated
  with cosmic rays of top-down origin}, Phys.\ Lett. B\,555 (2003) 22.
\newblock \href {http://arxiv.org/abs/hep-ph/0205230} {\path{
  arXiv:hep-ph/0205230}}.

\bibitem{Auger-photons}
Pierre Auger Coll., J.\,Abraham \etal, {\it Upper limit on the cosmic-ray
  photon flux above $10^{19}\ev$ using the surface detector of the {Pierre
  Auger Observatory}}, Astropart.\ Phys. 29 (2008) 243.
\newblock \href {http://arxiv.org/abs/0712.1147 [astro-ph]} {\path{
  arXiv:0712.1147 [astro-ph]}}.

\bibitem{Silvestri-2007}
A.\,Silvestri, {\it Limit on ultra high energy neutrino flux}, Ph.D. thesis,
  University of California, Irvine, {ISBN} 978-0-549-48228-4 (2007).
\newblock Available from: \url{http://www.ps.uci.edu/\til
  silvestr/DISSERTATION/DISSERTATION.pdf}.

\bibitem{icecube-2004}
IceCube Coll., J.\,Ahrens \etal, {\it Sensitivity of the {IceCube} detector to
  astrophysical sources of high energy muon neutrinos}, Astropart.\ Phys. 20
  (2004) 507.
\newblock \href {http://arxiv.org/abs/astro-ph/0305196} {\path{
  arXiv:astro-ph/0305196}}.

\bibitem{DM-Cotta-2009}
R.C.\,Cotta, J.S.\,Gainer, J.L.\,Hewett and T.G.\,Rizzo, {\it Dark matter in
  the {MSSM}}, New J.\ Phys. 11 (2009) 105026.
\newblock \href {http://arxiv.org/abs/0903.4409 [hep-ph]} {\path{
  arXiv:0903.4409 [hep-ph]}}.

\bibitem{DM-Gilmore-2007}
R.\,Gilmore, {\it Mass limits on neutralino dark matter}, Phys.\ Rev. D\,76
  (2007) 043520.
\newblock \href {http://arxiv.org/abs/0705.2610 [hep-ph]} {\path{
  arXiv:0705.2610 [hep-ph]}}.

\bibitem{DM-Olive-2010}
K.\,Olive, {\it Dark matter in {SuperGUT} unification models}, J.\ Phys.\
  Conf.\ Ser. 315 (2011) 012021.
\newblock \href {http://arxiv.org/abs/1009.0232 [hep-ph]} {\path{
  arXiv:1009.0232 [hep-ph]}}.

\bibitem{DM-Jungman-1996}
G.\,Jungmann, M.\,Kamionkowski and K.\,Griest, {\it Supersymmetric dark
  matter}, Phys.\ Rep. 267 (1996) 195.
\newblock \href {http://arxiv.org/abs/hep-ph/9506380} {\path{
  arXiv:hep-ph/9506380}}.

\bibitem{DM-Halzen-2009}
F.\,Halzen and D.\,Hooper, {\it The indirect search for dark matter with
  {IceCube}}, New J.\ Phys. 11 (2009) 105019.
\newblock \href {http://arxiv.org/abs/0910.4513 [astro-ph.HE]} {\path{
  arXiv:0910.4513 [astro-ph.HE]}}.

\bibitem{DM-Bertone-2005}
G.\,Bertone, D.\,Hooper and J.\,Silk, {\it Particle dark matter: {Evidence},
  candidates and constraints}, Phys.\ Rep. 405 (2005) 279.
\newblock \href {http://arxiv.org/abs/hep-ph/0404175} {\path{
  arXiv:hep-ph/0404175}}.

\bibitem{DM-Bergstroem-2009}
L.\,Bergström, {\it Dark matter candidates}, New J.\ Phys. 11 (2009) 105006.
\newblock \href {http://arxiv.org/abs/0903.4849 [hep-ph]} {\path{
  arXiv:0903.4849 [hep-ph]}}.

\bibitem{Ellis-2010}
J.\,Ellis, {\it New light on dark matter from the {LHC}}, in: N.S.M.\,Bor\v
  stnik, H.B.\,Nielsen and D.\,Lukman (Eds.), Proc.~13th Workshop What Comes
  Beyond the Standard Models (2010), Bled, Slovenia, 2010, p.~33.
\newblock \href {http://arxiv.org/abs/1012.0222 [hep-ph]} {\path{
  arXiv:1012.0222 [hep-ph]}}.

\bibitem{Dirac-1931}
P.A.M.\,Dirac, {\it Quantized singularities in the electromagnetic field},
  Proc.\ Royal Soc.\ Lond. A\,133 (1931) 60.

\bibitem{Giacomelli-2008}
G.\,Giacomelli, S.\,Manzoor, E.\,Medinaceli and L.\,Patrizii, {\it Searches for
  magnetic monopoles, nuclearites and {Q-balls}}, J.\ Phys.\ Conf.\ Ser. 116
  (2008) 012005.
\newblock \href {http://arxiv.org/abs/hep-ex/0702050} {\path{
  arXiv:hep-ex/0702050}}.

\bibitem{Wick-2000}
S.D.\,Wick, T.W.\,Kephart, T.J.\,Weiler and P.L.\,Biermann, {\it Signatures for
  a cosmic flux of magnetic monopoles}, Astropart.\ Phys. 18 (2003) 663.
\newblock \href {http://arxiv.org/abs/astro-ph/0001233} {\path{
  arXiv:astro-ph/0001233}}.

\bibitem{Ryu-1998}
D.\,Ryu, H.\,Kang and P.L.\,Biermann, {\it Cosmic magnetic fields in large
  scale filaments and sheets}, Astron.\,\&\,Astrophys. 335 (1998) 19.
\newblock \href {http://arxiv.org/abs/astro-ph/9803275} {\path{
  arXiv:astro-ph/9803275}}.

\bibitem{Rubakov-1981}
V.A.\,Rubakov, {\it Superheavy magnetic monopoles and proton decay}, JETP Lett.
  33 (1981) 644.

\bibitem{Rujula-1984}
A.\,De Rujula and S.L.\,Glashow, {\it Nuclearites: {A} novel form of cosmic
  radiation}, Nature 312 (1984) 734.

\bibitem{Bakari-2000}
D.\,Bakari \etal, {\it Magnetic monopoles, nuclearites, {Q-balls}: {A}
  qualitative picture}.
\newblock \href {http://arxiv.org/abs/hep-ex/0004019} {\path{
  arXiv:hep-ex/0004019}}.

\bibitem{Kusenko-1997}
A.\,Kusenko, {\it Solitons in the supersymmetric extensions of the standard
  model}, Phys.\ Lett. B\,405 (1997) 108.

\bibitem{Fujii-2002}
M.\,Fujii and K.\,Hamaguchi, {\it Non-thermal dark matter via {Afflek-Dine}
  baryogenesis and its detection possibility}, Phys.\ Rev. D\,66 (2002) 083501.
\newblock \href {http://arxiv.org/abs/hep-ph/0205044} {\path{
  arXiv:hep-ph/0205044}}.

\bibitem{Dine-2004}
M.\,Dine and A.\,Kusenko, {\it The origin of matter-antimatter asymmetry},
  Rev.\ Mod.\ Phys. 76 (2004) 1.
\newblock \href {http://arxiv.org/abs/hep-ph/0303065} {\path{
  arXiv:hep-ph/0303065}}.

\bibitem{Buchmueller-2005}
W.\,Buchmüller, R.D.\,Peccei and T.\,Yanagida, {\it Leptogenesis as the origin
  of matter}, Ann.\ Rev.\ Nucl.\ Part.\ Sci. 55 (2005) 311.
\newblock \href {http://arxiv.org/abs/hep-ph/0502169} {\path{
  arXiv:hep-ph/0502169}}.

\bibitem{Wolfenstein-1978}
L.\,Wolfenstein, {\it Neutrino oscillations in matter}, Phys.\ Rev. D\,17
  (1978) 2369.

\bibitem{Mikheev-Smirnov-1985}
S.P.\,Mikheev and A.Y.\,Smirnov, {\it Resonance amplification of oscillations
  in matter and spectroscopy of solar neutrinos}, Sov.\ J.\ Nucl.\ Phys. 42
  (1985) 913.

\bibitem{PDG}
Particle Data Group, K.\,Nakamura \etal, {\it Review of particle physics}, J.\
  Phys. G\,37 (2010) 075021.

\bibitem{Grant-2009}
D.\,Grant, D.J.\,Koskinen and C.\,Rott for the IceCube Coll., {\it Fundamental
  neutrino measurements with {IceCube DeepCore}}, contribution to 31st Int.\
  Cosmic Ray Conf., \Lodz, Poland (2009).
\newblock Available from:
  \url{http://icrc2009.uni.lodz.pl/proc/pdf/icrc1336.pdf}.

\bibitem{Learned-1995}
J.J.\,Learned and S.\,Pakvasa, {\it Detecting tau-neutrino oscillations at
  {$\Pev$} energies}, Astropart.\ Phys. 3 (1995) 267.
\newblock \href {http://arxiv.org/abs/hep-ph/9405296, hep-ph/9408296} {\path{
  arXiv:hep-ph/9405296, hep-ph/9408296}}.

\bibitem{Beacom-2003}
J.F.\,Beacom \etal, {\it Measuring flavor ratios of high-energy astrophysical
  neutrinos}, Phys.\ Rev. D\,68 (2003) 093005.
\newblock \href {http://arxiv.org/abs/hep-ph/0307025} {\path{
  arXiv:hep-ph/0307025}}.

\bibitem{Mena-2008}
O.\,Mena, I.\,Mocioiu and S.\,Razzaque, {\it Neutrino mass hierarchy extraction
  using atmospheric neutrinos in ice}, Phys.\ Rev. D\,78 (2008) 093003.
\newblock \href {http://arxiv.org/abs/0803.3044 [hep-ph]} {\path{
  arXiv:0803.3044 [hep-ph]}}.

\bibitem{Fernandez-2010}
E.\,Fernandez-Martinez, G.\,Giordano, O.\,Mena and I.\,Mocioiu, {\it
  Atmospheric neutrinos in ice and measurement of neutrino oscillation
  parameters}, Phys.\ Rev. D\,82 (2010) 093011.
\newblock \href {http://arxiv.org/abs/1008.4783 [hep-ph]} {\path{
  arXiv:1008.4783 [hep-ph]}}.

\bibitem{Coleman-1999}
S.\,Coleman and S.L.\,Glashow, {\it High-energy tests of {Lorentz} invariance},
  Phys.\ Rev. D\,59 (1999) 116008.
\newblock \href {http://arxiv.org/abs/hep-ph/9812418} {\path{
  arXiv:hep-ph/9812418}}.

\bibitem{icecube-2009d}
IceCube Coll., R.\,Abbasi \etal, {\it Determination of the atmospheric neutrino
  flux and searches for new physics with {AMANDA-II}}, Phys.\ Rev. D\,79 (2009)
  102005.
\newblock \href {http://arxiv.org/abs/0902.0675 [astro-ph.HE]} {\path{
  arXiv:0902.0675 [astro-ph.HE]}}.

\bibitem{Morgan-2007}
D.\,Morgan, E.\,Winstanley, J.\,Brunner and L.F.\,Thompson, {\it Neutrino
  telescope modelling of {Lorentz} invariance violation in oscillations of
  atmospheric neutrinos}, Astropart.\ Phys. 29 (2008) 345.
\newblock \href {http://arxiv.org/abs/0705.1897 [astro-ph]} {\path{
  arXiv:0705.1897 [astro-ph]}}.

\bibitem{nemo-sn1}
P.\,Favali \etal, {\it {NEMO-SN-1} the first ``real-time'' seafloor observatory
  of {ESONET}}, Nucl.\ Inst.\ Meth. A\,567 (2006) 462.

\bibitem{nestor-laertis}
NESTOR Coll., G.\,Aggouras \etal, {\it {LAERTIS}, a multidisciplinary station},
  Nucl.\ Inst.\ Meth. A\,567 (2006) 468.

\bibitem{Favali-2011}
P.\,Favali, A.\,de~Santis and L.\,Beranzoli, {\it Sea floor observatories},
  Springer praxis books in geophysical sciences, Springer, Berlin, 2011, {ISBN
  978-3-642-11373-4}.

\bibitem{km3net-tdr}
KM3NeT Coll., P.\,Bagley \etal, {\it {Technical Design Report}}, {ISBN
  978-90-6488-033-9} (2010).
\newblock Available from: \url{www.km3net.org}.

\bibitem{emso-web}
EMSO homepage.
\newblock Available from: \url{http://www.emso-eu.org}.

\bibitem{Amanda-shallow-1995}
AMANDA Coll., P.\,Askebjer \etal, {\it Optical properties of the {South Pole}
  ice at depths between 0.8\,km and 1\,km}, Science 267 (1995) 1147.

\bibitem{Amanda-ice-2006}
AMANDA Coll., M.\,Ackermann \etal, {\it Optical properties of deep glacial ice
  at the {South Pole}}, Journ.~Geophys.~Res. 111 (2006) D13203.

\bibitem{icecube-geology}
N.E.\,Bramall \etal, {\it A deep high-resolution optical log of dust, ash, and
  stratigraphy in {South Pole} glacial ice}, Geophys.~Res.~Lett. 32 (2005)
  L21815.

\bibitem{microbiology-2007}
P.B.\,Price, {\it Microbial life in glacial ice and implications for a cold
  origin of life}, FEMS Microbiol.~Ecol. 59 (2007) 217.

\bibitem{Katz-2000}
U.F.~Katz, {\it Deep-Inelastic Positron--Proton Scattering in the
  High-Momentum-Transfer Regime of {HERA}}, Vol. 168 of Springer Tracts in
  Modern Physics, Springer, Berlin, Heidelberg, 2000.

\bibitem{Forte-2010}
S.\,Forte, {\it Parton distributions at the dawn of the {LHC}}, Acta Phys.\
  Pol. B\,41 (2010) 2859.
\newblock \href {http://arxiv.org/abs/1011.5247 [hep-ph]} {\path{
  arXiv:1011.5247 [hep-ph]}}.

\bibitem{Connolly-2011}
A.\,Connolly, R.S.\,Thorne and D.\,Waters, {\it Calculation of high energy
  neutrino-nucleon cross sections and uncertainties using the {MSTW} parton
  distribution functions and implications for future experiments}, Phys.\ Rev.
  D\,83 (2011) 113009.
\newblock \href {http://arxiv.org/abs/1102.0691 [hep-ph]} {\path{
  arXiv:1102.0691 [hep-ph]}}.

\bibitem{Gandhi-1998}
R.\,Gandhi, C.\,Quigg, M.H.\,Reno and I.\,Sarcevic, {\it Neutrino interactions
  at ultrahigh energies}, Phys.\ Rev. D\,58 (1998) 093009.
\newblock \href {http://arxiv.org/abs/hep-ph/9807264} {\path{
  arXiv:hep-ph/9807264}}.

\bibitem{Cooper-Sarkar-2008}
A.\,Cooper-Sarkar and S.~Sarkar, {\it Predictions for high energy neutrino
  cross-sections from the {ZEUS} global {PDF} fits} 0801 (2008) 075.
\newblock \href {http://arxiv.org/abs/0710.5303 [hep-ph]} {\path{
  arXiv:0710.5303 [hep-ph]}}.

\bibitem{antares-proposal}
ANTARES Coll., E.\,Aslanides \etal, {\it A deep sea telescope for high-energy
  neutrinos} (1999).
\newblock \href {http://arxiv.org/abs/astro-ph/9907432} {\path{
  arXiv:astro-ph/9907432}}.

\bibitem{Albuquerque-2002}
I.F.M\,Albuquerque, J.\,Lamoureux and G.F.\,Smoot, {\it Astrophysical neutrino
  event rates and sensitivity for neutrino telescopes}, Astrophys.\ J.\ Suppl.
  141 (2002) 195.
\newblock \href {http://arxiv.org/abs/hep-ph/0109177} {\path{
  arXiv:hep-ph/0109177}}.

\bibitem{Gandhi-1996}
R.\,Gandhi, C.\,Quigg, M.H.\,Reno and I.\,Sarcevic, {\it Ultrahigh-energy
  neutrino interactions}, Astropart.\ Phys. 5 (1996) 81.
\newblock \href {http://arxiv.org/abs/hep-ph/9512364} {\path{
  arXiv:hep-ph/9512364}}.

\bibitem{Spiering-Handbook-2010}
C.\,Spiering, {\it Neutrino detectors under water and ice}, in: C.W.\,Fabjan
  and H.\,Schopper (Eds.), Landolt Börnstein, Vol. I/21/B/2/6.2, Springer,
  Heidelberg, New York, 2011.

\bibitem{UChicago-PMTs}
{\it Photomultiplier handbook} (1989).
\newblock Available from:
  \url{http://psec.uchicago.edu/links/Photomultiplier\usc Handbook.pdf}.

\bibitem{Hamamatsu-PMTs}
Hamamatsu, {\it Photomultiplier tubes -- basics and applications} (2006).
\newblock Available from:
  \url{http://sales.hamamatsu.com/assets/applications/ETD/pmt\usc handbook\usc
  complete.pdf}.

\bibitem{Groom-2001}
D.E.\,Groom, N.V.\,Mokhov and S.I.\,Striganov, {\it Muon stopping power and
  range tables $10\mev$--$100\tev$}, Atom.~Data Nucl.~Data Tabl. 78 (2001) 183.

\bibitem{icecube-2011f}
IceCube Coll., R.\,Abbasi \etal, {\it A search for a diffuse flux of
  astrophysical muon neutrinos with the {IceCube} 40-string detector}, Phys.\
  Rev. D\,84 (2011) 082001.
\newblock \href {http://arxiv.org/abs/1104.5187 [astro-ph.HE]} {\path{
  arXiv:1104.5187 [astro-ph.HE]}}.

\bibitem{Whipple-1989}
Whipple Coll., T.C.\,Weekes \etal, {\it Observation of {$\Tev$} gamma rays from
  the {Crab} nebula using the atmospheric {Cerenkov} imaging technique},
  Astrophys.\ J. 342 (1989) 379.

\bibitem{Hegra-1996}
HEGRA Coll., D.\,Petry \etal, {\it Detection of {VHE} gamma-rays from {Mkn-421}
  with the {HEGRA Cherenkov} telescopes}, Astron.\,\&\,Astrophys. 311 (1996)
  L13.
\newblock \href {http://arxiv.org/abs/astro-ph/9606159} {\path{
  arXiv:astro-ph/9606159}}.

\bibitem{amanda-2004b}
AMANDA Coll., J.\,Ahrens \etal, {\it Muon track reconstruction and data
  selection techniques in {AMANDA}}, Nucl.\ Inst.\ Meth. A\,524 (2004) 169.
\newblock \href {http://arxiv.org/abs/astro-ph/0407044} {\path{
  arXiv:astro-ph/0407044}}.

\bibitem{Antares-fastreco}
ANTARES Coll., J.A.\,Aguilar \etal, {\it A fast algorithm for muon track
  reconstruction and its application to the {ANTARES} neutrino telescope},
  Astropart.\ Phys. 34 (2011) 652.

\bibitem{Baikal-mureco-1999}
Baikal Coll., V.A.\,Balkanov \etal, {\it Registration of atmospheric neutrinos
  with the {BAIKAL} neutrino telescope {NT-96}}, Astropart.\ Phys. 12 (1999)
  75.
\newblock \href {http://arxiv.org/abs/astro-ph/9903341} {\path{
  arXiv:astro-ph/9903341}}.

\bibitem{Amanda-2004c}
AMANDA Coll., M.\,Ackermann \etal, {\it Search for neutrino-induced cascades
  with {AMANDA}}, Astropart.\ Phys. 22 (2004) 127.
\newblock \href {http://arxiv.org/abs/astro-ph/0405218} {\path{
  arXiv:astro-ph/0405218}}.

\bibitem{Antares-casc-2006}
B.\,Hartmann, {\it Reconstruction of neutrino-induced hadronic and
  electromagnetic showers with the {ANTARES} experiment}, Ph.D. thesis,
  Univ.~Erlangen (2006).
\newblock \href {http://arxiv.org/abs/astro-ph/0606697} {\path{
  arXiv:astro-ph/0606697}}.

\bibitem{Baikal-showerreco-2009}
Baikal Coll., A.V\,Avrorin \etal, {\it Search for high-energy neutrinos in the
  {Baikal} neutrino experiment} 35 (2009) 651.

\bibitem{DUMAND-Roberts}
A.\,Roberts, {\it The birth of high-energy neutrino astronomy: {A} personal
  history of the {DUMAND} project}, Rev.\ Mod.\ Phys. 64 (1992) 259.

\bibitem{DUMAND-Project}
DUMAND Coll., P.~Bosetti \etal, {\it {DUMAND II}: {Proposal} to construct a
  deep-ocean laboratory for the study of high energy neutrino astrophysics and
  particle physics}, Tech. Rep. HDC-2-88, Hawaii DUMAND Center, University of
  Hawaii (1988).

\bibitem{DUMAND-Babson}
DUMAND Coll., E.\,Babson \etal, {\it Cosmic ray muons in the deep ocean},
  Phys.\ Rev. D\,42 (1990) 3613.

\bibitem{Baikal-Web}
Baikal homepage.
\newblock Available from: \url{http://baikalweb.jinr.ru/}.

\bibitem{Baikal-principle-Belolaptikov-1997}
Baikal Coll., I.A.\,Belolaptikov \etal, {\it The {Baikal} underwater neutrino
  telescope: {Design}, performance, and first results}, Astropart.\ Phys. 7
  (1997) 263.

\bibitem{Baikal-1984}
Baikal Coll., L.B.\,Bezrukov \etal, {\it Progress report on {Lake Baikal}
  neutrino experiment: {Site} studies and stationary string}, in:
  Proc.~XI.\,Conf.\ on Neutrino Physics and Astrophysics, Nordkirchen, Germany,
  1984, p. 550.

\bibitem{Baikal-1986}
G.V.\,Domogatsky \etal, {\it Present status of {Baikal} deep underwater
  experiment}, in: Proc.~XII.\,Conf.\ on Neutrino Physics and Astrophysics,
  Sendai, Japan, 1986, p. 737.

\bibitem{Baikal-1990}
L.B.\,Bezrukov \etal, {\it Search for superheavy magnetic monopoles in deep
  underwater experiments at {Lake Baikal} (in {Russian})}, Sov.\ J.\ Nucl.\
  Phys. 52 (1990) 54.

\bibitem{Baikal-atm-Balkanov-1999}
Baikal Coll., R.V.\,Balkanov \etal, {\it Reconstruction of atmospheric
  neutrinos with the {Baikal} neutrino telescope {NT-96}}, proc.~25th Int.\
  Cosmic Ray Conf., Durban, South Africa (1997).
\newblock \href {http://arxiv.org/abs/astro-ph/9705244} {\path{
  arXiv:astro-ph/9705244}}.

\bibitem{Baikal-OM-Bagduev-1999}
R.\,Bagduev \etal, {\it The optical module of the {Baikal} deep underwater
  neutrino telescope}, Nucl.\ Inst.\ Meth. A\,420 (1999) 138.
\newblock \href {http://arxiv.org/abs/astro-ph/9903347} {\path{
  arXiv:astro-ph/9903347}}.

\bibitem{Baikal-diff-Aynutdinov-2006}
Baikal Coll., V.\,Aynutdinov \etal, {\it Search for a diffuse flux of
  high-energy extraterrestrial neutrinos with the {NT200} neutrino telescope},
  Astropart.\ Phys. 25 (2006) 140.
\newblock \href {http://arxiv.org/abs/astro-ph/0508675} {\path{
  arXiv:astro-ph/0508675}}.

\bibitem{Baikal-NT200+}
Baikal Coll., V.\,Aynutdinov \etal, {\it The {BAIKAL} neutrino experiment:
  {From NT200} to {NT200+}}, Nucl.\ Inst.\ Meth. A\,567 (2006) 433.
\newblock \href {http://arxiv.org/abs/astro-ph/0609743} {\path{
  arXiv:astro-ph/0609743}}.

\bibitem{amanda-2000}
AMANDA Coll., E.\,Andrés \etal, {\it The {AMANDA} neutrino telescope:
  {Principle} of operation and first results}, Astropart.\ Phys. 13 (2000) 1.
\newblock \href {http://arxiv.org/abs/astro-ph/9906203} {\path{
  arXiv:astro-ph/9906203}}.

\bibitem{amanda-2001}
AMANDA Coll., E.\,Andrés \etal, {\it Observation of high-energy neutrinos using
  {\v cerenkov} detectors embedded deep in {Antarctic} ice}, Nature 410 (2001)
  441.

\bibitem{antares-web}
{ANTARES} homepage.
\newblock Available from: \url{http://antares.in2p3.fr}.

\bibitem{antares-demo1}
F.\,Blondeau for the ANTARES Coll., {\it The {ANTARES} demonstrator: {Towards}
  a high-energy undersea neutrino telescope}, Prog.\ Part.\ Nucl.\ Phys. 40
  (1998) 413.

\bibitem{antares-demo2}
F.\,Feinstein for the ANTARES Coll., {\it The {ANTARES} demonstrator towards an
  undersea neutrino telescope}, Nucl.\ Phys.\ Proc.\ Suppl. 70 (1999) 445.

\bibitem{antares-light}
ANTARES Coll., P.\,Amram \etal, {\it Background light in potential sites for
  the {ANTARES} undersea neutrino telescope}, Astropart.\ Phys. 13 (2000) 127.

\bibitem{antares-sedi}
ANTARES Coll., P.\,Amram \etal, {\it Sedimentation and fouling of optical
  surfaces at the {ANTARES} site}, Astropart.\ Phys. 19 (2003) 253.
\newblock \href {http://arxiv.org/abs/astro-ph/0206454} {\path{
  arXiv:astro-ph/0206454}}.

\bibitem{antares-detector}
ANTARES Coll., M.\,Ageron \etal, {\it {ANTARES: The} first undersea neutrino
  telescope}, Nucl.\ Inst.\ Meth. A\,656 (2011) 11.
\newblock \href {http://arxiv.org/abs/1104.1607 [astro-ph.IM]} {\path{
  arXiv:1104.1607 [astro-ph.IM]}}.

\bibitem{antares-line0}
ANTARES Coll., M.\,Ageron \etal, {\it Studies of a full-scale mechanical
  prototype line for the {ANTARES} neutrino telescope and tests of a prototype
  instrument for deep-sea acoustic measurements}, Nucl.\ Inst.\ Meth. A\,581
  (2007) 695.

\bibitem{antares-om}
ANTARES Coll., P.\,Amram \etal, {\it The {ANTARES} optical module}, Nucl.\
  Inst.\ Meth. A\,484 (2002) 369.
\newblock \href {http://arxiv.org/abs/astro-ph/0112172} {\path{
  arXiv:astro-ph/0112172}}.

\bibitem{antares-pmt}
ANTARES Coll., J.A.\,Aguilar \etal, {\it Study of large hemispherical
  photomultiplier tubes for the {ANTARES} neutrino telescope}, Nucl.\ Inst.\
  Meth. A\,555 (2005) 132.
\newblock \href {http://arxiv.org/abs/physics/0510031 [physics.ins-det]}
  {\path{ arXiv:physics/0510031 [physics.ins-det]}}.

\bibitem{antares-water}
ANTARES Coll., J.A.\,Aguilar \etal, {\it Transmission of light in deep sea
  water at the site of the {ANTARES} neutrino telescope}, Astropart.\ Phys. 23
  (2005) 131.
\newblock \href {http://arxiv.org/abs/astro-ph/0412126} {\path{
  arXiv:astro-ph/0412126}}.

\bibitem{antares-status-brunner}
J.\,Brunner for the ANTARES Coll., {\it The {ANTARES} neutrino telescope --
  status and first results}, Nucl.\ Inst.\ Meth. A\,626--627 (2011) S19.

\bibitem{antares-daq}
ANTARES Coll., J.A.\,Aguilar \etal, {\it The data acquisition system for the
  {ANTARES} neutrino telescope}, Nucl.\ Inst.\ Meth. A\,570 (2007) 107.
\newblock \href {http://arxiv.org/abs/astro-ph/0610029} {\path{
  arXiv:astro-ph/0610029}}.

\bibitem{antares-timecal}
ANTARES Coll., J.A.\,Aguilar \etal, {\it Time calibration of the {ANTARES}
  neutrino telescope}, Astropart.\ Phys. 34 (2011) 539.
\newblock \href {http://arxiv.org/abs/1012.2204 [astro-ph.IM]} {\path{
  arXiv:1012.2204 [astro-ph.IM]}}.

\bibitem{antares-beacon}
ANTARES Coll., M.\,Ageron \etal, {\it The {ANTARES} optical beacon system},
  Nucl.\ Inst.\ Meth. A\,578 (2007) 498.
\newblock \href {http://arxiv.org/abs/astro-ph/0703355} {\path{
  arXiv:astro-ph/0703355}}.

\bibitem{antares-pos1}
M.\,Ardid for the ANTARES Coll., {\it Positioning system of the {ANTARES}
  neutrino telescope}, Nucl.\ Inst.\ Meth. A\,602 (2009) 174.

\bibitem{antares-pos2}
A.M.\,Brown for the ANTARES Coll., {\it Positioning system of the {ANTARES}
  neutrino telescope}, contribution to 31st Int.\ Cosmic Ray Conf., \Lodz,
  Poland (2009).
\newblock Available from:
  \url{http://icrc2009.uni.lodz.pl/proc/pdf/icrc0178.pdf}, \href
  {http://arxiv.org/abs/0908.0814 [astro-ph.IM]} {\path{
  arXiv:0908.0814 [astro-ph.IM]}}.

\bibitem{antares-acou}
ANTARES Coll., J.A.\,Aguilar \etal, {\it {AMADEUS} -- {The} acoustic neutrino
  detection test system of the {ANTARES} deep-sea neutrino telescope}, Nucl.\
  Inst.\ Meth. A\,626--627 (2011) 128.
\newblock \href {http://arxiv.org/abs/1009.4179 [astro-ph.IM]} {\path{
  arXiv:1009.4179 [astro-ph.IM]}}.

\bibitem{nemo-web}
{NEMO} homepage.
\newblock Available from: \url{http://nemoweb.lns.infn.it}.

\bibitem{nemo-status-2004}
NEMO Coll., E.\,Migneco \etal, {\it {NEMO: Status} of the project}, Nucl.\
  Phys.\ Proc.\ Suppl. 136 (2004) 61.

\bibitem{nemo-status-2006}
NEMO Coll., E.\,Migneco \etal, {\it Status of {NEMO}}, Nucl.\ Inst.\ Meth.
  A\,567 (2006) 444.

\bibitem{nemo-status-2009}
NEMO Coll., A.\,Capone \etal, {\it Recent results and perspectives of the
  {NEMO} project}, Nucl.\ Inst.\ Meth. A\,602 (2009) 47.

\bibitem{nemo-atmuons}
NEMO Coll., S.\,Aiello \etal, {\it Measurement of the atmospheric muon flux
  with the {NEMO Phase-1} detector}, Astropart.\ Phys. 33 (2010) 263.
\newblock \href {http://arxiv.org/abs/0910.1269 [astro-ph.IM]} {\path{
  arXiv:0910.1269 [astro-ph.IM]}}.

\bibitem{nemo-latest}
NEMO Coll., M.\,Taiuti \etal, {\it The {NEMO} project: {A} status report},
  Nucl.\ Inst.\ Meth. A\,626--627 (2011) S25.

\bibitem{nemo-acou}
NEMO Coll., G.\,Riccobene \etal, {\it Long-term measurements of acoustic
  background noise in very deep sea}, Nucl.\ Inst.\ Meth. A\,604 (2009) S149.

\bibitem{nemo-whales}
F.\,Bénard-Coudal, P.\,Giraudet and H.\,Glautin, {\it Whale {3D} monitoring
  using astrophysic {NEMO ONDE} two meters wide platform with state optimal
  filtering by {Rao-Blackwell Monte Carlo} data association}, App.~Acou. 71
  (2010) 994.

\bibitem{nestor-web}
{NESTOR} homepage.
\newblock Available from: \url{http://www.nestor.noa.gr/}.

\bibitem{nestor-1994}
NESTOR Coll., L.K.\,Resvanis \etal, {\it {NESTOR: A} neutrino particle
  astrophysics underwater laboratory for the {Mediterranean}}, Nucl.\ Phys.\
  Proc.\ Suppl. 35 (1994) 294.

\bibitem{nestor-test}
NESTOR Coll., G.\,Aggouras \etal, {\it Operation and performance of the
  {NESTOR} test detector}, Nucl.\ Inst.\ Meth. A\,552 (2005) 420.

\bibitem{nestor-atmuons}
NESTOR Coll., G.\,Aggouras \etal, {\it A measurement of the cosmic-ray muon
  flux with a module of the {NESTOR} neutrino telescope}, Astropart.\ Phys. 23
  (2005) 377.

\bibitem{icecube-web}
IceCube homepage.
\newblock Available from: \url{http://www.icecube.wisc.edu/}.

\bibitem{icecube-sensi-2004}
IceCube Coll., J.\,Ahrens \etal, {\it Sensitivity of the {IceCube} detector to
  astrophysical sources of high energy muon neutrinos}, Astropart.\ Phys. 20
  (2004) 507.
\newblock \href {http://arxiv.org/abs/astro-ph/0305196} {\path{
  arXiv:astro-ph/0305196}}.

\bibitem{IceCube-PDD-2001}
IceCube Collaboration, {\it {IceCube Preliminary Design Document}} (2001).
\newblock Available from:
  \url{http://www.icecube.wisc.edu/science/publications/pdd/}.

\bibitem{icecube-daq-2009}
IceCube Coll., R.\,Abbasi \etal, {\it The {IceCube} data acquisition system:
  {Signal} capture, digitization, and timestamping}, Nucl.\ Inst.\ Meth. A\,601
  (2009) 294.
\newblock \href {http://arxiv.org/abs/0810.4930 [physics.ins-det]} {\path{
  arXiv:0810.4930 [physics.ins-det]}}.

\bibitem{Wiebusch-2009}
C.\,Wiebusch for the IceCube Coll., {\it Physics capabilities of the {IceCube
  DeepCore} detector}, contribution to 31st Int.\ Cosmic Ray Conf., \Lodz,
  Poland (2009).
\newblock Available from:
  \url{http://icrc2009.uni.lodz.pl/proc/pdf/icrc1352.pdf}, \href
  {http://arxiv.org/abs/0907.2263 [astro-ph.IM]} {\path{
  arXiv:0907.2263 [astro-ph.IM]}}.

\bibitem{IceCube-IceTop-2008}
T.\,Waldenmaier for the IceCube Coll., {\it {IceTop} -- cosmic ray physics with
  {IceCube}}, Nucl.\ Inst.\ Meth. A\,588 (2008) 130.
\newblock \href {http://arxiv.org/abs/0802.2540 [astro-ph]} {\path{
  arXiv:0802.2540 [astro-ph]}}.

\bibitem{amanda-sn-2002}
AMANDA Coll., J.\,Ahrens \etal, {\it Search for supernova neutrino-bursts with
  the {AMANDA} detector}, Astropart.\ Phys. 16 (2002) 345.
\newblock \href {http://arxiv.org/abs/astro-ph/0105460} {\path{
  arXiv:astro-ph/0105460}}.

\bibitem{icecube-sn-2009}
T.\,Kowarik, T.\,Griesel and A.\,Piégsa for the IceCube Coll., {\it Supernova
  search with the {AMANDA\,/\, IceCube} detectors}, contribution to 31st Int.\
  Cosmic Ray Conf., \Lodz, Poland (2009).
\newblock Available from:
  \url{http://icrc2009.uni.lodz.pl/proc/pdf/icrc1251.pdf}, \href
  {http://arxiv.org/abs/0908.0441 [astro-ph.HE]} {\path{
  arXiv:0908.0441 [astro-ph.HE]}}.

\bibitem{SNEWS-2004}
P.\,Antonioli \etal, {\it {SNEWS: The} supernova early warning system}, New J.\
  Phys. 6 (2004) 114.
\newblock \href {http://arxiv.org/abs/astro-ph/0406214} {\path{
  arXiv:astro-ph/0406214}}.

\bibitem{henap-2002}
The High Energy Neutrino Astrophysics Panel, E.\,Feernandez \etal, {\it High
  energy neutrino observatories} (2002).
\newblock Available from: \url{www.lngs.infn.it/lngs\usc
  infn/contents/docs/pdf/panagic/henap2002.pdf}.

\bibitem{km3net-cdr}
KM3NeT Coll., P.\,Bagley \etal, {\it {Conceptual Design Report}}, {ISBN
  978-90-6488-031-5} (2008).
\newblock Available from: \url{www.km3net.org}.

\bibitem{esfri-2006}
European Strategy Forum on Reasearch Infrastructures (ESFRI), {\it European
  roadmap for research infrastructures, report 2006} (2006).
\newblock Available from: \url{ftp://ftp.cordis.europa.eu/pub/
  esfri/docs/esfri-roadmap-report-26092006\usc en.pdf}.

\bibitem{esfri-2008}
European Strategy Forum on Reasearch Infrastructures (ESFRI), {\it European
  roadmap for research infrastructures, update 2008} (2008).
\newblock Available from: \url{ftp://ftp.cordis.europa.eu/pub/
  esfri/docs/esfri\usc roadmap\usc update\usc 2008.pdf}.

\bibitem{Kavatsyuk-2009}
O.~Kavatsyuk for the KM3NeT Coll., {\it Photo-sensor characteristics for a
  multi-{PMT} optical module in {KM3NeT}}, contribution to 31st Int.\ Cosmic
  Ray Conf., \Lodz, Poland (2009).
\newblock Available from:
  \url{http://icrc2009.uni.lodz.pl/proc/pdf/icrc0767.pdf}.

\bibitem{Baikal-GVD-Aynutdinov-2007}
Baikal Coll., V.\,Aynutdinov \etal, {\it The prototype string for the km3-scale
  {Baikal} neutrino telescope}, Nucl.\ Inst.\ Meth. A\,602 (2009) 227.
\newblock \href {http://arxiv.org/abs/0811.1110 [astro-ph]} {\path{
  arXiv:0811.1110 [astro-ph]}}.

\bibitem{Baikal-GVD-Avrorin-2010}
Baikal Coll., A.\,Avrorin \etal, {\it The {Baikal} experiment -- from {Megaton
  to Gigaton}}, J.\ Phys.\ Conf.\ Ser. 203 (2010) 012123.

\bibitem{Gabici-2008}
S.\,Gabici \etal, {\it The diffuse neutrino flux from the inner {Galaxy:
  Constraints} from very high energy gamma-ray observations}, Astropart.\ Phys.
  30 (2008) 180.
\newblock \href {http://arxiv.org/abs/0806.2459 [astro-ph]} {\path{
  arXiv:0806.2459 [astro-ph]}}.

\bibitem{Barr-2006}
G.D.\,Barr, T.\,Gaisser, S.\,Robbins and T.\,Stanev, {\it Uncertainties in
  atmospheric neutrino fluxes}, Phys.\ Rev. D\,74 (2006) 094009.
\newblock \href {http://arxiv.org/abs/astro-ph/0611266} {\path{
  arXiv:astro-ph/0611266}}.

\bibitem{Daum-1995}
K.\,Daum \etal, {\it Determination of the atmospheric neutrino spectra with the
  {Fréjus} detector}, Z.\ Phys. C\,66 (1995) 417.

\bibitem{SK-atmnu}
M.C.\,González-García, M.\,Maltoni and J.\,Rojo, {\it Determination of the
  atmospheric neutrino fluxes from experimental data}, Astrophys.\ Space Sci.
  309 (2007) 447.

\bibitem{icecube-2010d}
IceCube Coll., R.\,Abbasi \etal, {\it First search for extremely high energy
  cosmogenic neutrinos with the {IceCube} neutrino observatory}, Phys.\ Rev.
  D\,82 (2010) 072003.
\newblock \href {http://arxiv.org/abs/1009.1442 [astro-ph.CO]} {\path{
  arXiv:1009.1442 [astro-ph.CO]}}.

\bibitem{icecube-2011c}
IceCube Coll., R.\,Abbasi \etal, {\it Measurement of the atmospheric neutrino
  energy spectrum from {$100\gev$ to $400\tev$} with {IceCube}}, Phys.\ Rev.
  D\,83 (2011) 012001.
\newblock \href {http://arxiv.org/abs/1010.3980 [astro-ph.HE]} {\path{
  arXiv:1010.3980 [astro-ph.HE]}}.

\bibitem{icecube-siderial}
IceCube Coll., R.\,Abbasi \etal, {\it Search for a {Lorentz}-violating siderial
  signal with atmospheric neutrinos in {IceCube}}, Phys.\ Rev. D\,82 (2010)
  112003.
\newblock \href {http://arxiv.org/abs/1010.4096 [astro-ph.HE]} {\path{
  arXiv:1010.4096 [astro-ph.HE]}}.

\bibitem{Kowalski-2005}
M.\,Kowalski, {\it Measuring diffuse neutrino fluxes with {IceCube}}, J.\
  Cosm.\ Astropart.\ Phys. 0505 (2005) 010.
\newblock \href {http://arxiv.org/abs/astro-ph/0505506} {\path{
  arXiv:astro-ph/0505506}}.

\bibitem{icecube-2007b}
IceCube Coll., A.\,Achterberg \etal, {\it Multi-year search for a diffuse flux
  of muon neutrinos with {AMANDA-II}}, Phys.\ Rev. D\,76 (2007) 042008, erratum
  ibid., D\,77~(2008)~089904(E).
\newblock \href {http://arxiv.org/abs/0705.1315 [astro-ph]} {\path{
  arXiv:0705.1315 [astro-ph]}}.

\bibitem{Antares-diffuse}
ANTARES Coll., J.A.\,Aguilar \etal, {\it Search for a diffuse flux of
  high-energy $\nu_\mu$ with the {ANTARES} neutrino telescope}, Phys.\ Lett.
  B\,696 (2011) 16.
\newblock \href {http://arxiv.org/abs/1011.3772 [astro-ph.HE]} {\path{
  arXiv:1011.3772 [astro-ph.HE]}}.

\bibitem{icecube-2011b}
IceCube Coll., R.\,Abbasi \etal, {\it Search for neutrino-induced cascades with
  five years of {AMANDA} data}, Astropart.\ Phys. 34 (2011) 420.

\bibitem{Baikal-diff-2009}
A.\,Kochanov for the Baikal Coll., {\it Search for a diffuse flux of
  high-energy neutrinos with the {Baikal} neutrino telescope {NT200}},
  contribution to 31st Int.\ Cosmic Ray Conf., \Lodz, Poland (2009).
\newblock Available from:
  \url{http://icrc2009.uni.lodz.pl/proc/pdf/icrc1093.pdf}, \href
  {http://arxiv.org/abs/0909.5562 [astro-ph.HE]} {\path{
  arXiv:0909.5562 [astro-ph.HE]}}.

\bibitem{IC-22-cascades}
IceCube Coll., R.\,Abbasi \etal, {\it First search for atmospheric and
  extraterrestrial neutrino-induced cascades with the {IceCube} detector}
  (2011).
\newblock \href {http://arxiv.org/abs/1101.1692 [astro-ph.HE]} {\path{
  arXiv:1101.1692 [astro-ph.HE]}}.

\bibitem{MACRO-diffuse}
MACRO Coll., M.\,Ambrosio \etal, {\it Search for diffuse neutrino flux from
  astrophysical sources with {MACRO}}, Astropart.\ Phys. 19 (2003) 1.
\newblock \href {http://arxiv.org/abs/astro-ph/0203181} {\path{
  arXiv:astro-ph/0203181}}.

\bibitem{icecube-EHE2011}
IceCube Coll., R.\,Abbasi \etal, {\it Constraints on the extremely-high energy
  cosmic neutrino flux with the {IceCube} 2008--2009 data}, Phys.\ Rev. D\,83
  (2011) 092003.
\newblock \href {http://arxiv.org/abs/1103.4250 [astro-ph.CO]} {\path{
  arXiv:1103.4250 [astro-ph.CO]}}.

\bibitem{RICE-2006}
RICE Coll., I.\,Kravchenko \etal, {\it {RICE} limits on the diffuse ultrahigh
  energy neutrino flux}, Phys.\ Rev. D\,73 (2006) 082002.
\newblock \href {http://arxiv.org/abs/astro-ph/0601148} {\path{
  arXiv:astro-ph/0601148}}.

\bibitem{Auger-2010}
P.\,Billoir for the Pierre Auger Coll., {\it Limit on the diffuse flux of ultra
  high energy neutrinos using the {Pierre Auger Observatory}}, J.\ Phys.\
  Conf.\ Ser. 203 (2010) 012125.

\bibitem{ANITA-2009}
ANITA Coll., P.W.\,Gorham \etal, {\it Observational constraints on the
  ultrahigh energy cosmic neutrino flux from the second flight of the {ANITA}
  experiment}, Phys.\ Rev. D\,82 (2010) 022004, {Erratum in
  arXiv:1011.5004 [astro-ph.HE]}.
\newblock \href {http://arxiv.org/abs/1003.2961 [astro-ph.HE]} {\path{
  arXiv:1003.2961 [astro-ph.HE]}}.

\bibitem{Kotera-2011}
K.\,Kotera and A.V.\,Olinto, {\it The astrophysics of ultrahigh energy cosmic
  rays}, Ann.\,Rev.\,Astron.\,Astrophys. 49 (2011) 1.
\newblock \href {http://arxiv.org/abs/1101.4256 [astro-ph.HE]} {\path{
  arXiv:1101.4256 [astro-ph.HE]}}.

\bibitem{Gelmini-2011}
G.B.\,Gelmini, O.\,Kalashev and D.V.\,Semikoz, {\it Gamma-ray constraints on
  maximum cosmogenic neutrino fluxes and {UHECR} source evolution models}, to
  appear in JHEP (2011).
\newblock \href {http://arxiv.org/abs/1107.1672 [astro-ph.CO]} {\path{
  arXiv:1107.1672 [astro-ph.CO]}}.

\bibitem{MACROpoint-2001}
MACRO Coll., M.\,Ambrosio \etal, {\it Neutrino astronomy with the {MACRO}
  detector}, Astrophys.\ J. 546 (2001) 1038.
\newblock \href {http://arxiv.org/abs/astro-ph/0002492} {\path{
  arXiv:astro-ph/0002492}}.

\bibitem{SuperK-point-2009}
Super-Kamiokande Coll., E.\,Thrane \etal, {\it Search for astrophysical
  neutrino point sources at {Super-Kamiokande}}, Astrophys.\ J. 704 (2009) 503.
\newblock \href {http://arxiv.org/abs/0907.1594 [astro-ph.HE]} {\path{
  arXiv:0907.1594 [astro-ph.HE]}}.

\bibitem{icecube-2009a}
IceCube Coll., R.\,Abbasi \etal, {\it Search for point sources of high energy
  neutrinos with final data from {AMANDA-II}}, Phys.\ Rev. D\,79 (2009) 062001.
\newblock \href {http://arxiv.org/abs/0809.1646 [astro-ph]} {\path{
  arXiv:0809.1646 [astro-ph]}}.

\bibitem{Braun-2008}
J.\,Braun \etal, {\it Methods for point source analysis in high energy neutrino
  telescopes}, Astropart.\ Phys. 29 (2008) 299.
\newblock \href {http://arxiv.org/abs/0801.1604 [astro-ph]} {\path{
  arXiv:0801.1604 [astro-ph]}}.

\bibitem{Antares-pointlims}
J.~Hernandez, {\it Northern hemisphere neutrino telescopes}, presentation at
  {\it Very Large Volume Neutrino Telescopes (VLVnT11)}, Erlangen, Germany,
  Oct.\,2011 (2011).
\newblock Available from:
  \url{https://indico.cern.ch/contributionDisplay.py?contribId=5\amp
  confId=143656}.

\bibitem{IC-40-point}
IceCube Coll., R.\,Abbasi \etal, {\it Time-integrated searches for point-like
  sources of neutrinos with the 40-string {IceCube} detector}, Astrophys.\ J.
  732 (2011) 18.
\newblock \href {http://arxiv.org/abs/1012.2137} {\path{ arXiv:1012.2137}}.

\bibitem{IceCube-2009e}
IceCube Coll., R.\,Abbasi \etal, {\it First neutrino point-source results from
  the 22 string {IceCube} detector}, Astrophys.\ J. 701 (2009) L47.
\newblock \href {http://arxiv.org/abs/0905.2253 [astro-ph.HE]} {\path{
  arXiv:0905.2253 [astro-ph.HE]}}.

\bibitem{icecube-2009g}
IceCube Coll., R.\,Abbasi \etal, {\it Extending the search for neutrino point
  sources with {IceCube} above the horizon}, Phys.\ Rev.\ Lett. 103 (2009)
  221102.
\newblock \href {http://arxiv.org/abs/0911.2338 [astro-ph.HE]} {\path{
  arXiv:0911.2338 [astro-ph.HE]}}.

\bibitem{AMANDAB10-2000}
Amanda Coll., E.\,Andres \etal, {\it Results from the {AMANDA} high-energy
  neutrino detector}, Nucl.\ Phys.\ Proc.\ Suppl. 91 (2001) 423.
\newblock \href {http://arxiv.org/abs/astro-ph/0009242} {\path{
  arXiv:astro-ph/0009242}}.

\bibitem{Morlino-2009}
G.\,Morlino, E.\,Amato and P.\,Blasi, {\it Gamma ray emission from {SNR
  RX\,J1713.7-3946} and the origin of galactic cosmic rays}, Mon.\ Not.\ Roy.\
  Astron.\ Soc. 392 (2009) 240.
\newblock \href {http://arxiv.org/abs/0810.0094 [astro-ph]} {\path{
  arXiv:0810.0094 [astro-ph]}}.

\bibitem{Koers-2008}
H.B.J.\,Koers and P.\,Tinyakov, {\it Relation between the neutrino flux from
  {Centaurus A} and the associated diffuse neutrino flux}, Phys.\ Rev. D\,78
  (2008) 083009.
\newblock \href {http://arxiv.org/abs/0802.2403 [astro-ph]} {\path{
  arXiv:0802.2403 [astro-ph]}}.

\bibitem{Braun-2009}
J.\,Braun \etal, {\it Time-dependent point source search methods in high energy
  neutrino astronomy}, Astropart.\ Phys. 33 (2010) 175.
\newblock \href {http://arxiv.org/abs/0912.1572 [astro-ph.IM]} {\path{
  arXiv:0912.1572 [astro-ph.IM]}}.

\bibitem{Alba-2009}
J.L.\,Bazo Alba, E.\,Bernardini and R.\,Lauer for the IceCube Coll., {\it
  Search for neutrino flares from point sources with {IceCube}}, contribution
  to 31st Int.\ Cosmic Ray Conf., \Lodz, Poland (2009).
\newblock Available from:
  \url{http://icrc2009.uni.lodz.pl/proc/pdf/icrc0960.pdf}, \href
  {http://arxiv.org/abs/0908.4209 [astro-ph.HE]} {\path{
  arXiv:0908.4209 [astro-ph.HE]}}.

\bibitem{Gora-2011}
D.\,Góra, E.\,Bernardini and A.H.\,Cruz Silva, {\it A method for untriggered
  time-dependent searches for multiple flares from neutrino point sources},
  Astropart.\ Phys. 35 (2011) 201.
\newblock \href {http://arxiv.org/abs/1103.2644 [astro-ph.IM]} {\path{
  arXiv:1103.2644 [astro-ph.IM]}}.

\bibitem{IceCube-2011-time-dependent}
IceCube Coll., R.\,Abbasi \etal, {\it Time-dependent searches for point sources
  of neutrinos with the 40-string and 22-string configurations of {IceCube}}
  (2011).
\newblock \href {http://arxiv.org/abs/1104.0075 [astro-ph.HE]} {\path{
  arXiv:1104.0075 [astro-ph.HE]}}.

\bibitem{Markus-thesis}
M.\,Ackermann, {\it Searches for signals from cosmic point-like sources of high
  energy neutrinos in 5 years of {AMANDA-II} data}, Ph.D. thesis, Humboldt
  University Berlin (2006).
\newblock Available from:
  \url{http://edoc.hu-berlin.de/docviews/abstract.php?lang=ger\amp id=27726}.

\bibitem{Bernardini-2005}
E.\,Bernardini for the IceCube Coll., {\it Multi-messenger studies with
  {AMANDA\,/\,IceCube}: {Observations} and strategies}, contribution to the
  Cherenkov 2005 Conference, Palaiseau, France (2005).
\newblock \href {http://arxiv.org/abs/astro-ph/0509396} {\path{
  arXiv:astro-ph/0509396}}.

\bibitem{Reimer-2005}
A.\,Reimer, M.\,Böttcher and S.\,Postnikov, {\it Neutrino emission in the
  hadronic synchrotron mirror model: the ``orphan'' {$\Tev$} flare from
  {1ES\;1959+650}}, Astrophys.\ J. 630 (2005) 186.
\newblock \href {http://arxiv.org/abs/astro-ph/0505233} {\path{
  arXiv:astro-ph/0505233}}.

\bibitem{Halzen-Hooper-2005}
F.\,Halzen and D.\,Hooper, {\it High energy neutrinos from the {$\Tev$} blazar
  {1ES\;1959+650}}, Astropart.\ Phys. 23 (2005) 537.
\newblock \href {http://arxiv.org/abs/astro-ph/0502449} {\path{
  arXiv:astro-ph/0502449}}.

\bibitem{icecube-2011d}
IceCube Coll., R.\,Abbasi \etal, {\it Constraints on high-energy neutrino
  emission from {SN\;2008D}}, Astron.\,\&\,Astrophys. 527 (2011) A28.
\newblock \href {http://arxiv.org/abs/1101.3942 [astro-ph.HE]} {\path{
  arXiv:1101.3942 [astro-ph.HE]}}.

\bibitem{Guetta-2004}
D.\,Guetta \etal, {\it Neutrinos from individual gamma-ray bursts in the
  {BATSE} catalog}, Astropart.\ Phys. 20 (2004) 429.
\newblock \href {http://arxiv.org/abs/astro-ph/0302524} {\path{
  arXiv:astro-ph/0302524}}.

\bibitem{icecube-2008a}
IceCube and the InterPlanetary Network Coll., A.\,Achterberg \etal, {\it The
  search for muon neutrinos from northern hemisphere gamma-ray bursts with
  {AMANDA}}, Astrophys.\ J. 674 (2008) 357.
\newblock \href {http://arxiv.org/abs/0705.1186 [astro-ph]} {\path{
  arXiv:0705.1186 [astro-ph]}}.

\bibitem{icecube-2011e}
IceCube Coll., R.\,Abbasi \etal, {\it Limits on neutrino emission from
  gamma-ray bursts with the 40 string {IceCube} detector}, Phys.\ Rev.\ Lett.
  106 (2011) 141101.
\newblock \href {http://arxiv.org/abs/1101.1448 [astro-ph.HE]} {\path{
  arXiv:1101.1448 [astro-ph.HE]}}.

\bibitem{icecube-2011h}
P.\,Redl for the IceCube Coll., {\it Limits on neutrino emission from gamma-ray
  bursts with the 59 string {IceCube} detector}, contribution to 32nd Int.\
  Cosmic Ray Conf., Beijing, China (2011).
\newblock Available from:
  \url{http://galprop.stanford.edu/elibrary/icrc/2011/papers/HE2.3/icrc0764.pd%
f}.

\bibitem{Kowalski-Mohr-2007}
M.\,Kowalski and A.\,Mohr, {\it Detecting neutrino transients with optical
  follow-up observations}, Astropart.\ Phys. 27 (2007) 533.
\newblock \href {http://arxiv.org/abs/astro-ph/0701618} {\path{
  arXiv:astro-ph/0701618}}.

\bibitem{Antares-OFU-2008}
S.\,Basa \etal, {\it Neutrino alert systems for {Gamma Ray Bursts} and
  transient astronomical sources}, Nucl.\ Inst.\ Meth. A\,602 (2009) 275.
\newblock \href {http://arxiv.org/abs/0810.1394 [astro-ph]} {\path{
  arXiv:0810.1394 [astro-ph]}}.

\bibitem{Antares-OFU-2011}
ANTARES Coll., M.\,Ageron \etal, {\it The {ANTARES} telescope neutrino alert
  system}, subm.\ to Astropart.~Phys.\href
  {http://arxiv.org/abs/1103.4477 [astro-ph.IM]} {\path{
  arXiv:1103.4477 [astro-ph.IM]}}.

\bibitem{NToO-2007}
IceCube and MAGIC Coll., M.\,Ackermann \etal, {\it Neutrino triggered target of
  opportunity {(NToO)} test run with {AMANDA-II} and {MAGIC}}, in: R.~Caballero
  \etal (Ed.), Proc.\ 30th Int.~Cosmic Ray Conf., Merida, Mexico, Vol.~3, 2008,
  p. 1257.
\newblock \href {http://arxiv.org/abs/0709.2640 [astro-ph]} {\path{
  arXiv:0709.2640 [astro-ph]}}.

\bibitem{icecube-SN-2010}
Th.\,Kowarik, T.\,Griesel and A.\,Piégsa for the IceCube Coll., {\it Supernova
  search with the {AMANDA\,/\,IceCube} detectors}, contribution to 31st Int.\
  Cosmic Ray Conf., \Lodz, Poland (2009).
\newblock Available from:
  \url{http://icrc2009.uni.lodz.pl/proc/pdf/icrc1251.pdf}, \href
  {http://arxiv.org/abs/0908.0441 [astro-ph.HE]} {\path{
  arXiv:0908.0441 [astro-ph.HE]}}.

\bibitem{Raffelt-2007}
G.\,Raffelt, {\it Supernova neutrino observations: {What} can we learn?},
  proc.\ 22nd Int.\ Conf.\ Neutrino Physics and Astrophysics, Santa Fe, USA
  (2007).
\newblock Available from:
  \url{http://lss.fnal.gov/conf/C0606131/Neutrino06-Proceedings.pdf}, \href
  {http://arxiv.org/abs/astro-ph/0701677} {\path{ arXiv:astro-ph/0701677}}.

\bibitem{amanda-2006}
AMANDA Coll., M.\,Ackermann \etal, {\it Limits to the muon flux from neutralino
  annihilations in the {Sun} with the {AMANDA} detector}, Astropart.\ Phys. 24
  (2006) 459.
\newblock \href {http://arxiv.org/abs/astro-ph/0508518} {\path{
  arXiv:astro-ph/0508518}}.

\bibitem{Lim-2009}
G.M.A.\,Lim for the ANTARES Coll., {\it First results on the search for dark
  matter in the {Sun} with the {ANTARES} neutrino telescope}, contribution to
  31st Int.\ Cosmic Ray Conf., \Lodz, Poland (2009).
\newblock Available from:
  \url{http://icrc2009.uni.lodz.pl/proc/pdf/icrc0031.pdf}.

\bibitem{Motz-2009}
H.\,Motz for the ANTARES Coll., {\it Indirect search for dark matter with the
  {ANTARES} neutrino telescope}, DOI 10.1142/9789814293792\usc0041, proc.\ 7th
  Int.\ Conf.\ on Dark Matter in Astrophysics and Particle Physics (Dark09),
  Christchurch, New Zealand (2009).

\bibitem{Lim-2011}
G.M.A.\,Lim, {\it Searching for dark matter with the {ANTARES} neutrino
  telescope}, Ph.D. thesis, University of Amsterdam, The Netherlands (2011).
\newblock Available from:
  \url{http://www.nikhef.nl/pub/services/biblio/theses\usc pdf/thesis\usc G\usc
  Lim.pdf}.

\bibitem{Rott-2009}
C.\,Rott for the IceCube Coll., {\it Search for dark matter from the {Galactic}
  halo with {IceCube}} (2009).
\newblock \href {http://arxiv.org/abs/0912.5183 [astro-ph.HE]} {\path{
  arXiv:0912.5183 [astro-ph.HE]}}.

\bibitem{Heros-2011}
C.\,de los Heros for the IceCube Coll., {\it Dark matter searches with
  {IceCube}}, proc.\ 8th Int.\ Workshop on Identification of Dark Matter
  (IDM2010), Montpellier, France (2010).
\newblock \href {http://arxiv.org/abs/1012.0184} {\path{ arXiv:1012.0184}}.

\bibitem{Conrad-2011}
J.\,Conrad, {\it Indirect detection of dark matter with gamma-rays -- status
  and perspectives}, proc.\ 8th Int.\ Workshop on Identification of Dark Matter
  (IDM2010), Montpellier, France (2011).
\newblock \href {http://arxiv.org/abs/1103.5638 [astro-ph.CO]} {\path{
  arXiv:1103.5638 [astro-ph.CO]}}.

\bibitem{Porter-2011}
T.A.\,Porter, R.P.\,Johnson and P.W.\,Graham, {\it Dark matter searches with
  astroparticle data}, submitted to Ann.~Rev.~Astron.~Astrophys (2011).
\newblock \href {http://arxiv.org/abs/1104.2836 [astro-ph.HE]} {\path{
  arXiv:1104.2836 [astro-ph.HE]}}.

\bibitem{Kappl-2011}
R.\,Kappl and M.W.\,Winkler, {\it New limits on dark matter from
  {Super-Kamiokande}}, Nucl.\ Phys. B\,850 (2011) 505.
\newblock \href {http://arxiv.org/abs/1104.0679 [hep-ph]} {\path{
  arXiv:1104.0679 [hep-ph]}}.

\bibitem{icecube-2010f}
IceCube Coll., R.\,Abbasi \etal, {\it Limits on a muon flux from {Kaluza-Klein}
  dark matter annihilations in the {Sun} from the {IceCube} 22-string
  detector}, Phys.\ Rev. D\,81 (2010) 057101.
\newblock \href {http://arxiv.org/abs/0910.4480 [astro-ph.CO]} {\path{
  arXiv:0910.4480 [astro-ph.CO]}}.

\bibitem{Baimon}
Baikal Coll., V.\,Aynutdinov \etal, {\it Search for relativistic magnetic
  monopoles with the {Baikal} neutrino telescope}, Astropart.\ Phys. 29 (2008)
  366.

\bibitem{icecube-2010e}
IceCube Coll., R.\,Abbasi \etal, {\it Search for relativistic magnetic
  monopoles with the {AMANDA-II} neutrino telescope}, Eur.\ Phys.\ J. C\,69
  (2010) 361.

\bibitem{antares-monopoles}
ANTARES Coll., S.\,Adrián-Martínez \etal, {\it Search for relativistic magnetic
  monopoles with the {ANTARES} neutrino telescope} (2011).
\newblock \href {http://arxiv.org/abs/1110.2656 [astro-ph.HE]} {\path{
  arXiv:1110.2656 [astro-ph.HE]}}.

\bibitem{icecube-mon-2011}
IceCube Coll., {\it Search for relativistic monopoles with {IceCube}}, to be
  subm.\ to Phys.\,Rev.\,Lett. (2011).

\bibitem{Turner-1982}
M.S.\,Turner, E.N.\,Parker and T.J.\,Bogdan, {\it Magnetic monopoles and the
  survival of {Galactic} magnetic fields}, Phys.\ Rev. D\,26 (1982) 1296.

\bibitem{Giacomelli-2007}
G.\,Giacomelli, S.\,Manzoor, E.\,Medinaceli and L.\,Patrizii, {\it Searches for
  magnetic monopoles, nuclearites and {Q-balls}}, J.\ Phys.\ Conf.\ Ser. 116
  (2008) 012005.
\newblock \href {http://arxiv.org/abs/hep-ex/0702050} {\path{
  arXiv:hep-ex/0702050}}.

\bibitem{baikal-1}
I.A.\,Belolpatikov \etal, {\it The experimental limits on {Q-ball} flux with
  the {Baikal} deep underwater array {`Gyrlyanda'}} (1998).
\newblock \href {http://arxiv.org/abs/astro-ph/9802223} {\path{
  arXiv:astro-ph/9802223}}.

\bibitem{baikal-2}
I.\,Sokalski for the Baikal Coll., {\it Search for magnetic monopoles with deep
  underwater {Cherenkov} detectors at {Lake Baikal}} (1995).
\newblock \href {http://arxiv.org/abs/9601160} {\path{ arXiv:9601160}}.

\bibitem{Arvid-Thesis}
A.\,Pohl, {\it Search for subrelativistic particles with the {AMANDA} neutrino
  telescope}, Ph.D. thesis, University of Uppsala, Sweden (2009).
\newblock Available from:
  \url{http://wwwiexp.desy.de/groups/astroparticle/pubs/Thesis.Arvid.090210.pd%
f}.

\bibitem{SPASE}
AMANDA and SPASE Coll., J.\,Ahrens \etal, {\it Measurement of the cosmic ray
  composition at the knee with the {SPASE-2/AMANDA-B10} detectors}, Astropart.\
  Phys. 21 (2004) 565.

\bibitem{SPASE-calib}
AMANDA and SPASE Coll., J.\,Ahrens \etal, {\it Calibration and survey of
  {AMANDA} with the {SPASE} detectors}, Nucl.\ Inst.\ Meth. A\,522 (2004) 347.

\bibitem{IceCube-Moon}
D.J.\,Boersma, L.\,Gladsstone and A.\,Karle for the IceCube Coll., {\it Moon
  shadow observation by {IceCube}}, contribution to 31st Int.\ Cosmic Ray
  Conf., \Lodz, Poland (2009).
\newblock Available from:
  \url{http://icrc2009.uni.lodz.pl/proc/pdf/icrc1173.pdf}, \href
  {http://arxiv.org/abs/1002.4900 [astro-ph.HE]} {\path{
  arXiv:1002.4900 [astro-ph.HE]}}.

\bibitem{antares-muon-zenith-2010}
ANTARES Coll., J.A.\,Aguilar \etal, {\it Zenith distribution and flux of
  atmospheric muons measured with the 5-line{ ANTARES} detector}, Astropart.\
  Phys. 34 (2010) 179.
\newblock \href {http://arxiv.org/abs/1007.1777 [astro-ph.HE]} {\path{
  arXiv:1007.1777 [astro-ph.HE]}}.

\bibitem{Bugaev-1998}
E.V.\,Bugaev \etal, {\it Atmospheric muon flux at sea level, underground and
  underwater}, Phys.\ Rev. D\,58 (1998) 054001.
\newblock \href {http://arxiv.org/abs/hep-ph/9803488} {\path{
  arXiv:hep-ph/9803488}}.

\bibitem{anisotropy-Milagro}
MILAGRO Coll., A.A.\,Abdo \etal, {\it The large-scale cosmic-ray anisotropy as
  observed with {Milagro}}, Astrophys.\ J. 698 (2009) 2121.
\newblock \href {http://arxiv.org/abs/0806.2293 [astro-ph]} {\path{
  arXiv:0806.2293 [astro-ph]}}.

\bibitem{anisotropy-Tibet}
Tibet-AS$\gamma$ Coll., M.\,Amenomori \etal, {\it Anisotropy and corotation of
  {Galactic} cosmic rays}, Science 314 (2006) 439.
\newblock \href {http://arxiv.org/abs/astro-ph/0610671} {\path{
  arXiv:astro-ph/0610671}}.

\bibitem{anisotropy-Argo}
S.\,Vernetto, Z.\,Guglielmotto and J.L.\,Zhang for the ARGO-YBJ Coll., {\it Sky
  monitoring with {ARGO-YBJ}}, contribution to 31st Int.\ Cosmic Ray Conf.,
  \Lodz, Poland (2009).
\newblock Available from:
  \url{http://icrc2009.uni.lodz.pl/proc/pdf/icrc0399.pdf}, \href
  {http://arxiv.org/abs/0907.4615 [astro-ph.HE]} {\path{
  arXiv:0907.4615 [astro-ph.HE]}}.

\bibitem{anisotropy-SK}
Super-Kamiokande Coll., G.\,Guillian \etal, {\it Observation of the anisotropy
  of {$10\tev$} primary cosmic ray nuclei flux with the {Super-Kamiokande-I}
  detector}, Phys.\ Rev. D\,75 (2007) 062003.
\newblock \href {http://arxiv.org/abs/astro-ph/0508468} {\path{
  arXiv:astro-ph/0508468}}.

\bibitem{Compton-Getting}
A.H.\,Compton and I.A.\,Getting, {\it An apparent effect of {Galactic} rotation
  on the intensity of cosmic rays}, Phys.\ Rev. 47 (1935) 817.

\bibitem{icecube-2010b}
IceCube Coll., R.\,Abbasi \etal, {\it Measurement of the anisotropy of cosmic
  ray arrival directions with {IceCube}}, Astrophys.\ J. 718 (2010) L194.
\newblock \href {http://arxiv.org/abs/1005.2960 [astro-ph.HE]} {\path{
  arXiv:1005.2960 [astro-ph.HE]}}.

\bibitem{icecube-2011g}
IceCube Coll., R.\,Abbasi \etal, {\it Observation of anisotropy in the arrival
  directions of {Galactic} cosmic rays at multiple angular scales with
  {IceCube}}.
\newblock \href {http://arxiv.org/abs/1105.2326 [astro-ph.HE]} {\path{
  arXiv:1105.2326 [astro-ph.HE]}}.

\bibitem{Drury-2008}
L.\,Drury and F.\,Aharonian, {\it The puzzling {MILAGRO} hot spots},
  Astropart.\ Phys. 29 (2008) 420.
\newblock \href {http://arxiv.org/abs/0802.4403 [astro-ph]} {\path{
  arXiv:0802.4403 [astro-ph]}}.

\bibitem{Lazari-2010}
A.\,Lazarian and P.\,Desiati, {\it Magnetic reconnection as the cause of cosmic
  ray excess from the heliospheric tail}, Astrophys.\ J. 722 (2010) 188.
\newblock \href {http://arxiv.org/abs/1008.1981 [astro-ph.CO]} {\path{
  arXiv:1008.1981 [astro-ph.CO]}}.

\bibitem{Auger-Web}
Pierre Auger Observatory homepage.
\newblock Available from: \url{http://www.auger.org}.

\bibitem{TA-2003}
M.\,Sasaki, Y.\,Asaoka and M.\,Jobashi, {\it Detecting very high energy
  neutrinos by the {Telescope Array}}, Astropart.\ Phys. 19 (2003) 37.
\newblock \href {http://arxiv.org/abs/astro-ph/0204167} {\path{
  arXiv:astro-ph/0204167}}.

\bibitem{Fargion-2002}
D.\,Fargion, {\it Discovering ultra-high-energy neutrinos by horizontal and
  upward $\tau$ air-showers: {Evidences} in terrestrial gamma flashes?},
  Astrophys.\ J. 570 (2002) 909.
\newblock \href {http://arxiv.org/abs/astro-ph/0002453} {\path{
  arXiv:astro-ph/0002453}}.

\bibitem{Fargion-2004}
D.\,Fargion, P.G.\,De Sanctis Lucentini and M.\,De Santis, {\it Tau air showers
  from {Earth}}, Astrophys.\ J. 613 (2004) 1285.
\newblock \href {http://arxiv.org/abs/hep-ph/0305128} {\path{
  arXiv:hep-ph/0305128}}.

\bibitem{Auger-nutau}
Pierre Auger Coll., J.\,Abraham \etal, {\it Upper limit on the diffuse flux of
  {UHE} tau neutrinos from the {Pierre Auger Observatory}}, Phys.\ Rev.\ Lett.
  100 (2008) 211101.
\newblock \href {http://arxiv.org/abs/0712.1909 [astro-ph]} {\path{
  arXiv:0712.1909 [astro-ph]}}.

\bibitem{EUSO-2002}
EUSO Coll., A.\,Petrolini \etal, {\it The {Extreme Universe Space Observatory
  (EUSO)} instrument}, Nucl.\ Phys.\ Proc.\ Suppl. 113 (2002) 329.

\bibitem{JEM-EUSO-Web}
JEM-EUSO homepage.
\newblock Available from: \url{http://jemeuso.riken.jp/en/index.html}.

\bibitem{JEM-EUSO-2010}
F.\,Kajino for the JEM-EUSO Coll., {\it The {JEM-EUSO} mission to explore the
  extreme universe}, Nucl.\ Inst.\ Meth. A\,623 (2010) 422, contribution to 1st
  Int.\ Conf.\ on Technology and Instrum.\ in Part.\ Phys.\ (TIPP09), Tsukuba,
  Japan.

\bibitem{Fargion-2002b}
D.\,Fargion, {\it Highest energy neutrino showers in {EUSO}} (2002).
\newblock \href {http://arxiv.org/abs/astro-ph/0212342} {\path{
  arXiv:astro-ph/0212342}}.

\bibitem{JEM-EUSO-2010b}
K.\,Shinozaki for the JEM-EUSO Coll., {\it Extreme energy gamma rays and
  neutrinos and their observation in {JEM-EUSO} mission}, AIP Conf.\ Proc. 1238
  (2010) 377, contribution to 7th Tours Symp.\ on Nucl.\ Phys.\ and Astrophys.,
  Kobe, Japan.

\bibitem{Kotera-2010}
K.\,Kotera, D.\,Allard and A.V.\,Olinto, {\it Cosmogenic neutrinos: {Parameter}
  space and detectabilty from {$\Pev$} to {$\Zev$}}, J.\ Cosm.\ Astropart.\
  Phys. 1010 (2010) 013.
\newblock \href {http://arxiv.org/abs/1009.1382 [astro-ph.HE]} {\path{
  arXiv:1009.1382 [astro-ph.HE]}}.

\bibitem{JEM-EUSO-2009}
K.\,Asano, K.\,Shinozaki and M.\,Teshima for the JEM-EUSO Coll., {\it
  Sensitivity of {JEM-EUSO} to {GRB} neutrinos}, contribution to 31st Int.\
  Cosmic Ray Conf., \Lodz, Poland (2009).
\newblock Available from:
  \url{http://icrc2009.uni.lodz.pl/proc/pdf/icrc0692.pdf}, \href
  {http://arxiv.org/abs/0908.0392 [astro-ph.HE]} {\path{
  arXiv:0908.0392 [astro-ph.HE]}}.

\bibitem{radio-Askaryan-1962}
G.A.\,Askaryan, {\it Excess negative charge of an electron-photon shower and
  its coherent radio emission}, Sov.\ Phys.\ JETP 14 (1962) 441.

\bibitem{radio-Saltzberg-2001}
D.\,Saltzberg \etal, {\it Observation of the {Askaryan} effect: {Coherent}
  microwave {Cherenkov} emission from charge asymmetry in high energy particle
  cascades}, Phys.\ Rev.\ Lett. 86 (2001) 2802.
\newblock \href {http://arxiv.org/abs/hep-ex/0011001} {\path{
  arXiv:hep-ex/0011001}}.

\bibitem{radio-Gorham-2007}
P.W.\,Gorham \etal, {\it Observations of the {Askaryan} effect in ice}, Phys.\
  Rev.\ Lett. 99 (2007) 171101.
\newblock \href {http://arxiv.org/abs/hep-ex/0611008} {\path{
  arXiv:hep-ex/0611008}}.

\bibitem{Price-1996}
B.\,Price, {\it Comparison of optical, radio, and acoustical detectors for
  ultrahigh-energy neutrinos}, Astropart.\ Phys. 5 (1996) 43.
\newblock \href {http://arxiv.org/abs/astro-ph/9510119} {\path{
  arXiv:astro-ph/9510119}}.

\bibitem{radio-RICE-2006}
I.\,Kravchenko \etal, {\it {RICE} limits on the diffuse ultra-high energy
  neutrino flux}, Phys.\ Rev. D\,73 (2006) 082002.
\newblock \href {http://arxiv.org/abs/astro-ph/0601148} {\path{
  arXiv:astro-ph/0601148}}.

\bibitem{radio-ANITA-2006}
ANITA Coll., S.\,Barwick \etal, {\it Constraints on cosmic neutrino fluxes from
  the {ANITA} experiment}, Phys.\ Rev.\ Lett. 96 (2006) 171101.
\newblock \href {http://arxiv.org/abs/astro-ph/0512265} {\path{
  arXiv:astro-ph/0512265}}.

\bibitem{radio-ARIANNA-2007}
S.\,Barwick, {\it {ARIANNA: A} new concept for {UHE} neutrino detection}, J.\
  Phys.\ Conf.\ Ser. 60 (2007) 278.
\newblock \href {http://arxiv.org/abs/astro-ph/0610631} {\path{
  arXiv:astro-ph/0610631}}.

\bibitem{radio-ARA-2011}
ARA Coll., P.\,Allison \etal, {\it Design and initial performance of the
  {Askaryan Radio Array} prototype {EeV} neutrino detector at the {South
  Pole}}, subm.\ to Astropart.~Phys.
\newblock \href {http://arxiv.org/abs/1105.2854 [astro-ph.IM]} {\path{
  arXiv:1105.2854 [astro-ph.IM]}}.

\bibitem{radio-GLUE-2004}
P.W.\,Gorham \etal, {\it Experimental limit on the cosmic diffuse ultrahigh
  energy neutrino flux}, Phys.\ Rev.\ Lett. 93 (2004) 041101.
\newblock \href {http://arxiv.org/abs/astro-ph/0310232} {\path{
  arXiv:astro-ph/0310232}}.

\bibitem{radio-NUMOON-2008}
O.\,Scholten \etal, {\it Improved flux limits for neutrinos with energies above
  $10^{22}\ev$ from observations with the {Westerbork Synthesis Radio
  Telescope}}, Phys.\ Rev.\ Lett. 103 (2009) 191301.
\newblock \href {http://arxiv.org/abs/0910.4745 [astro-ph.HE]} {\path{
  arXiv:0910.4745 [astro-ph.HE]}}.

\bibitem{LOFAR-2010}
A.\,Horneffer \etal, {\it Cosmic ray and neutrino measurements with {LOFAR}},
  Nucl.\ Inst.\ Meth. A\,617 (2010) 482.

\bibitem{acoustic-Askaryan-1957}
G.A.\,Askaryan, {\it Hydrodynamic radiation from the tracks of ionizing
  particles in stable liquids}, Sov.\ J.\ Atom.\ Energy 3 (1957) 921.

\bibitem{acoustic-Learned-1979}
J.G.\,Learned, {\it Acoustic radiation by charged atomic particles in liquids:
  {An} analysis}, Phys.\ Rev. D\,19 (1979) 3293.

\bibitem{acoustic-SPATS-2009}
IceCube Coll., R.\,Abbasi \etal, {\it Measurement of sound speed vs.\ depth in
  {South Pole} ice for neutrino astronomy}, Astropart.\ Phys. 33 (2010) 277.
\newblock \href {http://arxiv.org/abs/0909.2629 [astro-ph.IM]} {\path{
  arXiv:0909.2629 [astro-ph.IM]}}.

\bibitem{acoustic-SPATS-2010}
IceCube Coll., R.\,Abbasi \etal, {\it Measurement of acoustic attenuation in
  {South Pole} ice}, Astropart.\ Phys. 34 (2010) 382.
\newblock \href {http://arxiv.org/abs/1004.1694 [astro-ph.IM]} {\path{
  arXiv:1004.1694 [astro-ph.IM]}}.

\bibitem{acoustic-SPATS-2011}
IceCube Coll., R.\,Abbasi \etal, {\it Background studies for acoustic neutrino
  detection at the {South Pole}} (2011).
\newblock \href {http://arxiv.org/abs/1103.1216 [astro-ph.IM]} {\path{
  arXiv:1103.1216 [astro-ph.IM]}}.

\bibitem{acoustic-Nahnhauer-2008}
R.\,Nahnhauer, A.A.\,Rostovtsev and D.\,Tosi, {\it Permafrost -- {An}
  alternative target material for ultra-high energy neutrino detection?},
  Nucl.\ Inst.\ Meth. A\,587 (2008) 29.
\newblock \href {http://arxiv.org/abs/0707.3757 [astro-ph]} {\path{
  arXiv:0707.3757 [astro-ph]}}.

\bibitem{ARENA-Zeuthen}
R.\,Nahnhauer and S.\,Boeser (Eds.), {\it {Proc.\ 1st Int.\ Workshop on
  Acoustic and Radio $\Eev$ Neutrino detection Activities (ARENA), Zeuthen,
  Germany}}, Int.\ J.\ Mod.\ Phys.~A, World Scientific, Singapore, 2006, {ISBN
  978-981-256-755-0}.

\bibitem{ARENA-Scotland}
L.\,Thompson and S.\,Danaher (Eds.), {\it {Proc.\ 2nd Int.\ Workshop on
  Acoustic and Radio $\Eev$ Neutrino detection Activities (ARENA\,2006),
  Northumbria, United Kingdom}}, Vol.~81 of J.\ Phys.: Conf.\ Ser., Institute
  of Physics Publ., Bristol, United Kingdom, 2007, {ISBN 978-160-560-262-2}.

\bibitem{ARENA-Rome}
F.\,Ameli \etal (Ed.), {\it {Proc.\ 3rd Int.\ Workshop on Acoustic and Radio
  $\Eev$ Neutrino detection Activities (ARENA\,2006), Rome, Italy}}, Vol.
  A\,604 of Nucl.\ Inst.\ Meth., Elsevier, 2009.

\bibitem{acoustic-Lehtinen-2002}
N.\,Lehtinen \etal, {\it Sensitivity of an underwater acoustic array to
  ultra-high energy neutrinos}, Astropart.\ Phys. 17 (2002) 279.
\newblock \href {http://arxiv.org/abs/astro-ph/0104033} {\path{
  arXiv:astro-ph/0104033}}.

\bibitem{acoustic-Vandenbroucke-2005}
J.\,Vandenbroucke, G.\,Gratta and N.\,Lehtinen, {\it Experimental study of
  acoustic ultrahigh-energy neutrino detection}, Astrophys.\ J. 621 (2005) 301.
\newblock \href {http://arxiv.org/abs/astro-ph/0406105} {\path{
  arXiv:astro-ph/0406105}}.

\bibitem{acoustic-SAUND-2006}
SAUND II Coll., N.\,Kurahashi \etal, {\it Study of acoustic ultra-high energy
  neutrino detection phase {II}}, Int.\ J.\ Mod.\ Phys. A\,21\;S\,1 (2006) 217.

\bibitem{acoustic-AMADEUS}
ANTARES Coll., J.A.\,Aguilar \etal, {\it {AMADEUS -- The} acoustic neutrino
  detection test system of the {ANTARES} deep-sea neutrino telescope} (2010).
\newblock \href {http://arxiv.org/abs/1009.4179 [astro-ph.IM]} {\path{
  arXiv:1009.4179 [astro-ph.IM]}}.

\end{thebibliography}
}

\end{document}